\documentclass[a4paper,11pt]{article}
\pdfoutput=1 

\usepackage{amssymb}
\usepackage{dsfont}
\usepackage[T1]{fontenc} 
\usepackage{physics}
\usepackage{comment}
\usepackage{amsmath}
\usepackage{graphicx}
\usepackage{array, makecell}
\usepackage{float}
\setcellgapes{3pt}
\usepackage{xcolor}
\usepackage{amssymb}
\usepackage[normalem]{ulem}
\usepackage[numbers,sort&compress]{natbib}
\usepackage{geometry}
\usepackage{pdflscape}
\usepackage{mathtools}
\usepackage{multirow}

\definecolor{azure}{rgb}{0.0, 0.5, 1.0}
\definecolor{darkblue}{rgb}{0.15,0.35,0.7}
\definecolor{reddish}{rgb}{0.65, 0.2, 0.2}
\definecolor{brandeisblue}{rgb}{0.0, 0.44, 1.0}
\definecolor{ceruleanblue}{rgb}{0.16, 0.32, 0.75}
\definecolor{indigo(dye)}{rgb}{0.0, 0.25, 0.42}
\definecolor{dgrey}{rgb}{0.3,0.3,0.3}
\definecolor{grey}{rgb}{0.9,0.9,0.9}

\usepackage[linktocpage=true]{hyperref}
\hypersetup{
colorlinks=true,
citecolor=ceruleanblue,
linkcolor=ceruleanblue,
urlcolor=ceruleanblue,
pdfauthor={},
pdftitle={},
pdfsubject={}
}

\usepackage{cleveref}
\usepackage{bm}
\crefname{lem}{lemma}{lemmas}
\crefname{thm}{theorem}{theorems}
\crefname{cor}{corollary}{corollaries}
\crefname{rem}{remark}{remarks}
\crefname{prop}{proposition}{propositions}

\usepackage{colortbl}
\definecolor{dgreen}{rgb}{0, 0.55, 0}
\definecolor{llightyellow}{rgb}{1.0, 0.95, 0.7}
\definecolor{llightblue}{rgb}{0.7, 0.9, 1.0}
\definecolor{llightpink}{rgb}{1.0, 0.85, 0.95}
\definecolor{llightgreen}{rgb}{0.7, 1.0, 0.4}
\colorlet{lightyellow}{llightyellow!50!white}
\colorlet{lightblue}{llightblue!50!white}
\colorlet{lightgreen}{llightgreen!50!white}
\colorlet{lightpink}{llightpink!50!white}

\usepackage[framemethod=TikZ]{mdframed} 
\usepackage{tikz-cd} 
\usetikzlibrary{arrows,snakes,shapes.arrows,decorations.markings}
     \tikzset{>=triangle 90}
     \tikzstyle{bbc}=[draw,circle,fill=black,scale=.75]
     \tikzstyle{rc}=[circle,fill=red,scale=.6]
     \tikzstyle{wc}=[draw,circle,scale=.75]

\usetikzlibrary{cd}

\tikzset{snake it/.style={decorate, decoration=snake}}

\tikzset{
	on each segment/.style={
		decorate,
		decoration={
			show path construction,
			moveto code={},
			lineto code={
				\path [#1]
				(\tikzinputsegmentfirst) -- (\tikzinputsegmentlast);
			},
			curveto code={
				\path [#1] (\tikzinputsegmentfirst)
				.. controls
				(\tikzinputsegmentsupporta) and (\tikzinputsegmentsupportb)
				..
				(\tikzinputsegmentlast);
			},
			closepath code={
				\path [#1]
				(\tikzinputsegmentfirst) -- (\tikzinputsegmentlast);
			},
		},
	},
	mid arrow/.style={postaction={decorate,decoration={
				markings,
				mark=at position .5 with {\arrow[#1]{stealth}}
	}}},
}

\usetikzlibrary{automata,positioning}
\usetikzlibrary{decorations.markings}

\tikzset{line/.style={line width=0.25mm},
curve/.style={line,smooth,tension=1},
->-/.style={decoration={
  markings,
  mark=at position #1 with {\arrow[>=stealth]{>}}},postaction={decorate}},
-<-/.style={decoration={
  markings,
  mark=at position #1 with {\arrow[>=stealth]{<}}},postaction={decorate}},
}

\tikzset{bg/.style={opacity=.5}}

\tikzset{
    partial ellipse/.style args={#1:#2:#3}{
        insert path={+ (#1:#3) arc (#1:#2:#3)}
    }
}

\makeatletter
\renewcommand\section{\@startsection {section}{1}{\z@}%
                               {-3.5ex \@plus -1ex \@minus -.2ex}
                               {2.3ex \@plus.2ex}%
                               {\normalfont\large\bfseries}}
\renewcommand\subsection{\@startsection{subsection}{2}{\z@}%
                                 {-3.25ex\@plus -1ex \@minus -.2ex}%
                                 {1.5ex \@plus .2ex}%
                                 {\normalfont\bfseries}}
\makeatother

\let\non\nonumber

\newfont{\goth}{ygoth.tfm scaled 1200}                   

\numberwithin{equation}{section}

\newcommand{\fsa}{\epsilon_{\CD}}

\newcommand{\be}{\begin{equation}}
\newcommand{\ee}{\end{equation}}
\newcommand{\bee}{\begin{equation} \begin{aligned}}
\newcommand{\eee}{\end{aligned} \end{equation}}

\newcommand{\CA}{\mathcal{A}}
\newcommand{\CC}{\mathcal{C}}
\newcommand{\CD}{\mathcal{D}}
\newcommand{\CE}{\mathcal{E}}

\newcommand{\CZ}{\mathcal{Z}}
\newcommand{\CH}{\mathcal{H}}
\newcommand{\CL}{\mathcal{L}}
\newcommand{\CN}{\mathcal{N}}
\newcommand{\CM}{\mathcal{M}}
\newcommand{\CO}{\mathcal{O}}

\newcommand\doubleA{\mathbb{A}}

\newcommand\doubleC{\mathbb{C}}

\newcommand\doubleZ{\mathbb{Z}}

\newcommand\scriptA{\mathcal{A}}
\newcommand\scriptB{\mathcal{B}}
\newcommand\scriptC{\mathcal{C}}
\newcommand\scriptD{\mathcal{D}}
\newcommand\scriptE{\mathcal{E}}

\newcommand\scriptL{\mathcal{L}}

\newcommand\scriptN{\mathcal{N}}
\newcommand\scriptO{\mathcal{O}}

\newcommand\scriptU{\mathcal{U}}
\newcommand\scriptV{\mathcal{V}}

\newcommand\scriptX{\mathcal{X}}

\newcommand\scriptZ{\mathcal{Z}}

\newcommand{\VEC}{\operatorname{Vec}}

\newcommand\Spin{\operatorname{Spin}}

\newcommand{\dsi}{\mathds{1}}
\newcommand{\IZ}{\mathbb{Z}}

\newcommand{\ii}{\mathsf{i}}

\newcommand\Rep{\operatorname{Rep}}
\newcommand\TY{\operatorname{TY}}
\newcommand\Hom{\operatorname{Hom}}

\newcommand\EqBr{\operatorname{EqBr}}
\newcommand\BrPic{\operatorname{BrPic}}
\newcommand\ind{\operatorname{ind}}
\newcommand\conj{\operatorname{conj}}
\newcommand\res{\operatorname{res}}

\newcommand{\Aut}{\operatorname{Aut}}
\newcommand\Inv{\operatorname{Inv}}
\newcommand\Cor{\operatorname{cor}}

\begin{document} 

\begin{titlepage}
\begin{center}

\hfill         \phantom{xxx}  

\vskip 2 cm {\Large \bf SymSETs and self-dualities under gauging non-invertible symmetries} 


\vskip 1.25 cm {\bf Da-Chuan Lu${}^{1,2,3}$, Zhengdi Sun${}^{4}$, Zipei Zhang${}^1$}\non \\

\vskip 0.2 cm
 {\it ${}^{1}$ Department of Physics, University of California, San Diego, CA 92093, USA}

\vskip 0.2 cm
 {\it ${}^{2}$ Department of Physics and Center for Theory of Quantum Matter, University of Colorado, Boulder, Colorado 80309, USA}
\vskip 0.2 cm
 {\it ${}^{3}$ Department of Physics, Harvard University, Cambridge, MA 02138, USA}
\vskip 0.2 cm

{\it ${}^{4}$ Mani L. Bhaumik Institute for Theoretical Physics, Department of Physics and Astronomy, University of California Los Angeles, CA 90095, USA
}

\end{center}
\vskip 1.5 cm

\begin{abstract}
\noindent 

The self-duality defects under discrete gauging in a categorical symmetry $\mathcal{C}$ can be classified by inequivalent ways of enriching the bulk SymTFT of $\mathcal{C}$ with $\mathbb{Z}_2$ 0-form symmetry. The resulting Symmetry Enriched Topological (SET) orders will be referred to as \textit{SymSETs} and are parameterized by choices of $\mathbb{Z}_2$ symmetries, as well as symmetry fractionalization classes and discrete torsions. In this work, we consider self-dualities under gauging \textit{non-invertible} $0$-form symmetries in $2$-dim QFTs and explore their SymSETs. Unlike the simpler case of self-dualities under gauging finite Abelian groups, the SymSETs here generally admit multiple choices of fractionalization classes. We provide a direct construction of the SymSET from a given duality defect using its \textit{relative center}. Using the SymSET, we show explicitly that changing fractionalization classes can change fusion rules of the duality defect besides its $F$-symbols. We consider three concrete examples: the maximal gauging of $\Rep H_8$, the non-maximal gauging of the duality defect $\mathcal{N}$ in $\Rep H_8$ and $\Rep D_8$ respectively. The latter two cases each result in 6 fusion categories with two types of fusion rules related by changing fractionalization class. In particular, two self-dualities of $\Rep D_8$ related by changing the fractionalization class lead to $\Rep D_{16}$ and $\Rep SD_{16}$ respectively. Finally, we study the physical implications such as the spin selection rules and the SPT phases for the aforementioned categories.

\baselineskip=18pt

\end{abstract}
\end{titlepage}

\tableofcontents

\flushbottom

\newpage

\section{Introduction}
For a given quantum field theory (QFT), it is often useful to first explore its global symmetries and their implications, before diving into harder dynamical questions, as the former generally imposes non-trivial constraints on possible answers to the latter question. In the past ten years, ordinary invertible symmetries that act on local operators have been generalized to non-invertible symmetries whose symmetry operator generically does not admit an inverse, and higher form symmetries that act on extended operators. These are collectively referred to as generalized symmetries or categorical symmetries. And the latter name arises from that such symmetries in $d$-dim are described by fusion $(d-1)$-categories. They provide powerful tools to study the universal behaviors of various theories and reveal connections between different phases, as well as even seemingly unrelated models. For recent reviews on the developments of non-invertible symmetry in various fields, see \cite{gaiotto:generalizedsym,McGreevy:2022oyu,Cordova:2022ruw,daniel2023review,sakura2023review,lakshya2023review,shuheng2023review,runkel2023review,Costa:2024wks}.

The self-duality defect $\CN$ (also known as the Kramers-Wannier duality defect) in 2d Ising CFT is among the first examples of non-invertible symmetries. It is non-invertible as this operator does not admit an inverse, and it characterizes the invariance of the theory under gauging $\mathbb{Z}_2$ symmetry, for recent developments of non-invertible duality symmetry in lattice model, see  \cite{seiberg2023maj,moradi2023sym,seifnashri2023lieb,seiberg2024non,mana2024nisym,Lu:2024ytl,sal2024mod,cao2024nisym,sal2024tdual}. This self-duality plays a very important role in quantum phase transitions \cite{Kramers1941}. Since the duality maps the ordered phase to the disordered phase, the critical point is invariant and pinned by the self-duality if assuming there is one continuous phase transition \cite{Kramers1941}. With the self-duality, the critical phases along the critical line can be analyzed \cite{Aasen:2016dop,Aasen:2020jwb,Fendley2018Susy,Lu2021SelfDualDQCP,Nahum2021SelfdualZ2,omer2024selfdual,arkya2024selfdual,sinha2024selfdual}. 

This simple example is then generalized in many different ways. In $2$-dim, one can consider the self-duality under gauging other finite Abelian symmetries, with the resulting symmetries described by the Tambara-Yamagami fusion category \cite{tambara1998tensor,tambara2000representations}, or consider the self-duality under gauging non-invertible symmetries which are studied recently in \cite{Choi:2023vgk,Perez-Lona:2023djo,Diatlyk:2023fwf}. Alternatively, one can consider gauging which is not of order $2$, but described by a generic finite group $G$. This leads to the $G$-ality defects studied in \cite{Thorngren:2019iar,Thorngren:2021yso,Lu:2022ver,Lu:2024ytl,Lu:2024lzf,Ando:2024hun}. These constructions can also be generalized to higher dimensions (for instance, see \cite{Kaidi:2021xfk,Choi:2021kmx,Choi:2022zal,Bhardwaj:2024xcx,Decoppet:2024moc,Decoppet:2023bay,Choi:2024rjm,Cui:2024cav}).

To fully explore the physical implications of these self-duality defects, the their categorical structures must be determined. The analog for the ordinary symmetry is the 't Hooft anomaly, and it is well-known that the same symmetry groups with different 't Hooft anomalies will lead to drastically distinct constraints on the physical system. Just as with a given symmetry group the possible 't Hooft anomaly can be classified, and the allowed categorical structures of the self-dualities can be classified using the theory of $G$-graded extension (for duality defects, $G = \mathbb{Z}_2$) of a given (higher)-fusion categories \cite{Etingof:2009yvg,Bhardwaj:2024xcx, Decoppet:2024moc,Decoppet:2023bay}. 

It turns out that the symmetry topological field theories (SymTFTs) provide useful tools and a natural framework to study those extensions. Generically, the symmetry and its properties (such as 't Hooft anomaly, topological sectors, allowed gapped and gapless phases) of a $d$-dim QFT can be captured by the $(d+1)$-dim topological field theory known as the Symmetry TFT (SymTFT) \cite{Kitaev:2011dxc,symtft2019XGW,symtft2020XGW2,symtft2021XGW3,symtft2023XGW4,symtft2022XGW5,symtft2021Gaiotto,symtft2021Sakura,moradi2023topoholo,symtft2022Apruzzi,symtft2022freed2,Kaidi:2022cpf,symtft2022Kulp,symtft2023kaidi2,Brennan:2024fgj,Bonetti:2024cjk,DelZotto:2024tae,Franco:2024mxa,Putrov:2024uor,Huang:2024ror,Freed:2022qnc,Lin:2022dhv,Bhardwaj:2024igy,Antinucci:2024ltv,Bhardwaj:2024qiv,Bhardwaj:2024qrf,Bhardwaj:2023bbf,Bhardwaj:2023ayw,Bhardwaj:2023wzd,Sun:2023xxv,Zhang:2023wlu,Copetti:2024onh,Argurio:2024oym,Antinucci:2023ezl,Choi:2024wfm,Choi:2024tri,Cordova:2023bja,Antinucci:2024zjp,Bhardwaj:2024ydc,Chen:2023qnv,Cui:2024cav}. Roughly speaking, for a given categorical symmetry $\CC$, one considers its SymTFT (which we will denote as $\CZ(\CC)$). Then, any $G$-extension of $\CC$ corresponds to enriching the SymTFT with a corresponding $G$ $0$-form symmetry in a certain way to form a symmetry enriched topological order(SET) and vice versa. For simplicity, we will refer to those SETs naturally as \textit{SymSETs}. The $G$-extensions of a given fusion category $\CC$ can be classified through the allowed SymSETs in the bulk. 

To be more concrete, we will focus on the fusion category symmetry $\CD$ in a $2$-dim bosonic QFT $\scriptX$. The symmetry operators are line operators invariant under local deformations, therefore are referred to as topological defect lines(TDLs) \cite{Chang:2018iay}. Mathematically, these TDLs are objects in the fusion category $\CD$, with local junctions between TDLs and the fusion of the TDLs captured respectively by morphisms between objects and tensor product structure in the fusion category $\CD$. It is possible that theory $\mathcal{X}$ actually admits additional symmetries arising from the invariance under the invertible topological manipulation applied to the theory $\mathcal{X}$. Such manipulations could be tensor autoequivalences of the symmetry $\mathcal{D}$ or discrete gauging some algebra object $\mathcal{A}$ in $\scriptD$, or some combinations of both. For the self-dualties, the topological manipulation is an order-$2$ discrete gauging of some algebra object. The TDLs describing the enhanced symmetries constructed by performing the topological manipulation on half-space. Often, the new topological defects together with the known one in $\scriptD$ form a bigger fusion category $\CC \equiv \mathcal{E}_{G}\mathcal{D}$, known as a $G$-extension of $\mathcal{D}$. $\CC$ naturally admits a $G$-grading, meaning there is a direct sum decomposition
\begin{equation}
    \CC = \bigoplus_{g\in G} \CC_g ~, \quad \CC_1 = \scriptD ~,
\end{equation}
and the tensor product respects the $G$-grading $\CC_g \boxtimes \CC_{h} \subset \CC_{gh}$. Each non-trivial grading component $\CC_g$ consists of simple objects generated by fusing simple TDL in $\scriptD$ with the defect constructed from the half-space topological manipulation.

As first pointed it out \cite{Etingof:2009yvg}, any $G$-extension can be understood in terms of SymTFT as follows:
\begin{equation}\label{eq:BB_relation}
    \begin{tikzpicture}[scale=0.8,baseline={([yshift=-.5ex]current bounding box.center)},vertex/.style={anchor=base,
    circle,fill=black!25,minimum size=18pt,inner sep=2pt},scale=0.50]
    \node[black] at (0,0) {\footnotesize $\CC_1$};
    
    \node[black, below] at (0,-4.5) {\footnotesize SymTFT};
    \node[black, below] at (0,-5.5) {\footnotesize $\mathcal{Z}(\CC_1)$};
    \draw[thick, black, -stealth] (0,-0.5) -- (0,-4.5);
    \node[black, left] at (0,-2.5) {\scriptsize center};
    
    \draw[thick, black, -stealth] (2,-5) -- (7,-5);
    \node[black, above] at (4.5,-5) {\scriptsize $G$-crossed};
    \node[black, below] at (4.5,-5) {\scriptsize extension};

    \node[black, below] at (9,-4.5) {\footnotesize SymSET};
    \node[black, below] at (9,-5.5) {\footnotesize $\mathcal{Z}(\CC_1)_G^\times$};

    \draw[thick, black, -stealth] (11,-5) -- (15.5,-5);
    \node[black, above] at (13.5,-5) {\scriptsize $G$-equiv.};

    \node[black, below] at (17.5,-4.5) {\footnotesize SymTFT};
    \node[black, below] at (17.5,-5.5) {\footnotesize $\mathcal{Z}(\CC)$};
    
    \draw[thick, black, -stealth] (1,0) -- (16,0);
    \node[black, above] at (8,0) {\scriptsize $G$-extension};
    
    \node[black, right] at (16,0) {\footnotesize $\displaystyle \mathcal{C} = \bigoplus_{g\in G} \CC_g$};

    \draw[thick, black, -stealth] (17,-0.5) -- (17,-4.5);
    \node[black, right] at (17,-2.5) {\scriptsize center};

\end{tikzpicture} ~.
\end{equation}
Mathematically, the SymTFT of $\CC$ is described its Drinfeld center $\CZ(\CC)$. The group of invertible topological manipulations correspond to a specific $0$-form symmetry $G$ in the SymTFT $\CZ(\CC_1)$. Enriching the SymTFT $\CZ(\CC_1)$ with this group $G$ leads to an SymSET, and mathematically it is described by a braided $G$-crossed category $\CZ(\CC_1)^\times_{G}$. Notice with a given $G$, SymSET is not uniquely determined, and there are different choices of the symmetry fractionalization classes (which form a $H^2(G,\mathbf{A})$-torsor where $\mathbf{A}$ is the group of Abelian anyons in $\CZ(\CC_1)$) and discrete torsions (which are parameterized by $H^3(G,U(1))$)\footnote{Notice that in order to consistently choose a symmetry fractionalization class, the obstruction measured by the twisted cohomology $H^3(G,\mathbf{A})$ must vanish. After choosing the fractionalization class, one must also check the obstruction to defectification, which depends on the fractionalization class and is measured by a class in $H^4(G,U(1))$, vanishes to consistently pick a choice discrete torsion. More precisely, the fractionalization classes and discrete torsions are cohomologically inequivalent $C^2(G,\mathbf{A})$-cochain and $C^3(G,U(1))$-cochain which trivialized the two obstructions respectively. Therefore, $H^2(G,\mathbf{A})$ and $H^3(G,U(1))$ parameterize inequivalent ways of trivializing the above two obstructions.} \cite{Barkeshli:2014cna}. On the boundary, these data specifies the fusion structures of TDLs describing enhanced symmetries. After choosing the corresponding SymSET, one can perform $G$-equivariantization to get the SymTFT $\CZ(\CC)$ of the extended category $\CC$. Naturally, different choices of the SymSET data will generically lead to distinct SymTFTs describing inequivalent $G$-extensions of $\CC_1$.

Let's consider the example of Tambara-Yamagami fusion categories $\TY(A,\chi,\epsilon)$, which is the $\mathbb{Z}_2$-extension of a finite Abelian group symmetry $\VEC_{A}$. The non-trivial grading component contains a unique simple TDL $\CN$, describing the self-duality under gauging the $\VEC_{A}$-symmetry. The fusion rule is uniquely determined, but the categorical structure depends on the choice of the non-degenerate symmetric bicharacter $\chi$ as well as the FS indicator $\epsilon = \pm 1$. From the bulk point of view, $\chi$ labels different choice of the bulk $\mathbb{Z}_2^{em}$ symmetry; and the FS indicator $\epsilon$ corresponds to different choices of the discrete torsion $H^3(\mathbb{Z}_2^{em},U(1)) = \mathbb{Z}_2$. 

While the choice of the symmetry fractionalization classes is unique in the well-known $\TY$-fusion categories \cite{Etingof:2009yvg}, as we observed in the example discussed in \cite{Lu:2024lzf} as well as by the examples we will study in this paper, $G$-symmetries in the bulk with non-trivial choices of the symmetry fractionalization classes are not uncommon. 

It is therefore natural to ask how the symmetry fractionalization classes affect the extended boundary fusion category symmetry. For the other piece of the bulk data--the discrete torsions, the answer is well-known and it changes the $F$-symbols of the boundary symmetry in a simple way \cite{Chang:2018iay}. Therefore, we would like to know that starting with a specific $G$-extension $\CC$, is it possible to work out all the other $G$-extensions related to $\CC$ by changing the symmetry fractionalization classes, and how does the physical implication change accordingly.

In this paper, we approach the above questions using two different tools. First, to systematically track changes in the categorical structure resulting from variations in the fractionalization class, we generalize the previous work to develop an approach to compute SymSET from the boundary fusion category symmetry $\CC$. This framework allows us to derive the categorical data in the resulting boundary symmetry after changing the fractionalization class. Once the new categorical structure is determined, one can then use all the familiar tools such as group-theoretical fusion categories, SymTFT, etc to study the properties of the fusion categories, such as the anomaly, spin selection rules, properties of the SPT phases. It would be interesting to see if one could access those physical implications directly using the SymSETs and we will leave this question for future study.

\

\subsection{Summary of the main results}
The main results of this paper are divided into two parts. The first part is to develop a generic approach to compute SymSET data from a given $G$-extension using relative center. The second part is to consider its applications in three examples.

\subsubsection*{Computing SymSET via relative center}
We provide an explicit construction of the SymSET directly from the $G$-extended category $\CC$, and mathematically this is known as the relative center $\CZ_{\CC_1}(\CC)$ \cite{Gelaki:2009blp}. Generically, for the fusion category $\CC$, and any of its bimodule categories $\CM$, one can define the center of $\CM$ denoted as $\CZ_{\CC}(\CM)$. Since the fusion category $\CC$ can be viewed as a bimodule category of any of its subcategory $\CD$, the center $\CZ_{\CD}(\CC)$ of $\CC$ can be defined in this way and is known as \textit{relative center}. In the case where $\CD. =\CC$, the relative center $\CZ_{\CC}(\CC)$ is the familiar Drinfeld center of $\CC$ (which physically describes the SymTFT of $\CC$). The $G$-extended category $\CC$ can also be viewed as a bimodule category over its trivial grading component $\CC_1$, and it is shown in \cite{Gelaki:2009blp} that the relative center $\CZ_{\CC_1}(\CC)$ is the SymSET describing the corresponding $G$-symmetry enriching the original SymTFT $\CZ_{\CC_1}(\CC_1)$.

In this work, we generalize the computation of the Drinfeld center $\CZ(\CC_1)$ in the physics literature (e.g. \cite{Teo:2015xla,Choi:2024tri}) to the construction of relative center $\CZ_{\CC_1}(\CC)$. The TDLs in $\CC_1$ lead to the bulk anyons in the SymTFT of $\CC_1$, while the TDLs in the non-trivial grading components lead to twist defects of the bulk $G$-symmetry. Considering the diagrams involving twist defects and various consistency conditions allows us to derive the action of $G$-symmetry on the simple anyons, the junction of simple anyons, and the symmetry fractionalization class. 

When changing the fractionalization class, Abelian anyons are inserted into the junctions of the $0$-form symmetry based on elements in $H^2(G,\mathbf{A})$, modifying the junctions of the boundary TDLs. These Abelian anyons, emitted from local junctions of TDLs, lead to changes in the $F$-symbols of the boundary fusion category. Furthermore, if these Abelian anyons are not condensed on the boundary, the fusion rules of the boundary symmetry category \textit{could} be changed as well. We will provide explicit formulas that enable the computation of the categorical structure of the transformed categories resulting from a change in the fractionalization class.

As another application of the SymSET, once the bulk $G$-symmetry is known, one can check if the SymTFT $\CZ(\CC_1)$ admits a Lagrangian algebra stable under $G$ or not. If so, this implies that the $G$-extended category $\CC$ is Morita equivalent to another $G$-extended category $\CC'$ where the enhanced symmetries are invertible\footnote{Notice that this only guarantees at least one invertible TDL in $\CC'_g$, because fusing the non-invertible TDL in $\CC'_1$ will generically turn this invertible TDL to an non-invertible one.}. When this Lagrangian algebra is isomorphic to the regular representation of some finite group $H$, then $\CC_1'$ itself can be made into $\VEC_H^{\omega_0}$ for some $\omega_0 \in H^3(H,U(1))$. The full fusion category $\CC$ is therefore group-theoretical and Morita equivalent to $\VEC_\Gamma^\omega$ where the $\Gamma$ fits into the extension 
\begin{equation}
    0 \rightarrow H \rightarrow \Gamma \rightarrow G \rightarrow 0 ~,
\end{equation}
where the group structure $\Gamma$ can be determined from the bulk $G$-symmetry as well as its fractionalization class; and $\omega$ can be determined from $\omega_0$, the symmetry fractionalization class, and the discrete torsion. 

Finally, we find that changing the fractionalization class in the SymSET not only allows us to relate different fusion categories, but can also be used to map between symmetric gapped phases of these fusion categories. We will demonstrate this in a simple example and leave a thorough investigation of this for future study.

\

In the second part of the paper, we apply the above approach to study three cases of self-duality under gauging non-invertible symmetries where in all cases the SymSETs admit non-trivial choices of the fractionalization classes. We summarize our results below. 

\subsubsection*{Example I: Self-duality under maximally gauging in $\Rep H_8$}
The first one is the self-duality under maximal gauging of $\Rep H_8$ studied in \cite{Choi:2023vgk,Diatlyk:2023fwf,Perez-Lona:2023djo}. Here, $\Rep H_8 = \TY(\mathbb{Z}_2 \times \mathbb{Z}_2, \chi_{d}, +1)$ is a $\TY$-fusion category containing a non-invertible duality defect $\CN$ besides the invertible symmetries $\mathbb{Z}_2^a \times \mathbb{Z}_2^b$. Here, $\chi_{d}$ denotes the diagonal bicharacter of $\mathbb{Z}_2^a \times \mathbb{Z}_2^b$. The category is self-dual under the maximal gauging with algebra object
\begin{equation}
    \CA = \dsi \oplus a \oplus b \oplus ab \oplus 2\CN ~.
\end{equation}
Including the extra duality defect $\CD$\footnote{To match the original notation used in \cite{Choi:2023vgk}, we will use $\CD$ to denote extra duality defects from now on, and the reader should be able to distinguish it from the fusion subcategory $\CD$ based on context.} leads to an $\mathbb{Z}_2$-extension of $\Rep H_8$, which we denote as $\underline{\CE}_{\mathbb{Z}_2}^{(i,\kappa_\CD,\epsilon_\CD)}\Rep H_8$ \cite{Choi:2023vgk}. Here, the superscripts $(i,\kappa_\CD,\epsilon_\CD),\  i = 1,2$ and $\kappa_\CD,\epsilon_\CD=\pm$ label inequivalent categorical structures in the classification acquired in \cite{Choi:2023vgk} by directly solving the pentagon equations. We revised the bulk interpretation in \cite{Choi:2023vgk} using our new approach. We compute the corresponding SymSET to explicitly show that there are $2$ choices of the bulk $\mathbb{Z}_2^{em}$ symmetry labeled by $i = 1,2$, and $\kappa_{\CD} = \pm 1$ instead labels two distinct choices of the symmetry fractionalization classes. We also demonstrate explicitly how to derive the $F$-symbols of the fusion category acquired by changing the symmetry fractionalization class.

\subsubsection*{Example II: Self-duality under gauging $\dsi \oplus ab \oplus \CN$ in $\Rep H_8$}
The second example is the self-duality under non-maximal gauging the non-invertible symmetry $\CN$ of $\Rep H_8$. This case is studied in \cite{Diatlyk:2023fwf,Perez-Lona:2023djo} in concrete theories such as $c=1$ compact bosons. In this case, we instead gauge a smaller algebra object containing the non-invertible symmetry $\CN$:
\begin{equation}
    \CA = 1 \oplus ab \oplus \CN ~,
\end{equation}
which leads to a (partial) duality defect $\CD_1$ with the fusion rule $\CD_1 \times \CD_1 = \CA$ by the usual half-space gauging argument. But because the $\mathbb{Z}_2^a$ (or equivalently $\mathbb{Z}_2^b$) subgroup of $\Rep H_8$ is not gauged, fusing $a$ (or $b$) with $\CD_1$ will lead to a new defect $\CD_2$. And it's not hard to work out the full fusion rules, which are given by
\begin{equation}\label{eq:intro_tI}
\begin{aligned}
    & \CD_i \times \CN = \CN \times \CD_i = \CD_1 + \CD_2 ~, \quad \CD_i \times g = g \times \CD_i = \prescript{g}{}{\CD_i} ~, \\
    & \CD_1 \times \CD_1 = \CD_2 \times \CD_2 = 1 + ab + \CN ~, \quad \CD_1 \times \CD_2  = \CD_2 \times \CD_1 = a + b + \CN ~,
\end{aligned}
\end{equation}
where $\prescript{g}{}{\CD_i}$ is defined as $\prescript{\dsi}{}{\CD_i} = \prescript{ab}{}{\CD_i} = \CD_i$ and $\prescript{a}{}{\CD_i}=\prescript{b}{}{\CD_i}$ exchanges $\CD_1$ and $\CD_2$. The classification of the above fusion categories is acquired in \cite{vercleyen2024low} by directly solving pentagon equations, where in total $4$ inequivalent categories are found. With the $F$-symbols explicitly known, we are ready to apply our method to derive the corresponding SymSETs. As expected, we find the four categories are labeled by two choices of the bulk $\mathbb{Z}_2$ symmetries, and two choices of the $\mathbb{Z}_2$ discrete torsions. Following the notation of \cite{Choi:2023vgk}, we will denote these categories as $\CE_{\mathbb{Z}_2,\text{I}}^{(i,\pm)}\Rep H_8$ where $i = 1,2$ labels two choices of the bulk symmetry and $\pm$ denotes two choices of the discrete torsion in the bulk. Here, the subscript I denotes that the extension has fusion rules \eqref{eq:intro_tI} (which we will refer as type I fusion rules) which corresponds a specific choice of symmetry fractionalization classes.

Notice that for each choice of the bulk $\mathbb{Z}_2$ symmetry, it similarly admits two symmetry fractionalization classes. Interestingly, when the fractionalization class is changed, the fusion rules of the boundary symmetry category will change into
\begin{equation}
\begin{aligned}
    & \CD_i \times \CN = \CN \times \CD_i = \CD_1 + \CD_2 ~, \quad \CD_i \times g = g \times \CD_i = \prescript{g}{}{\CD_i} ~, \\
    & \CD_1 \times \CD_2 = \CD_2 \times \CD_1 = 1 + ab + \CN ~, \quad \CD_1 \times \CD_1  = \CD_2 \times \CD_2 = a + b + \CN ~.
\end{aligned}
\end{equation}
We see that with this new fusion rule (which we will refer to as \textit{type II} and the previous fusion one will be referred to as \textit{type I}), the $\CD_i$'s are no longer unoriented, but are orientation reversal to each other $\overline{\CD}_1 = \CD_2$. We want to highlight this is the analog of non-trivial $[w] \in H^2(G,A)$ class in the central extension of finite groups
\begin{equation}
    0 \rightarrow A \rightarrow \Gamma \rightarrow G \rightarrow 0 ~,
\end{equation}
where activating a non-trivial $[w]$ modify the product in $\Gamma$ as
\begin{equation}
    (a_1, g_1) \times (a_2, g_2) = (a_1 a_2, g_1 g_2) \rightarrow (a_1, g_1) \times (a_2, g_2) = (a_1 a_2 w(g_1,g_2), g_1 g_2) ~,
\end{equation}
where we parameterize the elements in $\Gamma$ as $(a,g) \in A\times G$\footnote{Notice that this extension can be interpreted using the SymSET as well. One can simply start with the SymTFT of the Abelian group $A$, and enrich it with a trivially acting $G$-symmetry. We will demonstrate this in a simple example in Appendix \ref{app:Z2_example}}. Furthermore, with the new fusion rules, the discrete torsion can be trivialized by the boundary symmetry, therefore there are only two inequivalent fusion categories labeled by two choices of the bulk $\mathbb{Z}_2^{pem}$ symmetry, which we will denote as $\CE^i_{\mathbb{Z}_2,\text{II}}\Rep H_8$ and $i=1,2$ correspond to the different bulk symmetries.

Using the $F$-symbols in \cite{vercleyen2024low}, we compute the spin selection rules of these fusion categories and find that these categories can be completely distinguished by the spin selection rules. Next, with the help of the SymSET, we show that all these categories are group-theoretical and provide an explicit group-theoretical construction for each category. For the type I fusion rules, the categories all arise from gauging $\mathbb{Z}_2^s$ in $D_{16} = \langle r,s|r^8 = s^2 = 1, srs = r^{-1}\rangle$ symmetry with suitable anomaly $\omega \in H^3(D_{16},U(1))$. For the type II fusion rules, the categories instead arise from gauging $\mathbb{Z}_2^s$ in the order-$16$ \textit{semi-dihedral} group $SD_{16} = \langle r,s|r^8 = s^2 = 1, srs = r^3\rangle$ with corresponding anomalies. The group-theoretical construction allows us to conclude that all these categories are anomalous as they do not admit trivially gapped phases (that is, gapped phases with a unique ground state).

Finally, we consider the examples of these duality defects in the $c = 1$ compact boson as discussed in \cite{Diatlyk:2023fwf,Perez-Lona:2023djo}, and our classification together with the spin selection rules allow us to determine the categorical structure of these duality defects.

\

\subsubsection*{Example III: Self-duality under gauging $\dsi \oplus ab \oplus \CN$ in $\Rep D_8$}
Our last example is the self-duality under gauging $\dsi \oplus ab \oplus \CN$ in $\Rep D_8 = \TY(\mathbb{Z}_2 \times \mathbb{Z}_2, \chi_{od}, +1)$, where $\chi_{od}$ stands for off-diagonal bicharacter. This case is also studied in \cite{Diatlyk:2023fwf,Perez-Lona:2023djo}. Notice that $\Rep D_8$ has the same fusion rule as $\Rep H_8$. Therefore, the corresponding duality defects $\CD_i$ will have the same fusion rules as the $\Rep H_8$ case in Example II as well. The classification is similar. For type I fusion rules, there are four inequivalent categories $\CE_{\mathbb{Z}_2,\text{I}}^{(i,\pm)}\Rep D_8$ labeled by two choices of the bulk symmetry $i = 1,2$ and two choices of the discrete torsion $\pm 1$. These two symmetries admit fractionalization classes labeled by $\mathbb{Z}_2$, and changing the fractionalization class leads to type II fusion rules. Again, the discrete torsion can be trivialized by the boundary, hence there are only two inequivalent fusion categories denoted by $\CE_{\mathbb{Z}_2,\text{II}}^{i}\Rep D_8$.

Next, we compute spin-selection rules and we find that unlike the case of $\Rep H_8$, the spin-selection rules can not distinguish between $\CE_{\mathbb{Z}_2,\text{I}}^{(1,+)}\Rep D_8$ and $\CE_{\mathbb{Z}_2,\text{I}}^{(1,-)}\Rep D_8$. These two categories can be distinguished by the numbers of SPT they admit, or by the numbers of the twisted sector with certain spin. All the six categories are group-theoretical, and their group theoretical construction is similar to the case of $\Rep H_8$ and the only difference is the anomaly of the corresponding groups $D_{16}$ and $SD_{16}$. Furthermore, some of the categories are anomaly free and admit (for some cases, more than one) SPT phase(s). We then study their SPT phases and interface algebras.

It is also interesting to notice that changing fractionalization class can potentially lead to an igSPT, which stands for intrinsically gapless SPT\cite{Ruben2021igSPT,huang2023igspt,wen2023igSPT,sakura2024igspt}. A well-known example is the group $\mathbb{Z}_4$, which fits into the non-trivial central extension 
\begin{equation}
    0 \rightarrow \mathbb{Z}_2 \rightarrow \mathbb{Z}_4 \rightarrow \mathbb{Z}_2 \rightarrow 0 ~.
\end{equation}
From the boundary point of view, $\mathbb{Z}_4$ admits the igSPT because the $H^3(\mathbb{Z}_2,U(1))$ of the quotient $\mathbb{Z}_2$ can be trivialized inside the $\mathbb{Z}_4$. From the bulk point of view, the $\mathbb{Z}_4$ is acquired from the trivial extension $\mathbb{Z}_2 \times \mathbb{Z}_2$ via changing fractionalization class. Given the similarities, it is natural to ask if the type II fusion category can similarly admit gapless SPT related to this. Notice that only $\CE_{\mathbb{Z}_2,\text{II}}^{1}\Rep D_8$ admits SPT phases, and one can show using \cite{Bhardwaj:2024qrf,Bhardwaj:2023bbf} that $\CE_{\mathbb{Z}_2,\text{II}}^{1}\Rep D_8$ does admit a gapless SPT related to changing the symmetry fractionalization class.

Finally, we mention some concrete physical examples of these symmetries. 

\

The paper is organized as follows. In Section \ref{sec:review}, we briefly review the fusion category symmetries in $2$-dim, as well as the setup to compute the SymTFT (the Drinfeld center $\CZ(\CC)$) from the fusion category $\CC$. In Section \ref{sec:RDC}, we discuss in general how to compute the SymSET (the relative center $\CZ_{\CC_1}(\CC)$), as well as how to derive the categorical structure of the resulting 2d fusion category symmetry when changing the symmetry fractionalization classes in the SymSET. In Section \ref{sec:TY_RC}, we demonstrate our approach in the simple example of $\TY$-fusion categories as a warm-up. In Section \ref{sec:max_RepH8}, we consider the maximal gauging in $\Rep H_8$ and revisit the bulk interpretation of the classifications in \cite{Choi:2023vgk}. In Section \ref{sec:non_max_RepH8}, we study the self-duality under gauging of $\dsi \oplus ab\oplus \CN$ in $\Rep H_8$. In Section \ref{sec:non_max_RepD8}, we study the self-duality under gauging of $\dsi \oplus ab\oplus \CN$ in $\Rep D_8$ instead. Finally, there are a few appendices on some technical details the readers may find useful. 

\section{Fusion category symmetry and its SymTFT}\label{sec:review}
In this section, we briefly review the fusion category symmetry $\CC$ formed by the topological defect lines(TDLs) in 2d bosonic CFTs and the corresponding 3d SymTFT $\CZ(\CC)$.

\subsection{Topological defect lines in 2d CFTs}\label{sec:convention}
Topological defect lines are line operators commuting with the stress-energy tensor, and we focus on the TDLs which mathematically form the unitary fusion categories \cite{Bhardwaj:2017xup, Fuchs:2002cm}. Those TDLs therefore generate finite categorical symmetries in $2$d. A generic TDL can be written as a direct sum of other TDLs, and the \textit{simple} TDLs are those which can't be further decomposed. Fusing two simple TDLs $a$ and $b$ by putting them close to each other will generically generate a finite sum of other simple TDLs
\begin{equation}
    a \otimes b = \bigoplus_{c} N_{ab}^c c ~, \quad N_{ab}^c \in \doubleZ_{\geq 0} ~.
\end{equation}
Here, the integer $N_{ab}^c$ is known as the fusion multiplicity. We denote the trivial TDL as $\dsi$.

When $N_{ab}^c > 0$, two simple TDLs $a$ and $b$ can fuse locally into the simple TDL $c$ at a trivalent junction. The set of such topological junction operators forms a $N_{ab}^c$-dim $\mathbb{C}$-vector space $V_{ab}^c$. Similarly, $c$ can also locally split into $a$ and $b$ locally, and the topological junction operators also form a $N_{ab}^c$-dim $\mathbb{C}$-vector space $V_{c}^{ab}$. We fix a choice of basis for the splitting and the fusion junctions
\begin{equation}\label{eq:simple_basis}
 ~, \quad \quad \quad \mu = 1,\cdots, N_{ab}^c ~, 
\end{equation}
such that the following identities are satisfied
\begin{equation}\label{eq:fus_spl_relation}
%
 ~,
\end{equation}
where $d_a$ is the quantum dimension of the TDL $a$ defined as
\begin{equation}
d_a = \quad %
 ~.
\end{equation}
Given a four-way junction of TDLs, there are two generically inequivalent ways to decompose it into two trivalent junctions. As the four-way junctions also form a $\doubleC$-vector spaces, the two decompositions are related to each other by a linear transformation, known as the crossing relations. With the basis explicitly chosen for all spaces of trivalent junctions, we can express these crossing relations in terms of their $\doubleC$-number components. We define the $F$-symbols to be the ones characterizing the crossing relation of the splitting junctions to match the convention used in most modular tensor category(MTC) literature
\begin{equation}\label{eq:F-symbols}
 ~.
\end{equation}
And these $F$-symbols are constrained by the so-called pentagon equations:
\begin{equation}
\begin{aligned}
    &\sum_{\delta }\left[ F_{e}^{fcd}\right] _{\left( g,\beta ,\gamma \right)
\left( l,\nu, \delta \right) }\left[ F_{e}^{abl}\right]
_{\left( f,\alpha ,\delta \right) \left( k,\mu, \lambda \right) }\nonumber\\
=&\sum_{h,\sigma ,\psi ,\rho }\left[ F_{g}^{abc}\right] _{\left( f,\alpha
,\beta \right) \left( h,\psi,\sigma \right) }\left[ F_{e}^{ahd}\right]
_{\left( g,\sigma ,\gamma \right) \left( k,\rho, \lambda \right) }%
\left[ F_{k}^{bcd}\right] _{\left( h,\psi ,\rho \right) \left(
l,\nu ,\mu \right) } ~.
\end{aligned}
\end{equation}

Once the $F$-symbols are specified, crossing relations of four-way junctions with different numbers of incoming and outgoing lines can be derived from it. For instance, the crossing move of the fusion junctions is characterized by
\begin{equation}\label{eq:G-symbols}
 ~,
\end{equation}
and the $G$-symbols are related to the $F$-symbols via 
\begin{equation}\label{eq:GF_relation}
    \sum_{f,\rho,\sigma} \left[G_{abc}^d\right]_{(e,\mu,\nu),(f,\rho,\sigma)} \left[F^{abc}_d\right]_{(e',\mu',\nu'),(f,\rho,\sigma)} = \delta_{ee'} \delta_{\mu\nu'} \delta_{\nu\nu'} ~.
\end{equation}
There are also two possible four-way junctions involving two incoming lines and two outgoing lines, and their crossings are given by
\begin{equation}\label{eq:other_two_crossings}
\begin{aligned}
    %
 ~.
\end{aligned} 
\end{equation}

Notice that in the recent paper \cite{Choi:2024tri} (also see \cite{Chang:2018iay}), a $\times$ is used to mark out a specific line in the three-way junction to unambiguously indicate the hom space. We want to point out here that such subtlety is due to the operation of simultaneously reversing the orientation (direction of the arrow) of a TDL $a$ and relabeling it as its orientation reversal $a^*$ in a three way junction. As a result, one can convert between fusion and splitting junctions up to certain coefficients (the $A$ and $B$ matrices in \cite{Choi:2024tri}), such that every crossing always involves three incoming lines. In contrast, we will not consider such an operation in this work, therefore even without a such marked line, the hom space is unambiguously specified by checking the incoming lines and the out going lines in a given trivalent junction. As demonstrated above, crossing moves of different types can still be expressed in terms of the $F$-symbols; and there are no essential differences between the two approaches.

\

Fusion category symmetries are ubiquitous in 2d QFTs. As an example, the category $\VEC_G^\omega$ describes the finite group $0$-form symmetry $G$ with the 't Hooft anomaly $[\omega] \in H^3(G,U(1))$ with simple lines and fusion rules given by the group elements and the group multiplication respectively. And the $G$-symbols are simply given by $G_{ghk} = \omega(g,h,k)$. One can further consider gauging an anomaly free subgroup $H \subset G$ with a choice of discrete torsion $[\psi] \in H^2(H,U(1))$. The resulting symmetry category is known as the \textit{group-theoretical} fusion category and we denote it as $\CC(G,\omega;H,\psi)$. The simple objects, their fusion rules, as well as their $F$-symbols can be straightforwardly derived (e.g. following the approach in \cite{Lu:2022ver}).

A TDL $a$ may end on a (non-local) operator $\CO$, and for a CFT, under the state-operator correspondence map, such non-local operator $\CO$ is mapped to a state in the Hilbert space $\CH_a$ quantized on $S^1$ with the boundary condition twisted by $a$:
\begin{equation}

\end{equation}
And $\CH_a$ is known as the defect/twisted Hilbert space.

The TDLs act on the (non-)local operators via the following \textit{lasso} diagram \cite{Chang:2018iay,Lin:2022dhv}:
\begin{equation}
    %
 ~.
\end{equation}
These lasso actions form a basis for the tube algebra $\operatorname{Tube}(\mathcal{C})$ with multiplication rules defined by the following diagrams:
\begin{equation}
   %
 ~.
\end{equation}
(Non-)Local operators form representations of $\operatorname{Tube}(\mathcal{C})$, and the irreducible representations $\operatorname{Tube}(\mathcal{C})$ are in $1$-to-$1$ correspondence with the simple anyons in the bulk SymTFT (mathematically this is known as the Drinfeld center $\CZ(\CC)$ of the fusion category $\CC$).

The twisted partition function $\Tr_{\mathcal{H}_{\mathcal{L}_1}}\left(\widehat{\mathcal{L}_2}_{(\mathcal{L}_3,\mu,\nu)} q^{L_0 - c/24} \overline{q}^{\overline{L}_0 - c/24}\right)$ is an important observable of the 2d CFT:
\begin{equation}
\Tr_{\mathcal{H}_{\mathcal{L}_1}}\left(\widehat{\mathcal{L}_2}_{(\mathcal{L}_3,\mu,\nu)} q^{L_0 - c/24} \overline{q}^{\overline{L}_0 - c/24}\right) \equiv Z[\mathcal{L}_1,\mathcal{L}_2,\mathcal{L}_3;\mu,\nu](\tau) = \begin{tikzpicture}[baseline={([yshift=+.5ex]current bounding box.center)},vertex/.style={anchor=base,
    circle,fill=black!25,minimum size=18pt,inner sep=2pt},scale=0.5]
    \filldraw[grey] (-2,-2) rectangle ++(4,4);
    \draw[thick, dgrey] (-2,-2) -- (-2,+2);
    \draw[thick, dgrey] (-2,-2) -- (+2,-2);
    \draw[thick, dgrey] (+2,+2) -- (+2,-2);
    \draw[thick, dgrey] (+2,+2) -- (-2,+2);
    \draw[thick, red, -stealth] (0,-2) -- (0.354,-1.354);
    \draw[thick, red] (0,-2) -- (0.707,-0.707);
    \draw[thick, blue, -stealth] (2,0) -- (1.354,-0.354);
    \draw[thick, blue] (2,0) -- (0.707,-0.707);
    \draw[thick, red, -stealth] (-0.707,0.707) -- (-0.354,1.354);
    \draw[thick, red] (0,2) -- (-0.707,0.707);
    \draw[thick, blue, -stealth] (-0.707,0.707) -- (-1.354,0.354);
    \draw[thick, blue] (-2,0) -- (-0.707,0.707);
    \draw[thick, dgreen, -stealth] (0.707,-0.707) -- (0,0);
    \draw[thick, dgreen] (0.707,-0.707) -- (-0.707,0.707);
    \filldraw[black] (0.707,-0.707) circle (2pt);
    \filldraw[black] (-0.707,0.707) circle (2pt);
    \node[red, below] at (0,-2) {\scriptsize $\mathcal{L}_1$};
    \node[blue, right] at (2,0) {\scriptsize $\mathcal{L}_2$};
    \node[dgreen, above] at (0.2,0) {\scriptsize $\mathcal{L}_3$};
    \node[black, below] at (0.9,-0.607) {\scriptsize $\mu$};
    \node[black, below] at (-0.707,0.707) {\scriptsize $\nu$};
\end{tikzpicture} ~.
\end{equation}
In the case of the fusion category being multiplicity free (that is, every fusion coefficient $N_{ab}^c = 0,1$), we will simply drop the $\mu,\nu$ indices, and write $Z[\mathcal{L}_1,\mathcal{L}_2,\mathcal{L}_3]$. 

\subsection{Gauging non-invertible symmetries and the dual categorical symmetry}
In this subsection, we briefly review the gauging of non-invertible symmetries and how to find the dual fusion category symmetry. Readers can check \cite{Choi:2023vgk, Diatlyk:2023fwf, Perez-Lona:2023djo}for more details. 

\subsubsection*{Algebra objects}
The gauging of the fusion category symmetry requires the notion of symmetric $\Delta$-separable Frobenius algebra object, which contains an object $\CA \in \CC$ together with several additional structures. In this work, we will assume that $\CA$ contains only one copy of identity object $\dsi$, and simply refer to it as an algebra object. In this case, all the additional structures can be uniquely determined by the object $\CA$ itself together with a multiplication morphism $\mu \in \Hom_{\CC}(\CA\otimes \CA, \CA)$ satisfying the following constraint:
\begin{equation}\label{eq:alg}
 ~.
\end{equation}
To describe this in more detail we consider expand $\CA$ in terms of simple objects $a$'s
\begin{equation}
    \CA = \sum_{a} m_a a ~, \quad \quad m_a \in \doubleZ_{>0} ~,
\end{equation}
where $m_a$ is the multiplicity of $a$. By choosing a basis for $\Hom(\CA,a)$ and $\Hom(a,\CA)$ for all simple objects $a$ in $\CA$,
\begin{equation}\label{eq:A_basis}
    %
 ~, \quad \alpha = 1, \dots, m_a ~,
\end{equation}
satisfying the normalization condition
\begin{equation}\label{eq:alg_norm}
    %
 ~.
\end{equation}
We can then characterize $\mu:\CA\otimes\CA \rightarrow \CA$ in terms of simple objects using the natural basis of local fusion junctions of the simple objects
\begin{equation}\label{eq:alg_basis}
%
 ~,
\end{equation}
where the two definitions are equivalent to each other and can be related to each other using \eqref{eq:alg_norm}. 

This then allows us to express the condition \eqref{eq:alg} in terms of the junction coefficients $t_{a\alpha,b\beta}^{e\epsilon,\rho}$:
\begin{equation}\label{eq:alg_junc_eq}
    \sum_{e,\epsilon,\mu,\nu} t_{a\alpha,b\beta}^{e\epsilon,\mu} t_{e\epsilon,c\gamma}^{d\delta,\nu} [G^{d}_{abc}]_{(e,\mu,\nu),(f,\rho,\sigma)} = \sum_{\varphi} t_{b\beta,c\gamma}^{f\varphi,\rho} t_{a\alpha,f\varphi}^{d\delta,\sigma} ~.
\end{equation}
Notice that one can consider a change of basis in \eqref{eq:A_basis} preserving the normalization condition \eqref{eq:alg_norm}:
\begin{equation}\label{eq:A_gauge}
 ~, 
\end{equation}
which changes the junction coefficients $t$'s as follows:
\begin{equation}
    t_{a\alpha,b\beta}^{c\gamma,\rho} \rightarrow \sum_{\tilde{\alpha},\tilde{\beta},\tilde{\gamma}} (S_a)_{\alpha\tilde{\alpha}} (S_b)_{\beta\tilde{\beta}} t_{a\tilde{\alpha},b\tilde{\beta}}^{c\tilde{\gamma},\rho} (S_c^{-1})_{\tilde{\gamma}\gamma} ~.
\end{equation}
algebra objects related by this gauge transformation are considered equivalent. 

To gauge the non-invertible symmetry given by $\CA$ is to insert a mesh of algebra $\CA$ along the dual edge of a given triangulation of the 2d spacetime. Generically, besides the fusion junction $\mu$, to define the mesh also requires a splitting junction $\mu^\vee: \CA\rightarrow \CA \otimes \CA$ which is determined from $\mu$ via the following relations:
\begin{equation}\label{eq:alg_fus_split_rel}
 ~.
\end{equation}
One can expand $\mu^\vee$ in terms of the splitting of the simple objects and the junction coefficients $t^\vee$ satisfy
\begin{equation}\label{eq:alg_sp_basis}
     ~, \quad \begin{array}{c}
\displaystyle \sum_{\gamma} (t^\vee)^{c\gamma,\rho}_{a\alpha,b\beta} t^{c\gamma,\tilde{\rho}}_{\tilde{a}\tilde{\alpha},\tilde{b}\tilde{\beta}} = \delta_{a\tilde{a}}\delta_{b\tilde{b}}\delta_{\alpha\tilde{\alpha}}\delta_{\beta\tilde{\beta}}\delta_{\rho\tilde{\rho}} \sqrt{\frac{d_c}{d_a d_b}} ~, \\ \displaystyle \sum_{a,\alpha,b,\beta,\rho} \sqrt{\frac{d_a d_b}{d_c}}  (t^\vee)_{a\alpha,b\beta}^{c\gamma,\rho} t_{a\alpha,b\beta}^{c\tilde{\gamma},\rho} = \delta_{\gamma\tilde{\gamma}} ~. \end{array} 
\end{equation}
Furthermore, it is straightforward to check $\mu^\vee$ and $\mu$ have trivial crossing relations and this together with \eqref{eq:alg_fus_split_rel} ensures that the gauging is independent of the choices of the triangulation.

For instance, on a torus, the gauging is described by
\begin{equation}
    %
 ~.
\end{equation}

\subsubsection*{Constructing dual fusion category symmetry}\label{sec:dual_symmetry}
The dual fusion category symmetry after gauging $(\CA,\mu)$ is characterized by the fusion category of the $\CA$-bimodule objects, denoted as ${}_\CA \CC_{\CA}$. An $\CA$-bimodule object is an object $M \in \CC$ together with two morphisms $\lambda\in \Hom_{\CC}(\CA\otimes M,M)$ and $\rho\in \Hom_{\CC}(M \otimes \CA, M)$ which describes the left and right action of $\CA$ on $M$ respectively:
\begin{equation}\label{eq:bimod_fusion}
%
.
\end{equation}
Generally, a bimodule object can be written as a direct sum of some bimodule objects. The ones which can not be further decomposed are known as the indecomposable bimodule objects. They are the simple objects in the dual fusion category ${}_\CA \CC_{\CA}$ and generate the quantum symmetry in the gauged theory. 
The intuition is that the conditions in \eqref{eq:bimodule_condition} allow the line $M$ to be inserted and deformed across the mesh of $\CA$ in a consistent way. 

For the rest of the paper, we are interested in computing certain twisted partition functions of the gauged theory. For that, we will need the splitting version $\lambda^\vee,\rho^\vee$ of the $\lambda,\rho$ and their relation to the fusion junctions:
\begin{equation}\label{eq:bimod_split}
 ~.
\end{equation}
As an example, the torus partition function capturing the action of the dual symmetry $(M,\lambda,\rho)$ in the gauged theory is given by the following diagram
\begin{equation}\label{eq:M_action}
    %
 ~.
\end{equation}
Furthermore, one can also consider twisted torus partition functions where two cycles both have non-trivial defects inserted. Calculations of these diagrams will require the construction of the fusion junctions of the bimodule objects. Suppose the bimodule $M$ and $N$ can locally fuse into the bimodule $Y$, and a local junction $\phi_{M,N}^Y:M\otimes N \rightarrow Y$
\begin{equation}
    %
 ~,
\end{equation}
where we abbreviate $\phi_{M,N}^Y$ as $\phi$. The numbers of inequivalent solutions correspond to the multiplicity of $Y$ in the decomposition $M\otimes N$. For practical calculations, one would expand $M$ and the junctions in terms of simple objects similar to what one would do in algebra objects, and solve the corresponding junction coefficients. We will skip the details here.  

\subsection{SPTs of fusion category symmetries}\label{sec:SPT_review}
Similar to the ordinary symmetry, the fusion category symmetry can have symmetry protected topological phases (SPTs), which is defined as the fusion category symmetric gapped phase with a unique ground state \cite{Thorngren:2019iar,shuheng2024cluster,Inamura:2024jke,gu2024nispt}. Different SPTs can be smoothly connected if no symmetry is required. The SPTs of fusion category symmetry $\scriptC$ is mathematically described by the $\scriptC$-module category with one simple object or equivalently the fiber functors of the fusion category symmetry. We say the fusion category $\CC$ is anomalous if it does not admit an SPT \cite{Thorngren:2019iar}(sometimes $\CC$-SPTs is also referred to as $\CC$-symmetric trivially gapped phases). Notice that this notion of the anomaly also relates to the generalization of the 't Hooft anomaly of ordinary symmetry, which measures the obstruction of gauging the symmetry. More precisely, the existence of SPTs of $\CC$ is equivalent to the ability to gauge the entire $\CC$ with the algebra object of the form \cite{Choi:2023xjw}
\begin{equation}
    \CA_{max} = \sum_{\text{simple} \, a} d_{a} a ~.
\end{equation}
Furthermore, there are also fusion category symmetry protected degenerate interface modes between different SPTs which are captured by the representation of interface algebra or the boundary tube algebra \cite{Inamura:2024jke,Choi:2024tri}. 

Below we will review the above three equivalently ways of characterizing the SPTs, with an emphasis on (a) how to construct the torus twisted partition functions of a given SPT and (b) the interface algebras between SPTs and their irreducible representations. 
\subsubsection*{Fiber functors and the SPT twisted partition functions}
Recall that a fiber functor $\phi$ of the fusion category $\CC$ is a tensor functor from $\CC$ to $\VEC$, the category of vector spaces. More concretely, it maps any simple object $a \in \CC$ to the $d_a$-dim vector space $\doubleC^{d_a}$. The vector space of local fusion junction $\Hom_{\CC}(a\otimes b,c)$ is then mapped to the space of linear maps $\Hom_{\doubleC}(\doubleC^{d_a}\otimes \doubleC^{d_b}, \doubleC^{d_c})$. Naturally, the basis in $\Hom(a\otimes b,c)$ defined in \eqref{eq:simple_basis} is mapped to a basis $\{\phi_{ab}^{c,\rho}\}_{\rho=1}^{N_{ab}^c}$ in $\Hom_{\doubleC}(\doubleC^{d_a}\otimes \doubleC^{d_b}, \doubleC^{d_c})$. Furthermore, upon choosing a basis in each $\doubleC^{d_{a}}$, each tensor $\phi_{ab}^{c,\rho}:\doubleC^{d_a}\otimes\doubleC^{d_b} \rightarrow \doubleC^{d_c}$ can be characterized by its $\doubleC$-number components: 
\begin{equation}
    (\phi_{ab}^{c,\rho})_{\alpha\beta}^\gamma \in \doubleC ~, \quad \rho = 1,\cdots,N_{ab}^c,~ \alpha = 1,\cdots, d_a,~ \beta = 1,\cdots, d_b,~ \gamma = 1,\cdots, d_c ~.
\end{equation}
A tensor functor $\phi$ must commute with the associativity map; and in terms of $(\phi_{ab}^{c,\rho})_{\alpha\beta}^\gamma$'s, this is equivalent to the following set of equations:
\begin{equation}\label{eq:fiber_functor}
    \sum_{e,\epsilon,\mu,\nu} (\phi_{ab}^{e,\mu})_{\alpha\beta}^\epsilon (\phi_{ec}^{d,\nu})_{\epsilon\gamma}^\delta [G^{d}_{abc}]_{(e,\mu,\nu),(f,\rho,\sigma)} = \sum_{\varphi} (\phi_{bc}^{f,\rho})_{\beta\gamma}^{\varphi} (\phi_{af}^{d,\sigma})_{\alpha\varphi}^{\delta} ~.
\end{equation}
A non-trivial solution $(\phi_{ab}^{c,\rho})_{\alpha\beta}^\gamma$'s to the above equations specify a fiber functor $\phi$. Notice that given a fiber functor $\phi$, $(\phi_{ab}^{c,\rho})_{\alpha\beta}^\gamma$ depends on the choices of the basis on $\doubleC^{d_a}$, therefore is not unique. Consider a change of basis on each $\doubleC^{d_a}$ parametrized by the matrices $S_a$, $(\phi_{ab}^{c,\rho})_{\alpha\beta}^\gamma$'s transform as
\begin{equation}
    (\phi_{ab}^{c,\rho})_{\alpha\beta}^\gamma \rightarrow \sum_{\tilde{\alpha},\tilde{\beta},\tilde{\gamma}} (S_a)_{\alpha\tilde{\alpha}} (S_b)_{\beta\tilde{\beta}} (\phi_{ab}^{c,\rho})_{\tilde{\alpha}\tilde{\beta}}^{\tilde{\gamma}} (S_c^{-1})_{\tilde{\gamma}\gamma} ~.
\end{equation}
And any two sets of solutions $(\phi_{ab}^{c,\rho})_{\alpha\beta}^\gamma$'s related by the above gauge transformation specify the same fiber functor. 

Notice that $\phi$ also maps the basis of the splitting junctions in \eqref{eq:simple_basis} to a set of $\doubleC$-numbers which we denote as $(\psi^{c,\rho}_{ab})_{\alpha\beta}^\gamma$. As the $\phi$ should preserve the normalization condition, one can easily determine $(\psi^{c,\rho}_{ab})_{\alpha\beta}^\gamma$ in terms of $(\phi_{ab}^{c,\rho})_{\alpha\beta}^\gamma$ (where we use the identity morphism $\operatorname{id}_a$ is mapped to the identity linear map on $\doubleC^{d_a}$)
\begin{equation}
    \delta_{\alpha \tilde{\alpha}} \delta_{\beta \tilde{\beta}} = \sum_{c,\rho,\gamma} \sqrt{\frac{d_c}{d_a d_b}} (\psi^{c,\rho}_{ab})_{\alpha\beta}^\gamma (\phi_{ab}^{c,\rho})_{\tilde{\alpha}\tilde{\beta}}^\gamma ~, \quad \sum_{\alpha\beta}(\psi^{c,\rho}_{ab})_{\alpha\beta}^\gamma (\phi_{ab}^{c,\tilde{\rho}})_{\alpha\beta}^{\tilde{\gamma}} = \delta_{\rho\tilde{\rho}} \delta_{\gamma\tilde{\gamma}} \sqrt{\frac{d_a d_b}{d_c}} ~.
\end{equation}

Physically, a fiber functor $\phi$ describes a trivially gapped phase with the fusion category symmetry $\CC$. We will refer to these phases as the SPT phases of $\CC$. The twisted partition function of the SPT $\phi$ can be constructed using $(\psi^{c,\rho}_{ab})_{\alpha\beta}^\gamma$ and $(\phi_{ab}^{c,\rho})_{\alpha\beta}^\gamma$ by replacing the fusion and the splitting junction with these tensors and have the vector space indices contracted accordingly. For instance, 
\begin{equation}
    \begin{tikzpicture}[baseline={([yshift=-1ex]current bounding box.center)},vertex/.style={anchor=base,circle,fill=black!25,minimum size=18pt,inner sep=2pt},scale=0.5]
    \filldraw[grey] (-2,-2) rectangle ++(4,4);
    \draw[thick, dgrey] (-2,-2) rectangle ++(4,4);
    \draw[thick, black, -stealth] (0,-2) -- (0.354,-1.354);
    \draw[thick, black] (0,-2) -- (0.707,-0.707);
    \draw[thick, black, -stealth] (2,0) -- (1.354,-0.354);
    \draw[thick, black] (2,0) -- (0.707,-0.707);
    \draw[thick, black, -stealth] (-0.707,0.707) -- (-0.354,1.354);
    \draw[thick, black] (0,2) -- (-0.707,0.707);
    \draw[thick, black, -stealth] (-0.707,0.707) -- (-1.354,0.354);
    \draw[thick, black] (-2,0) -- (-0.707,0.707);
    \draw[thick, black, -stealth] (0.707,-0.707) -- (0,0);
    \draw[thick, black] (0.707,-0.707) -- (-0.707,0.707);
    
    \filldraw[black] (+0.707,-0.707) circle (2pt);
    \filldraw[black] (-0.707,+0.707) circle (2pt);
    
    \node[black, below] at (0,-2) {\footnotesize $a$};
    \node[black, right] at (2,0) {\footnotesize $b$};
    \node[black, above] at (0,2) {\footnotesize $a$};
    \node[black, left] at (-2,0) {\footnotesize $b$};
    \node[black, above] at (0,0) {\footnotesize $c$};

    \node[black, below] at (+0.807,-0.707) {\scriptsize $\rho$};
    \node[black, above] at (-0.807,+0.707) {\scriptsize $\tilde{\rho}$};
    
\end{tikzpicture} = \sum_{\alpha,\beta,\gamma} (\phi_{a,b}^{c,\rho})_{\alpha\beta}^{\gamma} (\psi_{b,a}^{c,\tilde{\rho}})_{\alpha\beta}^\gamma ~.
\end{equation}

\subsubsection*{Algebra objects of the maximal gauging}
There is a special case of the algebra object with $\CA_{max}$ where every simple TDL $a$ appears in $\CA$ with multiplicity being its quantum dimension $m_{a} = d_{a}$:
\begin{equation}
    \CA_{max} = \sum_{\text{simple} \, a} d_{a} a ~.
\end{equation}
We will refer to the gauging associated with this type of algebra object as maximal gauging. Notice that given this form of $\CA_{max}$, in general the allowed multiplication maps $\mu$ are not unique (up to the equivalent relation \eqref{eq:A_gauge}). And inequivalent $(\CA_{max},\mu)$'s are in $1$-to-$1$ correspondence with the fiber functors of the fusion category symmetry $\CC$. To see this, we simply notice that the junction coefficients $t_{a\alpha,b\beta}^{c\gamma,\rho}$ satisfying the algebraic condition \eqref{eq:alg_junc_eq} will also satisfy the equations \eqref{eq:fiber_functor} which define the fiber functor (under the identification $t_{a\alpha,b\beta}^{c\gamma,\rho} = (\phi_{a,b}^{c,\rho})_{\alpha\beta}^\gamma$). Physically, this means that the notion of the anomaly (in terms of the obstruction to the $\CC$-SPT phases) is equivalent to the obstruction to maximally gauging the fusion category symmetry $\CC$.

\subsubsection*{Module categories and interfaces of SPTs}
Generically, $\CC$-symmetric gapped phases are classified by (right) module categories $\CM$'s over $\CC$. A generic object in $\CM$ can be expanded as a direct sum of simple objects; and the number of simple objects in $\CM$ equals the number of ground states on $S^1$. Instead of describing the structure of a generic module category, below we restrict ourselves to the case where $\CM$ contains a unique simple object $m$. Therefore, $\CM$ describes a trivially gapped $\CC$-symmetric phase, or an $\CC$-SPT. 

Being a right module category over $\CC$ implies that there is a fusion structure characterizing $\CC$ action on $\CM$ from the right. We can describe this structure by considering the fusion between the unique simple object $m \in \CM$ and the simple objects $c \in \CC$, and the fusion is given by
\begin{equation}
    m \times c = d_c m ~,
\end{equation}
where $d_c$ is the quantum dimension of $c$. Naturally, one can consider a set of local fusion and splitting junctions
\begin{equation}\label{eq:mod_cat_junc}
 ~, \quad \gamma = 1,\cdots, d_c ~.
\end{equation}
Notice that we will always use the blue line to denote the unique simple object $m \in \CM$, and we will drop the label for simplicity. The fusion and splitting junctions satisfy the following normalization conditions:
\begin{equation}
    %
 ~.
\end{equation}

The crossing relations of the local fusion junctions are characterized by the $M$-symbols
\begin{equation}
    %
 ~,
\end{equation}
and they satisfy the following pentagon equations
\begin{equation}\label{eq:M_pentagon}
    \sum_{e,\epsilon,\mu,\nu} [M_{ab}]_{(\alpha,\beta),(e, \mu, \epsilon)} [M_{ec}]_{(\epsilon,\gamma),(d,\nu,\delta)} [G^{d}_{abc}]_{(e,\mu,\nu),(f,\rho,\sigma)} = \sum_{\varphi} [M_{bc}]_{(\beta,\gamma),(f,\rho,\varphi)} [M_{af}]_{(\alpha,\varphi),(d,\sigma,\delta)} ~,
\end{equation}
where we again see that $M$-symbols satisfy exactly the same defining equations as the fiber functors. Therefore, for the rest of the paper, we will use fiber functors and module categories (with a unique simple object) interchangeably.

Similarly, the crossing relations of the splitting junctions are characterized by
\begin{equation}
    \begin{tikzpicture}[baseline={([yshift=-1ex]current bounding box.center)},vertex/.style={anchor=base,circle,fill=black!25,minimum size=18pt,inner sep=2pt},scale=1]
        \draw[blue, line width = 0.4mm, ->-=0.6] (0,0) -- (0,1);
        \draw[blue, line width = 0.4mm, ->-=0.6] (0,1) -- (0,2);
        \draw[blue, line width = 0.4mm, ->-=0.6] (0,2) -- (0,3);
        \draw[black, line width = 0.4mm, ->-=0.6] (0,1) -- (2,3);
        \draw[black, line width = 0.4mm, ->-=0.6] (0,2) -- (1,3);
        \filldraw[blue] (0,1) circle (1.5pt);
        \filldraw[blue] (0,2) circle (1.5pt);

        \node[black, above] at (1,3) {\footnotesize $a$};
        \node[black, above] at (2,3) {\footnotesize $b$};
        \node[black, left] at (0,2) {\scriptsize $\alpha$};
        \node[black, left] at (0,1) {\scriptsize $\beta$};
    \end{tikzpicture} = \sum_{c,\rho,\gamma} [M_{ab}^\vee]_{(\alpha,\beta),(c,\rho,\gamma)} \quad \begin{tikzpicture}[baseline={([yshift=-1ex]current bounding box.center)},vertex/.style={anchor=base,circle,fill=black!25,minimum size=18pt,inner sep=2pt},scale=1]
        \draw[blue, line width = 0.4mm, ->-=0.6] (0,0) -- (0,1);
        \draw[blue, line width = 0.4mm, ->-=0.6] (0,1) -- (0,3);
        \draw[black, line width = 0.4mm, ->-=0.6] (0,1) -- (1,2);
        \draw[black, line width = 0.4mm, ->-=0.6] (1,2) -- (1,3);
        \draw[black, line width = 0.4mm, ->-=0.6] (1,2) -- (2,3);
        \node[black, left] at (0,1) {\scriptsize $\gamma$};
        \filldraw[blue] (0,1) circle (1.5pt);
        \filldraw[black] (1,2) circle (1.5pt);
        \node[above, black] at (0.5,1.5) {\footnotesize $c$};
        \node[below, black] at (1,2) {\footnotesize $\rho$};
        \node[black, above] at (1,3) {\footnotesize $a$};
        \node[black, above] at (2,3) {\footnotesize $b$};

    \end{tikzpicture} ~.
\end{equation}
The relation between $[M_{ab}^\vee]_{(\alpha,\beta),(c,\rho,\gamma)}$ and $[M_{ab}]_{\alpha\beta,c\gamma}$ can be derived similarly to the relation between $F$-symbols and $G$-symbols, which is given by
\begin{equation}
    \sum_{c,\rho,\gamma} [M_{ab}]_{(\alpha,\beta),(c,\rho,\gamma)} [M^\vee_{ab}]_{(\tilde{\alpha},\tilde{\beta}),(c,\rho,\gamma)} = \delta_{\alpha\tilde{\alpha}}\delta_{\beta\tilde{\beta}} ~.
\end{equation}

Notice that in choosing the basis for the local fusion junction \eqref{eq:mod_cat_junc}, one can consider a change of basis
\begin{equation}
 ~.
\end{equation}
This will change the $M$-symbols as
\begin{equation}
    [M_{ab}]_{(\alpha,\beta),(c,\rho,\gamma)} \rightarrow \sum_{\tilde{\alpha},\tilde{\beta},\tilde{\gamma}} (S_a)_{\alpha\tilde{\alpha}} (S_b)_{\beta\tilde{\beta}} [M_{ab}]_{(\tilde{\alpha},\tilde{\beta}),(c,\rho,\tilde{\gamma)}} (S_c^{-1})_{\tilde{\gamma}\gamma} ~,
\end{equation}
and two sets of $M$-symbols are considered equivalent module categories if they are related by the above gauge transformation.

Let's now take two SPTs specified by $M$ and $\widetilde{M}$, and consider the fusion category symmetry $\CC$ acting on their interface. The symmetry operators are defined as
\begin{equation}
    \begin{tikzpicture}[baseline={([yshift=-1ex]current bounding box.center)},vertex/.style={anchor=base,circle,fill=black!25,minimum size=18pt,inner sep=2pt},scale=1]
    \draw[blue, line width = 0.4mm, ->-=0.3] (0,0) -- (0,2);
    \draw[blue, line width = 0.4mm, ->-=0.8] (0,2) -- (0,4);
    \node[blue, below] at (0,0) {\footnotesize $m$};
    \node[blue, above] at (0,4) {\footnotesize $\widetilde{m}$};
    \filldraw[blue] (0,2) circle (1.5pt);
    \draw[black, line width = 0.4mm, ->-=0.5] (0,1) arc(-90:90:0.8 and 1);
    \filldraw[black] (0,1) circle (1.5pt);
    \filldraw[black] (0,3) circle (1.5pt);
    \node[black, right] at (0.8,2) {\footnotesize $a$};
    \node[black, left] at (0,3) {\footnotesize $\mu$};
    \node[black, left] at (0,1) {\footnotesize $\nu$};
    \end{tikzpicture} \equiv \scriptV[a,\mu,\nu] ~, \quad \mu,\nu = 1, \cdots, d_a ~,
\end{equation}
where we use $m$ and $\tilde{m}$ to denote the corresponding simple object respectively, and the blue dot in the middle to denote the interface between the two SPT phases. The symmetry operators $\scriptV[a,\mu,\nu]$'s satisfy the following algebraic relations
\begin{equation}
    \scriptV[b,\alpha,\beta] \scriptV[a,\mu,\nu] = \sum_{c,\rho,\gamma,\gamma'} \sqrt{\frac{d_a d_b}{d_c}}[\widetilde{M}_{ab}]_{(\mu,\alpha),(c,\rho,\gamma)} [M^\vee_{ab}]_{(\beta,\nu),(c,\rho,\gamma')} \scriptV[c,\gamma,\gamma'] ~,
\end{equation}
which are derived from
\begin{equation}
 ~.
\end{equation}
Generically, this algebra may not admit $1$-dimensional irreducible representations. As shown in \cite{Inamura:2024jke}, there exists $1$-dim irrep if and only if the two SPT phases are equivalent to each other. Physically, the non-existence of 1-dim irrep of the interface algebra implies that the interface between two distinct SPTs will have non-trivial degeneracies. We want to emphasize here that the existence of irrep with dim $> 1$ itself does not guarantee the inequivalence between two SPTs. The self-interface algebra of a given SPT necessarily has 1-dim irreps but may also contain higher dimensional irreps. For instance, the self-interface algebra of $\Rep H_8$ contains one 2-dim irrep \cite{Inamura:2024jke}; and in the examples considered in this paper later, we also find there are some SPTs of $\Rep D_{16}$ and $\Rep SD_{16}$ admitting $2$-dim irreps. 

Notice that the 1-dim irreps of the self-interface algebra are in 1-1 correspondence with the automorphism of an SPT. In particular, the equivalence class of the 1-dim irreps can be used to define an invariant of an $S^1$-parameterized family of injective MPSs as discussed in \cite{Inamura:2024jke}.

To construct representations of the interface algebra, one can start with the regular representation of this algebra, and block diagonalize the matrices to get all the irreducible representations. Namely, if we start with some semi-simple algebras $\mathcal{U}_a$ such that
\begin{equation}
    \mathcal{U}_a \mathcal{U}_b = f_{ab}^c \mathcal{U}_c ~,
\end{equation}
then since the algebra is semi-simple, due to the Wedderburn–Artin theorem, it must be a direct sum of matrix algebras. Using the associativity of $\mathcal{U}_a$'s, one can directly check that $(\mathcal{U}_a)_{cb} = f_{ab}^c$ is a matrix representation of the algebra (analog to the regular representation of a finite group), which we can then decompose into irreducible representations.

\subsubsection*{SPTs of group-theoretical fusion categories}
To conclude, we quickly review the classification of the SPTs of a group-theoretical fusion category $\CC(G,\omega;H,\psi)$, which is the dual symmetry one gets from gauging an anomaly-free subgroup $H$ with discrete torsion $[\psi] \in H^2(H,U(1))$ in $\VEC_G^\omega$. It is known from \cite{ostrik2002module} that any symmetric gapped phase of $\CC(G,\omega;H,\psi)$ can be acquired from a symmetric gapped phase of $\VEC_G^\omega$. It is straightforward to characterize the symmetric gapped phases of $\VEC_G^\omega$ by the pairs $(K,\psi_K)$ where $K$ is the unbroken subgroup of $G$ (therefore anomaly free) and $[\psi_K] \in H^2(K,U(1))$ is the $K$-SPT realized on a given ground state. Gauging $H$ subgroup with discrete torsion $\psi$ maps this phase of $\VEC_G^\omega$ to a phase of $\CC(G,\omega;H,\psi)$, which contains a unique ground state if and only if $(K,\psi_K)$ satisfies $G \subset HK$ and the 2-cocycle represented by
\begin{equation}\label{eq:group_theoretical_SPT}
    \frac{\psi_K|_{H\cap K}}{\psi|_{H\cap K}} \in Z^2(H\cap K,U(1))
\end{equation}
is non-degenerate (i.e., there is a unique projective irrep of $H\cap K$) \cite{ostrik2002module}. 

We want to point out an important caveat here: generically $(K,\psi_K)$ and $(K,\psi'_K)$ can describe equivalent partial SSB phases of $\VEC_G^\omega$ even if $\psi_K$ and $\psi'_K$ are inequivalent cohomologically. As pointed out, $\psi_K$ specifies the $K$-SPT of a given ground state, and the $K$-SPTs on other ground states are uniquely determined by the $\VEC_G^\omega$-symmetry. However, this does not necessarily mean that the other ground states will have the same $K$-SPTs as $\psi_K$. This may lead to the identification of some of the partial SSB phases. 

Let's consider a concrete example where $G = \doubleZ_2^a \times \doubleZ_2^b \times \mathbb{Z}_2^c$ with $\omega$ being the type III mixed anomaly between the three $\mathbb{Z}_2$'s. Say $K = \mathbb{Z}_2^b \times \mathbb{Z}_2^c$, then the two choices of $\psi_K$ in $H^2(\mathbb{Z}_2^b\times \mathbb{Z}_2^c,U(1)) = \mathbb{Z}_2$ actually label the same partial SSB phase. This is because, due to the type III mixed anomaly, the $\mathbb{Z}_2^c$ action which maps one ground state to the other will also stack a non-trivial $\mathbb{Z}_2^b \times \mathbb{Z}_2^c$-SPT phase, therefore the two ground states must realize different SPT phases. Thus it doesn't matter which ground state realizes the trivial SPT of the two. Hence, there is only one partial SSB phase with $\mathbb{Z}_2^a \times \mathbb{Z}_2^b$ unbroken. 

If $(K,\psi_K)$ and $(K,\psi_K')$ actually describes the same $\VEC_G^\omega$ phases, then gauging $(H,\psi)$ will lead to identical phases of $\CC(G,\omega;H,\psi)$. One must be careful with this subtlety when classifying the SPT phases of $\CC(G,\omega;H,\psi)$. We will encounter an example of this later. 

\subsection{Computing Drinfeld center from the boundary fusion category}\label{sec:DC}
In this subsection, we review how to construct the SymTFT $\mathcal{Z}(\CC)$ for the 2d fusion category symmetry $\CC$, following \cite{Teo:2015xla}. Specifically, we will give a skeletal setup on how to compute the Drinfeld center with a given fusion category $\CC$, which we will then generalize to the relative center to compute the SET corresponding to a given $G$-extension. After discussing the general formalism, we will consider $\CC = \TY(\doubleA,\chi,\epsilon)$ as an example, which is also useful when we discuss the self-duality under gauging non-invertible symmetries. Notice that the same result has been worked out in numerous places via similar formalism, including \cite{Gelaki:2009blp,Zhang:2023wlu}. We include this simple example to simply demonstrate our convention.

\subsubsection{Generalities}\label{subsection:Drinfeld Center}
A fusion category symmetry $\mathcal{C}$ can be realized on the corresponding gapped boundary of its SymTFT (its Drinfeld center) $\mathcal{Z}(\mathcal{C})$. Notice that there are two relations between the bulk SymTFT and the boundary fusion category symmetries which are useful for different purposes. First, given a bulk SymTFT and a choice of Lagrangian algebra $\mathcal{L}$, one can ask what fusion category symmetry $\CC_\CL$ is realized on the boundary. This can be done using anyon condensation theory following the approach described in \cite{Lou:2020gfq}. Second, given a boundary fusion category symmetry $\CC$, one can ask what is the corresponding bulk SymTFT. This can be computed by simply using the definition of the Drinfeld center in a skeletal formalism (for more details, see \cite{Teo:2015xla,Zhang:2023wlu,Choi:2024tri}). We will focus on the second relation in this work.

If one pushes a simple bulk anyon $X$ to the boundary, it will become a (generically non-simple) boundary TDL which we denote $x$. This implies that there is a local junction between the bulk anyon $X$ and the boundary TDL $x$. Furthermore, one can decompose this junction in terms of simple TDLs on the boundary
\begin{equation}\label{eq:boundary_bulk_junction}
 ~.
\end{equation}
The above junctions provide a natural basis for the local fusion junction between bulk anyon $X$ and the boundary TDL $a$:
\begin{equation}
\begin{aligned}
    & %
 ~.
\end{aligned}
\end{equation}
Generically, two distinct simple anyons $X$ and $X'$ in the bulk may become the same TDL $x$ when placed on the boundary; therefore, the data of $x$ itself is not sufficient to uniquely determine the bulk anyon. There is an additional piece of data allowing us to distinguish between $X$ and $X'$ by the boundary simple TDLs. This is the half-braiding between the anyon $X$ and the TDLs on the boundary:
\begin{equation}
    %
 ~,
\end{equation}
which can be interpreted as the action of the boundary TDL on the junction between $X$ and $(c,\mu)$ using the above basis
\begin{equation}\label{eq:R_symbol_basis}
    %
 ~.
\end{equation}
On the other hand, not every pair $X = (x,R^{*,X}_*)$ corresponds to an anyon in the bulk. They are subjected to the following constraint known as the hexagon equations, derived from matching two sequences of $F$- and $R$-moves shown in Figure \ref{fig:DC_hex}:
\begin{figure}[H]
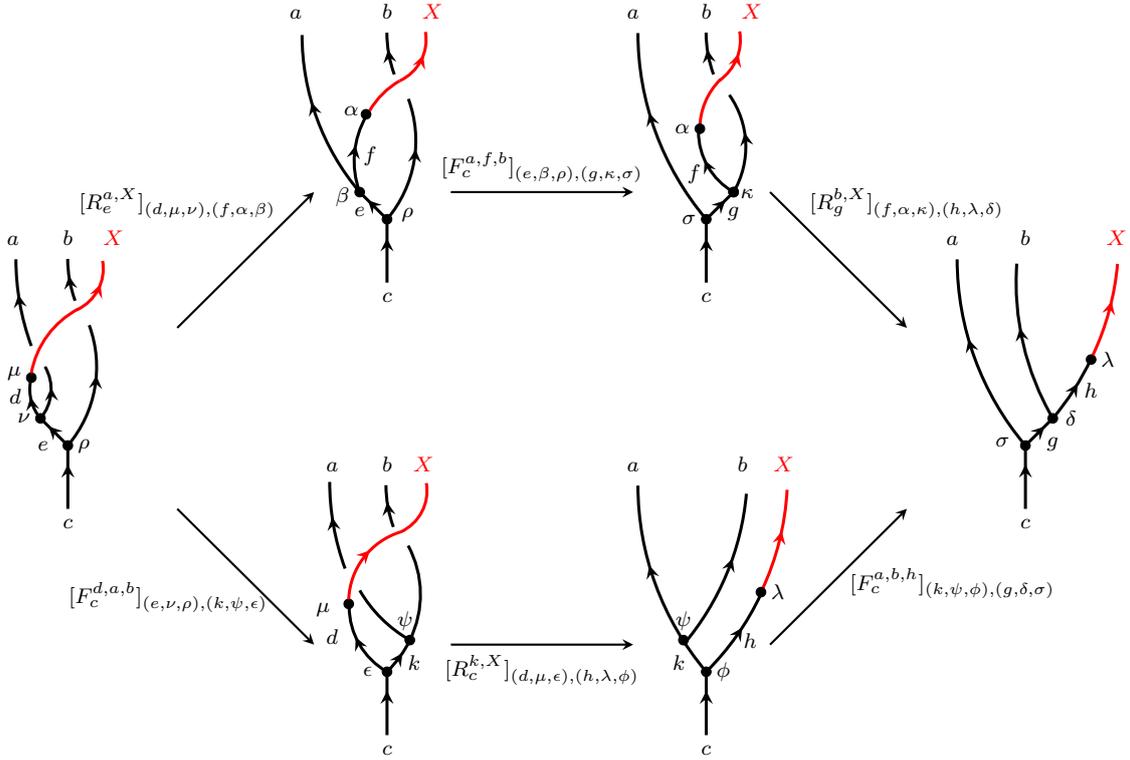

    \centering
    %
    \caption{Hexagon equation that computes the half braiding of anyons.}
    \label{fig:DC_hex}
\end{figure}
\begin{equation}\label{eq:bbhex}
\begin{aligned}
 & \sum_{\alpha,\beta,\kappa,f}\left[R^{a,X}_e\right]_{(d,\mu,\nu),(f,\alpha,\beta)} \left[F^{a,f,b}_{c}\right]_{(e,\beta,\rho),(g,\kappa,\sigma)} \left[R^{b,X}_g\right]_{(f,\alpha,\kappa),(h,\lambda,\delta)} \\
    = & \sum_{\psi,\epsilon,\phi,k}\left[F^{d,a,b}_{c}\right]_{(e,\nu,\rho),(k,\psi,\epsilon)} \left[R^{k,X}_c\right]_{(d,\mu,\epsilon),(h,\lambda,\phi)} \left[F^{a,b,h}_{c}\right]_{(k,\psi,\phi),(g,\delta,\sigma)} ~.
\end{aligned}
\end{equation}
This then allows us to construct the SymTFT using only the data of the fusion category symmetry $\mathcal{C}$. More concretely, to find all the simple anyons $(x,R^{*,X}_*)$ in the bulk, one would first start with the simple TDLs $x \in \mathcal{C}$, then find all the inequivalent solutions to \eqref{eq:bbhex}. The corresponding bulk anyon has the quantum dimension
\begin{equation}
    d_{X \equiv (x,R^{*,X}_*)} = d_x~.
\end{equation}
And the total quantum dimension $d_{\mathcal{Z}(\mathcal{C})}$ of the $\mathcal{Z}(\mathcal{C})$ should satisfy
\begin{equation}
    d_{\mathcal{Z}(\mathcal{C})}=\sqrt{\sum_{\text{simple}\, X \in \mathcal{Z}(\mathcal{C})} d_X^2} = \sum_{\text{simple} \, x\in \mathcal{C}} d_x^2~.
\end{equation}
If all the simple bulk anyons of the form $(x,R^{*,X}_*)$ where $x$ is simple lead to a smaller total quantum dimension, then one must gradually increase the number of simple objects in $X$ and continue to find solutions to the hexagon equation \eqref{eq:bbhex}, until the total quantum dimension is saturated.

Notice that in this bulk-boundary setup, every bulk computation can be computed by pushing part of the bulk anyons to the boundary. To do that, besides the junctions on the RHS of \eqref{eq:boundary_bulk_junction}, we also need another sets of basis junctions
\begin{equation}\label{eq:bulk_boundary_junction}
 ~.
\end{equation}
This then allows us to compute the bulk observables such as the $S$ and $T$ matrices by pushing the anyons to the boundary, where we find
\begin{equation}
\begin{aligned}
    \theta_X &= \sum_{a,b,\mu,\alpha}\frac{d_b}{d_X}[R^{a,X}_{b}]_{(a,\mu,\alpha),(a,\mu,\alpha)} ~, \\
    S_{XY} &= \sum_{a,b,\mu,\nu,\alpha,\beta}\frac{d_c}{d_{\mathcal{Z}(\mathcal{C})}}[R^{b,X}_c]_{(a,\mu,\alpha),(a,\mu,\beta)}[R^{a,Y}_c]_{(b,\nu,\beta),(b,\nu,\alpha)} ~,
\end{aligned}
\end{equation}
where the details can be found in \eqref{eq:ST_matrices_cal}.

\subsubsection{Example I: $\CC = \VEC_{A}$}\label{sec:VecADC}
As the first example, let's consider $\CC = \VEC_{A}$. In this case, if we take $X = g \in A$, then the potentially non-trivial $R$-symbols are given by
\begin{equation}
 ~.
\end{equation}
Then, the hexagon equations simply become
\begin{equation}
    R^{a,X}_{ag} R^{b,X}_{bg} = R^{ab,X}_{abg} ~.
\end{equation}
Immediately we see that there are $|A|^2$ anyons of this type, parameterized by $(g,R^{*,X} \equiv \hat{g})$ where $g \in A$ and $\hat{g} \in \Hom(A,U(1)) \simeq \widehat{A}$. Since these anyons already saturate the quantum dimension, we conclude that the Drinfeld center is parameterized by $A \times \widehat{A}$. And we denote these Abelian anyons as $X_{(a,\widehat{b})}$ where $a \in A, \widehat{b} \in \widehat{A}$.

\subsubsection{Example II: $\CC = \TY(A,\chi,\epsilon)$}\label{sec:TYDC}
As another example, let's consider $\CC = \TY(A,\chi,\epsilon)$. Here, $\chi$ is a symmetric non-degenerate bicharacter of $A$ and $\epsilon = \pm 1$ is known as the FS indicator. Besides the group elements of $A$, it also admits a duality defect $\CN$ with the fusion rules
\begin{equation}
    a \times \CN = \CN \times a = \CN ~, \quad a \in A ~, \quad \CN \times \CN = \sum_{a\in A} a ~.
\end{equation}
The non-trivial $F$-symbols are given by
\begin{equation}
    F^{a\CN b}_{\CN} = F^{\CN a \CN}_b = \chi(a,b) ~, \quad [F^{\CN\CN\CN}_\CN]_{a,b} = \frac{1}{\sqrt{|A|}} \frac{1}{\chi(a,b)} ~,
\end{equation}
where $|A|$ is the order of the group $A$.

There are three types of anyons in the Drinfeld center which we denoted as $X$-type, $Y$-type, and $Z$-type following \cite{Gelaki:2009blp}. We will not explain in detail how to find these solutions here, as the interpretation will soon be clear as we explain the relative center later in Section \ref{sec:TY_RC}.

\textbf{$X$-type anyons:}

The $X$-type anyons are Abelian anyons with $X = g \in A$, and the potentially non-trivial $R$-symbols are given by
\begin{equation}
 ~.
\end{equation}
The hexagon equations are given by
\begin{equation}
\begin{aligned}
    R^{h,X} R^{k,X} &= R^{hk,X} ~, \\
    R^{h,X} R^{\CN,X} &= \chi(g,h) R^{\CN,X} ~, \\
    R^{\CN,X} \chi(h,g) R^{\CN,X} &= R^{g^{-1}h,X} ~,
\end{aligned}
\end{equation}
where we see $2|A|$ independent solutions given by
\begin{equation}
    X = g ~, \quad R^{h,X} = \chi(g,h) ~, \quad R^{\CN,X} = \pm \frac{1}{\sqrt{\chi(g,g)}} ~.
\end{equation}

\textbf{$Y$-type anyons}

\

$Y$-type anyons correspond to composite TDL where $Y = g_1 \oplus g_2$ where $g_1 \neq g_2$. The quantum dimension of a $Y$-type anyon is therefore $d_Y = 2$. Since $Y$ is a composite object, its fusion rules with simple TDLs are 
\begin{equation}
    Y \times h = g_1 h + g_2 h ~, \quad Y \times \CN = 2\CN ~,
\end{equation}
and the local fusion junction has a canonical choice of basis via the junction between $Y$ and its component simple object $g_i$:
\begin{equation}
 ~.
\end{equation}
The $R$-symbols and the hexagon equations can be conveniently expressed using the above basis. The potentially non-trivial $R$-symbols are given by
\begin{equation}
%
 ~.
\end{equation}
The four hexagon equations are given by
\begin{equation}
\begin{aligned}
    R^{h,Y}_{g_i} R^{k,Y}_{g_i} &= R_{g_i}^{hk,Y} ~, \\
    R^{h,Y}_{g_i} (R^{\CN,Y}_{\CN})_{ij} &= \chi(h,g_j) (R^{\CN,Y}_{\CN})_{ij} ~, \\
    (R^{\CN,Y}_{\CN})_{ij} R^{h,Y}_{g_j} &= \chi(h,g_i) (R^{\CN,Y}_{\CN})_{ij} ~, \\
    \sum_{j=1}^2 (R^{\CN,Y}_{\CN})_{ij} \chi(h,g_j) (R^{\CN,Y}_{\CN})_{jk} &= R^{h,Y}_{g_i} \delta_{ik} ~.
\end{aligned}
\end{equation}
Corresponding to each distinct choice of $Y = g_1 \oplus g_2$, there is a unique inequivalent solution given by
\begin{equation}\label{eq:Y_type_R_symbol}
    R^{h,Y}_{g_1} = \chi(g_2, a) ~, \quad R^{h,Y}_{g_2} = \chi(g_1,a) ~, \quad (R^{\CN,Y}_{\CN})_{ij} = \sqrt{\chi(g_1,g_2)}
 ~.
\end{equation}
\textbf{$Z$-type anyons}

\

The $Z$-type anyons are anyons with $Z = \CN$, therefore with quantum dimension $d_Z = \sqrt{|A|}$. And the potentially non-trivial $R$-symbols are given by
\begin{equation}
%
 ~.
\end{equation}
The hexagon equations are given by
\begin{equation}
\begin{aligned}
    R^{g,Z}_\CN R^{h,Z}_{\CN} \chi(g,h) &= R^{gh,Z}_\CN ~, \\
    R^{\CN,Z}_g R^{h,Z}_{\CN} &= \chi(h,gh) R^{\CN,Z}_{gh} ~, \\
    R^{\CN,Z}_g R^{\CN,Z}_h &= \frac{1}{\sqrt{|A|}} \sum_{k \in A} R^{k,Z}_{\CN} \frac{\chi(g,h)}{\chi(gh,k)} ~,
\end{aligned}
\end{equation}
where we find $2|A|$ inequivalent solutions. First, since $\chi(g,h)$ is cohomologically trivial, there are $|A|$ distinct solutions to the first equations. Furthermore, the last two equations relate $R^{\CN,Z}_g$ to $R^{g,Z}_{\CN}$ via
\begin{equation}
    R^{\CN,Z}_g = \pm \frac{1}{\chi(g,g)} R^{g,Z}_{\CN} ~.
\end{equation}
Hence, in total there are $2|A|$ distinct $Z$-type anyons with quantum dimension $\sqrt{|A|}$ in the bulk.

\section{Constructing SETs from $G$-graded extension}\label{sec:RDC}
The $G$-graded extension of a fusion category $\CC$ is an important construction in studying the generalized symmetries \cite{Kaidi:2021xfk,Choi:2021kmx,Choi:2022zal,Bhardwaj:2024xcx,Decoppet:2024moc,Decoppet:2023bay,Choi:2024rjm}, as it provides a systematically way of exploring non-invertible symmetries which can be easily generalized to higher dimensions. 

A $G$-graded fusion category $\CC$ is a fusion category with a direct sum decomposition
\begin{equation}
    \CC = \bigoplus_{g\in G}\CC_g ~, 
\end{equation}
such that the tensor product preserving the $G$-grading $\CC_g \otimes \CC_h \subset \CC_{gh}$. For a fusion category $\CD$, we say that a $G$-graded fusion category $\CC$ is a $G$-graded extension of $\CD$ if the trivial grading component $\CC_1 = \CD$. As an example, the $\TY$-fusion category mentioned previously is a $\mathbb{Z}_2$-graded fusion category. The trivial grading component is $\VEC_{A}$ while the non-trivial grading component is $\{\CN\}$ consisting of the non-invertible duality symmetry $\CN$. Generically, $G$-extension $\CC$ arises from including the enhanced symmetry describing the invariance under invertible topological manipulations in $\CC_1$. And when such invertible topological manipulation involves gauging\footnote{It is important to notice that finite gauging is an invertible topological manipulation, as one can always gauge the dual symmetry to recover the original theory.}, the enhanced symmetry is non-invertible. For more details, see \cite{Choi:2023vgk}. This is the reason $G$-extensions are important in the study of non-invertible symmetries. 

In the remarkable work \cite{Etingof:2009yvg}, it is shown that the $G$-extension can be understood and classified by the bulk SymSETs. More concretely, this is described by the following diagram \cite{Etingof:2009yvg,Gelaki:2009blp,Bhardwaj:2024qiv}:
\begin{equation}
    \begin{tikzpicture}[scale=0.8,baseline={([yshift=-.5ex]current bounding box.center)},vertex/.style={anchor=base,
    circle,fill=black!25,minimum size=18pt,inner sep=2pt},scale=0.50]
    \node[black] at (0,0) {\footnotesize $\CC_1$};
    
    \node[black] at (0,-5) {\footnotesize $\mathcal{Z}(\CC_1)$};
    \draw[thick, black, -stealth] (0,-0.5) -- (0,-4.5);
    \node[black, left] at (0,-2.5) {\scriptsize center};
    
    \draw[thick, black, -stealth] (1.5,-5) -- (5.5,-5);
    \node[black, above] at (3.5,-5) {\scriptsize $G$-crossed};
    \node[black, below] at (3.5,-5) {\scriptsize extension};

    \node[black] at (7,-5) {\footnotesize $\mathcal{Z}(\CC_1)_G^\times$};

    \draw[thick, black, -stealth] (8.5,-5) -- (12.5,-5);
    \node[black, above] at (10.5,-5) {\scriptsize $G$-equiv.};

    \node[black, right] at (12.5,-5) {\footnotesize $\mathcal{Z}(\CC) = [\CZ(\CC_1)_G^\times]^G$};
    
    \draw[thick, black, -stealth] (1.0,0) -- (12.5,0);
    \node[black, above] at (6,0) {\scriptsize $G$-extension};
    
    \node[black, right] at (12.5,0) {\footnotesize $\displaystyle \mathcal{C} = \bigoplus_{g\in G} \CC_g$};

    \draw[thick, black, -stealth] (13,-0.5) -- (13,-4.5);
    \node[black, right] at (13,-2.5) {\scriptsize center};

    \draw[thick, red, -stealth] (12.5,-0.5) -- (7,-4.5);
    \node[red, left] at (9.5, -2.5) {\scriptsize relative center};
    
\end{tikzpicture} ~.
\end{equation}
That is, the bulk correspondence of a $G$-extension is a $2$-step process: one first construct the SET phase $\mathcal{Z}(\CC_1)_G^\times$ which enriches the SymTFT $\mathcal{Z}(\CC_1)$ with the corresponding $G$-symmetry in a certain way, and then perform the $G$-equivariantization to finish gauging $G$ to get the SymTFT $\mathcal{Z}(\CC)$ for the boundary symmetry $\CC$. As there is no choice left in the final $G$-equivariantization step, the $G$-extension $\CC$ is completely specified by which $0$-form symmetry $G$ we use to enrich the SymTFT $\CZ(\CC_1)$ and how we enrich it. As we will review shortly, fixing a concrete $0$-form symmetry $G$, an SET is characterized by the symmetry fractionalization class labeled by $H^2(G,\mathbf{A})$ where $\mathbf{A}$ is the group of Abelian anyons in $\CZ(\CC_1)$, and the defectification class labeled by $H^3(G,U(1))$ (which can also be interpreted as the discrete torsion of the $0$-form symmetry $G$) \cite{Barkeshli:2014cna}, provided that the obstruction $[\mathfrak{O}_3] \in H^3(G,\mathbf{A})$ to symmetry fractionalization vanishes and the 't Hooft anomaly $[\mathfrak{O}_4] \in H^4(G,U(1))$ of $G$-symmetry vanishes. From the boundary point of view, choosing the $G$-symmetry in the bulk fixes which invertible topological manipulations in $\CC_1$, while the fractionalization classes and the discrete torsions determine how those enhanced symmetry lines would fuse each other to form a larger fusion category $\CC$.

The goal of our paper is to extract the SymSET data from a given $G$-extension $\CC$, and to study how the categorical structure of the boundary $G$-extension category changes when the fractionalization class changes from the SymSET. To derive the SymSET, one way to start with the SymTFT $\CZ(\CC)$ and using the anyon condensation theory to condense all the $G$-Wilson lines to acquire the SymSET $\CZ_{\CC_1}(\CC)$. This is a $2$-step procedure, and despite being complicated, it is also unclear how to extract the data described in Section \ref{sec:SET data} except the fusion rules in a concrete setup. One would hope to find a single-step calculation to reconstruct all the SET data from the $G$-graded fusion category $\CC$. 

Indeed, such an approach is provided in \cite{Gelaki:2009blp} and the corresponding bulk category is known as the relative center $\mathcal{Z}_{\CC_1}(\CC)$. After briefly reviewing some basic notions of the SETs in Section \ref{sec:SET data}, we will first explain the relative center in the categorical language quickly in Section \ref{sec:Relative_center} and then dive into the concrete skeletal setup which allows us to compute the SET data in Section \ref{sec:SymSET_computation}. Finally, in Section \ref{sec:SymSET_application}, we will describe how to extract the change in the categorical structure of the boundary $G$-extension when changing the fractionalization classes in the SymSET, as well as some other applications.

\subsection{Basic of the SETs}\label{sec:SET data}
We briefly review some basic facts about SET where more detailed discussion is available in \cite{Barkeshli:2014cna,Etingof:2009yvg,Gelaki:2009blp,teo2015theory}\footnote{Here we only discuss bosonic unitary symmetries in a bosonic TQFT described by Unitary Modular Tensor Categories(UMTCs). More general discussions can be found in \cite{Barkeshli:2014cna}.}. The intrinsic symmetry of a topological order (a unitary modular tensor category $\scriptB$) forms a group $\Aut(\scriptB)$ (and mathematically this is also known as the group $\EqBr(\scriptB)$ of braided autoequivalences of $\scriptB$). A generic group element $\varphi$ permutes the anyons 
\begin{equation}
    \varphi(a) = a'
\end{equation}
as well as linearly transforms the local splitting(fusion) spaces:
\begin{equation}
    \varphi(\ket{a,b;c,\mu}) = \sum_{\mu'}[U(a',b';c')]_{\mu\mu'} \ket{a',b';c',\mu'}~, \quad\mu=1, \cdots,N^{c}_{ab}
\end{equation}
where $
\ket{a,b;c,\mu}$ is a basis vector for the splitting junction as defined in \eqref{eq:simple_basis}. Diagrammatically, this is captured by 
\begin{equation}
 ~.
\end{equation}
As a result, $\varphi$ transforms the $F$-symbols and $R$-symbols in a non-trivial way:
\begin{equation}
\begin{aligned}
    \varphi([R^{ab}_c]_{(\mu,\nu)}) &= \sum_{\mu',\nu'}[U(b',a';c')]_{\mu\mu'}[R^{a'b'}_{c'}]_{\mu'\nu'}[U(a',b';c')^{-1}]_{\nu'\nu} ~, \\
    \varphi([F^{abc}_d]_{(e,\alpha,\beta),(f,\mu,\nu)}) &= \sum_{\alpha',\beta',\mu'\nu'}[U(a',b';e')]_{\alpha\alpha'}[U(e',c';d')]_{\beta\beta'}[F^{a'b'c'}_{d'}]_{(e',\alpha',\beta'),(f',\mu',\nu')} \\
    & \quad \quad \quad \quad \quad \quad \quad \quad \quad \quad \quad \quad \quad \quad \quad \quad \quad \quad \quad \quad 
        [U(b',c';f')^{-1}]_{\mu'\mu}[U(a',f';d')^{-1}]_{\nu'\nu} ~,
\end{aligned}
\end{equation}
and the transformed $F$-symbols and the $R$-symbols are required to be identical as the original ones:
\begin{equation}
   \varphi([F^{abc}_d]_{(e,\alpha,\beta),(f,\mu,\nu)}) = [F^{abc}_d]_{(e,\alpha,\beta),(f,\mu,\nu)} ~, \quad \varphi([R^{ab}_c]_{(\mu,\nu)}) = [R^{ab}_c]_{\mu,\nu} ~.
\end{equation}
Furthermore, the gauge invariant data such as the fusion multiplicity, quantum dimensions, $S$ and $T$-matrices are therefore invariant under the $\varphi$-transformations.

The following transformations (known as the natural transformations), while trivially satisfying the above requirement, are considered trivial elements in $\Aut(\scriptB)$:
\begin{equation}\label{eq:gauge redundancy}
    \beta(a) = a ~, \quad \beta(\ket{a,b;c,\mu}) =\frac{\beta_a\beta_b}{\beta_c}\ket{a,b;c,\mu} ~, \quad \beta_i \in U(1) ~.
\end{equation}
Physically, such transformation can be thought of as dressing the junction between $0$-form symmetry and the anyon with contact terms, therefore are not considered as an actual symmetry. Naturally, any two $\varphi$ and $\varphi'$ related by such a $\beta$ are considered equivalent bulk symmetries.

It is important to mention that while generically the action on the anyons do not uniquely determine the symmetry (for an example, see \cite{Davydov:2013xov,Kobayashi:2025ykb}), it is shown in \cite{Benini:2018reh} that for if $\scriptB$ is multiplicity free (that is, $N_{ab}^c = 0,1$), the anyon permutation action does uniquely determine the symmetry.

A physical symmetry $G$ is realized through a homomorphism from $G$ to $\Aut(\scriptB)$:
\begin{equation}
    [\rho]: G \rightarrow \Aut(\scriptB) ~,
\end{equation}
where for each $g \in G$, we associate a $\rho_g \in \Aut(\scriptB)$ specified by 
\begin{equation}
    \rho_g(a) = {}^g a ~, \quad \rho_g(\ket{a,b;c,\mu}) = \sum_{\mu'}[U_g(a',b';c')]_{\mu\mu'}\ket{a',b';c',\mu'} ~.
\end{equation}
Notice that because of the equivalence relations due to natural transformations, the multiplications of $\rho_g$'s may differ from the group multiplication up to natural transformations:
\begin{equation}\label{eq:comp_rho_g}
    \kappa_{g,h} \circ \rho_g \circ \rho_h = \rho_{gh} ~,
\end{equation}
where $\kappa_{g,h}$'s are natural transformations specified by $\eta_a(g,h) \in U(1)$:
\begin{equation}
    \kappa_{g,h}(a) = a ~, \quad \kappa_{g,h}(\ket{a,b;c,\mu}) = \frac{\eta_a(g,h)\eta_b(g,h)}{\eta_c(g,h)} (\ket{a,b;c,\mu}) ~.
\end{equation}
The composition laws \eqref{eq:comp_rho_g} then requires
\begin{equation}
    \frac{\eta_a(g,h)\eta_b(g,h)}{\eta_c(g,h)}\delta_{\mu\nu} = \sum_{\alpha,\beta}[U_{g}(a,b;c)^{-1}]_{\mu\alpha}[U_{h}({}^{g^{-1}}a,{}^{g^{-1}}b,{}^{g^{-1}}c)^{-1}]_{\alpha\beta}[U_{gh}(a,b;c)]_{\beta\nu} ~.
\end{equation}
As pointed it out in \cite{Barkeshli:2014cna,Benini:2018reh}, $\eta_a(g,h)$ appears when one try to move the anyon $a$ across the junction between $g$ and $h$
\begin{equation}
 ~.
\end{equation}
Notice that the two fusion channels of three $0$-form symmetry $g,h,k$ can change into each other locally in $3$-dim, therefore leading to a consistency condition from pulling the anyon $a$ across the junctions of $0$-form symmetries in two different fusion channels. Using the property of the UMTC, it can be shown that the obstruction to satisfy this consistency condition is captured the cohomology class $[\mathfrak{O}_3] \in H^3_\rho(G,\mathbf{A})$ where $\mathbf{A}$ is the group of Abelian anyons in $\scriptB$. This obstruction is known as the obstruction to the symmetry fractionalization, and in the context of $2$-group symmetry, this is known as the Postnikov class. We will not discuss them in detail, as in all the cases we study, such obstruction vanishes. For more details, see \cite{Barkeshli:2014cna,Benini:2018reh}.

In the case $\mathfrak{O}_3$ is cohomologically trivial, we have $\mathfrak{O}_3 = d\mathfrak{M}$, where $\mathfrak{M}_2 \in C^2_\rho(G,\mathbf{A})$. Physically, this means we can dress the fusion junction of the $0$-form symmetries with Abelian anyons specified by $\mathfrak{M}_2$ to effectively change $\eta_a(g,h)$:
\begin{equation}
 ~,
\end{equation}
where $\mathfrak{M}(g,h)(a)$ denotes the $1$-form symmetry action of the Abelian anyon $\mathfrak{M}(g,h)$ on the simple anyon $a$. With the new effective $\eta$'s, the consistency conditions are satisfied. Here, the cohomology class of $\mathfrak{M}$ specifies a symmetry fractionalization class. Inequivalent symmetry fractionalization classes therefore form a $H^2_\rho(G,\mathbf{A})$-torsor. Namely, to change one symmetry fractionalization to another, we take a corresponding 2-cocycle in $[\mathfrak{t}] \in H^2_\rho(G,\mathbf{A})$, and dress the current fusion junctions of $0$-form symmetry by Abelian anyons specified by $\mathfrak{t}$.

The next step in defining a SET is the introduction of twist $g$-defects, which reside at the boundaries of $g$-surface operator. Anyon lines can be thought of as living on the boundary of a trivial surface operator. The fusion of twist defects must respect the group structure of $G$: fusing a twist $g$-defect with an $h$-defect results in a $gh$-defect. This fusion rule can be represented diagrammatically as follows: 
\begin{equation} \label{eq:twist fusion junction}
 ~,  \quad \quad \quad \mu = 1,\cdots, N_{a_gb_h}^{c_{gh}} ~.
\end{equation}
Here, the subscripts $g$ and $h$ denote the attached surface operator extending out of the plane of the paper. Similarly, braiding of twist defect can be represented diagrammatically as 
\begin{equation}\begin{aligned}
    %
 ~.
\end{aligned}
\end{equation}
In this process, $b_h$ pierces the surface operator associated with $a_g$, causing it to transform into a new twist defect residing at the boundary of a $ghg^{-1}$ surface:
\begin{equation}
    {}^{g}b_h=\tilde{\rho}_g(b_h)=b'_{ghg^{-1}} ~.
\end{equation}
In particular, if we take $h = 1$ to be the identity operator, then this crossed braiding will capture the action of the $0$-form symmetry $g$ on the anyon $b$.

Naturally, one can expect that the $U$-symbols and $\eta$-symbols are also extended to include the twist defects; we will skip those details as we will not explicitly use them in this work and refer the reader to \cite{Barkeshli:2014cna} for more details.

Mathematically all the data forms the so-called $G$-crossed braided fusion category, which we will denote as $\scriptB_G^\times$. $\scriptB_G^\times$ is also called an $G$-crossed extension of $\scriptB$. It is important to notice that after fixing the symmetry $G$ and a fractionalization class, the SET $\scriptB_G^\times$ does not always exist and the obstruction is nothing but the 't Hooft anomaly $[\mathfrak{O}_4] \in H^4(G,U(1))$ of the $0$-form symmetry $G$ (also known as the obstruction to defectification). It is important to notice that this 't Hooft anomaly generically depends on the choice of the fractionalization class \cite{Brennan:2022tyl,Delmastro:2022pfo}. In the case where the 't Hooft anomaly vanishes, different choices of the SETs are parameterized by classes in $H^3(G,U(1))$, which are nothing but the discrete torsions of the $0$-form symmetry $G$.

Despite the additional data involving the twist defects, we want to emphasize here that the SET is completely specified by the choice of the symmetry $G$, the fractionalization class, and the discrete torsion. Thus, we only focus on those in this work, although the additional data involving twist defects can be computed by straightforwardly generalizing some of our setups.

\subsection{The relative center of a fusion category}\label{sec:Relative_center}
Here, we briefly review the notion of relative center of a fusion category and we will refer the readers to \cite{Gelaki:2009blp} for more details. 

To start, let's consider a fusion category $\CD$ together with a $\CD$-bimodule category $\CM$. The center $\CZ_{\CD}(\CM)$ of $\CM$ is the category of $\CD$-bimodule functors from $\CD$ to $\CM$. More concretely, the objects of More concretely, objects in $\CZ_{\CD}(\CM)$ are pairs $(m,\gamma)$, where $m$ is an object in $\CM$ and 
\begin{equation}
    \gamma = {\gamma_x: x \otimes m \xrightarrow{\sim} m \otimes x} ~, \quad x \in \CD ~,
\end{equation}
such that the following diagram commutes (where $x,y \in \CD$):
\begin{equation}
    \begin{tikzpicture}
    \node[thick] at (0,0) {$x \otimes(y \otimes m)$};
    \draw[thick, black, -stealth] (0.75, 0.35) to (2.75,1.35);
    \node[thick] at (1.4, 1.1) {\footnotesize $id_x \otimes \gamma_y$};
    \node[thick, right] at (2.75,1.35) {$x\otimes(m\otimes y)$};
    \draw[thick, black, -stealth] (5.25, 1.35) to (7.75,1.35);
    \node[thick, above] at (6.50,1.35) {\footnotesize $\alpha^{-1}_{x,m,y}$};
    \node[thick, right] at (7.75,1.35) {$(x\otimes m) \otimes y$};
    \draw[thick, -stealth] (10.25,1.35) to (12,0.35);
    \node[thick] at (11.75,1.1) {\footnotesize $\gamma_x \otimes id_y$};
    \node[thick] at (12.75,0) {$(m\otimes x)\otimes y$}; 
    
    \draw[thick, black, -stealth] (0.75, -0.35) to (2.75,-1.35);
    \node[thick] at (1.4, -1.1) {\footnotesize $\alpha^{-1}_{x,y,m}$};
    \node[thick, right] at (2.75,-1.35) {$(x\otimes y)\otimes m$};
    \draw[thick, black, -stealth] (5.25, -1.35) to (7.75,-1.35);
    \node[thick, below] at (6.50,-1.35) {\footnotesize $\gamma_{x\otimes y}$};
    \node[thick, right] at (7.75,-1.35) {$m \otimes (x \otimes y)$};
    \draw[thick, -stealth] (10.25,-1.35) to (12,-0.35);
    \node[thick] at (11.75,-1.1) {\footnotesize $\alpha_{m,x,y}^{-1}$};
\end{tikzpicture} ~,
\end{equation}
where $\alpha$'s are the associativity map from the bimodule category structure $\CM$. One can immediately see that this generalizes the definition of the Drinfeld center (which is the special case of taking $\CM = \CD$ itself). If the bimodule category $\CM$ can be written as a direct sum of other bimodule categories, then its center also admits a similar direct sum decomposition
\begin{equation}
    \CM = \bigoplus_i \CM_i \implies \CZ_{\CD}(\CM) = \bigoplus_{i} \CZ_{\CD}(\CM_i) ~. \
\end{equation}

When $\CM$ is an invertible $\CD$-bimodule category, $\CM$ corresponds to an invertible $0$-form symmetry in the SymTFT $\CZ(\CD)$. And its center $\scriptZ_{\CD}(\CM)$ contains all the twist defects ending on the corresponding bulk $0$-form symmetry operator. We want to emphasize that at this stage, we have not equipped those twist defects with a fusion structure. Generally, it is possible that a consistent fusion structure does not exist; and when it does exist, generally such structures are not unique.

Let's now consider the case where $\CD$ is a subcategory $\CC$. Because $\CC$ can be viewed as a bimodule category of any of its subcategories, the center $\CZ_{\CD}(\CC)$ of $\CC$ is well-defined and is known as the \textit{relative center} of $\CC$. Furthermore, when $\CC$ is a $G$-extension of $\CD$, it is shown in \cite{Gelaki:2009blp} that when $\CC$ is a $G$-extension of $\CD$, then the relative center $\CZ_{\CD}(\CC)$ is a braided $G$-crossed fusion category, which exactly describes the SymSET we are looking for.

More concretely, the $G$-grading on $\CC$ will lead to a $G$-grading on the relative center $\CZ_{\CD}(\CC)$:
\begin{equation}
    \CC = \bigoplus_{g\in G} \CC_g \implies \CZ_{\CD}(\CC) = \bigoplus_{g\in G} \CZ_{\CD}(\CC_g) ~.
\end{equation}
Here, the trivial grading component $\CZ_{\CD}(\CC_1)$ is nothing but the Drinfeld center of $\CD$, therefore describes its SymTFT. Since each grading component $\CC_g$ in $\CC$ is an invertible $\CD$-bimodule category, they each specify an invertible symmetry in the SymTFT of $\CD$, and together form a finite group $G$. Furthermore, the corresponding twist defects are described by the center $\CZ_{\CD}(\CC_g)$. Finally, additional structures can be equipped to $\CZ_{\CD}(\CC)$ in a unique way specified by $\CC$ to make it a braided $G$-crossed fusion category. We will skip the mathematical description here and refer the readers to the original work \cite{Gelaki:2009blp}. Below, we will describe in a more physics-friendly setup on how to extract the data.

It is worth noticing that once a bulk symmetry action is fixed, there are two places where inequivalent choices of the SET structure can arise. First, the $0$-form symmetry can fractionalize non-trivially on the Abelian anyons (which are invertible $1$-form symmetry operators) in the bulk; second, one can stack a $G$-SPT to the existing SET. The second piece is well-understood in the context of $G$-extension. For instance, for $\TY$-fusion categories, this choice of $G=\mathbb{Z}_2$-SPT in the bulk corresponds to the FS indicator of $\CN$ on the boundary corresponds, which is a choice of $\pm$ sign in $F^{\CN\CN\CN}_{\CN}$. The first piece is less acknowledged as for $\TY$-fusion categories, Shapiro's lemma guarantees that there is a unique choice of symmetry fractionalization. In the following, we will derive an explicit relation relating inequivalent $F$-symbols related by changing the fractionalization classes. This will play an important role in the bulk understanding of the self-duality defect under gauging non-invertible symmetries. 

As a final remark, even using the approach below, computing the full SET data is still a non-trivial task in any non-trivial examples. But this is due to the large amount of data one has in the SET phase. For any specific piece of data, one wants to compute in a concrete example, it is not hard to isolate these data and compute them. In this work, since we are only interested in understanding the data related to the classification of the SET phases and its effect on the boundary extended fusion category symmetries, we only need to compute the SET data for the Abelian anyons in many occasions. This is a great simplification.

\subsection{Computation of the SymSETs}\label{sec:SymSET_computation}

\subsubsection{Spectrum of twist defects}
The setup is nearly identical to that described in Section  \ref{subsection:Drinfeld Center}, with the key difference being that the boundary fusion category $\scriptC$ is now a $G$-extension of $\scriptD$. Specifically,
\begin{equation}
    \scriptC= \bigoplus_{g\in G} \scriptC_g ~, \quad \scriptD=\scriptC_1 ~.
\end{equation}
Consequently, the relative center $\mathcal{Z}_{\scriptD}(\scriptC)$ inherits a $G$-grading
\begin{equation}
    \mathcal{Z}_{\scriptD}(\scriptC)=\underset{g\in G}{\bigoplus}\mathcal{Z}_{\mathcal{\scriptD}}(\scriptC_g) ~.
\end{equation}
Here, the simple bulk anyon lines in the SymTFT $\CZ(\CD)$ are in the trivial grading component $\CZ_{\CD}(\CC_1)$. Bulk twist $g$-defects in $\mathcal{Z}_{\mathcal{\scriptD}}(\scriptC_g)$ become (generically non-simple) boundary TDL $x_g \in \scriptC_g$ when pushed to the boundary. The local junction between the bulk twist $g$-defect and boundary TDL $x_g$ can be decomposed using simple TDLs in $\CC$
\begin{equation}\label{eq:frac_bb_setup}

\end{equation}
The above junctions provide a natural basis for the local fusion junction between bulk twist $g$-defect $X_g$ and the boundary TDL $a$:
\begin{equation}
\begin{aligned}
    & %
 ~,
\end{aligned}
\end{equation}
where we use the green lines to denote the twist defects in the bulk and black lines to denote the TDLs on the boundary. The surface $g$-symmetry operator extend out of the paper. The data which distinguishes different bulk twist $g$-defect $X_g$ corresponding to the same TDL $x_g$ is the half-braiding:
\begin{equation}
    %
 ~,
\end{equation}
which can be interpreted as the action of the boundary TDL $a$ on the junction between $X_g$ and $(c_g,\mu)$ using the above basis
\begin{equation}
    %
 ~.
\end{equation}
On the other hand, not every pair $X_g = (x_g,R^{*,X_g}_*)$ correspond to a twist $g$-defect in the bulk. They are subjected to the following constraint between the boundary TDLs and the bulk twist $g$-defect $(x_g,R^{*,x_g}_*)$, derived from matching two sequences of $F$- and $R$-moves shown in Figure \ref{fig:twist defect_hex}:
\begin{figure}[H]
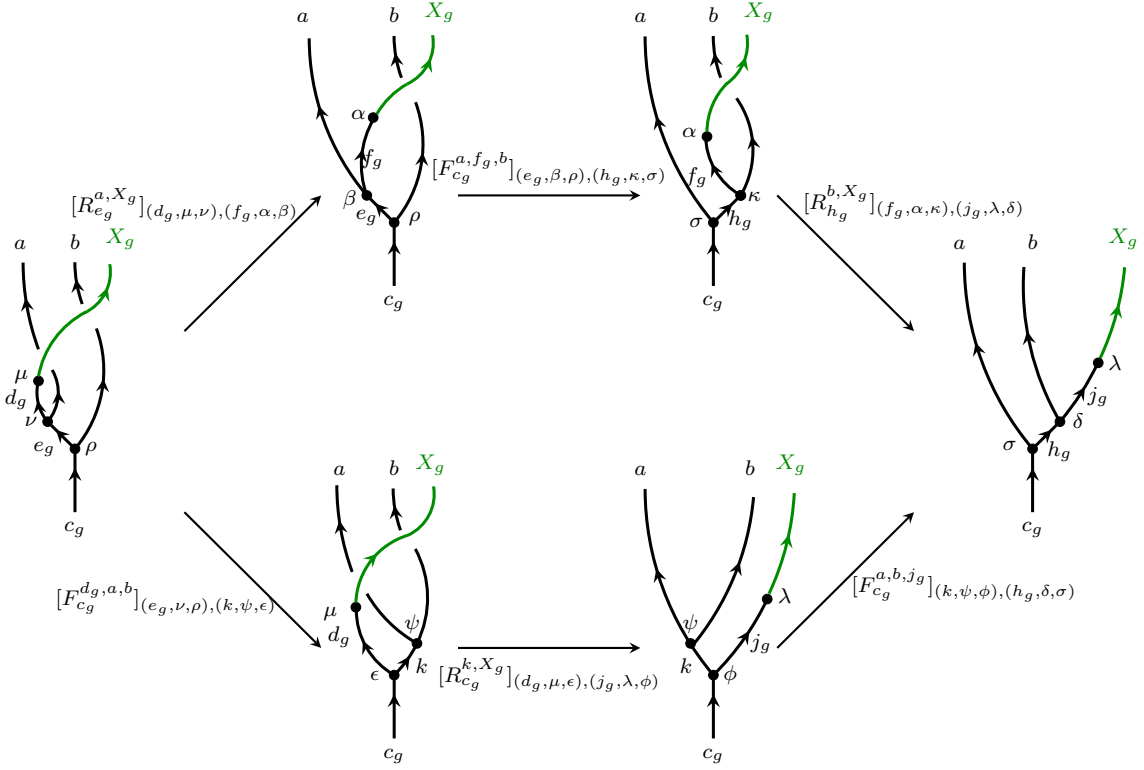

    \centering
%
    \caption{Hexagon equation that allows us to compute the half braiding of twist $g$-defect.}
    \label{fig:twist defect_hex}
\end{figure}
\begin{equation}\label{eq:hexagon for twist}
\begin{aligned}
    & \sum_{\alpha,\beta,\kappa,f_g}\left[R^{a,X_g}_{e_g}\right]_{(d_g,\mu,\nu),(f_g,\alpha,\beta)} \left[F^{a,f_g,b}_{c_g}\right]_{(e_g,\beta,\rho),(h_g,\kappa,\sigma)} \left[R^{b,X_g}_{h_g}\right]_{(f_g,\alpha,\kappa),(j_g,\lambda,\delta)} \\
    = & \sum_{\psi,\epsilon,\phi,k}\left[F^{d_g,a,b}_{c_g}\right]_{(e_g,\nu,\rho),(k,\psi,\epsilon)} \left[R^{k,X_g}_{c_g}\right]_{(d_g,\mu,\epsilon),(j_g,\lambda,\phi)} \left[F^{a,b,j_g}_{c_g}\right]_{(k,\psi,\phi),(h_g,\delta,\sigma)} ~.
\end{aligned}
\end{equation}
This allows us to construct bulk twist $g$-defect using the data of boundary graded fusion category $\scriptC$. More concretely, to find all the simple twist $g$-defect $(x_g,R^{*,x_g}_*)$, one needs to find all the inequivalent solutions to \eqref{eq:hexagon for twist}. The corresponding bulk twist $g$-defect has quantum dimension
\begin{equation}
    d_{X_g \equiv (x_g,R^{*,x_g}_*)} = d_{x_g}~.
\end{equation}
And the total quantum dimension of the $\mathcal{Z}_{\scriptD}(\mathcal{C}_g)$ should be the same as that of the SymTFT $\mathcal{Z}(\scriptD)$
\begin{equation}
    \sum_{\text{simple}\, X_g \in \mathcal{Z}_{\scriptD}(\mathcal{C}_g)} d_{X_g}^2 = \left(\sum_{\text{simple} \, x_g\in \mathcal{C}_g} d_{x_g}^2\right)^2 = \left(\sum_{\text{simple} \, x\in \mathcal{C}_0} d_x^2\right)^2~.
\end{equation}

\subsubsection{Symmetry action on the anyons}\label{sec:sym_action}
Knowing the spectrum of anyons and twist defects, we are now ready to equip it with a $G$-crossed braiding structure to turn it into a SET. First, we would like to know the $G$-symmetry actions on the anyons and twist defects. For simplicity, we will describe how to compute the action on anyons only, but it can be straightforwardly generalized to compute the action on the twist defects as well.

Let's consider the following crossed braiding configurations
\begin{equation}
 ~.
\end{equation}
where we use $a_g,b_g$ to denote TDLs in the non-trivial grading component $\scriptC_g$, and $X$ is an anyon in the SymTFT $\CZ(\CD)$ of $\scriptD$. Notice that in the SymSET setup, $a_g$ and $b_g$ extend into the bulk as the symmetry defect of $g \in G$, therefore, when the bulk anyon $X$ going through it as depicted on the L.H.S., it will be transformed by the symmetry $g$ and we denote the resulting anyon as ${}^g X$. We also use a red dot to denote the intersection between $X$ and $g$ symmetry defect in the bulk.

To find out the symmetry action, we consider the hexagon equations acquired from the diagrams in Figure \ref{fig:hex_g_action}:
\begin{figure}
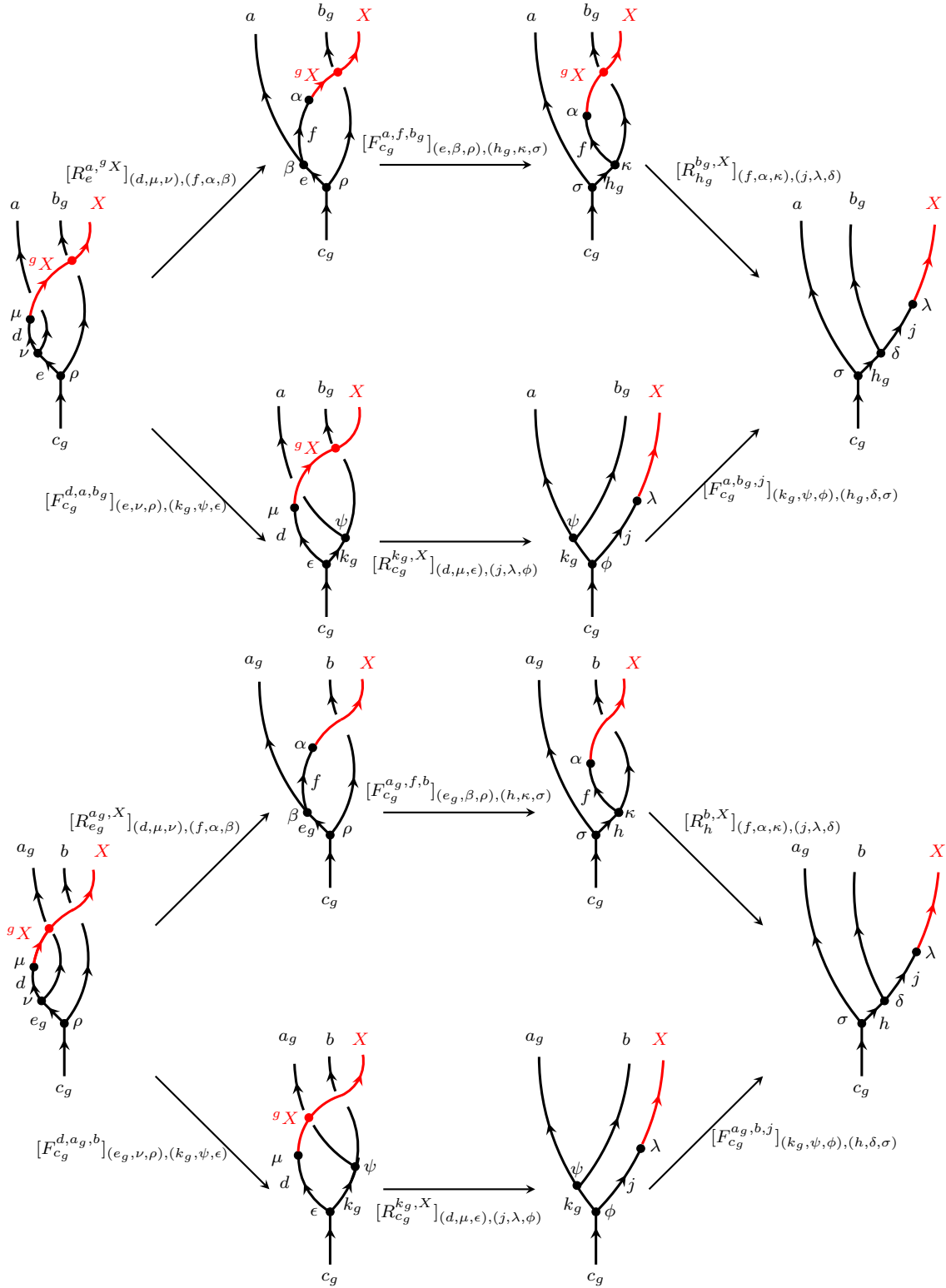

    \centering

    \caption{The hexagon equations which allows us to compute the $g$ action on the bulk anyon $X$.}
    \label{fig:hex_g_action}
\end{figure}
\begin{equation}\label{eq:hexagon for bulk action}
\begin{aligned}
     & \sum_{\alpha,\beta,\kappa,f}\left[R^{a,{}^{g}X}_{e}\right]_{(d,\mu,\nu),(f,\alpha,\beta)} \left[F^{a,f,b_g}_{c_g}\right]_{(e,\beta,\rho),(h_g,\kappa,\sigma)} \left[R^{b_g,X}_{h_g}\right]_{(f,\alpha,\kappa),(j,\lambda,\delta)} \\
    = & \sum_{\psi,\epsilon,\phi,k_g}\left[F^{d,a,b_g}_{c_g}\right]_{(e,\nu,\rho),(k_g,\psi,\epsilon)} \left[R^{k_g,X}_{c_g}\right]_{(d,\mu,\epsilon),(j,\lambda,\phi)} \left[F^{a,b_g,j}_{c_g}\right]_{(k_g,\psi,\phi),(h_g,\delta,\sigma)} ~,\\
      & \sum_{\alpha,\beta,\kappa,f}\left[R^{a_g,X}_{e_g}\right]_{(d,\mu,\nu),(f,\alpha,\beta)} \left[F^{a_g,f,b}_{c_g}\right]_{(e_g,\beta,\rho),(h,\kappa,\sigma)} \left[R^{b,X}_{h}\right]_{(f,\alpha,\kappa),(j,\lambda,\delta)} \\
    = & \sum_{\psi,\epsilon,\phi,k_g}\left[F^{d,a_g,b}_{c_g}\right]_{(e_g,\nu,\rho),(k_g,\psi,\epsilon)} \left[R^{k_g,X}_{c_g}\right]_{(d,\mu,\epsilon),(j,\lambda,\phi)} \left[F^{a_g,b,j}_{c_g}\right]_{(k_g,\psi,\phi),(h,\delta,\sigma)} ~.
\end{aligned}
\end{equation}
Notice that only when $X$ and ${}^g X$ are actually related by the symmetry $g$, the above equations admit non-zero solutions to the $R$-symbols. Notice that $X$ and ${}^g X$ must have the same quantum dimensions and the topological spins, therefore we only need to test such pairs. This is how we determines $g$-symmetry actions on the bulk anyon $X$.

\subsubsection{$U$-symbols and $\eta$-symbols}
\begin{figure}
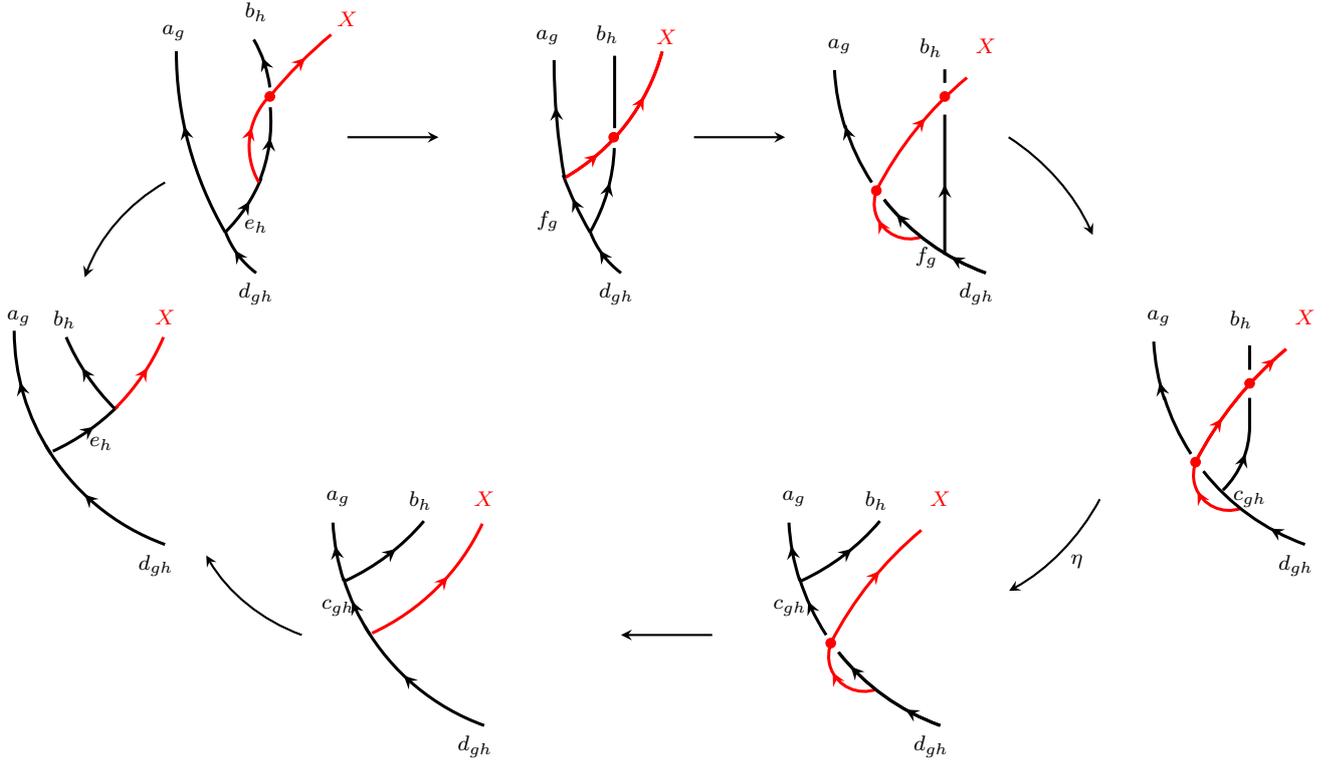

    \centering

    \caption{The heptagon equation which allows us to compute the projective action $\eta$ of $g$ and $h$ on bulk anyon $X$. Notice that we have omitted the indices on the fusion junctions for simplicity.}
    \label{fig:eta_symbol_cal}
\end{figure}

Once the equations are solved and the action of $g$ is known, we are ready to compute the action of bulk symmetry on the junction of the anyons. To do so, consider the heptagon equations given by the Figure \ref{fig:hep_U_symbol}:
\begin{figure}[H]
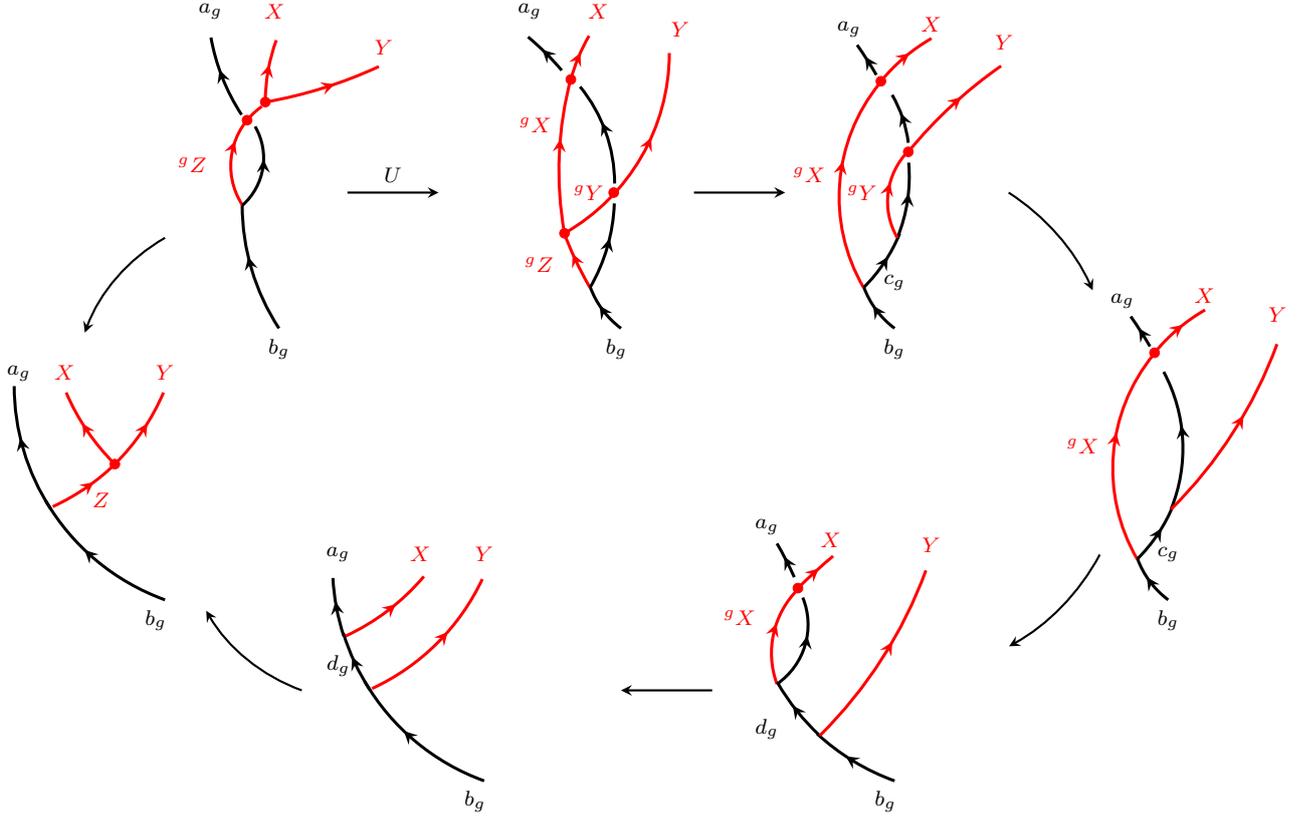

    \centering

    \caption{The heptagon equations which allows us to compute the $g$ action on splitting junction of the bulk anyon $X$ and $Y$. Notice that we have omitted the indices on the fusion junctions for simplicity.}
    \label{fig:hep_U_symbol}
\end{figure}
Notice that the only unknown variables in the equations are the $U$-symbols, hence we acquire an expression for $U$-symbols in the bulk SET. However, to write down a close form equation for the $U$-symbols requires us to construct explicitly the local splitting junctions of the bulk anyons in terms of the boundary TDLs explicitly, which of course can be done straightforwardly. Similarly, once the crossed braiding is known, one can compute the $\eta$-symbols via heptagon equations given by the diagram in Figure \ref{fig:eta_symbol_cal}, as it is the only unknown in the equation. Knowing the two allows us to unambiguously determine the choice of symmetry fractionalization classes in the corresponding SET.

But in this work, however, we will not need the expression of the $U$-symbols and the $\eta$-symbols explicitly, as we are only interested in  tracking how the \textit{difference} between fractionalization classes changes the categorical structure of the boundary fusion categories. As a result, we will skip the details of these equations.

\subsection{Applications}\label{sec:SymSET_application}
\subsubsection{How symmetry fractionalization classes change the boundary categorical structure?}\label{sec:transformed_F_symbols}
We are now ready to derive how changing the symmetry fractionalization classes changes the boundary extended fusion category. Recall that to change the symmetry fractionalization class, one would take a non-trivial class $[\mathfrak{t}] \in H^2_{[\rho]}(G,\mathbf{A})$ where $\mathbf{A}$ denotes the group of Abelian anyons in the SymTFT, and dress the Abelian anyons $\mathfrak{t}_{g,h}$ to the fusion junctions of $g,h$ surface operators. Notice that in our convention, $\mathfrak{t}$ satisfies
\begin{equation}
    {}^{\overline{k}} \mathfrak{t}_{g,h} \mathfrak{t}_{gh,k} = \mathfrak{t}_{h,k} \mathfrak{t}_{g,hk} ~,
\end{equation}
where we used $\overline{k}$ to denote the inverse of $k$ and notice that while this is not the conventional twisted $2$-cocycle conditions, they are equivalent to the often used one. In the bulk-boundary setup, naively we expect to modify the local splitting junction
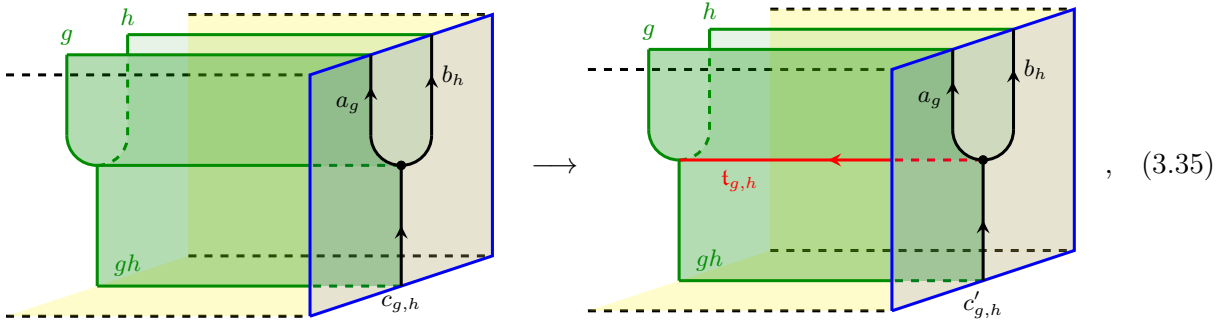
\begin{equation}\label{eq:frac_bb_modif}
 \quad ,
\end{equation}
where the Abelian anyon $\mathfrak{t}_{g,h}$ inserted at the junction of the bulk $0$-form symmetry will fuse into the local splitting junction of the fusion $a_g \times b_h \rightarrow c_{gh}$. Notice that in this modification, if the Abelian anyon condenses on the symmetry boundary (namely it can terminate on the boundary), the incoming TDL $c_{gh}$ will not be changed. Otherwise $c_{gh}$ is modified into a different anyon $c'_{gh}$ which is the fusion product of the boundary image of $\mathfrak{t}_{g,h}$ and $c_{gh}$. Notice that since $\mathfrak{t}_{g,h}$ becomes an invertible TDL on the boundary, $c'_{gh}$ remains a simple TDL with the same quantum dimension (since $c_{gh}$ is assumed to be simple to begin with). In general, $\mathfrak{t}_{g,h}|_{bdy}$ is not a central element in the boundary extended fusion category and we need to make a choice whether $\mathfrak{t}_{g,h}|_{bdy}$ fuses with $c_{gh}$ from the left or from the right. Importantly, these two choices are equivalent since they are related by a braiding in the bulk; and we will choose the convention where it fuses with $c_{gh}$ from the right to match with our definition of $R$-symbols.

The symmetry fractionalization provides a useful way of generalizing the $2$-cocycle twist in the central extension to the generic $G$-extension of a fusion category symmetry. In the well-known story of central extension, given a finite group $G$, one can consider extending $G$ by a finite Abelian group $A$ characterized by the following short exact sequence
\begin{equation}
    0 \rightarrow A \rightarrow \Gamma \rightarrow G \rightarrow 0 ~,
\end{equation}
where, assuming $G$ acts trivially on $A$, inequivalent extensions are labeled by elements $[w] \in H^2(G,A)$. The group multiplication structure on $\Gamma \equiv A \times_{w_2} G$ is given by
\begin{equation}
    (a,g) \times (a',g') = (a a' w(g,g'), gg') ~.
\end{equation}
This example admits a simple understanding of from the bulk. The SymSET is simply given by $A$-gauge theory extended by a trivially acting $G$ symmetry. A choice of the fractionalization class is specified by $[\mathfrak{t}] \in H^2(G,A\times \widehat{A})$. Notice that whether the fractionalization class will lead to non-trivial $w$ depends on if $\mathfrak{t}_{g,h}$ contains non-trivial anyons in $A$. This is because if all $\mathfrak{t}_{g,h}$ is condensed on the boundary (we assume the symmetry boundary is acquired by condensing all the anyons labeled by $\widehat{A}$), then $\mathfrak{t}$ will not modify the fusion rule at all (but this will generically modify the 't Hooft anomaly of the resulting group). On the other hand, if one takes $\mathfrak{t}$ to be the lift of $w$ from $H^2(G,A)$ to $H^2(G,A\times \widehat{A})$, then this would give us the extension $A\times_{w}G$ with trivial 't Hooft anomaly. We discuss a concrete example with $A = \mathbb{Z}_2$ and $G = \mathbb{Z}_2$ in Appendix \ref{app:Z2_example}.

The situation becomes a bit complicated when generalizing to non-invertible symmetries. Naively, we expect when $\mathfrak{t}_{g,h}$ is not completely trivial when pushing to the boundary, it will modify the boundary fusion rule exactly as what $w$ does
\begin{equation}
    a_g \times b_h = \sum_{c_{gh}} N_{a_g, b_h}^{c_{gh}} c_{gh} \longrightarrow a_g \times b_h = \left(\sum_{c_{gh}} N_{a_g, b_h}^{c_{gh}} c_{gh}\right) (\mathfrak{t}_{g,h}|_{bdy}) ~,
\end{equation}
where we use $\mathfrak{t}_{g,h}|_{bdy}$ to denote the boundary image of $\mathfrak{t}_{g,h}$. However, even if $\mathfrak{t}_{g,h}|_{bdy}$ is non-trivial, since $c_{gh}$ can be non-invertible, it can certainly be the case that $c_{gh} \times \mathfrak{t}_{g,h}|_{bdy} = c_{gh}$ such that the fusion rule is unchanged. A generic case like this is the $G$-ality defects under which the full fusion category is gauged. Since there is a unique simple $G$-ality defect $\CN_g$ in every non-trivial grading component, it then must be true that $\CN_{gh} \times t_{g,h}|_{bdy} = \CN_{gh}$. Thus, in order to realize a case in the non-invertible symmetries similar to $2$-cocycle $w$ in the central extension, $c_{gh}$ can only be the defect where the fusion category symmetry is non-maximally gauged. We will see examples of both cases later when discussing the self-duality under gauging non-invertible symmetries in $\Rep D_8$ and $\Rep H_8$.

\

Let's now describe how to compute the modified $F$-symbols. To proceed, we start by pointing out that this intuitive physical picture in \eqref{eq:frac_bb_modif} introduces a four-way local junction of the lines on the boundary, and we want to resolve it into two three-way junction. Generically, this is not a unique procedure and one needs to pick a convention and stick with it. We choose our convention to match our definition of the $R$-symbols, and the local splitting junctions are modified to 
\begin{equation}

\end{equation}
where we use subscript $g,h\in G$ to track the $G$-grading of the TDLs. The new $F$-symbols, which we denote as $\widetilde{F}$'s, are defined as
\begin{equation}
    %
 ~,
\end{equation}
where we have dropped the labels of the line which are determined from the fusion rule and will use $\mathfrak{t}_{g,h}$ to denote its boundary image for ease of notation. It is then straightforward to show that
\begin{equation}\label{eq:tilde_F}
\begin{aligned}
    & \left[\widetilde{F}^{a_g,b_h,c_k}_{d_{ghk}}\right]_{(e_{gh},\mu,\nu),(f_{hk},\alpha,\beta)} \\
    = & \sum_{\rho,\sigma,\gamma} \left[F^{e_{gh}\mathfrak{t}_{g,h}^{-1},\mathfrak{t}_{g,h},c_k}_{d_{ghk} \mathfrak{t}^{-1}_{gh,k}}\right]_{\nu\rho} R^{c_k, {}^{\overline{k}}\mathfrak{t}_{g,h}}_{\mathfrak{t}_{g,h}c_k} \left[F^{e_{gh}\mathfrak{t}_{g,h}^{-1},c_k,{}^{\overline{k}}\mathfrak{t}_{g,h}}_{d_{ghk} \mathfrak{t}^{-1}_{gh,k}}\right]^{-1}_{\rho\sigma} F^{d_{ghk}\mathfrak{t}_{gh,k}^{-1} {}^{\overline{k}}\mathfrak{t}_{g,h}^{-1} ,{}^{\overline{k}}\mathfrak{t}_{g,h},\mathfrak{t}_{gh,k}}_{d_{ghk}} \\
     & \left[F^{a_g,b_h,c_k}_{d_{ghk} \mathfrak{t}_{gh,k}^{-1}{}^{\overline{k}}\mathfrak{t}_{g,h}^{-1}}\right]_{(e_{gh}\mathfrak{t}_{g,h}^{-1},\mu,\sigma),(f_{hk}\mathfrak{t}_{h,k}^{-1},\alpha,\gamma)} [F^{d_{ghk}\mathfrak{t}_{g,hk}^{-1}\mathfrak{t}_{h,k}^{-1},\mathfrak{t}_{h,k},\mathfrak{t}_{g,hk}}_{d_{ghk}}]^{-1} \left[F^{a_g,f_{hk} \mathfrak{t}^{-1}_{h,k},\mathfrak{t}_{h,k}}_{d_{ghk} \mathfrak{t}^{-1}_{g,hk}}\right]_{\gamma\beta} ~.
\end{aligned}
\end{equation}
An important caveat here is that the transformed $F$-symbols $\widetilde{F}$ may not satisfy the pentagon equations. Naively, one might want to appeal to the Mac-Lane coherence theorem, but carefully tracking the moves \cite{Etingof:2009yvg,Barkeshli:2019vtb} shows that the violation of the pentagon equations is characterized by phases made of $U$ and $\eta$ symbols in the SET data. Since the phase arises due to the crossing and the braiding among $\mathfrak{t}_{g,h}$ themselves and moving them through the junctions of the bulk $0$-form symmetries, these phases only depend on the grading of the four outgoing lines in the pentagon equations. These phases are nothing but the obstruction class $[\mathfrak{O}_4]$ valued in $H^4(G,U(1))$, and concretely it is given by
\begin{equation}\label{eq:O4}
    \begin{aligned}
        \mathfrak{O}_4(g,h,k,l) = &U_l({}^{\bar{k}}\mathfrak{t}_{g,h},\mathfrak{t}_{gh,k})U^{-1}_l(\mathfrak{t}_{h,k},\mathfrak{t}_{g,hk})\eta_{\mathfrak{t}_{g,h}}(k,l)[R^{-1}]^{{}^{\overline{kl}}\mathfrak{t}_{g,h},\mathfrak{t}_{k,l}}\\
        &G_{{}^{\overline{kl}}\mathfrak{t}_{g,h}{}^{\bar{l}}\mathfrak{t}_{gh,k}\mathfrak{t}_{ghk,l}}G^{-1}_{{}^{\overline{kl}}\mathfrak{t}_{g,h}\mathfrak{t}_{k,l}\mathfrak{t}_{gh,kl}}G_{\mathfrak{t}_{k,l}{}^{\overline{kl}}\mathfrak{t}_{g,h}\mathfrak{t}_{gh,kl}}G^{-1}_{\mathfrak{t}_{k,l}\mathfrak{t}_{h,kl}\mathfrak{t}_{g,hkl}} G_{{}^{\overline{l}}\mathfrak{t}_{h,k}\mathfrak{t}_{hk,l}\mathfrak{t}_{g,hkl}}G^{-1}_{{}^{\overline{l}}\mathfrak{t}_{h,k}{}^{\bar{l}}\mathfrak{t}_{g,hk}\mathfrak{t}_{ghk,l}} ~,
    \end{aligned}
\end{equation}
where $\overline{g}$ denotes the inverse of $g$. From the bulk point of view, this is nothing but the 't Hooft anomaly of the $0$-form symmetry $G$ in the SET, while from the boundary point of view, this is the obstruction to the existence of the corresponding $G$-extension. 

It is interesting to notice that \eqref{eq:O4} is a normalized cochain, that is, $\omega(g,h,k,l) = 1$ when any one of $g,h,k,l$ equals to identity, as long as we choose $\mathfrak{t}_{g,h}$ to be a normalized $2$-cocycle. This means that when $\mathfrak{O}_4$ is cohomologically trivial, the $3$-cochain $\omega$ we use to trivialize it is also normalized. As a result, the actual $F$-symbols $\omega \widetilde{F}$ differs from the naive one $\widetilde{F}$ only when the three incoming TDLs are all in the non-trivial grading components. This allows us to use \eqref{eq:tilde_F} to show explicitly that $F$-symbols of the form $F^{a_g, b, c}_{d_g},F^{a, b_g, c}_{d_g},F^{a, b, c_g}_{d_g}$ does not transform under changing fractionalization class, matching the general statement in \cite{Choi:2023vgk}.

As a result, one can see that the changing of the fractionalization class changes the spin selection rules in an interesting way. Notice that the spin selection rules of the defect Hilbert space $\CH_{a_g}$ is derived via relating the spins to the symmetry actions of $\CC_1$ on $\CH_{a_g}$, and the latter depends only on the bimodule structure specified by $F^{a_g, b, c}_{d_g},F^{a, b_g, c}_{d_g},F^{a, b, c_g}_{d_g}$. Thus, changing the fractionalization class will not change the algebra relations and the representation of $\CC_1$, but will change the relation between the spin and the representation to change the spin selection rules. 

To eventually acquire a consistent set of $F$-symbols, we must compute $\mathfrak{O}_4$ explicitly and check that it is cohomologically trivial and pick a trivialization of it when this is the case. While one can certainly compute $\mathfrak{O}_4$ in terms of $U$ and $\eta$ (notice that since it only depends on the $U$ and $\eta$ symbols of Abelian anyons only, this is not too hard) practically it is even simpler to just extract those phases from the violations of the pentagon equations. This is the approach we will take in this paper. And when $\mathfrak{O}_4$ is indeed cohomologically trivial, inequivalent choices are then labeled by $H^3(G,U(1))$ and are known as the discrete torsions.

\

To conclude, we summarize the procedure to answer the first question we ask. When a specific $G$-extension $\CC$ of the fusion category $\CC_1$ is given, one can derive all the rest of the fusion categories related to $\CC$ by changing the fractionalization classes and discrete torsions via the following procedure:
\begin{enumerate}
    \item Compute the group of Abelian anyons $\mathbf{A}$ in the SymTFT of $\CC_1$ following Section \ref{sec:DC};
    \item Compute the action of bulk $G$-symmetry action $\rho$ on the Abelian anyons following Section \ref{sec:sym_action}, and then compute the cohomology group  $H^2_\rho(G,\mathbf{A})$ whose elements labels the choices of symmetry fractionalization;
    \item Fix a choice of $[\mathfrak{t}] \in H^2_\rho(G,\mathbf{A})$, compute $\widetilde{F}$ using equation $(\ref{eq:tilde_F})$, and extract $\CO_4$ from the violation of the pentagon equations;
    \item Check if there exists $\omega \in C^3(G,U(1))$ such that $\CO_4 = d\omega$, and the transformed $F$-symbols are given by $\omega \widetilde{F}$.
\end{enumerate}

\subsubsection{Coupling SymSET to physical boundary $\mathcal{B}_{phys}$ and a map of gapped phases}
Previously, we focused on the symmetry boundary of the SymSET. Here, we briefly consider placing the SymSET on an interval with one physical boundary $\scriptB_{phys}$ and one symmetry boundary $\scriptB_{sym}$, and we will briefly discuss how some basic notions in the SymTFT generalizes to the case of SymSET.

To begin, let's briefly discuss the SymTFT subsector of the SymSET. For now, let us assume the 2d physical theory after interval compactification realizes the symmetry faithfully. In this case, the anyons in the SymTFT will terminate on local operators on $\scriptB_{phys}$:
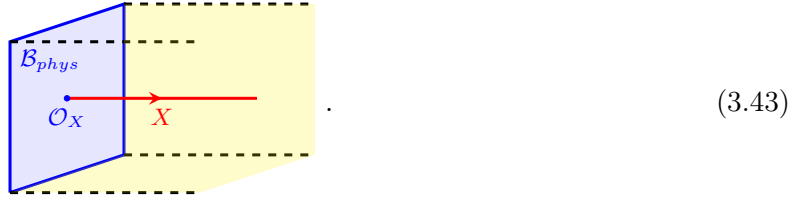
\begin{equation}
    \begin{tikzpicture}[baseline={([yshift=-1ex]current bounding box.center)},vertex/.style={anchor=base,
    circle,fill=black!25,minimum size=18pt,inner sep=2pt},scale=0.5]
    \draw[blue, line width = 0.4mm] (-5,0) -- (-5,4) -- (-2,5) -- (-2,1) -- (-5,0);
    \draw[fill = blue, opacity = 0.1] (-5,0) -- (-5,4) -- (-2,5) -- (-2,1) -- (-5,0);
    
    \draw[black, line width = 0.4mm, dashed] (-5,0) -- (0,0);
    \draw[black, line width = 0.4mm, dashed] (-5,4) -- (0,4);
    \draw[black, line width = 0.4mm, dashed] (-2,5) -- (3,5);
    \draw[black, line width = 0.4mm, dashed] (-2,1) -- (3,1);
    \fill[yellow, opacity = 0.2] (-5,0) -- (0,0) -- (3,1) -- (-2,1) -- (-5,0);
    \fill[yellow, opacity = 0.2] (-2,5) -- (3,5) -- (3,1) -- (-2,1) -- (-2,5);

    \draw[red, line width = 0.4mm, ->-=0.5] (-3.5,2.5) -- (1.5,2.5);
    \filldraw[blue] (-3.5,2.5) circle (2pt);
    
    \node[blue, right] at (-5,3.5) {\footnotesize $\scriptB_{phys}$};
    \node[red, below] at (-1,2.5) {\footnotesize $X$};
    \node[blue, below] at (-3.5,2.5) {\footnotesize $\mathcal{O}_X$};
\end{tikzpicture} ~.
\end{equation}
Notice that all the local operators $\mathcal{O}_X$'s terminating on the same anyon $X$ transform under the same irreducible representation of the non-invertible symmetries. To get a genuine 2d QFT, one would pair the physical boundary $\scriptB_{phys}$ with a symmetry boundary $\scriptB_{sym}$. If the anyon $X$ is condensed on $\scriptB_{sym}$, that is, there exists a local junction between the anyon $X$ and the identity line $\dsi$ on $\scriptB_{sym}$ (or equivalently, $X$ can terminate on $\scriptB_{sym}$), after the interval compactification, $\scriptO_X$ corresponds to a local operator in the 2d theory; while if $X$ can only terminate on some non-trivial TDL $x$ on $\scriptB_{sym}$, then after reduction along the interval, $\scriptO_X$ becomes a non-local operator ending on the TDL $x$:
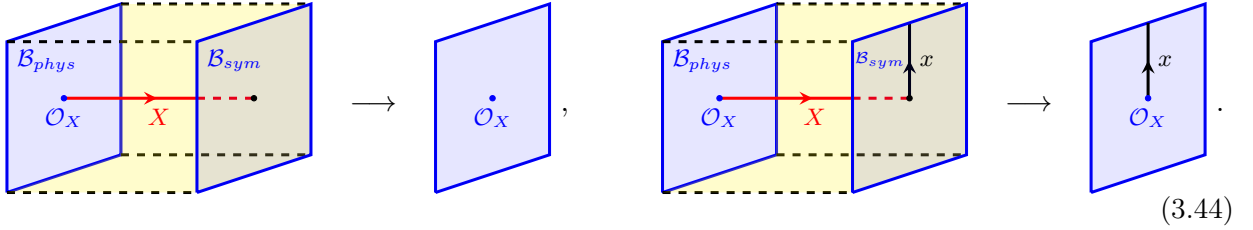
\begin{equation}
 ~.
\end{equation}

Let us now consider the $0$-form symmetry operator $g \in G$ in the SymSET. To proceed, we will assume that the $\scriptB_{phys}$ when pairing with $\scriptB_{sym}$ leads to a 2d theory which admits the $G$-extended symmetry $\scriptC$. We already discussed that $g$ terminates on some $a_g \in \CC_g$ TDL on $\scriptB_{sym}$; we propose that $g$ terminates on $\scriptB_{phys}$ as an invertible symmetry operator $U_g$. Namely, moving the local operator $\scriptO_X$ together with the bulk anyon $X$ across the symmetry defect $g$, $X$ will be transformed into ${}^g X$ and the local operator $\scriptO_X$ will be transformed into another operator $U_g(\scriptO_X)$ with the same physical property as $\scriptO_X$, as shown below
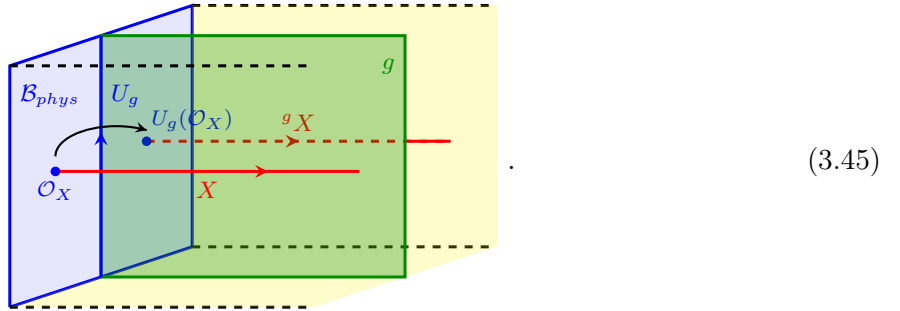
\begin{equation}
    \begin{tikzpicture}[baseline={([yshift=-1ex]current bounding box.center)},vertex/.style={anchor=base,
    circle,fill=black!25,minimum size=18pt,inner sep=2pt},scale=0.8]
    \draw[blue, line width = 0.4mm] (-5,0) -- (-5,4) -- (-2,5) -- (-2,1) -- (-5,0);
    \draw[fill = blue, opacity = 0.1] (-5,0) -- (-5,4) -- (-2,5) -- (-2,1) -- (-5,0);
    
    \draw[black, line width = 0.4mm, dashed] (-5,0) -- (0,0);
    \draw[black, line width = 0.4mm, dashed] (-2,5) -- (3,5);
    \draw[black, line width = 0.4mm, dashed] (-2,1) -- (3,1);
    \fill[yellow, opacity = 0.2] (-5,0) -- (0,0) -- (3,1) -- (-2,1) -- (-5,0);
    \fill[yellow, opacity = 0.2] (-2,5) -- (3,5) -- (3,1) -- (-2,1) -- (-2,5);

    \draw[red, line width = 0.4mm, ->-=0.5, dashed] (-2.75,2.75) -- (2.25,2.75);
    \draw[red, line width = 0.4mm] (1.5,2.75) -- (2.25,2.75);

    \node[red, above] at (-0.25,2.75) {\footnotesize ${}^g X$};
    \filldraw[blue]  (-2.75,2.75) circle (2pt);
    \node[blue, above] at (-2,2.75) {\scriptsize $U_g(\mathcal{O}_X)$};

    \fill[dgreen, opacity = 0.3] (-3.5,0.5) -- (-3.5,4.5) -- (1.5,4.5) -- (1.5,0.5) -- (-3.5,0.5);
    \draw[dgreen, line width = 0.4mm] (-3.5,0.5) -- (-3.5,4.5) -- (1.5,4.5) -- (1.5,0.5) -- (-3.5,0.5);

    \draw[black, line width = 0.4mm, dashed] (-5,4) -- (0,4);
    \draw[blue, line width = 0.4mm, ->-=0.6] (-3.5,0.5) -- (-3.5,4.5);

    \draw[red, line width = 0.4mm, ->-=0.7] (-4.25,2.25) -- (0.75,2.25);
    \filldraw[blue] (-4.25,2.25) circle (2pt);

    \draw[black, thick, ->-=1] (-4.25,2.5) arc(180:60:1 and 0.5);
    
    \node[blue, right] at (-5,3.5) {\footnotesize $\scriptB_{phys}$};
    \node[red, below] at (-1.75,2.25) {\footnotesize $X$};
    \node[blue, below] at (-4.25,2.25) {\scriptsize $\mathcal{O}_X$};
    \node[blue, right] at (-3.5,3.5) {\footnotesize $U_g$};
    \node[dgreen, left] at (1.5,4) {\footnotesize $g$};
\end{tikzpicture} ~.
\end{equation}
Notice that this picture is valid only when $\scriptB_{phys}$ satisfies the previous assumption. In the most general case where $\scriptB_{phys}$ is not invariant under the $G$-action, then $g \in G$ will map $\scriptB_{phys}$ to an inequivalent physical boundary ${}^g \scriptB_{phys}$, and $U_g$ should be considered as an interface between two physical boundary conditions. 

Under the dimensional reduction, the $g$-defect stretching between $\scriptB_{phys}$ and $\scriptB_{sym}$ becomes the TDL $a_g$ in the non-trivial grading component:
\begin{equation}\label{eq:g_action_X_OX}
 ~.
\end{equation}
Similarly, dimensionally reducing \eqref{eq:g_action_X_OX} implies the following action on the operators in the 2d theory:
\begin{equation}
    %
 ~,
\end{equation}
where we use $x$ and ${}^g x$ to denote the boundary TDLs that $X$ and ${}^g X$ can terminate on respectively. We see that when reducing to the 2d theory, $a_g$ may act non-invertibly on local operators if, for instance, $X$ is condensed on $\scriptB_{sym}$ ($x = \dsi$) such that $\scriptO_X$ is a local operator and ${}^g X$ is not condensed on $\scriptB_{sym}$ such that $U_g(\scriptO_X)$ is a non-local operator. 

Finally, we notice that the twist defect can also terminate on local operators $\scriptO_{X_g}$ on $\scriptB_{phys}$, and after reducing to 2d, it becomes the non-local operator terminates on the TDL $x_g$:
\begin{equation}
 ~.
\end{equation}

With the above setup, we immediately see that changing the symmetry fractionalization class is not a topological manipulation. This is because in the heuristic picture \eqref{eq:frac_bb_modif}, we need to insert an Abelian anyon $\mathfrak{t}_{g,h}$ into the junction of $g \times h \rightarrow gh$:
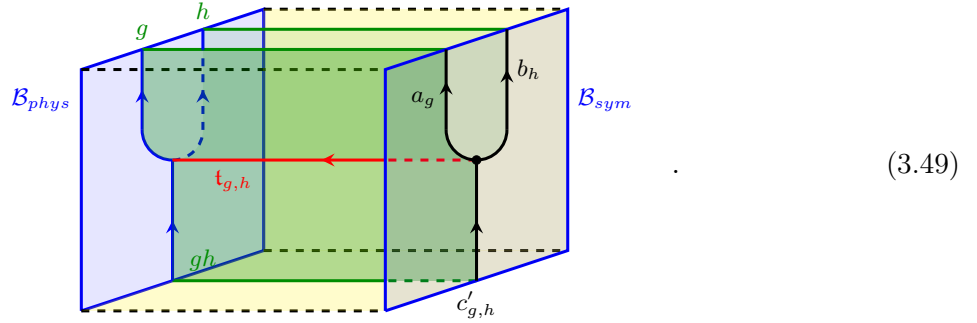
\begin{equation}
%
 \quad .
\end{equation}
With our assumption of $\scriptB_{phys}$ realizing a 2d QFT with a unique ground state and $\scriptC$ faithfully realized, only the trivial anyon can end topologically on $\scriptB_{phys}$ (where it ends on the identity local operator). After reducing, this amounts to modifying the fusion category by inserting a non-topologically (non)-local operators on the corresponding local fusion junction $a_g \times b_h \rightarrow c_{gh}$, which is obviously not a topological manipulation. This explains why we only find the $\underline{\CE}_{\doubleZ_2}\Rep H_8$'s relating to each other by changing the bulk symmetry at the $\operatorname{Ising}^2$ point of the $c = 1$ compact boson in \cite{Choi:2023vgk}.

Before concluding this section, we want to point out an intriguing construction. Notice that we can consider the case where $\scriptB_{phys}$ describes a gapped phase of $\scriptB_{sym}$. Then, the anyon $\mathfrak{t}_{g,h}$ becomes a topological line operator through a topological junction on $\scriptB_{phys}$. Therefore, in this setup, changing the symmetry fractionalization class is a topological manipulation. Furthermore, if there is no obstruction in gauging the $G$ $0$-form symmetry in this bulk-boundary system, then gauging $G$ and performing the dimensional reduction would lead to a gapped phase of the categorical symmetry acquired from changing fractionalization class. This would allow one to define a map of gapped phases between different $G$-extensions related by changing fractionalization classes. 

As a quick example, one can consider the SymSET describes the $\mathbb{Z}_2 \times \mathbb{Z}_2$ symmetry extended from $\doubleZ_2$. As mentioned in Appendix \ref{app:Z2_example}, the SymSET is acquired by enriching the $\doubleZ_2$-gauge theory with a trivially acting $\mathbb{Z}_2$ symmetry. Let's consider the SPT phase of $\mathbb{Z}_2\times \mathbb{Z}_2$-symmetry, and in the interval setup, on $\scriptB_{sym}$ the electric line $e$ is condensed while on $\scriptB_{phys}$ the magnetic line $m$ is condensed. Furthermore, the bulk $0$-form symmetry operator $b$ can terminate on both $\scriptB_{sym}$ and $\scriptB_{phys}$ as a trivially acting $\doubleZ_2$ line operator which we also denote as $b$. Let's now change the fractionalization class by inserting the electric line $e$ to the junction of the $0$-form symmetry (without stacking an $\mathbb{Z}_2^b$-SPT phase which would change the discrete torsion). Since $e$ is not condensed on $\scriptB_{phys}$, the fusion rule of $b \times b = 1$ is modified to $b \times b = e$. As a result, gauging $\mathbb{Z}_2^b$ symmetry in the entire system would lead to a partial SSB phase. This is because, to maintain the symmetry structure on $\scriptB_{sym}$ the $b$ lines can not be condensed on $\scriptB_{sym}$; and we will choose to condense $b$-lines on $\scriptB_{phys}$ instead of on $\scriptB_{sym}$. However, because of the fusion rule $b$, one can not condense the $b$-lines on $\scriptB_{phys}$ without also condensing the $e$-lines on $\scriptB_{phys}$ as well. This then implies that $e$-lines are condensed on both $\scriptB_{phys}$ and $\scriptB_{sym}$, which indicates ground state degeneracy. Notice that this is consistent with the fact that fractionalizing on $e$ introduces a mixed anomaly between two $\mathbb{Z}_2$ symmetries, which obstructs the existence of SPT phases. As another example, one can consider fractionalizing on the magnetic line instead, and since $m$ is now condensed on $\scriptB_{phys}$, the fusion rule of $b$ is no longer modified therefore we expect to get a trivially gapped phase after gauging $\mathbb{Z}_2^b$. This is also consistent with the fact that fractionalizing on $m$ will change the symmetry from $\mathbb{Z}_2\times \mathbb{Z}_2$ to $\mathbb{Z}_4$ without introducing anomalies, which admits a SPT phase. Describing this map completely requires one to understand all the possible choices of the boundary conditions of the dynamical $G$-gauge field on $\scriptB_{phys}$ as well as how the bulk discrete torsion interplays with $\scriptB_{phys}$, and we leave it for future studies.


\section{Example $0$: Self-duality under gauging $\VEC_{A}$}\label{sec:TY_RC}
As a trivial example, let's consider the self-duality under gauging $\VEC_{A}$ where $A$ is a finite Abelian group. The corresponding fusion category capturing this duality is of course the $\TY(A,\chi,\epsilon)$ fusion categories. We want to demonstrate in this simplest setup how we could reconstruct the SET which enriches the $A$-gauge theory with a $\mathbb{Z}_2^{em}$ symmetries specified by the bicharacter $\chi$ and derive the action of the bulk $\mathbb{Z}_2^{em,\chi}$.

\subsection{Spectrum of anyons and twist defects}
The anyons in the SymSET are simply the Abelian anyons in the $A$-gauge theory discussed in Section \ref{sec:VecADC}. Let's then compute the twist defects, which are given by a pair $(\sigma, R^{a,\sigma}_{\CN})$ where $\sigma$ is a (potentially non-simple) object built out of the TDL $\CN$ from non-trivial grading component. We will start with $\sigma = \scriptN$ and see if the twist defects we get will saturate the quantum dimension.

For possible twist defects of the form $\sigma = (\scriptN, R^{c,\sigma}_{\scriptN})$, the $R$-symbol characters the following diagram
\begin{equation}
 ~,
\end{equation}
which are solutions to the following hexagon equations
\begin{equation}
    R^{a,\sigma}_\CN R^{b,\sigma}_{\CN} F^{a,\CN,b}_{\CN}  = F^{\CN,a,b}_{\CN} R^{ab,\sigma}_\CN F^{a,b,\CN}_{\CN} ~,
\end{equation}
which, using the $F$-symbols of $\TY$, becomes
\begin{equation}
    R^{a,\sigma}_\CN R^{b,\sigma}_{\CN} \chi(a,b) = R^{ab,\sigma}_{\CN} ~.
\end{equation}
Since $\chi(a,b)$ is cohomologically trivial as $Z^2(A,U(1))$, we conclude that there are $|A|$ solutions to the above equations. And we denote the corresponding anyons as $\sigma_{\rho}$ where $\rho:A \rightarrow U(1)$ satisfying $\rho(a) \rho(b) \chi(a,b) = \rho(ab)$. Since each twist defect has quantum dimension $\sqrt{|A|}$, the total quantum dimensions of all twist defects of the form $\sigma_{\rho}$ are $|A|^2$ which already saturate the total quantum dimension. Thus, there are no other twist defects besides $\sigma_{\rho}$. Then orientation reversal of $\sigma_\rho$ is denoted as $\sigma_{\overline{\rho}}$ where
\begin{equation}
    \overline{\rho}(a) := \rho(a^{-1}) ~,
\end{equation}
where $\chi(a,b) = \chi(a^{-1},b^{-1})$ implies $\overline{\rho}(a)\overline{\rho}(b) \chi(a,b) = \overline{\rho}(ab)$. Notice that generically $\overline{\rho}$ is not the complex conjugate of $\rho$.

It is also straightforward to work out the fusion rules in the SET using the fact that the half-braiding phases multiplies when fusing two lines (anyons or twist defects) in the bulk:
\begin{equation}
\begin{aligned}
    & \sigma_{\rho} \times \phi_{(a,\widehat{b})} = \phi_{(a,\widehat{b})} \times \sigma_{\rho} = \sigma_{\rho \widehat{b}} ~, \\
    & \sigma_{\rho_1} \times \sigma_{\rho_2} = \sum_{a \in A} \phi_{(a,\widehat{b})} ~, \quad where \quad \widehat{b} = \frac{\rho_2}{\overline{\rho}_1} = \frac{\rho_1}{\overline{\rho}_2} ~.
\end{aligned}
\end{equation}

\subsection{Action of the $\doubleZ_2^{em}$-symmetry in the bulk}
Let us first compute the $\doubleZ_2^{em}$ invariant anyons. For an Abelian anyon $\phi_{(a,\widehat{b})}$, for it to be invariant under $\doubleZ_2$, it must admit the following non-zero $R$-symbol
\begin{equation}
 ~,
\end{equation}
satisfies the following equations from Figure \ref{fig:hex_g_action}
\begin{equation}
\begin{aligned}
    R^{c,\phi_{(a,\widehat{b})}}_{ca} F^{c,a,\CN}_{\CN} R^{\CN,\phi_{(a,\widehat{b})}}_{\CN} &= F^{a,c,\CN}_{\CN} R_{\CN}^{\CN,\phi_{(a,\widehat{b})}} F^{c,\CN,a}_{\CN} ~, \\
    R^{\CN,\phi_{(a,\widehat{b})}}_{\CN} F^{\CN,a,c}_{\CN} R^{c,\phi_{(a,\widehat{b})}}_{ca} &= F^{a,\CN,c}_{\CN} R^{\CN,\phi_{(a,\widehat{b})}}_{\CN} F^{c,a,\CN}_{\CN} ~,
\end{aligned}
\end{equation}
where $c \in A$ and $R^{c,\phi_{(a,\widehat{b})}}_{ca} = \widehat{b}(c)$. Using the $F$-symbols of the $\TY$ fusion category, we find
\begin{equation}
    \widehat{b}(c) R^{\CN,\phi_{(a,\widehat{b})}}_{\CN} = \chi(c,a) R^{\CN,\phi_{(a,\widehat{b})}}_{\CN} ~, \quad R^{\CN,\phi_{(a,\widehat{b})}} \widehat{b}(c) = \chi(a,c) R^{\CN,\phi_{(a,\widehat{b})}}_{\CN} ~,
\end{equation}
which doesn't have a non-trivial solution unless $\widehat{b}(c) = \chi(a,c)$ for all $c\in A$ (in other words, $\widehat{b} = \chi(a,\cdot) \equiv \chi(a)$). This implies that only the Abelian anyon of the form $\phi_{(a,\chi(a))}$ is $\doubleZ_{2}^{em}$-invariant.

\

Then, let's compute the non-trivial orbits of $\doubleZ_2^{em}$ which consists of two anyons $\phi_{(a_1,\widehat{b}_1)}$ and $\phi_{(a_2,\widehat{b}_2)}$ related by $\doubleZ_2^{em}$ symmetry. Then, the crossed braidings are encoded by two $R$-symbols:
\begin{equation}
 ~,
\end{equation}
satisfying \footnote{Notice that there will be another two equations one can get from swapping $\phi_{(a_1,\widehat{b}_1)}$ with $\phi_{(a_2,\widehat{b}_2)}$, but the two equations we have here are already sufficient to determine the $\doubleZ_2^{em}$-action.}
\begin{equation}
\begin{aligned}
    R^{c,\phi_{(a_2,\widehat{b}_2)}}_{ca_2} F^{c,a_2,\CN}_{\CN} R^{\CN,\phi_{(a_1,\widehat{b}_1)}}_{\CN} &= F^{a_2,c,\CN}_{\CN} R_{\CN}^{\CN,\phi_{(a_1,\widehat{b}_1)}} F^{c,\CN,a_1}_{\CN} ~, \\
    R^{\CN,\phi_{(a_1,\widehat{b}_1)}}_{\CN} F^{\CN,a_1,c}_{\CN} R^{c,\phi_{(a_1,\widehat{b}_1)}}_{ca_1} &= F^{a_2,\CN,c}_{\CN} R^{\CN,\phi_{(a_1,\widehat{b}_1)}}_{\CN} F^{c,a_1,\CN}_{\CN} ~, \\
\end{aligned}
\end{equation}
which, after plugging in the $F$-symbols, become
\begin{equation}
\begin{aligned}
    & \widehat{b}_2(c) R^{\CN,\phi_{(a_1,\widehat{b}_1)}}_{\CN} = R^{\CN,\phi_{(a_1,\widehat{b}_1)}}_{\CN} \chi(c,a_1) ~, \quad R^{\CN,\phi_{(a_1,\widehat{b}_1)}}_{\CN} \widehat{b}_1(c) = \chi(a_2,c) R^{\CN,\phi_{(a_1,\widehat{b}_1)}}_{\CN} ~.
\end{aligned}
\end{equation}
The above equations of the two $R$-symbols have non-zero solutions if and only if
\begin{equation}
    \widehat{b}_1(c) = \chi(a_2,c) ~, \quad \widehat{b}_2(c) = \chi(a_1,c) ~.
\end{equation}
Then, we conclude that the $\doubleZ_2^{em}$ symmetry acts on the Abelian anyons as
\begin{equation}
    \doubleZ_2^{em}: \phi_{(a,\chi(b))} \longleftrightarrow \phi_{b,\chi(a))} ~,
\end{equation}
and we see explicitly that the symmetric non-degenerate bicharacter $\chi$ specifies a specific bulk $\doubleZ_2^{em}$ symmetries. 

\,

As another example, let's compute the action of $\doubleZ_2^{em}$ on the twist defects on $\sigma_\rho$. Let us first search for $\doubleZ_2^{em}$-invariant twist defects, which admits a non-zero $R$-symbol
\begin{equation}
 ~,
\end{equation}
satisfying the following hexagon equations
\begin{equation}
\begin{aligned}
    R^{a,\sigma_\rho}_{\CN} F^{a\CN\CN}_b R^{\CN,\sigma_\rho}_{a^{-1}b} &= F^{\CN a\CN}_b R^{\CN,\sigma_\rho}_{b} F^{a\CN\CN}_b ~, \\
    R^{\CN,\sigma_\rho}_{ba^{-1}} F^{\CN\CN a}_b R^{a,\sigma_\rho}_{\CN} &= F^{\CN\CN a}_b R^{\CN,\sigma_\rho}_b F^{\CN a \CN}_b ~,
\end{aligned}
\end{equation}
which are solved by
\begin{equation}
    R^{\CN,\sigma_\rho}_{a} = \frac{1}{\rho(a^{-1})} R^{\CN,\sigma_{\rho}}_{\dsi} ~,
\end{equation}
where $R^{\CN,\sigma_{\rho}}_{\dsi}$ is a undetermined coefficient. Hence, we conclude that the $\doubleZ_2^{em}$ symmetry in the SET acts trivially on all twist defects.

\

Since the action of $\doubleZ_2^{em}$ is known, one can compute the possible choices of the symmetry fractionalization classes. It then follows from Shapiro's lemma that there is a unique choice of the symmetry fractionalization class in the SET. Furthermore, stacking $\doubleZ_2^{em}$-SPT in the bulk will change the choice of the FS indicator on the boundary.

Before concluding this section, we want to quickly point out the connection between the relative center and the full Drinfeld center. Essentially there will be a few additional hexagon equations one needs to solve for the bulk anyon. For the $\doubleZ_2^{em}$-invariant anyons such as $\phi_{(a,\chi(a))}$ and $\sigma_\rho$, the additional equations will fix their $R$-symbols with the duality defect $\CN$ up to a sign. This sign from the bulk point of view determines the $\doubleZ_2^{em}$ flux one could attach; and the pure flux is simply the dual $1$-form symmetry from the$\doubleZ_2^{em}$ gauging. For the $\doubleZ_2^{em}$ non-invariant anyons, after gauging only the orbit survives which is given by the composite $\phi_{(a,\chi(b))} + \phi_{(b,\chi(a))}$, which explains the form of the $R$-symbols \eqref{eq:Y_type_R_symbol} for $Y$-type objects in the full Drinfeld center.  

\section{Example I: Self-duality under maximal gauging in $\Rep H_8$}\label{sec:max_RepH8}
In this section, we revisit the self-duality under the maximal gauging in $\Rep H_8$ fusion categories. The main point of this section is to derive the bulk interpretation of the classification given in \cite{Choi:2023vgk}, and demonstrate explicitly that among the $8$ inequivalent fusion categories, there are only two choices of the bulk $\mathbb{Z}_2^{em}$ symmetry and but there are two choices of symmetry fractionalization classes (and the last factor of $2$ is from two choices of the FS indicator). 

We begin by listing the anyons contents of the SymTFT of $\Rep H_8$ in Section \ref{sec:RepH8_SymTFT}. Then, we compute the bulk symmetry action in Section \ref{sec:RepH8_bulk_symmetry} and show  that there are only two inequivalent choices of the bulk $\mathbb{Z}_2^{em}$ symmetry. Finally, in Section \ref{sec:RepH8_cf_revisited} we show explicitly that the $(i,+,\pm)$ categories\footnote{Here, we change the notation such that $i = 1,2$ corresponds to the first $\pm$ in the three labels specifying the fusion categories in \cite{Choi:2023vgk}.} are indeed related to $(i,-,\pm)$ categories via the change of symmetry fractionalization classes in the bulk following the general procedure in Section \ref{sec:transformed_F_symbols}.

\subsection{SymTFT of $\Rep H_8$}\label{sec:RepH8_SymTFT}
Let us summarize the anyons in the SymTFT of $\Rep H_8$ symmetry. There are $3$ types of anyons which we call $X$-type, $Y$-type, and $Z$-type respectively.

The spectrum of the SymTFT of the $\TY$ fusion categories has been worked out in Section \ref{sec:TYDC}. In this case, we find $8$ Abelian anyons $X_{(g,s)}$, $6$ $Y$-type anyons $Y_{\{g,h\}}$ with quantum dimensions $2$, and $8$ $Z$-type anyons with quantum dimensions $2$. 

In the following, we denote the $X$-type anyons as $X_{(g,s)}$ where $g = \dsi,a,b,ab$ and the $R$-symbols are given by
\begin{equation}
    R^{h,X_{(g,s)}}_{hg} = \chi(h,g) ~, \quad R^{\CN,X_{(g,s)}}_{\CN} = s ~.
\end{equation}
And the $8$ $X$-type anyons are given by
\begin{equation}
\begin{array}{rll}
    X_{(\dsi,\pm 1)}: & R^{g,X_{(\dsi,\pm 1)}}_{g} = (1,+1,+1,+1) ~, & R^{\CN,X_{(\dsi,\pm 1)}}_{\CN} = \pm 1 ~, \\
    X_{(a,\pm i)}: & R^{g,X_{(a,\pm i)}}_{ga} = (1,-1,+1,-1) ~, & R^{\CN,X_{(a,\pm i)}}_{\CN} = \pm i ~, \\
    X_{(b,\pm i)}: & R^{g,X_{(b,\pm i)}}_{gb} = (1,+1,-1,-1) ~, & R^{\CN,X_{(b,\pm i)}}_{\CN} = \pm i ~, \\
    X_{(ab,\pm 1)}: & R^{g,X_{(ab,\pm 1)}}_{gab} = (1,-1,-1,+1) ~, & R^{\CN,X_{(ab,\pm 1)}}_{\CN} = \pm 1 ~. \\
\end{array}
\end{equation}
The $6$ $Y$-type anyons $Y_{\{g,h\}}$ together with their $R$-symbols are given by
\begin{equation}
\begin{array}{rlll}
    Y_{\{\dsi,a\}}: & R^{g,Y_{\{\dsi,a\}}}_{g} = (1,-1,+1,-1) ~, & R^{g,Y_{\{\dsi,a\}}}_{ga} = (1,+1,+1,+1) ~, & [R^{\CN,Y_{\{\dsi,a\}}}_{\CN}]_{\mu\nu} = (\sigma_x)_{\mu\nu} ~, \\
    Y_{\{\dsi,b\}}: & R^{g,Y_{\{\dsi,b\}}}_{g} = (1,+1,-1,-1) ~, & R^{g,Y_{\{\dsi,b\}}}_{gb} = (1,+1,+1,+1) ~, & [R^{\CN,Y_{\{\dsi,b\}}}_{\CN}]_{\mu\nu} = (\sigma_x)_{\mu\nu} ~, \\
    Y_{\{\dsi,ab\}}: & R^{g,Y_{\{\dsi,ab\}}}_{g} = (1,-1,-1,+1) ~, & R^{g,Y_{\{\dsi,ab\}}}_{gab} = (1,+1,+1,+1) ~, & [R^{\CN,Y_{\{\dsi,ab\}}}_{\CN}]_{\mu\nu} = (\sigma_x)_{\mu\nu} ~, \\
    Y_{\{a,b\}}: & R^{g,Y_{\{a,b\}}}_{ga} = (1,+1,-1,-1) ~, & R^{g,Y_{\{a,b\}}}_{gb} = (1,-1,+1,-1) ~, & [R^{\CN,Y_{\{a,b\}}}_{\CN}]_{\mu\nu} = (\sigma_x)_{\mu\nu} ~, \\
    Y_{\{a,ab\}}: & R^{g,Y_{\{a,ab\}}}_{ga} = (1,-1,-1,+1) ~, & R^{g,Y_{\{a,ab\}}}_{gab} = (1,-1,+1,-1) ~, & [R^{\CN,Y_{\{a,ab\}}}_{\CN}]_{\mu\nu} = i (\sigma_x)_{\mu\nu} ~, \\
    Y_{\{b,ab\}}: & R^{g,Y_{\{b,ab\}}}_{gb} = (1,-1,-1,+1) ~, & R^{g,Y_{\{b,ab\}}}_{gab} = (1,+1,-1,-1) ~, & [R^{\CN,Y_{\{b,ab\}}}_{\CN}]_{\mu\nu} = i (\sigma_x)_{\mu\nu} ~. \\
\end{array}
\end{equation}
We denote the $8$ type anyons $Z_{(i,\epsilon)}$ where $i=1,2,3,4$ and $\epsilon = \pm 1$, and the corresponding $R$-symbols are given by
\begin{equation}
\begin{aligned}
    Z_{(1,\pm 1)}&: R^{g,Z_{(1,\pm 1)}}_{\CN} = (1,+i,+i,-1) ~, \quad R^{\CN,Z_{(1,\pm 1)}}_{g} = \pm \sqrt{+i}(1,-i,-i,-1) ~, \\
    Z_{(2,\pm 1)} &: R^{g,Z_{(2,\pm 1)}}_{\CN} = (1,-i,+i,+1) ~, \quad R^{\CN,Z_{(2,\pm 1)}}_{g} = \pm (1,+i,-i,+1) ~, \\
    Z_{(3,\pm 1)} &: R^{g,Z_{(3,\pm 1)}}_{\CN} = (1,+i,-i,+1) ~, \quad R^{\CN,Z_{(3,\pm 1)}}_{g} = \pm (1,-i,+i,+1) ~, \\
    Z_{(4,\pm 1)} &: R^{g,Z_{(4,\pm 1)}}_{\CN} = (1,-i,-i,-1) ~, \quad R^{\CN,Z_{(4,\pm 1)}}_{g} = \pm \sqrt{-i}(1,+i,+i,-1) ~.
\end{aligned}
\end{equation}
Before proceed, notice that $\Rep H_8$ is group-theoretical and can be realized via starting with $D_8 = \langle r,s| r^4 = s^2 = 1 ~, srs = r^{-1} \rangle$ symmetry with an anomaly $[\gamma] \in H^3(D_8,U(1))$
\begin{equation}\label{eq:3cocyclegamma}
    \gamma(r^{i_1}s^{j_1},r^{i_2}s^{j_2},r^{i_3}s^{j_3}) = \exp(\frac{4\pi i}{4^2}(-1)^{j_1} i_1(i_2 + (-1)^{j_2} i_3 - [i_2 + (-1)^{j_2}i_3]_4)) 
\end{equation}
and gauge the $\mathbb{Z}_2 = \langle s\rangle$ subgroup. Thus, the SymTFT $\CZ(\Rep H_8)$ can also be described by $D_8$ gauge theory with the twist $\gamma$. Recall that a generic anyon in a $G$-gauge theory with twist $\omega$ is labeled by $([g],\pi_g)$, where $[g]$ is a conjugacy class of $G$ and $\pi_g$ is an irrep of the centralizer subgroup $C_g := \{h\in G:hg=gh\}$ twisted by some $2$-cocycle $\beta_g$ depending on $\omega$. In Appendix \ref{app:RepH8_VECD8gamma_match}, we give an explicit map between the anyon spectrum in the SymTFT computed in these two different ways. For the rest of the paper, however, we will only make use of the map between Abelian anyons. In the $\CZ(\VEC_{D_8}^\omega)$ calculation, the $8$ Abelian anyons are $(\dsi, \chi_i)$ and $(r^2, \chi_i)$ where $i = 1,2,3,4$ are four $1$-dimension irreps of $D_8$ given in \eqref{eq:D8CTab}, and the mapping between two calculations are given by
\begin{equation}\label{eq:AA_RepH8_match}
\begin{array}{|c|c|c|c|c|c|c|c|}
\hline
 (\dsi,\chi_1) & (r^2,\chi_3) & (r^2,\chi_4) &  (\dsi,\chi_4) &  (\dsi,\chi_3) &  (r^2,\chi_1) &  (r^2,\chi_2) & (\dsi,\chi_2) \\
\hline
 X_{(\dsi,+1)} & X_{(a,i)} & X_{(b,i)} & X_{(ab,1)} & X_{(\dsi,-1)} & X_{(a,-i)} & X_{(b,-i)} & X_{(ab,-1)} \\\hline
\end{array} ~.
\end{equation}

\subsection{$\doubleZ_2^{em}$ symmetry in the bulk}\label{sec:RepH8_bulk_symmetry}
In this subsection, we derive the corresponding  $\doubleZ_2^{em}$-symmetries for $\underline{\CE}_{\doubleZ_2}^{(i,\kappa_D,\epsilon_D)}\Rep H_8$. In summary, we find that there are only two order-$2$ anyon-permutation actions, which we denote as $\eta_i$. One might wonder if it is possible that there are additional bulk $\doubleZ_2$ symmetry, distinguished by different actions on the fusion junctions of the anyons. However, $\CZ(\Rep H_8)$ is a multiplicity-free MTC, and as explicitly shown in \cite{Benini:2018reh}, for a multiplicity-free MTC, its symmetry is completely determined by its permutation action on the anyons.

Before proceeding, in order to match the classification data computed in \cite{Choi:2023vgk}, we will take the $F$-symbols defined in terms of fusion junctions computed there and rename them as $G$-symbols as in Section \ref{sec:convention}, and compute the $F$-symbols defined in terms the splitting junctions using \eqref{eq:GF_relation}.

\subsubsection{$\doubleZ_2^{em}$ action on the Abelian anyons}
Let us first compute the action on the Abelian anyons $X_{(g,s)}$. The two Abelian anyons $X_{(g_1,s_1)}$ and $X_{(g_2,s_2)}$ exchanged by the bulk $\doubleZ_2^{em}$-symmetry admit crossed braiding
\begin{equation}
 ~,
\end{equation}
satisfying the following hexagon equations
\begin{equation}
\begin{aligned}
    R^{\CD,X_{(g_1,s_1)}}_{\CD} F^{\CD g_1 h}_{\CD} R^{h,X_{(g_1,s_1)}}_{hg_1} &= F^{g_2\CD h}_{\CD} R^{\CD,X_{(g_1,s_1)}}_{\CD} F^{\CD h g_1}_{\CD} ~, \\
    R^{h,X_{(g_2,s_2)}}_{hg_2} [F^{hg_2\CD}_{\CD}] R^{\CD,X_{(g_1,s_1)}}_{\CD} &= F^{g_2 h\CD}_{\CD} R^{\CD,X_{(g_1,s_1)}}_{\CD} F^{h\CD g_1}_{\CD} ~, \\
    R^{\CD,X_{(g_1,s_1)}}_{\CD} [F^{\CD g_1\CN}_{\CD}]_{\mu\nu} R^{\CN,X_{(g_1,s_1)}}_{\CN} &= \sum_{\rho}[F^{g_2\CD\CN}_{\CD}]_{\mu\rho} R^{\CD,X_{(g_1,s_1)}}_{\CD} [F^{\CD\CN g_1}_{\CD}]_{\rho\nu} ~, \\
    R^{\CN,X_{(g_2,s_2)}}_{\CN} [F^{\CN g_2 \CD}_{\CD}]_{\mu\nu} R^{\CD,X_{(g_1,s_1)}}_{\CD} &= \sum_{\rho} [F^{g_2\CN\CD}_{\CD}]_{\mu\rho} R^{\CD,X_{(g_1,s_1)}}_{\CD} [F^{\CN\CD g_1}_{\CD}]_{\rho\nu} ~.
\end{aligned}
\end{equation}
Notice that the special case of the above equations is $X_{(g_1,s_1)} = X_{(g_2,s_2)} \equiv X_{(g,s)}$. In this case, $\mathbb{Z}_2^{em}$ acts trivially on the $X_{(g,s)}$.

We find two different $\doubleZ_2^{em}$-actions (which only depends on $i = \pm$ in the parametrization $(i,\kappa_D,\epsilon_D)$), and the non-trivial actions are given by
\begin{equation}
\begin{aligned}
    & \eta_1: X_{(\dsi,-1)} \longleftrightarrow X_{(ab,1)} ~, \quad X_{(a,-i)} \longleftrightarrow X_{(b,-i)} ~, \\
    & \eta_2: X_{(\dsi,-1)} \longleftrightarrow X_{(ab,1)} ~, \quad X_{(a,+i)} \longleftrightarrow X_{(b,+i)} ~.
\end{aligned}    
\end{equation}
Given the $\doubleZ_2^{em}$ action on the Abelian anyons, one can compute the possible symmetry fractionalization classes. Notice that the Postnikov class obstruction automatically vanishes, as we already know the $\doubleZ_2^{em}$ symmetry is gaugable, which follows from the existence of $\underline{\CE}^{(i,\kappa_D,\epsilon_D)}_{\doubleZ_2}\Rep H_8$. Using the well-known result (see e.g. \cite{brown2012cohomology})
\begin{equation}
    H^2_\rho(\doubleZ_N, \mathbf{A}) \simeq \frac{\operatorname{Ker}(\sum_{i=0}^{N-1} P^i)}{\operatorname{Im}(\dsi - P)} ~, 
\end{equation}
where $P$ is the generator of $\doubleZ_N$ action on $\mathbf{A}$, it is straightforward to check that 
\begin{equation}
    H^2_{\eta_1}(\doubleZ_2,\mathbf{A}) = H^2_{\eta_2}(\doubleZ_2,\mathbf{A}) = \doubleZ_2 ~.
\end{equation}
Furthermore, they are generated by the two following normalized $2$-cocycles respectively
\begin{equation}
    \mathfrak{t}^1_{\eta_1,\eta_1} = X_{(a,+i)} ~, \quad \mathfrak{t}^2_{\eta_2,\eta_2} = X_{(a,-i)} ~.
\end{equation}
\subsubsection{$\doubleZ_2^{em}$ action on the non-Abelian anyons}
Since both $Y$-type defects and $Z$-type defects both have quantum dimension $2$ in $\CZ(\Rep H_8)$, then it is possible that the bulk $\doubleZ_2^{em}$ symmetry exchanges the $Y$ and $Z$ anyons. Notice that if a given anyon has a unique spin, then the bulk $\mathbb{Z}_2^{em}$ symmetry must preserve this anyon, as any anyon permutation symmetry will preserve $S,T$ matrices, therefore, the topological spin. These anyons are $Z_{(1,\pm 1)}, Z_{(4,\pm 1)}$.

Let's first consider $\doubleZ_2^{em}$ permute two $Y$-type anyons. This is characterized by the following crossed braiding
\begin{equation}
 ~,
\end{equation}
satisfying
\begin{equation}
\begin{aligned}
    [R^{\CD,Y_{\{g_1,g_2\}}}_{\CD}]_{\mu\nu} F^{\CD g_\nu k}_{\CD} R^{k,Y_{\{g_1,g_2\}}}_{kg_\nu} &= F^{h_\mu \CD k}_{\CD} [R^{\CD,Y_{\{g_1,g_2\}}}_{\CD}]_{\mu\nu} F^{\CD k g_\nu}_{\CD} ~, \\
    R^{k,Y_{\{h_1,h_2\}}}_{kh_\mu} F^{kh_\mu\CD}_{\CD} [R^{\CD,Y_{\{g_1,g_2\}}}_\CD]_{\mu\nu} &= F^{h_\mu k\CD}_{\CD} [R^{\CD,Y_{\{g_1,g_2\}}}_{\CD}]_{\mu\nu} F^{k\CD g_\nu}_{\CD} ~, \\
    \sum_{\rho}[R^{\CD,Y_{\{g_1,g_2\}}}_{\CD}]_{\mu\rho} [F^{\CD g_\rho \CN}_{\CD}]_{\nu\sigma} [R^{\CN,Y_{\{g_1,g_2\}}}_{\CN}]_{\rho \kappa} &= \sum_{\rho} [F^{h_\mu \CD \CN}_{\CD}]_{\nu\rho} [R^{\CD,Y_{\{g_1,g_2\}}}_{\CD}]_{\mu\kappa} [F^{\CD\CN g_\kappa}_{\CD}]_{\rho\sigma} ~, \\
    \sum_\rho [R^{\CN,Y_{\{h_1,h_2\}}}_{\CN}]_{\mu\rho} [F^{\CN h_\rho \CD}_{\CD}]_{\nu\sigma} [R^{\CD,Y_{\{g_1,g_2\}}}_{\CD}]_{\rho\kappa} &= \sum_{\rho} [F^{h_\mu \CN \CD}_{\CD}]_{\nu\rho} [R^{\CD,Y_{\{g_1,g_2\}}}_{\CD}]_{\mu\kappa} [F^{\CN \CD g_\kappa}_{\CD}]_{\rho\sigma} ~.
\end{aligned}
\end{equation}
For all $\underline{\CE}_{\doubleZ_2}^{(i,\kappa_\CD,\epsilon_\CD)}\Rep H_8$, we find that $Y_{\{\dsi,ab\}}$ is exchanged with $Y_{\{a,b\}}$. This is the only pair of $Y$ type anyons that are exchanged by $\eta_i$ and there is no $\mathbb{Z}_2^{em}$-invariant $Y$-type anyon.

Finally, let's consider the case where the $Y$-type anyon and the $Z$-type anyon are exchanged. In this case, the crossed braiding is given by
\begin{equation}
 ~,   
\end{equation}
where $[R^{\CD,Y_{\{g_1,g_2\}}}_{\CD}]_{\mu\nu}$ satisfy the following sets of equations
\begin{equation}
\begin{aligned}
    [R^{\CD,Y_{\{g_1,g_2\}}}_\CD]_{\mu\rho} F^{\CD g_\rho k}_{\CD} R^{k,Y_{\{g_1,g_2\}}}_{kg_\rho} &= \sum_{\nu} [F^{\CN\CD k}_{\CD}]_{\mu\nu} [R^{\CD,Y_{\{g_1,g_2\}}}_\CD]_{\nu\rho} F^{\CD k g_\rho}_\CD ~, \\
    \sum_\nu R^{k,Z_{(i,\epsilon)}}_{\CN} [F^{k \CN \CD}_{\CD}]_{\mu\nu} [R^{\CD,Y_{\{g_1,g_2\}}}_\CD]_{\nu\rho} &= \sum_\nu [F^{\CN k\CD}_{\CD}]_{\mu\nu} [R^{\CD,Y_{\{g_1,g_2\}}}_\CD]_{\nu\rho} F^{k\CD g_\rho}_\CD ~, \\
    \sum_{\rho} [R^{\CD,Y_{\{g_1,g_2\}}}_\CD]_{\mu\rho} [F^{\CD g_\rho\CN}_{\CD}]_{\nu\sigma} [R^{\CN,Y_{\{g_1,g_2\}}}_\CN]_{\rho\kappa} &= \sum_{\rho,\delta} [F^{\CN\CD\CN}_\CD]_{\mu\nu,\rho\delta} [R^{\CD,Y_{\{g_1,g_2\}}}_\CD]_{\delta\kappa} [F^{\CD\CN g_\kappa}_\CD]_{\rho\sigma} ~, \\
    \sum_{\mu} R^{\CN,Z_{(i,\epsilon)}}_k [F^{\CN\CN\CD}_{\CD}]_{k,\mu\nu} [R^{\CD,Y_{\{g_1,g_2\}}}_\CD]_{\mu\rho} &= \sum_{\alpha,\beta}[F^{\CN\CN\CD}_\CD]_{k,\alpha\beta} [R^{\CD,Y_{\{g_1,g_2\}}}_\CD]_{\beta\rho} [F^{\CN\CD g_\rho}_{\CD}]_{\alpha\nu} ~,
\end{aligned}
\end{equation}
while $[R^{\CD,Z_{(i,\epsilon)}}_{\CD}]_{\mu\nu}$ satisfy
\begin{equation}
\begin{aligned}
    \sum_{\nu} [R^{\CD,Z_{(i,\epsilon)}}_{\CD}]_{\mu\nu} [F^{\CD\CN k}_{\CD}]_{\nu\rho} R^{k,Z_{(i,\epsilon)}}_{\CN} &= \sum_\nu F^{g_\mu \CD k}_{\CD} [R^{\CD,Z_{(i,\epsilon)}}_{\CD}]_{\mu\nu} [F^{\CD k \CN}_{\CD}]_{\nu\rho} ~, \\
    R^{k,Y_{\{g_1,g_2\}}}_{kg_\mu} F^{kg_\mu\CD}_{\CD} [R^{\CD,Z_{(i,\epsilon)}}_{\CD}]_{\mu\rho} &= \sum_\nu F^{g_\mu k \CD}_{\CD} [R^{\CD,Z_{(i,\epsilon)}}_{\CD}]_{\mu\nu} [F^{k\CD\CN}_{\CD}]_{\nu\rho} ~, \\
    \sum_{\rho} [R^{\CD,Z_{(i,\epsilon)}}_{\CD}]_{\mu\rho} [F^{\CD\CN\CN}_{\CD}]_{\rho\nu,k} R^{\CN,Z_{(i,\epsilon)}}_k &= \sum_{\rho,\sigma} [F^{g_\mu\CD\CN}_\CD]_{\nu\rho} [R^{\CD,Z_{(i,\epsilon)}}_{\CD}]_{\mu\sigma} [F^{\CD\CN\CN}_{\CD}]_{\rho\sigma,k} ~, \\
    \sum_{\rho} [R^{\CN,Y_{\{g_1,g_2\}}}_{\CN}]_{\mu\rho} [F^{\CN g_\rho \CD}_{\CD}]_{\nu\sigma} [R^{\CD,Z_{(i,\epsilon)}}_{\CD}]_{\rho\kappa} &= \sum_{\rho,\alpha} [F^{g_\mu \CN\CD}_{\CD}]_{\nu\rho} [R^{\CD,Z_{(i,\epsilon)}}_{\CD}]_{\mu\alpha} [F^{\CN\CD\CN}_{\CD}]_{\rho\alpha,\kappa\sigma} ~.
\end{aligned}
\end{equation}
Looking for pairs of $Y$-anyon and $Z$-anyon which admits non-zero crossed braiding, again we find that the $\mathbb{Z}_2^{em}$ action only depends on $i$ in the classification $\underline{\CE}_{\doubleZ_2}^{(i,\kappa_\CD,\epsilon_\CD)}\Rep H_8$, and we summarize the action together with the Abelian anyons as
\begin{equation}\label{eq:max_act}
    \eta_1: \begin{pmatrix} X_{(ab,1)} \\ X_{(a,-i)} \\ Y_{\{\dsi,a\}} \\ Y_{\{\dsi,b\}} \\ Y_{\{\dsi,ab\}} \\ Y_{\{a,ab\}} \\ Y_{\{b,ab\}}\end{pmatrix} \longleftrightarrow \begin{pmatrix}
        X_{(\dsi,-1)} \\ X_{(b,-i)} \\ Z_{(3,1)} \\ Z_{(2,1)} \\ Y_{\{a,b\}} \\ Z_{(2,-1)} \\ Z_{(3,-1)}
    \end{pmatrix} ~, \quad \eta_2: \begin{pmatrix} X_{(ab,1)} \\ X_{(a,+i)} \\ Y_{\{\dsi,a\}} \\ Y_{\{\dsi,b\}} \\ Y_{\{\dsi,ab\}} \\ Y_{\{a,ab\}} \\ Y_{\{b,ab\}}\end{pmatrix} \longleftrightarrow \begin{pmatrix}
        X_{(\dsi,-1)} \\ X_{(b,+i)} \\ Z_{(2,1)} \\ Z_{(3,1)} \\ Y_{\{a,b\}} \\ Z_{(3,-1)} \\ Z_{(2,-1)}
    \end{pmatrix} ~.
\end{equation}
One might wonder if it is possible to have two different $\mathbb{Z}_2^{em}$ symmetry, which while permuting the anyon in the same way, but acts on the fusion junctions of anyons in an inequivalent way. However, this possibility can be ruled out by combining the result in \cite{Cordova:2018cvg}, which states that in any multiplicity-free MTC (meaning the fusion coefficient $\leq 1$), the permutation action on the anyon type completely determines the symmetry, as well as the fact that $\CZ(\Rep H_8)$ is multiplicity free (which can be checked using the Verlinde formula and the explicit $S$-matrix given in \cite{Gelaki:2009blp}).

Therefore, we conclude that we find two $\doubleZ_2^{em}$ symmetries which can lead to the self-duality under the maximal gauging on the boundary. 

To conclude we want to point out that the symmetry action \eqref{eq:max_act} suggests that the self-duality under gauging $\CN$ symmetry in the $\TY$ category is a rather rare phenomenon, as it requires exchanging the $Z$-type anyon with $Y$-type anyon. But or a generic $\TY(A,\chi,\epsilon)$ category, they have distinct quantum dimensions where $d_Y = 2$ and $d_Z = \sqrt{|A|}$.

\subsection{Classification revisited}\label{sec:RepH8_cf_revisited}
Given the result in the previous subsection, we can now revisit the interpretation of the classification of $F$-symbols in \cite{Choi:2023vgk}. Among the $8$ inequivalent fusion categories, a factor of $2$ comes from two different choices of the FS indicator, but the rest of the factor of $4$ should be interpreted as two inequivalent choices of the bulk $\mathbb{Z}_2^{em}$ symmetry, and for each choice of $\mathbb{Z}_2^{em}$, there are two choices of the symmetry fractionalization classes. More concretely, for the 8 fusion categories $\underline{\CE}_{\mathbb{Z}_2}^{(i,\kappa_\CD,\epsilon_{\CD})}\Rep H_8$ labeled by $i = 1,2$, $\kappa_{\CD} = \pm 1$, and $\epsilon_{\CD} = \pm 1$, the bulk interpretations are summarized as below:  

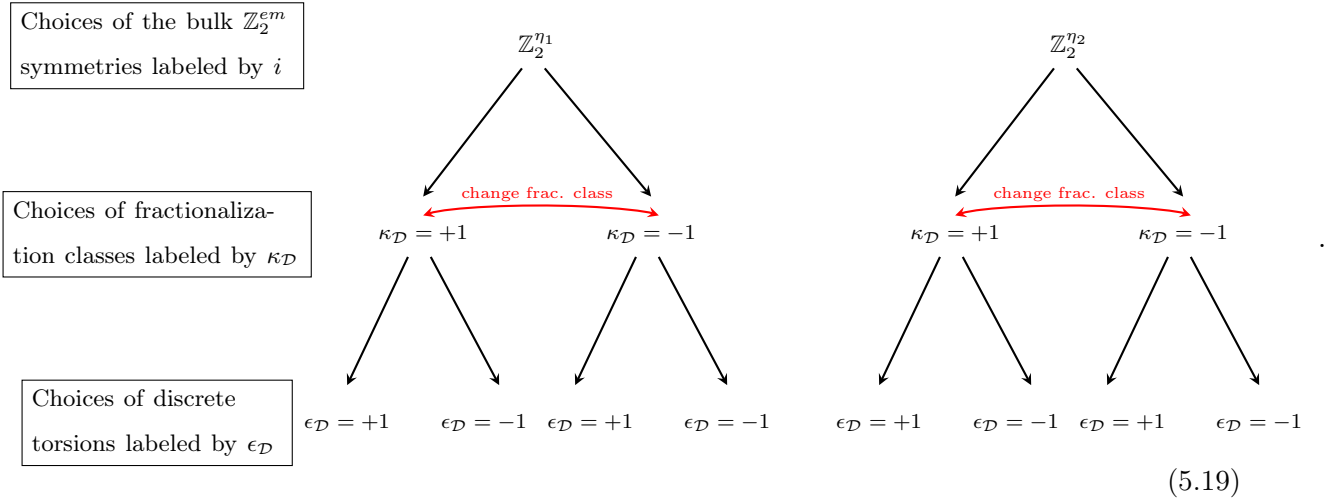
\begin{equation}
 ~.
\end{equation}
Namely, one first pick a choice of the bulk $\mathbb{Z}_2$ symmetries out of $\mathbb{Z}_2^{\eta_i}$ and this determines the $i$ label in $\underline{\CE}_{\mathbb{Z}_2}^{(i,\kappa_\CD,\epsilon_{\CD})}\Rep H_8$; after that one picks a choice of the fractionalization classes which then determines the $\kappa_{\CD} = \pm 1$ label in $\underline{\CE}_{\mathbb{Z}_2}^{(i,\kappa_\CD,\epsilon_{\CD})}\Rep H_8$; finally, the choice of the discrete torsions then determines the FS indicator $\epsilon_{\CD} = \pm 1$ in $\underline{\CE}_{\mathbb{Z}_2}^{(i,\kappa_\CD,\epsilon_{\CD})}\Rep H_8$. Changing the symmetry fractionalization would swap between $\kappa_{\CD} = \pm 1$ in the second layer in the above branching diagram, which means if we start with some specific $\underline{\CE}_{\mathbb{Z}_2}^{(i,\kappa_{\CD},\epsilon_{\CD})}\Rep H_8$ category and consider the change in the fractionalization class, we will have access to both $\underline{\CE}_{\mathbb{Z}_2}^{(i,-\kappa_{\CD},\epsilon_{\CD})}\Rep H_8$ and $\underline{\CE}_{\mathbb{Z}_2}^{(i,-\kappa_{\CD},-\epsilon_{\CD})}\Rep H_8$ categories. 

Recall that in \cite{Choi:2023vgk}, it is pointed out that the $F$-symbols for the $G$-extended fusion category relates to bulk classification data as given in Table \ref{tab:clsf_F_symbol}. 
\begin{table}[h]
    \centering
    \begin{tabular}{c|c|c}
    \hline \hline
        Abstract structure & $F$-symbols & Classification data  \\
    \hline
        Left $\Rep H_8$-module category structure & $F^{r_1 r_2 \mathcal{D}}_{\mathcal{D}}$ & \multirow{3}*{$\rho:\doubleZ_2 \rightarrow \BrPic(\Rep H_8)$} \\
    \cline{1-2}
        Right $\Rep H_8$-module category structure & $F^{\mathcal{D}r_1 r_2}_{\mathcal{D}}$ & \\
    \cline{1-2}
        \makecell{Bimodule structure which glues \\ the left/right module structure together} & $F^{r_1\mathcal{D}r_2}_{\mathcal{D}}$ &  \\
    \hline
        \makecell{A choice of the equivalence functor \\ $\mathcal{C}_\eta \boxtimes_{\Rep H_8} \mathcal{C}_\eta \simeq \Rep H_8$} & $F^{\mathcal{D}\mathcal{D} r_1}_{r_2}, F^{\mathcal{D}r_1 \mathcal{D}}_{r_2}, F^{r_1 \mathcal{D}\mathcal{D}}_{r_2}$ & $[\mathfrak{t}] \in H^2_{[\rho]}(\doubleZ_2, A)$ \\
    \hline
        A choice of the FS indicator & the $\pm$ sign in  $\left[F^{\mathcal{D}\mathcal{D}\mathcal{D}}_{\mathcal{D}}\right]_{\dsi \dsi}$ & $\fsa \in H^3(\doubleZ_2, U(1))$ \\
    \hline\hline
    
    \end{tabular}
    \caption{The correspondence between the abstract structure appearing in the classification analysis and the concrete $F$-symbols, where $r_i \in \Rep H_8$. We drop the labels for the internal channel of the $F$-symbol for simplicity.}
    \label{tab:clsf_F_symbol}
\end{table}
However, when checking this relation one must be very careful on the gauge transformation of the $F$-symbols. As we show, certain two sets (e.g. $(i,+,\pm)$ and $(i,-,\pm)$) of $F$-symbols given there have different but gauge-equivalent $F^{r_1 r_2 \CD}_{\CD}, F^{r_1 \CD r_2}_{\CD}, F^{\CD r_1 r_2}_{\CD}$, thus corresponds to the same bulk symmetry. 

\

To further check this new interpretation and provide an explicit example of how the calculation in Section \ref{sec:transformed_F_symbols} works, we will start with the $F$-symbols of the $(1,+,+)$ category, and consider a change of symmetry fractionalization class to get a new fusion category. Depending on the further choices of the FS indicator, we will see that the new $\widetilde{F}$-symbols are equivalent to $(1,-,\pm)$ respectively, since they have the same spin selection rules. Indeed, the new $F$-symbols have the same $\widetilde{F}^{r_1 r_2 \CD}_{\CD},\widetilde{F}^{r_1 \CD r_2}_{\CD},\widetilde{F}^{\CD r_1 r_2}_{\CD}$ as those of the $(1,+,+)$ category, therefore they correspond to the same bulk $\mathbb{Z}_2^{em}$ symmetry as the $(1,+,+)$ category, thus confirm our new interpretation. 

Let's demonstrate explicitly how this works. In this case, given the symmetry action of $\eta_1$ in \eqref{eq:max_act}, a representative of the non-trivial class in $H^2(\mathbb{Z}_2^{\eta_1},\mathbf{A})$ is
\begin{equation}
    \mathfrak{t}_{1,1} = \mathfrak{t}_{1,\eta_1} = \mathfrak{t}_{\eta_1,1} = \dsi ~, \quad \mathfrak{t}_{\eta_1,\eta_1} = X_{(a,+i)} ~.
\end{equation}
Hence, we are only modifying the following junctions
\begin{equation}
 ~.
\end{equation}
Notice that we have used the fact that $g^{-1} = g$ for $g \in \mathbb{Z}_2 \times \mathbb{Z}_2$ to simplify the notation. Apparently, this transformation only modifies the $F$-symbols with more than one outgoing $\CD$ defects. Let's first consider $\widetilde{F}^{\CD\CD g}_h$, which is the coefficient between
\begin{equation}
    %
 ~.
\end{equation}
We find
\begin{equation}
    \widetilde{F}^{\CD\CD g}_h = F^{ahg,a,g}_h R^{g,X_{(a,+i)}}_{ga} [F^{hga,g,a}_h]^{-1} F^{\CD\CD g}_{ha} = R^{g,X_{(a,+i)}}_{ga} F^{\CD\CD g}_{ha} ~.
\end{equation}
The following $\widetilde{F}$-symbols can be derived similarly:
\begin{equation}
\begin{aligned}
    [\widetilde{F}^{\CD\CD g}_{\CN}]_{\mu\nu} & = R^{g,X_{(a,+i)}}_{ga} [F^{\CD\CD g}_{\CN}]_{\mu\nu} ~, \\
    [\widetilde{F}^{\CD\CD\CN}_g]_{\mu\nu} & = F^{\CN a \CN}_g R^{\CN,X_{(a,+i)}}_\CN [F^{\CD\CD\CN}_{ga}]_{\mu\nu} ~, \\ [\widetilde{F}^{\CD\CD\CN}_{\CN}]_{g,\mu\nu} & = R^{\CN,X_{(a,+i)}}_\CN [F^{ga,\CN,a}_{\CN}]^{-1} [F^{\CD\CD\CN}_{\CN}]_{ga,\mu\nu} ~.
\end{aligned}
\end{equation}
Next, for $\widetilde{F}^{\CD g\CD}_h$ given by the following diagram,
\begin{equation}

\end{equation}
it is obvious that it remains unchanged
\begin{equation}
    \widetilde{F}^{\CD g\CD}_h = F^{\CD g \CD}_{ha} ~,
\end{equation}
and the following $\widetilde{F}$-symbols are not changed for the same reason
\begin{equation}
    [\widetilde{F}^{\CD g\CD}_{\CN}]_{\mu\nu} = [F^{\CD g \CD}_{\CN}]_{\mu\nu} ~, \quad [\widetilde{F}^{\CD \CN \CD}_g]_{\mu\nu} = [F^{\CD \CN \CD}_{ga}]_{\mu\nu} ~, \quad [\widetilde{F}^{\CD \CN \CD}_{\CN}]_{\mu\nu,\alpha\beta} = [F^{\CD\CN\CD}_{\CN}]_{\mu\nu,\alpha\beta} ~.
\end{equation}
Then let's consider $\widetilde{F}^{g \CD \CD}_h$, which is the phase between
\begin{equation}

\end{equation}
and is given by
\begin{equation}
    \widetilde{F}^{g\CD\CD}_h = F^{g\CD\CD}_{ha} F^{g,gha,a}_h = F^{g\CD\CD}_{ha} ~.
\end{equation}
And the following $\widetilde{F}$-symbols can be computed similarly:
\begin{equation}
    [\widetilde{F}^{g\CD\CD}_{\CN}]_{\mu\nu} = F^{g\CN a}_{\CN} [F^{g\CD\CD}_\CN]_{\mu\nu}  ~, \quad [\widetilde{F}^{\CN\CD\CD}_g]_{\mu\nu} = [F^{\CN\CD\CD}_{ga}]_{\mu\nu} ~, \quad [\widetilde{F}^{\CN\CD\CD}_{\CN}]_{\mu\nu,g} = [F^{\CN\CD\CD}_{\CN}]_{\mu\nu,ga} ~.
\end{equation}
Finally, let's consider $[\widetilde{F}^{\CD\CD\CD}_{\CD}]_{gh}$ and $[\widetilde{F}^{\CD\CD\CD}_{\CD}]_{g,\mu\nu}$, which are coefficients in the following relation
\begin{equation}
\begin{aligned}
 ~,
\end{aligned}
\end{equation}
which can be straightforwardly computed,
\begin{equation}
\begin{aligned}
    [\widetilde{F}^{\CD\CD\CD}_{\CD}]_{gh} &= F^{ga,a,\CD}_{\CD} R^{\CD, X_{(a,+i)}}_{\CD} [F^{\CD\CD\CD}_{\CD}]_{ga,h} [F^{\CD\CD a}_h]^{-1} ~, \\
    [\widetilde{F}^{\CD\CD\CD}_{\CD}]_{g,\mu\nu} &= \sum_\alpha F^{ga,a,\CD}_{\CD} R^{\CD, X_{(a,+i)}}_{\CD} [F^{\CD\CD\CD}_{\CD}]_{ga,\alpha\nu} [F^{\CD\CD a}_{\CN}]_{\alpha\mu}^{-1} ~.
\end{aligned}
\end{equation}
And the final $\widetilde{F}$-symbols can be derived similarly,
\begin{equation}
\begin{aligned}
    [\widetilde{F}^{\CD\CD\CD}_{\CD}]_{\mu\nu,g} &= \sum_{\rho} R^{\CD,X_{(a,+i)}}_\CD [F^{\CD\CD a}_g]^{-1} [F^{\CN a \CD}_{\CD}]_{\nu\rho} [F^{\CD\CD\CD}_{\CD}]_{\mu\rho,g} ~, \\
    [\widetilde{F}^{\CD\CD\CD}_{\CD}]_{\mu\nu,\alpha\beta} &= \sum_{\rho,\sigma} R^{\CD,X_{(a,+i)}}_\CD [F^{\CN a \CD}_{\CD}]_{\nu\rho} [F^{\CD\CD\CD}_{\CD}]_{\mu\rho,\sigma\beta} [F^{\CD\CD a}_\CN]^{-1}_{\sigma\alpha} ~.
\end{aligned}
\end{equation}
Taking the new $\widetilde{F}$-symbols can plug-in the pentagon equations explicitly, we find the only constraint is\footnote{Notice that here we have been a bit sloppy--in principle, one can solve a few more equations in the SymSET to determine $R^{\CD,X_{(a,+i)}}_{\CD}$, which is actually not $\pm 1$. But then, we are allowed to use a $\omega \in H^3(\mathbb{Z}_2,U(1))$ to trivialize the obstruction and the full equation should be $(R^{\CD,X_{(a,+i)}}_{\CD})^2 \omega(\eta,\eta,\eta)^2 = 1$. And because only $\widetilde{F}^{\CD\CD\CD}_{\CD}$'s depend on $R^{\CD,X_{(a,+i)}}_{\CD}$ and depends on it linearly, and are shifted as $\omega(\eta,\eta,\eta) \widetilde{F}^{\CD\CD\CD}_{\CD}$, we can simply treat $R^{\CD,X_{(a,+i)}}_{\CD}$ (which is actually $R^{\CD,X_{(a,+i)}}_{\CD}\omega(\eta,\eta,\eta)$) as a free parameter by dropping $\omega(\eta,\eta,\eta)$ for the purpose of determining $F$-symbols.}
\begin{equation}
    (R^{\CD,X_{(a,+i)}}_{\CD})^2 = 1 ~.
\end{equation}
This implies that choosing $R^{\CD,X_{(a,+i)}}_{\CD} = \pm 1$ in $\widetilde{F}$'s leads to a consistent set of $\widetilde{F}$-symbols with the FS indicator being $\pm 1$ respectively. To determine which fusion category it corresponds to, we only need to compute the spin selection rules using the formula listed in Appendix \ref{app:max_spin_selection}\footnote{We revisit the calculation \cite{Choi:2023vgk} and carefully distinguish the fusion and the splitting junctions following Section \ref{sec:RepH8_nm_ssr}. This leads to identical results as one can check explicitly.}, and we find
\begin{equation}
    \begin{array}{|c|c|}
    \hline
        R^{\CD,X_{(a,+i)}}_{\CD} & s \mod \frac{1}{2} \\
    \hline
    +1 & \frac{1}{16},~ \frac{3}{32},~ \frac{11}{32},~ \frac{7}{16} \\ \hline
    -1 & \frac{3}{32},~ \frac{3}{16},~ \frac{5}{16},~ \frac{11}{32} \\ \hline
\end{array} ~,
\end{equation}
which agrees with the spin selection rules of the $(1,-,\pm)$ categories respectively, therefore they must be equivalent. And it is not hard to find the explicit gauge transformation. Finally, it is straightforward to repeat the same exercise for the $(2,+,\cdot)$ category and show that changing fractionalization classes map it to the $(2,-,\cdot)$ categories, and the $\cdot$ denotes different choices of FS indicators and can be fixed by choosing appropriate $\omega$'s.


\section{Example II: Self-duality under non-maximal gauging in $\Rep H_8$}\label{sec:non_max_RepH8}
In this section, we consider non-maximal gauging in $\Rep H_8$ associates with the algebra object 
\begin{equation}
    \CA = \dsi + ab + \scriptN ~,
\end{equation}
and there is a unique choice of the multiplication map leading to self-duality. Since the gauging is non-maximal as $a$ (or $b$) is not gauged, we expect there are two duality defects $\scriptD_1$ and $\scriptD_2$, related by fusing with $a$ (or $b$). Notice that fusing $\CD_i$ with $ab$ does not change the defect type as $ab$ is gauged inside $\CA$.

It is then straightforward to write down the following fusion rules for the defects:
\begin{equation}\label{eq:nm_fus_type_I}
\begin{aligned}
    & \CD_i \times \CN = \CN \times \CD_i = \CD_1 + \CD_2 ~, \quad \CD_i \times g = g \times \CD_i = \prescript{g}{}{\CD_i} ~, \\
    & \CD_1 \times \CD_1 = \CD_2 \times \CD_2 = 1 + ab + \CN ~, \quad \CD_1 \times \CD_2  = \CD_2 \times \CD_1 = a + b + \CN ~,
\end{aligned}
\end{equation}
where $\prescript{g}{}{\CD_i}$ is defined as $\prescript{\dsi}{}{\CD_i} = \prescript{ab}{}{\CD_i} = \CD_i$ and $\prescript{a}{}{\CD_i}=\prescript{b}{}{\CD_i}$ exchanges $\CD_1$ and $\CD_2$. The categorification of this set of fusion rules is worked out in \cite{vercleyen2024low}, where $4$ inequivalent unitary fusion categories are discovered. Working out the corresponding bulk interpretation, we find the four fusion categories correspond to two distinct choices of $\mathbb{Z}_2$ symmetries and two choices of the FS indicator, which we denote as $\CE_{\doubleZ_2,\text{I}}^{(i,\pm)}\Rep H_8$. Notice that the lack of the underline in $\CE$ denotes the gauging is non-maximal.

What is interesting is that each of the bulk symmetry admits two distinct choices of the symmetry fractionalization classes. Therefore, one would expect there are more fusion categories to describe this duality. Since across the defect $\CD_i$, not every TDL is gauged, thus changing the symmetry fractionalization class can change the fusion rule. Indeed, we find there is an alternative set of fusion rules corresponding to the non-trivial fractionalization class:
\begin{equation}\label{eq:nm_fus_type_II}
\begin{aligned}
    & \CD_i \times \CN = \CN \times \CD_i = \CD_1 + \CD_2 ~, \quad \CD_i \times g = g \times \CD_i = \prescript{g}{}{\CD_i} ~, \\
    & \CD_1 \times \CD_1 = \CD_2 \times \CD_2 = a + b + \CN ~, \quad \CD_1 \times \CD_2  = \CD_2 \times \CD_1 = 1 + ab + \CN ~.
\end{aligned}
\end{equation}
These new fusion rules can be naturally interpreted as non-trivially extending the original fusion rules \eqref{eq:nm_fus_type_I} by multiplying $a$ (or equivalently $b$) to the fusion of $\CD_i \times \CD_j$. As we will argue (with the help of group-theoretical constructions), this non-trivial extension has the effect of trivializing the bulk $\mathbb{Z}_2$ SPT in the bulk, therefore we are left only with two choices of bulk $\mathbb{Z}_2$ symmetry, which we will denote as $\CE_{\doubleZ_2,\text{II}}^{i}\Rep H_8$. This matches with the explicit $F$-symbol classification studied in \cite{vercleyen2024low}.

We begin by discussing the action of the bulk partial electric-magnetic $\mathbb{Z}_2^{pem}$ symmetry which leads to the partial duality defects and the classification of these duality defects in Section \ref{sec:bulk_RepH8_nm}. We then compute the spin selection rules of these duality defects and find they are sufficient to distinguish categorically inequivalent duality defects in Section \ref{sec:RepH8_nm_ssr}. Next, we use the SymTFT to show that all the duality defects are group-theoretical and provide explicit group-theoretical constructions in Section \ref{sec:gpt_RepH8}, which also allows us to determine that all the duality defects are anomalous. Finally, in Section \ref{sec:RepH8_nm_example}, we consider some examples.

\subsection{Action of the bulk $\mathbb{Z}_2$ symmetry and classification}\label{sec:bulk_RepH8_nm}
Recall that the electric Lagrangian algebra of $\Rep H_8$ is given by 
\begin{equation}
    \CL_e = X_{(\dsi,+1)} + X_{(\dsi,-1)} + Y_{\{\dsi,a\}} + Y_{\{\dsi,b\}} + Y_{\{\dsi,ab\}} ~,
\end{equation}
we learn that the following two bulk $\mathbb{Z}_2$ symmetries will implement the non-maximal gauging\footnote{Notice that the symmetries of $\CZ(\Rep H_8)$ is $\mathbb{Z}_2 \times \mathbb{Z}_2 \times \mathbb{Z}_2$ as pointed it out in \cite{2016arXiv160304318M}, and the concrete symmetries can be determined by looking for anyon permutations which preserves $S$ and $T$-matrices and those transformations precisely form $\mathbb{Z}_2 \times \mathbb{Z}_2 \times \mathbb{Z}_2$.}:
\begin{equation}\label{eq:non_max_act}
    \rho_1: \begin{pmatrix} X_{(ab,1)} \\ X_{(a,-i)} \\ Y_{\{\dsi,a\}} \\ Y_{\{\dsi,b\}} \\ Y_{\{a,ab\}} \\ Y_{\{b,ab\}}\end{pmatrix} \longleftrightarrow \begin{pmatrix}
        X_{(\dsi,-1)} \\ X_{(b,-i)} \\ Z_{(3,1)} \\ Z_{(2,1)} \\ Z_{(2,-1)} \\ Z_{(3,-1)}
    \end{pmatrix} ~, \quad 
       \rho_2: \begin{pmatrix} X_{(ab,1)} \\ X_{(a,i)} \\ Y_{\{\dsi,a\}} \\ Y_{\{\dsi,b\}} \\ Y_{\{a,ab\}} \\ Y_{\{b,ab\}}\end{pmatrix} \longleftrightarrow \begin{pmatrix}
        X_{(\dsi,-1)} \\ X_{(b,i)} \\ Z_{(2,1)} \\ Z_{(3,1)} \\ Z_{(3,-1)} \\ Z_{(2,-1)}
    \end{pmatrix} ~.
\end{equation}
Alternatively, one can get this result by taking the explicit $F$-symbols given in \cite{vercleyen2024low} and use the similar approach in Section  \ref{sec:RepH8_bulk_symmetry}. 

For both choices of the $\mathbb{Z}_2^{pem}$ symmetry, we see that there are two non-trivial choices of the symmetry fractionalization classes. Notice that the action on the anyons uniquely determines the choice of the symmetries exactly as the case of maximal gauging in $\Rep H_8$. 

Then, we proceed to study the symmetry fractionalization. Unlike the case of maximal gauging, changing the fractionalization class will change the fusion rule of the fusion category. For instance, let's consider $\rho_1$, and again changing the fractionalization class can be acquired from taking $[\mathfrak{t}] \in H^2(\mathbb{Z}_2^{\rho_1}, \mathbf{A})$ 
\begin{equation}
    \mathfrak{t}_{1,1} = \mathfrak{t}_{1,\rho_1} = \mathfrak{t}_{\rho_1,1} = \dsi ~, \quad \mathfrak{t}_{\rho_1,\rho_1} = X_{(a,+i)} ~.
\end{equation}
However, because the boundary image of $X_{(a,+i)}$ is the TDL $a$. This will modify the fusion rule
\begin{equation}
    \CD_i \times \CD_j \mapsto a (\CD_i \times \CD_j) ~,
\end{equation}
which will exchange between type I and type II fusion rules. Once the fractionalization class is chosen, there are two choices of $F$-symbols related by the choice of the discrete torsion in the bulk. Then, naively one would expect that there should be $8$ inequivalent fusion categories. 

However, there is an important caveat. For the type II fusion rule, the duality defect $\CD_i$s are no longer unoriented. As a result, its FS indicator is no longer gauge invariant, and one can not use this to argue that changing discrete torsion necessarily leads to inequivalent fusion categories. In the simplest example of the central extension
\begin{equation}
    0 \rightarrow \mathbb{Z}_2^{c} \rightarrow \Gamma \rightarrow \mathbb{Z}_2^{\eta} \rightarrow 0 ~,
\end{equation}
when the non-trivial $[w] \in H^2(\mathbb{Z}_2^{\eta},\mathbb{Z}_2^{c})$ is activated, $\Gamma = \mathbb{Z}_4$ and the analog of the changing discrete torsion which sits inside $H^3(\mathbb{Z}_2^\eta,U(1))$ is trivialized, in the sense that the change of the $F$-symbols can be removed by a change of basis in the fusion junctions. Notice that this argument follows from the fusion rule itself and does not requires the explicit values of the $F$-symbols.

Similar phenomena happen for the type II fusion rules but in a more subtle way. The trivialization does not directly follow from the fusion rule itself. That is, if we start with $[F^{a,b,c}_{d}]_{e,f}$ and consider
\begin{equation}
    [\widetilde{F}^{a,b,c}_{d}]_{e,f} = \omega(a,b,c) [F^{a,b,c}_{d}]_{e,f}
\end{equation}
where $\omega(a,b,c) = -1$ if $a,b,c \in \{\CD_1,\CD_2\}$ and $\omega(a,b,c) = 1$ otherwise. It is straightforward to check that $\widetilde{F}$'s also satisfy the pentagon equations, but there is no gauge transformation that will relate $F$ and $\widetilde{F}$ like the above case of the extension of $\mathbb{Z}_2$. Namely, there is no solution to 
\begin{equation}
    \frac{\phi(a,b,e)\phi(e,c,f)}{\phi(b,c,f)\phi(a,f,d)} = \omega(a,b,c) ~.
\end{equation}
To check that stacking $\omega$ does not change the equivalence class of the $F$-symbols one must use the explicit form of \cite{vercleyen2024low}\footnote{Notice that it is possible to derive this without using the explicit $F$-symbols in this particular case. As we will see shortly, gauging the invertible symmetry $\mathbb{Z}_2^{ab}$ leads to the dual invertible symmetry $SD_{16}$ with certain anomalies, and as we will see shortly that desired anomaly $SD_{16}$ which would change the FS indicator is actually trivialized. Then, we can reach the same conclusion.}. Furthermore, to relate $F$'s and $\widetilde{F}$'s, besides the gauge transformation, one must also do a relabel of the lines which will exchange $(a,b)$ and $(\CD_1,\CD_2)$. Hence, we see that while stacking of the anomalous $\mathbb{Z}_2$ symmetry can be trivialized by the non-trivial extension in the context of the non-invertible symmetries, it happens in a much more subtle way.

\subsection{The spin selection rules}\label{sec:RepH8_nm_ssr}
Using the $F$-symbols, it is straightforward to compute the spin selection rules of the defect Hilbert space $\CH_{\CD_i}$. Notice that the Drinfeld center contains more information than the spin selection rules. But the latter has the advantage of being much easier to compute, and often the spin selection rules themselves already allow us to completely determine the fusion category with a given fusion rules.

While this is a rather standard calculation, we notice an important subtlety--even all the line defects in this fusion category is orientation reversal invariant, it is important to label the orientation of the lines such that we can distinguish the fusion and splitting junctions. Naively treating those diagrams as if all the lines have no arrow on them will sometimes lead to wrong results. Because of this, we will demonstrate explicitly how to simplify the diagrams explicitly in the Appendix \ref{app:spin_selection}.

To proceed, we start with the torus partition function of the defect Hilbert space. Applying a $T^{-2}$ transformation results in a partition sum with an extra weight of $e^{-4\pi is}$. As shown in \eqref{eq:der_ss_tp1}, the same twisted partition function is related to the action of the symmetry operators as
\begin{equation}
 ~.
\end{equation}
This allows us to relate the spin to the symmetry operators via
\begin{equation}\label{eq:RepH8_nm_spin_ssr}
    e^{-4\pi is} = \sum_{\CL_1 \in \Rep H_8 } \frac{1}{\sqrt{d_{\CL_1}}} \left[F^{\CD_i \CD_i \CD_i}_{\CD_i}\right]_{\CL_1 \CL_1}^{-1} \scriptU[\CL_1,\CD_i] ~,
\end{equation}
where the symmetry operator $\scriptU[a,b]$ is defined via the lasso action
\begin{equation}
    \scriptU[a,b] := \quad %
 ~, \quad a \in \Rep H_8 ~, b \in \{\CD_1,\CD_2\} ~.
\end{equation}
The algebra of these $\scriptU$ operators can be derived by simplifying the following diagram via crossing relations
\begin{equation}
    %
\end{equation}
and as shown in details in \eqref{eq:tube_alge_der}, we find
\begin{equation}
    \scriptU[a,b] \scriptU[c,d] = \sum_{e,f} \sqrt{\frac{d_f d_{\CD_i}}{d_b d_d}} \sqrt{\frac{d_a d_c}{d_{e}}} \left[F^{c\CD_i a}_f\right]_{d,b} \left[F^{ca\CD_i}_f\right]_{b,e}^{-1} \left[F^{\CD_i c a}_f\right]^{-1}_{e,d} \scriptU[e,f] ~.
\end{equation}
We find the allowed spin by first solving the irreducible representations of the tube algebra, and then using \eqref{eq:RepH8_nm_spin_ssr}. The result is listed in the following Table \ref{tab:spin_repH8}
\begin{table}[H]
    \centering
    \begin{tabular}{c|c}
    \hline
     Fusion categories & $s \mod \frac{1}{2}$ \\
    \hline
    \hline
       $\mathcal{F}_1 = \CE_{\doubleZ_2,\text{I}}^{(1,+)}\Rep H_8$  & $0,~ \frac{1}{32},~ \frac{5}{32},~ \frac{9}{32},~ \frac{13}{32}$ \\ \hline
       $\mathcal{F}_2 = \CE_{\doubleZ_2,\text{I}}^{(2,+)}\Rep H_8$  & $0,~ \frac{3}{32},~ \frac{7}{32},~ \frac{11}{32},~ \frac{15}{32}$ \\ \hline
       $\mathcal{F}_3 = \CE_{\doubleZ_2,\text{I}}^{(1,-)}\Rep H_8$  & $\frac{1}{4},~ \frac{1}{32},~ \frac{5}{32},~ \frac{9}{32},~ \frac{13}{32}$ \\ \hline
       $\mathcal{F}_4 = \CE_{\doubleZ_2,\text{I}}^{(2,-)}\Rep H_8$  & $\frac{1}{4},~ \frac{3}{32},~ \frac{7}{32},~ \frac{11}{32},~ \frac{15}{32}$ \\
    \hline
    \hline
        $\widetilde{\mathcal{F}}_1 = \CE_{\doubleZ_2,\text{II}}^{2}\Rep H_8$ & $\frac{1}{8},~ \frac{3}{8},~ \frac{3}{32},~ \frac{7}{32},~ \frac{11}{32},~ \frac{15}{32}$ \\
    \hline
        $\widetilde{\mathcal{F}}_2 = \CE_{\doubleZ_2,\text{II}}^{1}\Rep H_8$ & $\frac{1}{8},~ \frac{3}{8},~ \frac{1}{32},~ \frac{5}{32},~ \frac{9}{32},~ \frac{13}{32}$ \\
    \hline
    \end{tabular}
    \caption{Spin selection rules of the defect Hilbert space $\mathcal{H}_{\CD_i}$ for the different fusion categories derived from the non-maximal gauging of $\Rep H_8$. Notice that for $\CD_1$ and $\CD_2$ the results are the same. Here, the $\mathcal{F}_i$'s and $\widetilde{\mathcal{F}}_i$'s denote the fusion categories with type I fusion rules and type II fusion rules find in \cite{vercleyen2024low} via epxlicitly solving $F$-symbols.}
    \label{tab:spin_repH8}
\end{table}

As one can see, the spin selection rules of the defect Hilbert space allow us to completely distinguish these fusion categories.

\subsection{Group-theoretical construction}\label{sec:gpt_RepH8}
SymTFT is a useful tool to check if a given duality defect can be constructed via gauging subgroups of invertible symmetry. Using the corresponding bulk symmetries given in \eqref{eq:non_max_act}, it is straightforward to check that the Lagrangian subcategory $\Rep D_8$ generated by
\begin{equation}\label{eq:ZRepH8_LSC}
    X_{(\dsi,1)} ~, \quad X_{(\dsi,-1)} ~, \quad X_{(ab,1)} ~, \quad X_{(ab,-1)} ~, \quad Y_{\{1,ab\}} ~,
\end{equation}
is stable under both $\mathbb{Z}_2$ symmetries. Therefore, we conclude that all the six fusion categories are group-theoretical by \cite{Gelaki:2009blp,Sun:2023xxv}.

To find the group-theoretical construction, we notice that under gauging the $\mathbb{Z}_2$ symmetry generated by $ab$, $\Rep H_8$ becomes $\VEC_{D_8}^\gamma$ where $[\gamma]$ is an element in $H^3(D_8,U(1))$ given by
\begin{equation}
    \gamma(r^{i_1}s^{j_1},r^{i_2}s^{j_2},r^{i_3}s^{j_3}) = \exp(\frac{4\pi i}{4^2}(-1)^{j_1} i_1(i_2 + (-1)^{j_2} i_3 - [i_2 + (-1)^{j_2}i_3]_4)) ~,
\end{equation}
where we parameterize $D_8 = \langle r,s| r^4 = s^2 = 1 ~, srs = r^{-1} \rangle$. This quantum symmetry corresponds to the boundary condition generated by the Lagrangian subcategory $\Rep D_8$ in the bulk. As a result, the $\mathbb{Z}_2$ extension becomes an invertible extension, therefore its generator must correspond to a tensor autoequivalence of $\VEC_{D_8}^\gamma$. Recall that for a generic $\VEC_G^\omega$, its group of tensor autoequivalences $\Aut(\VEC_G^\omega)$ fit into the following short exact sequence
\begin{equation}
    0 \rightarrow H^2(G,U(1)) \rightarrow \Aut(\VEC_G^\omega) \rightarrow \operatorname{Stab}(\omega) \rightarrow 0 ~,
\end{equation}
where $\operatorname{Stab}(\omega)$ is the subgroup of the group of automorphisms of the group $G$ which preserves the cohomology class of $\omega$. And a generic element in $\Aut(\VEC_G^\omega)$ parameterized by $\rho \in \operatorname{Stab}(\omega)$ and $\mu \in C^2(G,U(1))$ acting on the fusion junction as 
\begin{equation}
    \begin{tikzpicture}[baseline={([yshift=-1ex]current bounding box.center)},vertex/.style={anchor=base,
    circle,fill=black!25,minimum size=18pt,inner sep=2pt},scale=0.5]
        \draw[black, thick, ->-=.5, line width = 0.4mm] (2,0) -- (2,1.333);

        \draw[black, line width = 0.4mm] (1.333,2) arc (180:270:0.667);

        \draw[black, line width  = 0.4mm, ->-=0.5] (2.667,2) -- (2.667,3);
        
        \draw[black, line width = 0.4mm, ->-=0.5] (1.333,2) -- (1.333,3);
        
        \draw[black, line width = 0.4mm] (2.667,2) arc (0:-90:0.667);

        \filldraw[black] (2,1.333) circle (3pt);

        \node[black, above] at (1.333,3) {\footnotesize $g$};
        \node[black, above] at (2.667,3) {\footnotesize $h$};
        \node[black, below] at (2,0) {\footnotesize $gh$}; 
    \end{tikzpicture} \rightarrow \mu(g,h) \begin{tikzpicture}[baseline={([yshift=-1ex]current bounding box.center)},vertex/.style={anchor=base,
    circle,fill=black!25,minimum size=18pt,inner sep=2pt},scale=0.5]
        \draw[black, thick, ->-=.5, line width = 0.4mm] (2,0) -- (2,1.333);

        \draw[black, line width = 0.4mm] (1.333,2) arc (180:270:0.667);

        \draw[black, line width  = 0.4mm, ->-=0.5] (2.667,2) -- (2.667,3);
        
        \draw[black, line width = 0.4mm, ->-=0.5] (1.333,2) -- (1.333,3);
        
        \draw[black, line width = 0.4mm] (2.667,2) arc (0:-90:0.667);

        \filldraw[black] (2,1.333) circle (3pt);

        \node[black, above] at (1.333,3) {\footnotesize $\rho(g)$};
        \node[black, above] at (2.667,3) {\footnotesize $\rho(h)$};
        \node[black, below] at (2,0) {\footnotesize $\rho(gh)$}; 
    \end{tikzpicture}
\end{equation}
where $\mu$ satisfies 
\begin{equation}
    d\mu (g,h,k) = \frac{{}^\rho \omega(g,h,k)}{\omega(g,h,k)} ~,
\end{equation}
where ${}^\rho\omega(g,h,k) = \omega(\rho(g),\rho(h),\rho(k))$. This tensor autoequivalence of $\VEC_{G}^\omega$ naturally induces a symmetry in the SymTFT $\CZ(\VEC_G^\omega)$, and the symmetry action on the anyons is described in detail in Appendix \ref{app:induction_map}. 

It is shown in \cite{2016arXiv160304318M} that the image of $\Aut(\VEC_{D_8}^\gamma)$ inside $\EqBr(\CZ(\VEC_{D_8}^\gamma))$ is $\mathbb{Z}_2\times \mathbb{Z}_2$ where the four corresponding boundary tensor autoequivalences are given by
\begin{equation}
    F_{1,1} ~, \quad F_{1,\mu} ~, \quad F_{\phi,\mu_0} ~, \quad F_{\phi, \mu_0 \mu} ~,
\end{equation}
where $1$ is the identity isomorphism of $D_8$ and $\phi: r \mapsto r, ~ s \mapsto r^3s$, and
\begin{equation}
    \mu(r^{i_1} s^{j_1}, r^{i_2}s^{j_2}) = \exp(\frac{2\pi}{4} (-1)^{j_1} i_1 j_2) ~, 
\end{equation}
and $\mu_0 \in C^2(D_8,U(1))$ such that
\begin{equation}
    d\mu_0(g,h,k) = \frac{\gamma(g,h,k)}{\gamma(\phi(g),\phi(h),\phi(k))} ~.
\end{equation}
It turns out that to determine which two out of the above four leads to the two bulk symmetries in \eqref{eq:non_max_act}, we only need to check their action on the Abelian anyons. Following the procedure outlined in Appendix \ref{app:induction_map} and using the correspondence \eqref{eq:AA_RepH8_match}, we find that $F_{1,\mu} : X_{(a,\pm i)} \longleftrightarrow X_{(b,\pm i)}$ which does not match with any choices in \eqref{eq:non_max_act}, therefore it is ruled out. Thus, the only options left are $F_{\phi, \mu_0}$ and $F_{\phi, \mu_0 \mu}$. Therefore, we conclude the corresponding boundary symmetries are $F_{\phi,\mu_0}$ and $F_{\phi,\mu_0 \mu}$. Notice that the precise matching depends on the choice of $\mu_0$; and we will use an alternative way to completely determine the group theoretical construction, therefore it doesn't really matter here. It is interesting to notice that $\phi$ is an order-$4$ group automorphism, and squares into an inner automorphism generated by either $r$ or $r^3$, therefore could form a $\mathbb{Z}_2$ extension as explained in Appendix \ref{app:extension} according to the general theory of group extension developed by Schreier \cite{schreier1927untergruppen}. The $\doubleZ_2$ extension introduces an additional element $\eta$, satisfying
\begin{equation}
    \eta^{-1}g\eta = \phi(g) ~, \quad \forall g \in D_8 ~.
\end{equation}
But because $\phi^2$ equals to the inner automorphism generated by $r$ or $r^3$, instead of having $\eta^2 = \dsi$, we now must have
\begin{equation}
    \eta^2 = r ~, \quad \text{or} \quad \eta^2 = r^3 ~.
\end{equation}
It is straightforward to check that both choices lead to a well-defined group, and if we introduce the following new variables $\tilde{r} = r^2 \eta$ and $\tilde{s} = s$, we find
\begin{equation}
\begin{aligned}
    \eta^2 = r & \implies \langle \tilde{r}, \tilde{s}| \tilde{r}^8 = \tilde{s}^2 = \dsi, ~ \tilde{s}\tilde{r}\tilde{s} = \tilde{r}^3 \rangle = SD_{16} ~, \\
    \eta^2 = r^3 & \implies \langle \tilde{r}, \tilde{s}| \tilde{r}^8 = \tilde{s}^2 = \dsi, ~ \tilde{s} \tilde{r} \tilde{s} = \tilde{r}^{-1} \rangle = D_{16} ~.
\end{aligned}
\end{equation}
Notice that the two different group structures can be understood as different choices of the symmetry fractionalization classes in the bulk. To see this, we simply notice that when changing the fractionalization class, we insert an Abelian anyon $X_{(a,\pm i)}$ (depending on which bulk $\mathbb{Z}_2$ symmetry we are taken). However, this anyon $X_{(a,\pm i)}$ is not among these \eqref{eq:ZRepH8_LSC} which are condensed; thus, changing the symmetry fractionalization class will change the fusion rules.

Next step would be finding out the corresponding $3$-cocycles of $D_{16}$ or $SD_{16}$. The generic option would be starting with the $3$-cocycle $\gamma$ of $D_{8}$, and then follow the approach outlined in the Appendix of \cite{Etingof:2009yvg} (where more explicit formulas are given in the Appendix \ref{app:extension}) to extend it to be a $3$-cocycle of $D_{16}$ (or $SD_{16}$). 

In this work, however, we will take a shortcut in the following way. Since $H^3(D_{16},U(1))$ and $H^3(SD_{16},U(1))$ can be computed explicitly, we can simply see which $3$-cocycles when restricting to the $D_{8}$ subgroup would match the $3$-cocycle $\gamma$ in \eqref{eq:3cocyclegamma}. Notice that the $D_8$ is the subgroup generated by $\mathbb{Z}_2^{\tilde{s}}$ and $\mathbb{Z}_4^{\tilde{r}^2}$ in both cases. The $3$-cocycle $\gamma$ has the feature that the two $\mathbb{Z}_2$ subgroups $\langle \tilde{r}^4\rangle$ and $\langle \tilde{s}\rangle$ are anomaly free, therefore, any $\omega \in H^3(D_{16},U(1))$ (or $H^3(SD_{16},U(1))$) which can restrict to $\gamma$ must share the same feature. 

For the case of $D_{16}$, we parameterize the anomaly valued in $\mathbb{Z}_{8} \times \mathbb{Z}_2 \subset H^3(D_{16},U(1)) = \mathbb{Z}_8 \times \mathbb{Z}_2 \times \mathbb{Z}_2$ as 
\begin{equation}\label{eq:D16_generic_anomaly}
\begin{aligned}
    \omega_{m,n}(g_1,g_2,g_3) =  \exp\bigg[\frac{2\pi m \ii}{8^2} (-1)^{j_1} i_1 &(i_2 + (-1)^{j_2} i_3 - [i_2 + (-1)^{j_2} i_3]_8) \\
    & + \frac{2\pi n \ii}{4} i_1 (j_2 + j_3 - [j_2 + j_3]_2)\bigg] ~, \quad m \in \doubleZ_8, \quad n \in \doubleZ_2 ~,
\end{aligned}
\end{equation}
where $g = \tilde{r}^i \tilde{s}^j$. The $\mathbb{Z}_2$ factor we dropped in $H^3(D_{16},U(1))$ is the self-anomaly of $\mathbb{Z}_2^{\tilde{s}}$ and it must be trivial as we intend to gauge this subgroup. The previous condition leaves us with four choices of $\omega_{2 + 4m,n}$ where $m,n = 0,1$. It can be checked explicitly using the formalism reviewed in Section \ref{sec:dual_symmetry} that gauging $\mathbb{Z}_2^{\tilde{s}}$ leads to the fusion category with type I fusion rules, and we summarize the identification of simple lines in Table \ref{tab:D16_group_theoretical}.
\begin{table}
    \centering
    \begin{tabular}{|c|c|c|c|}
    \hline
        Dual Symmetry & Double Coset & Little Group & Irrep \\
    \hline
        $\dsi$  & \multirow{2}{*}{$\{1,\tilde{s}\}$} & \multirow{2}{*}{$\mathbb{Z}_2$} & $\mathbf{1}$ \\
    \cline{1-1} \cline{4-4}
        $ab$  & & & $\mathbf{1}_s$ \\
    \hline
        $a$  & \multirow{2}{*}{$\{\tilde{r}^4,\tilde{r}^4\tilde{s}\}$} & \multirow{2}{*}{$\mathbb{Z}_2$} & $\mathbf{1}$  \\
    \cline{1-1} \cline{4-4}
        $b$  & & & $\mathbf{1}_s$  \\
    \hline
        $\CN$  & $\{\tilde{r}^2,\tilde{r}^{-2},\tilde{r}^2\tilde{s},\tilde{r}^{-2}\tilde{s}\}$ & $\doubleZ_1$ & $\mathbf{1}$\\
    \hline
        $\CD_1$  & $\{\tilde{r},\tilde{r}^{-1},\tilde{r}\tilde{s},\tilde{r}^{-1}\tilde{s}\}$ & $\doubleZ_1$ & $\mathbf{1}$ \\
    \hline
        $\CD_2$  & $\{\tilde{r}^3,\tilde{r}^{-3},\tilde{r}^3\tilde{s},\tilde{r}^{-3}\tilde{s}\}$ & $\doubleZ_1$ & $\mathbf{1}$ \\
    \hline
    \end{tabular}
    \caption{Identifying the simple lines in terms of the group-theoretical construction with gauging $\mathbb{Z}_2^{\tilde{s}}$ in $\VEC_{D_{16}}^\omega$. Recall from the work of \cite{ostrik2002module}, simple lines in a group theoretical fusion category $\CC(G,\omega;H,\psi)$ are labeled by a double coset of $H$ as well as a (projective) irrep of the little group of any representative in the double coset. Notice that there is no canonical identification of the lines $a$ and $b$. As one can confirm by exchanging $a$ and $b$ in the $\CE_{\doubleZ_{2},\text{I}}^{(i,\pm)} \Rep H_8$ leads to equivalent fusion categories.}
    \label{tab:D16_group_theoretical}
\end{table}
To match the $3$-cocycle with the fusion category, we can use the spin selection rules. To derive the spin-selection rules of $\CD_i$ constructed via the group-theoretical construction, we notice that the following relation between twisted partition functions of two theories 
\begin{equation}
 ~,
\end{equation}
where $\scriptX$ is a theory with $\VEC_{D_{16}}^{\omega_{m,n}}$ symmetry, and $\scriptX/\doubleZ_2^{\tilde{s}}$ is the theory $\scriptX$ with $\mathbb{Z}_2^{\tilde{s}}$ gauged. This implies the spin $s\mod 1$ of the defect Hilbert space of $\CH_{\CD_1}$ ($\CH_{\CD_2}$) contains all the spin $s\mod 1$ in the defect Hilbert spaces of $\CH_{g}$ where $g \in \{\tilde{r},\tilde{r}^{-1},\tilde{r}\tilde{s}, \tilde{r}^{-1} \tilde{s}\}$ ($g \in \{\tilde{r}^3,\tilde{r}^{-3},\tilde{r}^3\tilde{s}, \tilde{r}^{-3} \tilde{s}\}$), and the later can be computed via the anyon spectrum in the DW theory as outlined in Appendix \ref{app:DW_spectrum}. We will skip the details of the calculation and list the result as well as the corresponding $\CE_{\doubleZ_2,\text{I}}^{(i,\pm)}\Rep H_8$ below:
\begin{equation}\label{eq:spin}
\begin{array}{|c|c|c|}
\hline
\text{Fusion Categories} & \text{TDL} & s \mod \frac{1}{2} \\ 
\hline
\hline
 \multirow{2}{*}{$\CC(D_{16},\omega_{2,0};\doubleZ_2^{\tilde{s}},1) = \CE^{(1,+)}_{\doubleZ_2,\text{I}}\Rep H_8 \equiv \mathcal{F}_1$} & \CD_1 & 0,~ \frac{1}{32},~ \frac{5}{32},~ \frac{9}{32},~ \frac{13}{32} \\

\cline{2-3} 

& \scriptD_2 & 0,~ \frac{1}{32},~ \frac{5}{32},~ \frac{9}{32},~ \frac{13}{32}  \\
\hline

\multirow{2}{*}{$\CC(D_{16},\omega_{2,1};\doubleZ_2^{\tilde{s}},1) = \CE^{(1,-)}_{\doubleZ_2,\text{I}}\Rep H_8 \equiv \mathcal{F}_3$} & \scriptD_1 & \frac{1}{4},~ \frac{1}{32},~ \frac{5}{32},~ \frac{9}{32},~ \frac{13}{32} \\

\cline{2-3}

& \scriptD_2 & \frac{1}{4},~ \frac{1}{32},~ \frac{5}{32},~ \frac{9}{32},~ \frac{13}{32} \\

\hline
\multirow{2}{*}{$\CC(D_{16},\omega_{6,0};\doubleZ_2^{\tilde{s}},1) =  \CE^{(2,+)}_{\doubleZ_2,\text{I}}\Rep H_8 \equiv \mathcal{F}_2$} & \scriptD_1 & 0,~ \frac{3}{32},~ \frac{7}{32},~ \frac{11}{32},~ \frac{15}{32} \\

\cline{2-3}

& \scriptD_2 & 0,~ \frac{3}{32},~ \frac{7}{32},~ \frac{11}{32},~ \frac{15}{32} \\

\hline
\multirow{2}{*}{$\CC(D_{16},\omega_{6,1};\doubleZ_2^{\tilde{s}},1) = \CE^{(2,-)}_{\doubleZ_2,\text{I}}\Rep H_8 \equiv \mathcal{F}_4$} & \scriptD_1 & \frac{1}{4},~ \frac{3}{32},~ \frac{7}{32},~ \frac{11}{32},~ \frac{15}{32} \\

\cline{2-3}

& \scriptD_2 & \frac{1}{4},~ \frac{3}{32},~ \frac{7}{32},~ \frac{11}{32},~ \frac{15}{32} \\

\hline

\end{array} ~.
\end{equation}

While for the case of $SD_{16}$, the space of anomaly is $\mathbb{Z}_8 \times \mathbb{Z}_2 = H^3(SD_{16},U(1))$ where the $\mathbb{Z}_2$ factor labels the self-anomaly of $\mathbb{Z}_2^{\tilde{s}}$. Again, this anomaly must be trivial as we intend to gauge it, so that we can focus on the $\mathbb{Z}_8$ subgroup of $H^3(SD_{16},U(1))$ and we denote a generic element in it as $\omega_m$ where $m = 0,\cdots,7$. Using the transfer map described in Appendix \ref{app:Deh_cohomology}, one can easily construct explicit forms of $\omega_{2n}$ which are given by
\begin{equation}\label{eq:SD16_generic_anomaly}
\begin{aligned}
    \omega_{2n}(g_1,g_2,g_3) = & \exp\Bigg\{\frac{2\pi n \ii}{8^2} i_1 \left([3^{j_1}i_2]_8 + [3^{j_1+j_2}i_3]_8 - [3^{j_1}i_2 + 3^{j_1+j_2}i_3]_8\right) \\
    & \quad \quad \quad \quad + \frac{2\pi n i}{8^2} [3i_1]_8 \left([3^{j_1+1}i_2]_8 + [3^{j_1+j_2 + 1}i_3]_8 - [3^{j_1 + 1}i_2 + 3^{j_1+j_2 + 1}i_3]_8\right) \Bigg\} ~.
\end{aligned}
\end{equation}
It happens that the requirement of matching the anomaly leaves us with only two options given by $\omega_2$ and $\omega_6$. Again, it is straightforward to check that gauging $\mathbb{Z}_2^{\tilde{s}}$ in $\VEC_{SD_{16}}^{\omega_n}$ leads to fusion categories with the type II fusion rule \eqref{eq:nm_fus_type_II}, and the matching with the $\CE_{\doubleZ_{2},\text{II}}^{i}\Rep H_8$ can be done easily via the spin selection rules, and we summarize the result as below:
\begin{equation}
\begin{array}{|c|c|c|}
\hline
\text{Fusion Categories} & \text{TDL} & s \mod \frac{1}{2} \\ 
\hline
\hline
 \multirow{2}{*}{$\CC(SD_{16},\omega_{2};\doubleZ_2^{\tilde{s}},1) = \CE_{\doubleZ_2,\text{II}}^{1} \Rep H_8 \equiv \widetilde{\mathcal{F}}_2 $} & \scriptD_1 & \frac{1}{8},~ \frac{3}{8},~ \frac{1}{32},~ \frac{5}{32},~ \frac{9}{32},~ \frac{13}{32} \\

\cline{2-3} 

& \scriptD_2 & \frac{1}{8},~ \frac{3}{8},~ \frac{1}{32},~ \frac{5}{32},~ \frac{9}{32},~ \frac{13}{32} \\
\hline

\multirow{2}{*}{$\CC(SD_{16},\omega_{6};\doubleZ_2^{\tilde{s}},1) =  \CE_{\doubleZ_2,\text{II}}^{2} \Rep H_8 \equiv \widetilde{\mathcal{F}}_1 $} & \scriptD_1 & \frac{1}{8},~ \frac{3}{8},~ \frac{3}{32},~ \frac{7}{32},~ \frac{11}{32},~ \frac{15}{32} \\

\cline{2-3}

& \scriptD_2 & \frac{1}{8},~ \frac{3}{8},~ \frac{3}{32},~ \frac{7}{32},~ \frac{11}{32},~ \frac{15}{32} \\

\hline

\end{array} ~.
\end{equation}

\subsubsection*{Anomaly of $\CE_{\doubleZ_2} \Rep H_8$}
In the group-theoretical construction, given $G = D_{16}$ and $H = \doubleZ_2^{\tilde{s}}$, in order to have $G \subset HK$, $K$ must contain $\tilde{r}$. But then $K$ must be anomalous. Hence, there are no symmetric gapped phases. Similarly for $G = SD_{16}$ and $H = \mathbb{Z}_2^{\tilde{s}}$, $K$ must contain $\tilde{r}$ or $\tilde{r}\tilde{s}$ and both leads to at least order $8$ subgroup of $G$. It is straightforward to check both of them are anomalous. Thus, we conclude that there are no trivially gapped phases that admit any type of partial duality defect under non-maximal gauging of $\CN$.

\subsection{Examples}\label{sec:RepH8_nm_example}
Let's consider the $\CD_i$-defects in $c=1$ compact boson found in \cite{Diatlyk:2023fwf}. The duality lines arise from the cosine lines $\CL_{\theta_m,\theta_w}$ on the orbifold branch. Recall that a general cosine line descends from a linear combination of $U(1)_m \times U(1)_w$ symmetry line $\CL^{S_1}_{(\theta_m,\theta_w)}$ on circle branch\cite{Chang:2020imq}:
\begin{equation}
    \CL_{(\theta_m,\theta_w)}=\CL^{S^1}_{(\theta_m,\theta_w)}+\CL^{S^1}_{(-\theta_m,-\theta_w)} ~,
\end{equation}
and has fusion rules
\begin{equation}\label{eq:fusion cosine lines}
    \CL_{(\theta_m,\theta_w)}\CL_{(\theta_m',\theta_w')}=\CL_{(\theta_m+\theta_m',\theta_w+\theta_w')}+\CL_{(\theta_m-\theta_m',\theta_w-\theta_w')} ~.
\end{equation}
Note that the following cosine lines are non-simple:
\begin{equation}
    \CL_{0,0} = 1 \oplus \eta ~, \quad \CL_{\pi,0} = \eta_m \oplus \eta\eta_m ~, \quad \CL_{0,\pi} = \eta_w \oplus \eta\eta_w ~, \quad \CL_{\pi,\pi} = \eta_m\eta_w \oplus \eta\eta_m\eta_w ~.
\end{equation}
where $\eta_m$ and $\eta_w$ descend from $\doubleZ_2^m \times \doubleZ_2^w \subset U(1)_m \times U(1)_w$ symmetry on the circle branch that survive the gauging the charge conjugation $\doubleZ_2^c$, and $\eta$ generates the dual magnetic $\doubleZ_2$ symmetry on orbifold branch. Together, they form the $D_8$ invertible symmetries on the orbifold branch.
At any point on the orbifold branch, the theory has a $\Rep H_8$ symmetry with $\doubleZ_2 \times \mathbb{Z}_2$ symmetry generated by $\eta$ and $\eta_m$ (and $ab$ is identified with $\eta$) and the duality defect $\CN$ given by the cosine line $\CL_{\frac{\pi}{2},\pi}$. As pointed out in \cite{Diatlyk:2023fwf}, one can consider the non-maximal gauging of $\dsi \oplus \eta \oplus \CL_{\frac{\pi}{2},\pi}$ and show the theory is self-dual under this. It is straightforward to check that the following pairs of cosine lines 
\begin{equation}
    (\CL_{\frac{\pi}{4},\frac{\pi}{2}},\CL_{\frac{5\pi}{4},\frac{\pi}{2}}) ~, \quad (\CL_{\frac{\pi}{4},\frac{3\pi}{2}},\CL_{\frac{3\pi}{4},\frac{\pi}{2}}) ~, 
\end{equation}
when identifying with $(\CD_1, \CD_2)$, all lead to type I fusion rules \eqref{eq:nm_fus_type_I} together with the lines $\Rep H_8$. To determine the categorical structure of these defects, we can use the spin selection rules given in  Table \ref{tab:spin_repH8}. To do so, notice that the twisted partition function computing the trace over defect Hilbert space of a generic cosine line is given in \cite{Chang:2020imq}:
\begin{equation} \label{eq:defect partition function}
      Z_{\CL_{(\theta_m,\theta_w)}}(\tau,\bar{\tau})=\sum_{m,w\in\doubleZ}\frac{q^{\frac{p^2_+}{4}}\bar{q}^{\frac{p_-^2}{4}}}{|\eta(\tau)|^2}+\frac{|\theta_2(\tau)\theta_3(\tau)|}{|\eta(\tau)|^2} ~, \quad p_{\pm}=\frac{m+\frac{\theta_w}{2\pi}}{R}\pm (w+\frac{\theta_m}{2\pi})R ~,
\end{equation}  
where $R$ is the radius of the compact boson and 
\begin{equation}
    |\theta_2(\tau)\theta_3(\tau)|=2\sum_{n,\bar{n}\in\doubleZ_{\ge0}} q^{\frac{(2n+1)^2}{16}} \bar{q}^{\frac{(2\bar{n}+1)^2}{16}} ~.
\end{equation}
One can easily check that the spin of operators in the defect Hilbert space of $(\CL_{\frac{\pi}{4},\frac{\pi}{2}},\CL_{\frac{5\pi}{4},\frac{\pi}{2}})$ and $(\CL_{\frac{\pi}{4},\frac{3\pi}{2}},\CL_{\frac{3\pi}{4},\frac{\pi}{2}})$ and show that they generate the category $\CE_{\doubleZ_2,\text{I}}^{(1,+)}\Rep H_8$ and $\CE_{\doubleZ_2,\text{I}}^{(2,+)}\Rep H_8$ respectively.

\section{Example III: Self-duality under non-maximal gauging in $\Rep D_8$}\label{sec:non_max_RepD8}
In this section, we study the self-duality under non-maximal gauging in $\Rep D_8$. In particular, we find 2 inequivalent bulk anyon permutation symmetries that correspond to the non-maximal gauging of $\CA = \dsi+ab+\CN$ in $\Rep D_8$\footnote{Notice that one can also consider gauging $\dsi + a + \CN$ or $\dsi +b + \CN$. But because the $\doubleZ_3$ permutation symmetry between $a,b,ab$, the three cases are equivalent.}. With different fractionalization classes, there are $6=4+2$ inequivalent fusion categories as given in Table \ref{tab:spin_repD8}. Unlike the case of $\Rep H_8$, the extended fusion categories may admit symmetric gapped phases with a unique ground state or symmetry protected topological phases (SPTs). Mathematically, the SPT corresponds to the module category with a simple object or equivalently the fiber functor of the fusion category. We solve the SPT data explicitly and further calculate the interface algebra between different SPTs. The irreducible representations' dimensions of the interface algebra indicate the doubly degenerate interface modes between different SPTs.

\subsection{SymSETs and the group-theoretical construction}
The derivation of the $\Rep D_8$ case is almost identical to the $\Rep H_8$ case, and we will simply list the result and highlight the difference.

Again, we find there are two types of fusion rules given by \eqref{eq:nm_fus_type_I} and \eqref{eq:nm_fus_type_II}. For the type I fusion rule, there are four inequivalent fusion categories $\CE_{\doubleZ_2,\text{I}}^{(i,\pm)}\Rep D_8$ where $i = 1,2$ labels two choices of the bulk symmetry which we will describe shortly and the $\pm$ labels the two choices of the discrete torsion in the bulk. For the type II fusion rule, there are only two inequivalent fusion categories $\CE_{\doubleZ_2,\text{II}}^{i}\Rep D_8$ specified by two choices of the bulk symmetry just like the $\Rep H_8$ case. 

Let us now write down the anyon spectrum for the SymTFT for $\Rep D_8$ so that we can describe the two choices of the symmetries. There are $8$ $X$-type anyons as
\begin{equation}
\begin{array}{rll}
    X_{(\dsi,\pm 1)} :& R^{g,X_{(\dsi,\pm 1)}}_{g} = (1,+1,+1,+1) ~, & R^{\CN,X_{(\dsi,\pm 1)}}_{\CN} = \pm 1 ~, \\
    X_{(a,\pm 1)} :& R^{g,X_{(a,\pm 1)}}_{ga} = (1,+1,-1,-1) ~, & R^{\CN,X_{(a,\pm 1)}}_{\CN} = \pm 1 ~, \\
    X_{(b,\pm 1)} :& R^{g,X_{(b,\pm 1)}}_{gb} = (1,-1,+1,-1) ~, & R^{\CN,X_{(b,\pm 1)}}_{\CN} = \pm 1 ~, \\
    X_{(ab,\pm 1)} :& R^{g,X_{(ab,\pm 1)}}_{gab} = (1,-1,-1,+1) ~, & R^{\CN,X_{(ab,\pm 1)}}_{\CN} = \pm 1 ~. \\
\end{array}
\end{equation}
The $6$ $Y$-type anyons $Y_{\{g,h\}}$ together with their $R$-symbols are given by
\begin{equation}
\begin{array}{rlll}
    Y_{\{\dsi,a\}} :& R^{g,Y_{\{\dsi,a\}}}_{g} = (1,+1,-1,-1) ~, & R^{g,Y_{\{\dsi,a\}}}_{ga} = (1,+1,+1,+1) ~, & [R^{\CN,Y_{\{\dsi,a\}}}_{\CN}]_{\mu\nu} = (\sigma_x)_{\mu\nu} ~, \\
    Y_{\{\dsi,b\}} :& R^{g,Y_{\{\dsi,b\}}}_{g} = (1,-1,+1,-1) ~, & R^{g,Y_{\{\dsi,b\}}}_{gb} = (1,+1,+1,+1) ~, & [R^{\CN,Y_{\{\dsi,b\}}}_{\CN}]_{\mu\nu} = (\sigma_x)_{\mu\nu} ~, \\
    Y_{\{\dsi,ab\}} :& R^{g,Y_{\{\dsi,ab\}}}_{g} = (1,-1,-1,+1) ~, & R^{g,Y_{\{\dsi,ab\}}}_{gab} = (1,+1,+1,+1) ~, & [R^{\CN,Y_{\{\dsi,ab\}}}_{\CN}]_{\mu\nu} = (\sigma_x)_{\mu\nu} ~, \\
    Y_{\{a,b\}} :& R^{g,Y_{\{a,b\}}}_{ga} = (1,-1,+1,-1) ~, & R^{g,Y_{\{a,b\}}}_{gb} = (1,+1,-1,-1) ~, & [R^{\CN,Y_{\{a,b\}}}_{\CN}]_{\mu\nu} = i(\sigma_x)_{\mu\nu} ~, \\
    Y_{\{a,ab\}} :& R^{g,Y_{\{a,ab\}}}_{ga} = (1,-1,-1,+1) ~, & R^{g,Y_{\{a,ab\}}}_{gab} = (1,+1,-1,-1) ~, & [R^{\CN,Y_{\{a,ab\}}}_{\CN}]_{\mu\nu} = i(\sigma_x)_{\mu\nu} ~, \\
    Y_{\{b,ab\}} :& R^{g,Y_{\{b,ab\}}}_{gb} = (1,-1,-1,+1) ~, & R^{g,Y_{\{b,ab\}}}_{gab} = (1,-1,+1,-1) ~, & [R^{\CN,Y_{\{b,ab\}}}_{\CN}]_{\mu\nu} = i(\sigma_x)_{\mu\nu} ~. \\
\end{array}
\end{equation}
The $8$ $Z$-type anyons with their $R$-symbols are given by
\begin{equation}
\begin{array}{rll}
    Z_{(1,\pm 1)} :& R^{g,Z_{(1,\pm 1)}}_{\CN} = (1,-1,+1,+1) ~, & R^{\CN,Z_{(1,\pm 1)}}_{g} = \pm (1,-1,+1,+1) ~, \\
    Z_{(2,\pm 1)} :& R^{g,Z_{(2,\pm 1)}}_{\CN} = (1,+1,-1,+1) ~, & R^{\CN,Z_{(2,\pm 1)}}_{g} = \pm (1,+1,-1,+1) ~, \\
    Z_{(3,\pm 1)} :& R^{g,Z_{(3,\pm 1)}}_{\CN} = (1,+1,+1,-1) ~, & R^{\CN,Z_{(3,\pm 1)}}_{g} = \pm (1,+1,+1,-1) ~, \\
    Z_{(4,\pm 1)} :& R^{g,Z_{(4,\pm 1)}}_{\CN} = (1,-1,-1,-1) ~, & R^{\CN,Z_{(4,\pm 1)}}_{g} = \pm (i,-i,-i,-i) ~.
\end{array}
\end{equation}
Using the explicit $F$-symbols from \cite{vercleyen2024low}, we find there are two inequivalent bulk $\mathbb{Z}_2^{pem}$ symmetry action by solving the same consistency equations in the SymSET as before:
\begin{equation}\label{eq:D8nm-z2}
    \rho_1: %
 ~. \quad 
\end{equation}
Once the symmetry is fixed, we find there are two inequivalent choices of the fractionalization classes which leads to type I and type II fusion rules respectively. For the type I fusion rule, $\CD_i$ is unoriented and the FS indicator is gauge invariant, therefore changing discrete torsion in the bulk necessarily leads to inequivalent fusion categories. In total, there are $4$ inequivalent fusion categories with the type I fusion rule, and we denote them as $\CE_{\doubleZ_2,\text{I}}^{(i,\pm)}\Rep D_8$. For the type II case, similar to the case of $\Rep H_8$, changing discrete torsion leads to equivalent categories, therefore in total, there are only $2$ inequivalent fusion categories with the type II fusion rule, denoted as $\CE_{\doubleZ_2,\text{II}}^{i}\Rep D_8$. All of these fusion categories admit group theoretical constructions of gauging $\mathbb{Z}_2^{\tilde{s}}$ subgroup in $D_{16}$ (or $SD_{16}$) using the same form of anomaly given by \eqref{eq:D16_generic_anomaly} and \eqref{eq:SD16_generic_anomaly}. 

We list their spin selection rules, the interpretation using the classification data in terms of the SymTFT, as well as the group-theoretical construction as in Table \ref{tab:spin_repD8} below.
\begin{table}[h]
    \centering
    \begin{tabular}{c|c}
    \hline
    Fusion Categories & $s \mod \frac{1}{2}$ \\
    \hline
    \hline
        $\mathcal{F}_1 = \CE_{\doubleZ_2,\text{I}}^{(1,+)}\Rep D_8 = \CC(D_{16},\omega_{0,0};\doubleZ_2^{\tilde{s}},1)$ & $0,~ \frac{1}{8},~ \frac{1}{4},~ \frac{3}{8}$  \\ \hline
        $\mathcal{F}_2 = \CE_{\doubleZ_2,\text{I}}^{(1,-)}\Rep D_8 = \CC(D_{16},\omega_{0,1};\doubleZ_2^{\tilde{s}},1) $ & $0,~ \frac{1}{8},~ \frac{1}{4},~ \frac{3}{8}$  \\ \hline
        $\mathcal{F}_3 = \CE_{\doubleZ_2,\text{I}}^{(2,+)}\Rep D_8 = \CC(D_{16},\omega_{4,0};\doubleZ_2^{\tilde{s}},1)$ & $0,~ \frac{1}{16},~ \frac{3}{16},~ \frac{5}{16},~ \frac{7}{16}$  \\ \hline
        $\mathcal{F}_4 = \CE_{\doubleZ_2,\text{I}}^{(2,-)}\Rep D_8 = \CC(D_{16},\omega_{4,1};\doubleZ_2^{\tilde{s}},1)$ & $\frac{1}{4},~ \frac{1}{16},~ \frac{3}{16},~ \frac{5}{16},~ \frac{7}{16}$  \\
    \hline
    \hline
        $\widetilde{\mathcal{F}}_1 = \CE_{\doubleZ_2,\text{II}}^{1}\Rep D_8 = \CC(SD_{16},\omega_0;\doubleZ_2^{\tilde{s}},1)$ & $0,~ \frac{1}{8},~ \frac{1}{4},~ \frac{3}{8}$ \\
    \hline
        $\widetilde{\mathcal{F}}_2 = \CE_{\doubleZ_2,\text{II}}^{2}\Rep D_8 = \CC(SD_{16},\omega_{4};\doubleZ_2^{\tilde{s}},1)$ & $0,~ \frac{1}{4},~ \frac{1}{16},~ \frac{3}{16},~ \frac{5}{16},~ \frac{7}{16}$ \\
    \hline
    \end{tabular}
    \caption{Spin selection rules for the defect Hilbert space $\CH_{\CD_i}$ in different fusion categories derived from non-maximal gauging of $\Rep D_8$. Notice that $\mathcal{F}_i$ and $\widetilde{\mathcal{F}}_i$ denote the fusion categories with explicit $F$-symbols acquired in \cite{vercleyen2024low}.}
    \label{tab:spin_repD8}
\end{table}

It is important to point out that for $\rho_1$ and the type I fusion rule, the spin selection rules do not distinguish the FS indicator. To match with the anomaly of the group-theoretical construction, one can either compute the FS indicator of the line $\CD_i$ for $\omega_{0,0}$ and $\omega_{0,1}$. Alternatively, one can match from the numbers of SPT phases allowed in the IR, as we will see shortly, $\CE_{\doubleZ_2,\text{I}}^{(1,\pm)}\Rep D_8$ admits different numbers of SPT phases in the IR. 

Another interesting fact one can check explicitly is that the fusion category $\CE_{\doubleZ_2,\text{I}}^{(1,+)}\Rep D_8$ is also equivalent to $\Rep D_{16}$ and the fusion category $\CE_{\doubleZ_2,\text{II}}^{1}\Rep D_8$ is also equivalent to $\Rep SD_{16}$.

To conclude, we want to point out that any $G$-extension of $\Rep D_8$ must be group-theoretical. This follows from the fact that its SymTFT admits a Lagrangian algebra which is the sum of all the Abelian anyons in the SymTFT; furthermore, all the Abelian anyons form a Lagrangian subcategory isomorphic to $\Rep(\doubleZ_2\times \doubleZ_2 \times \doubleZ_2)$. Since any invertible symmetries in the bulk must preserve the quantum dimension, they must also preserve this Lagrangian subcategory. Then, the result of \cite{Gelaki:2009blp,Sun:2023xxv} implies that the corresponding $G$-extension is group-theoretical. This suggests that all the $\CE_{\doubleZ_2,\text{I}/\text{II}}\Rep D_8$ admit alternative group-theoretical construction where the partial duality defects arising from certain autoequivalence of the finite group $\mathbb{Z}_2 \times \mathbb{Z}_2 \times \mathbb{Z}_2$ with type III mixed anomaly. We will not go into the details here.

\subsection{SPT phases}
The fusion categories that are derived from non-maximal gauging of $\Rep D_8$ can be anomaly free, therefore admit SPT phases, the result is summarized below,
\begin{equation}\label{eq:numspt}
\begin{array}{c|c|c|c|c}
\hline
\hline
     \text{Fusion category}& \CE_{\mathbb{Z}_2,\operatorname{I}}^{(1,+)}\Rep D_8 = \Rep D_{16}  & \CE_{\mathbb{Z}_2,\operatorname{I}}^{(1,-)}\Rep D_8 & \CE_{\mathbb{Z}_2,\operatorname{I}}^{(2,+)}\Rep D_8 & \CE_{\mathbb{Z}_2,\operatorname{I}}^{(2,-)}\Rep D_8  \\ \hline
    \text{Num. of SPTs} &  3& 1 & 1 & 0   \\ \hline \hline
    \text{Fusion category}&\CE_{\mathbb{Z}_2,\operatorname{II}}^{1} \Rep D_8   = \Rep SD_{16} &\CE_{\mathbb{Z}_2,\operatorname{II}}^{2} \Rep D_8 \\ \hline
    \text{Num. of SPTs} &  2 & 0 \\\hline
\end{array}
\end{equation}
In the SymTFT perspective, the SPT phases correspond to Lagrangian algebras which only overlap with the electric Lagrangian algebra on the trivial anyon. These Lagrangian algebras are known as magnetic Lagrangian algebras. 

For fusion category symmetry that admits multiple SPTs, the interface between different SPTs necessarily hosts degenerate modes, resulting from the representation of the interface algebra (boundary tube algebra). The multiplication of the interface algebra is determined by the data of the two SPTs as demonstrated in Section \ref{sec:SPT_review}.

In the following subsections, we first use the SymTFT and the group-theoretical fusion categories to determine which fusion category admits SPT and how many are there, the details of the SymTFT conventions are listed in App \ref{app:ZD16}. There are two fusion categories that admits more than one SPT: $\CE_{\mathbb{Z}_2,\operatorname{I}}^{(1,+)}\Rep D_8 = \Rep D_{16}$ has 3 SPTs while $\CE_{\mathbb{Z}_2,\operatorname{II}}^{1} \Rep D_8 = \Rep SD_{16}$ has 2 SPTs. We further solve the junction data and determine the interface algebra between different SPTs only has $2$-dim irreducible representations.

\subsubsection{$\CE_{\mathbb{Z}_2,\operatorname{I}}^{(1,+)}\Rep D_8 = \Rep D_{16}$}
Let's first study $\Rep D_{16} = \CE_{\mathbb{Z}_2,\text{I}}^{(1,+)}\Rep D_8$. Using its group-theoretical construction of gauging $\mathbb{Z}_2^{\tilde{s}}$ in $\VEC_{D_{16}}$, we find it admits three SPT phases via \eqref{eq:fiber_functor}:
\begin{enumerate}
    \item The first SPT phase is acquired from the partial SSB phase of $\VEC_{D_{16}}$ where the unbroken subgroup is $\langle r^2, rs \rangle = D_8 \subset D_{16}$\footnote{For the following discussion, we will parameterize the $D_{16}$ as $\langle r,s|r^8 = s^2 = 1, srs = r^{-1}\rangle$ by dropping the tilde in the previous notation. We will do the same for $SD_{16}$.} and the trivial SPT of $D_8$ is realized on each ground state. The corresponding magnetic Lagrangian algebra in the SymTFT is
    \begin{equation}
        \CL_m^1 = (1,\chi _1)+(1,\chi _4)+(r^4,\chi _1)+(r^4,\chi _4)+(r^2,\chi_0^{r})+(r^2,\chi _4^{r})+2 (r s,\chi _{0,0}^{rs}) ~,
    \end{equation}
    where we have chosen the electric Lagrangian algebra to be:
    \begin{equation}
        \CL_e =(1,\chi _1)+(1,\chi _3)+(1,\chi _5)+(1,\chi _6)+(1,\chi _7)+(s,\chi^s _{0,0})+(s,\chi^s _{0,1}) ~.
    \end{equation}

    \item The second SPT phase is acquired from the partial SSB phase of $\VEC_{D_{16}}$ where the unbroken subgroup is $\langle r^2, rs \rangle = D_8 \subset D_{16}$ and but with the non-trivial SPT of $D_8$ is realized on each ground state (recall that $H^2(D_8,U(1)) = \mathbb{Z}_2$). The corresponding magnetic Lagrangian algebra in the SymTFT is
    \begin{equation}
        \CL_m^2 = (1,\chi _1)+(1,\chi _4)+(r^4,\chi _2)+(r^4,\chi _3)+(r^2,\chi_0^{r})+(r^2,\chi_4^{r})+2 (r s,\chi^{rs}_{0,1})~.
    \end{equation}

    \item The third SPT phase is acquired from the partial SSB phase of $\VEC_{D_{16}}$ where the unbroken subgroup is $\langle r \rangle = \mathbb{Z}_8 \subset D_{16}$. The corresponding magnetic Lagrangian algebra in the SymTFT is
    \begin{equation}
        \CL_m^3 =(1,\chi _1)+(1,\chi _2)+(r^4,\chi _1)+(r^4,\chi _2)+2(r,\chi_0^{r})+2(r^2,\chi_0^{r})+2(r^3,\chi_0^{r})~.
    \end{equation}
\end{enumerate}

Notice that each SPTs of $\Rep D_{16}$ is automatically an SPT of the subgroup category $\Rep D_{8}$, and in fact, they reduce to the same SPT of $\Rep D_8$. This simply follows from that there is only one SPT of $\Rep D_8$ invariant under the gauging of $\CA = 1 \oplus ab \oplus \CN$.

Next, we aim to study the interface algebras and their irreducible representations. To do so, we first explicitly solve the SPT data using the $F$-symbols in \cite{vercleyen2024low}, and we list the complete solutions in the supplementary Mathematica file \cite{MMAfile}\footnote{Notice that we have performed some gauge transformations of the $F$-symbols in \cite{vercleyen2024low}, so that the $F$-symbols of the $\Rep D_8$ subcategory match the $F$-symbols in the $\TY$-fusion category literature.}. We will not go through the details of calculation, but highlight some features of the solutions. 

Let us first consider the three SPT phases of $\Rep D_8$, which are studied in \cite{Thorngren:2021yso}. First, a $\Rep D_8$ must be an SPT of $\mathbb{Z}_2\times \mathbb{Z}_2$ which is self-dual under the corresponding $\mathbb{Z}_2\times \mathbb{Z}_2$-gauging. And only the non-trivial SPT satisfies this requirement, which fixes
\begin{equation}
    t_{g,h}^{gh}=(-1)^{g_1 h_2} ~.
\end{equation}
Next, consider the duality-twisted sector where the $\CN$ line is placed along the time-like direction, then there is a left and a right (projective) $\mathbb{Z}_2 \times \mathbb{Z}_2$ action on the sector, given by
\begin{equation}
 ~.
\end{equation}
The crossing relations imply an isomorphism between the left and the right action, which is pictorially captured by
\begin{equation}
    %
 ~,
\end{equation}
where $\nu(g)$'s are phases trivializing the difference between two cohomologically equivalent $2$-cocycles characterizing the (projective) left and right $\mathbb{Z}_2\times \mathbb{Z}_2$-actions. Notice that generically $\sigma$ can be an automorphism of the Abelian group in the $\TY$-fusion category. This phase factor $\nu(g)$ distinguishes the three $\Rep D_8$ SPTs. More concretely, gauge transformation allows us to fix the concrete form of $t_{\CN\mu,g}^{\CN \nu}$'s capturing the right $\mathbb{Z}_2\times \mathbb{Z}_2$-action, and $t_{g,\CN\mu}^{\CN\nu}$'s capturing the left action relate to $t_{\CN\mu,g}^{\CN \nu}$ via
\begin{align}
     \quad t_{\CN \mu,g}^{\CN \nu}  =\nu(g)^{-1}t_{g,\CN \mu}^{\CN \nu}=\{\sigma^0,\sigma^1,\sigma^3,-\ii \sigma^2\} ~.
\end{align}
And the 3 fiber functors differ by $\nu_I(a)=\nu_{II}(b)=\nu_{III}(ab)=-1$, while other $\nu(g)=1$, with the rest of the $t$'s given by
\begin{align}
    \nu_{\text{I}}(a)=-1,\quad t_{\CN\mu,\CN\nu}^g = \{\sigma^3,\ii \sigma^2,\sigma^0,-\sigma^1\} ~, \\
    \nu_{\text{II}}(b)=-1,\quad t_{\CN\mu,\CN\nu}^g = \{\sigma^1,\sigma^0,-\ii\sigma^2,\sigma^3\} ~, \\
    \nu_{\text{III}}(ab)=-1,\quad t_{\CN\mu,\CN\nu}^g = \{\sigma^0,\sigma^1,\sigma^3,-\ii \sigma^2\}  ~.
\end{align}

Next, let's consider the three SPTs of $\Rep D_{16}$. It is straightforward to check that only the $\Rep D_8$-SPT of $\nu_{\text{III}}(ab) = -1$ is self-dual under the $\dsi \oplus ab\oplus \CN$-gauging. This fixes the $t$'s with all external lines in $\Rep D_8$ subcategory in the $\Rep D_{16}$-SPTs. Generalizing the above approach, we should consider the $\CD_i$-twisted sectors, and study the left and the right action of the $\Rep D_8$ subcategory. In practice, it is useful to simply consider the left and the right action of $\mathbb{Z}_2 \times \mathbb{Z}_2 \subset \Rep D_8$ instead. And we notice that the $F$-symbols involving $\CD_i$ and $g$ can be fixed to a form which resembles a similar structure as the $\Rep D_8$ case\footnote{More concretely, recall that the $G$-symbols of $\Rep D_8$ is 
\begin{equation}
\begin{aligned}
     &G^{g,h,k}_{ghk}=1 ~, \quad G^{gh\CN}_{\CN} = G^{\CN gh}_{\CN} = 1 ~, \quad G^{\CN \CN g}_h = G^{g \CN \CN}_h = 1 ~, \\
     &G^{g\CN h}_{\CN} = G^{\CN g \CN}_{h} = \chi_{od}(g,h) ~, \quad [G^{\CN\CN\CN}_{\CN}]_{g,h} = \frac{1}{2}\chi_{od}(g,h) ~,
\end{aligned}
\end{equation}
where $\chi_{od}(g,h)=(-1)^{g_1 h_2 +g_2 h_1}$. And we can gauge fix the $G$-symbols (acquired from the $F$-symbols given in \cite{vercleyen2024low}) such that 
\begin{equation}\label{eq:fsymbol_D}
\begin{aligned}
    &G^{gh\CD_i}_{{}^{gh}\CD_i} = G^{\CD_i gh}_{{}^{gh}\CD_i} = 1 ~, \quad G^{g \CD_i {}^{gh}\CD_i}_h = G^{\CD_i {}^{gh}\CD_i g}_h = 1 ~, \\
    &G^{g\CD_i h}_{{}^{gh}\CD_i} = G^{\CD_i,g,{}^{gh}\CD_i}_{h} =\chi_{od}(g,h) ~.
\end{aligned}
\end{equation}}. Because of this, the junctions with $\CD_i$ and $g$ have a similar structure as those with $\CN$ and $g$. In particular, $t_{\CD_{i}\mu,g}^{{}^g\CD_{i}\nu}$ is the projective representation of $\IZ_2\times \IZ_2$, and $t_{g,\CD_{i}\mu}^{{}^g\CD_{i}\nu} = \nu^\CD (g) t_{\CD_{i}\mu,g}^{{}^g\CD_{i}\nu}$, where $\nu^\CD(g)$ are the analog of $\nu(g)$ relates the left and the right action in the $\Rep D_{16}$ case. More concretely, we can take 
\begin{equation}
    t_{\CD_{i}\mu,g}^{{}^g\CD_{i}\nu} =\nu^\CD_{i}(g)^{-1}t_{g,\CD_{i} \mu}^{{}^g\CD_{i} \nu}=\{\sigma^0,\sigma^1,\sigma^3,-\ii \sigma^2\} ~,
\end{equation}
and the three SPTs phases of $\Rep D_{16}$ can be similarly distinguished by the choices of $\nu^\CD(g)$'s:
\begin{equation}
\begin{aligned}
    \nu^\CD_{\text{I}}(a)=-1,\quad t_{\CD_{i}\mu,{}^g\CD_{i}\nu}^g = \{\sigma^3,\ii \sigma^2,\sigma^0,-\sigma^1\} ~, \\
    \nu^\CD_{\text{II}}(b)=-1,\quad t_{\CD_{i}\mu,{}^g\CD_{i}\nu}^g = \{\sigma^1,\sigma^0,-\ii\sigma^2,\sigma^3\} ~, \\
    \nu^\CD_{\text{III}}(ab)=-1,\quad t_{\CD_{i}\mu,{}^g\CD_{i}\nu}^g = \{\sigma^0,\sigma^1,\sigma^3,-\ii \sigma^2\}  ~,
\end{aligned}
\end{equation}
where unlisted $\nu^\CD(g) = 1$. The complete solutions are listed in the supplementary Mathematica file \cite{MMAfile}.

What is left is to identify the solutions with the SPT classification discussed previously in terms of the partial SSB phases of $\VEC_{D_{16}}$. Notice that this is actually a very subtle procedure. One option is to start with the twisted partition functions of the three SPT phases and gauge the dual symmetry $\mathbb{Z}_2^{ab}$ (of the $\mathbb{Z}_2^{\tilde{s}}$ symmetry), and check whether the invertible symmetry $r$ is spontaneously broken or not. That is, we only need to compute the twisted partition function, 
\begin{equation}\label{eq:r_action}
 \right) = \begin{cases} 0 \quad \text{for I and II} \\ 2 \quad \text{for III} \end{cases}\\
 ~,
\end{equation}
where we have explicitly chosen the junction $\rho,\lambda$ of the bimodule $M_r$ corresponds to $r$ as
\begin{equation}
    %
 ~.
\end{equation}

This step is completely unambiguously, and we find that the $\text{SPT-III}$ is mapped to the partial SSB phase with $\langle r \rangle$ unbroken; and we also learn that the $\text{SPT-I}$ and $\text{SPT-II}$ are mapped to the partial SSB phase with $\langle r^2,rs\rangle$ unbroken. But to determine exactly which fiber functors would realize the non-trivial SPT of the unbroken subgroup is ambiguous as it depends on certain choices one can make in the gauging process. 

More concretely, to see which SPTs are realized, one can try to compute the twisted torus partition function, which requires the construction of the fusion junctions of bimodule objects explicitly. However, each junction can only be determined up to an overall phase by the bimodule conditions. Say for the SPT-I, different choices of these phases related by stacking the non-trivial SPT phases of $D_{16}$ will change the $D_8$ SPT realized on the ground state, as the non-trivial SPT phase of $D_{16}$ remains non-trivial when restricting to the $\langle r^2,rs\rangle$ subgroup. In the following, we will choose the convention such that the first solution corresponding to $\nu_{\text{I}}^\CD(g)$ is mapped to the partial SSB phase realizing the trivial $D_8$-SPT phase.

It is then straightforward to construct the interface algebra and compute its irreducible representations. The dimensions of the irreducible representations of the interface algebra are summarized below,
\begin{equation}
    \begin{array}{|c|c|c|c|}
    \hline
         & \operatorname{SPT-I} & \operatorname{SPT-II} & \operatorname{SPT-III}  \\
    \hline
        \operatorname{SPT-I} & \substack{8 \, \text{$\mathbf{1}$-dim irreps}  \\ 2 \, \text{$\mathbf{2}$-dim irreps}}& 4 \, \text{$\mathbf{2}$-dim irreps} & 4 \, \text{$\mathbf{2}$-dim irreps} \\
    \hline
        \operatorname{SPT-II} &  & \substack{8 \, \text{$\mathbf{1}$-dim irreps}  \\ 2 \, \text{$\mathbf{2}$-dim irreps}} & 4 \, \text{$\mathbf{2}$-dim irreps} \\
    \hline
        \operatorname{SPT-III} &  & & 16 \, \text{$\mathbf{1}$-dim irreps} \\
    \hline
    \end{array} ~.
\end{equation}
We see that all the self-interface admits $1$-dim irreducible representations; and all interfaces between different SPTs only admit $2$-dim irreducible representations, which implies that the spectrum on the interface must be doubly degenerated.

\subsubsection{$\CE_{\mathbb{Z}_2,\operatorname{I}}^{(1,-)} \Rep D_8$}
For $\CE_{\mathbb{Z}_2,\text{I}}^{(1,-)} \Rep D_8$, using its group-theoretical construction of gauging $\mathbb{Z}_2^s$ in $\VEC_{D_{16}}^{\omega_{0,1}}$, we find it only admits a single SPT phase where it acquires from the partial SSB phase of $\VEC_{D_{16}}^{\omega_{0,1}}$ with $\mathbb{Z}_8^r$ unbroken. Notice that the subgroup $D_8 = \langle r^2,rs\rangle$ acquires a 't Hooft anomaly from $\omega_{0,1}$, therefore can not be unbroken. 

Similar to the previous case of $\CE_{\mathbb{Z}_2,\text{I}}^{(1,+)} \Rep D_8$, we have
\begin{equation}
    t_{\CD_{i}\mu,g}^{{}^g\CD_{i}\nu} =\nu^\CD(g)^{-1}t_{g,\CD_{i} \mu}^{{}^g\CD_{i} \nu}=\{\sigma^0,\sigma^1,\sigma^3,-\ii \sigma^2\} ~,
\end{equation}
where $\nu^\CD(g)=(1,-1,-1,-1)$. Since $\CE_{\mathbb{Z}_2,\text{I}}^{1,-} \Rep D_8$ differs from $\CE_{\mathbb{Z}_2,\text{I}}^{(1,+)} \Rep D_8$ by the Frobenius-Schur indicator of $\CD_i$, this fiber functor is also reminiscent of that of $\Rep Q_8$, where $\Rep Q_8$ differs from $\Rep D_8 $ by the Frobenius-Schur indicator of $\CN$ line. The complete solution of the fiber functor is presented in the supplementary Mathematica file \cite{MMAfile}. The irreducible representations of the self-interface algebra are all 1-dimensional.

In the SymTFT, the electric Lagrangian is fixed as the untwisted case \eqref{eq:RepD16eL}, the magnetic Lagrangian algebra is given by,
\begin{equation}
    \CL_m =(1,\chi _1)+(1,\chi _2)+(r^4,\chi _1)+(r^4,\chi _2)+2 (r,\chi _2^r)+2 (r^2,\chi _2^r) +2 (r^3,\chi _2^r)~.
\end{equation}

\subsubsection{$\CE_{\mathbb{Z}_2,\operatorname{I}}^{(2,+)} \Rep D_8$}
For $\CE_{\mathbb{Z}_2,\text{I}}^{(2,+)} \Rep D_8$, using its group-theoretical construction of gauging $\mathbb{Z}_2^s$ in $\VEC_{D_{16}}^{\omega_{4,0}}$, we find it only admits a single SPT phase. Opposite to the case of $\CE_{\mathbb{Z}_2,\text{I}}^{(1,-)} \Rep D_8$, $\langle r^2, rs\rangle \simeq D_8$ is anomaly free while $\mathbb{Z}_8^r$ is anomalous. While it seems we may find two partial SSB phases with $\langle r^2, rs\rangle \simeq D_8$ unbroken labeled by $H^2(D_8,U(1)) = \mathbb{Z}_2$, explicitly solving the corresponding module categories shows that the two $D_8$-SPTs realized on the two ground states are actually distinct. As a result of this, there is a unique partial SSB phase with $\langle r^2, rs\rangle \simeq D_8$ unbroken. Therefore, $\CE_{\mathbb{Z}_2,\text{I}}^{(2,+)} \Rep D_8$ only admits a SPT phase. 

This is also confirmed by the SymTFT analysis. The electric Lagrangian algebra is the same as \eqref{eq:RepD16eL},
\begin{equation}
    \CL_e =(1,\chi _1)+(1,\chi _3)+(1,\chi _5)+(1,\chi _6)+(1,\chi _7)+(s,\chi^s _{0,0})+(s,\chi^s _{0,1}) ~,
\end{equation}
and one can show that there is a unique magnetic Lagrangian algebra given by
\begin{equation}
    \CL_m = (1,\chi _1)+(1,\chi _4)+(r^4,\chi _5)+(r^2,\chi^{r} _3)+(r^2,\chi^{r} _7)+(rs,\chi^{rs}_{1,0})+(rs,\chi^{rs}_{1,1})~.
\end{equation}
We explicitly solved the fiber functor, and the complete solution is listed in the supplementary Mathematica file \cite{MMAfile}. The self-interface algebra only has 1-dimensional irreducible representations.

\subsubsection{$\CE_{\mathbb{Z}_2,\operatorname{I}}^{(2,-)} \Rep D_8$}
For $\CE_{\mathbb{Z}_2,\text{I}}^{(2,-)} \Rep D_8$, using its group-theoretical construction of gauging $\mathbb{Z}_2^s$ in $\VEC_{D_{16}}^{\omega_{4,1}}$, we find it does not admit any SPT phases because both $\mathbb{Z}_8^r$ and $\langle r^2, rs\rangle \simeq D_8$ are anomalous.

\subsubsection{$\CE_{\mathbb{Z}_2,\operatorname{II}}^{1} \Rep D_8 = \Rep SD_{16}$}
For $\CE_{\mathbb{Z}_2,\text{II}}^{1} \Rep D_8 = \Rep SD_{16}$, following the group-theoretical analysis described previously, we find there are two SPT phases:
\begin{enumerate}
    \item The first SPT phase is acquired from the partial SSB phase of $\VEC_{SD_{16}}$ with $\langle r^2, rs\rangle \simeq Q_8$ unbroken. Notice that, unlike $D_8$, $Q_8$ does not admit non-trivial SPT. The corresponding magnetic Lagrangian algebra is,
    \begin{equation}
        \CL_m^1=(1,\chi _1)+(1,\chi _2)+(r^4,\chi _1)+(r^4,\chi _2)+(r^2,\chi _0^{r})+(r^2,\chi _4^{r})+2 (r s,\chi _0^{rs})
    \end{equation}
    where the electric Lagrangian algebra is fixed as,
    \begin{equation}
        \CL_e =(1,\chi _1)+(1,\chi _3)+(1,\chi _5)+(1,\chi _6)+(1,\chi _7)+(s,\chi^s _{0,0})+(s,\chi^{s}_{0,1})~.
    \end{equation}
    The details of $\CZ(SD_{16})$ are listed in App. \ref{app:ZSD16}.
    \item The second SPT phase is acquired from the partial SSB phase of $\VEC_{SD_{16}}$ with $\langle r \rangle \simeq \doubleZ_8$ unbroken.
    \begin{equation}
        \CL_m^2=(1,\chi _1)+(1,\chi _4)+(r^4,\chi _1)+(r^4,\chi _4)+2(r,\chi _0^{r})+2(r^2,\chi _0^{r})+2(r^5,\chi _0^{r})~.
    \end{equation}
\end{enumerate}
Similarly to the previous cases, we solve the SPT data explicitly. More concretely, we gauge fix the $F$-symbols and the SPTs of $\Rep SD_{16}$ can be solved similarly, we can take 
\begin{equation}
    t_{\CD_{i}\mu,g}^{{}^g\CD_{i}\nu} =\nu^\CD_{i}(g)^{-1}t_{g,\CD_{i} \mu}^{{}^g\CD_{i} \nu}=\{\sigma^0,\sigma^1,\sigma^3,-\ii \sigma^2\} ~,
\end{equation}
and the two SPTs phases of $\Rep SD_{16}$ can be similarly distinguished by the choices of $\nu^\CD(g)$'s:
\begin{equation}
\begin{aligned}
    \nu^\CD_{\text{I}}(a)=-1,\quad t_{\CD_{i}\mu,{}^g\CD_{i}\nu}^g = \{\sigma^3,\ii \sigma^2,\sigma^0,-\sigma^1\} ~, \\
    \nu^\CD_{\text{II}}(ab)=-1,\quad t_{\CD_{i}\mu,{}^g\CD_{i}\nu}^g = \{\sigma^0,\sigma^1,\sigma^3,-\ii \sigma^2\}  ~,
\end{aligned}
\end{equation}
where unlisted $\nu^\CD(g) = 1$. The complete solutions are listed in the supplementary Mathematica file \cite{MMAfile}. We further compute the interface algebras and their representations. We find
\begin{equation}
    \begin{array}{|c|c|c|}
    \hline
         & \operatorname{SPT-I} & \operatorname{SPT-II}   \\
    \hline
        \operatorname{SPT-I} & \substack{8 \, \text{$\mathbf{1}$-dim irreps}  \\ 2 \, \text{$\mathbf{2}$-dim irreps}}& 4 \, \text{$\mathbf{2}$-dim irreps} \\
    \hline
        \operatorname{SPT-II} &  & 16 \, \text{$\mathbf{1}$-dim irreps}  \\ 
    \hline
    \end{array} ~.
\end{equation}
Again, we find that the interface algebra between different fiber functions can only have 2-dimensional representations for the interface between different fiber functors.

It is worth mentioning that since $\Rep(SD_{16})$ can be obtained from $\Rep D_{16}$ by changing the symmetry fractionalization class as discussed around \eqref{eq:D8nm-z2}, $\Rep SD_{16}$ admit an intrinsic gapless SPT phase corresponding to the condensable algebra $\scriptA_\text{igSPT}=(1,\chi_1)+(r^4,\chi_{3})$\cite{Ruben2021igSPT,huang2023igspt,wen2023igSPT,sakura2024igspt}. The $\scriptA_\text{igSPT}$ is not maximal, therefore it yields a gapless state. Furthermore, because $\scriptA_\text{igSPT}\cap \scriptL_e=\{(1,\chi_1)\}$ and $\scriptA_\text{igSPT}$ is not the subalgebra of any magnetic Lagrangian algebra, $\scriptA_\text{igSPT}$ describes an intrinsic gapless SPT phase \cite{sakura2024igspt}. Notice that this igSPT relates to the changing of the fractionalization class, as the anyon $(r^4,\chi_{3})$ arises from the one we use to change the fractionalization class. 

\subsubsection{$\CE_{\mathbb{Z}_2,\operatorname{II}}^{2} \Rep D_8$}
For $\CE_{\mathbb{Z}_2,\text{II}}^{2} \Rep D_8$, using its group-theoretical construction of gauging $\mathbb{Z}_2^s$ in $\VEC_{SD_{16}}^{\omega_{4}}$, we find it does not admit SPT phases because both subgroups $\mathbb{Z}_8^r$ and $\langle r^2,rs\rangle \simeq Q_8$ are anomalous. 

\subsection{Examples}
As a first example, notice that the $c=1$ compact boson at orbifold branch also admits a $\Rep D_8$ symmetry with $\mathbb{Z}_2\times \mathbb{Z}_2$ symmetry generated by $\eta,\eta_m$ and the duality line $\CN = \CL_{\frac{\pi}{2},0}$. It is straightforward to check that the theory is self-dual under gauging $\dsi \oplus \eta \oplus \CL_{\frac{\pi}{2},0}$. The dualities $(\CD_1,\CD_2)$ are $(\CL_{\frac{\pi}{4},0}, \CL_{\frac{3\pi}{4},0})$, and together they generate the $\CE_{\doubleZ_2,\text{I}}^{(1,+)}\Rep D_{8}$ fusion category. This is because gauge $\doubleZ_2^\eta$ maps us back to the circle branch and the dual symmetry $D_{16}$ is a subgroup of $U(1)_m \rtimes \doubleZ_2^C$ therefore free of anomaly. Hence, from the group-theoretical construction, we see that the $\CD_i$'s generate the $\Rep D_{16}$ category.

On the other hand, it is not so easy to find interesting examples of the fusion categories with type II fusion rules. Tautological examples can be found by embedding $SD_{16}$ into $S_8$ and potentially to a bigger group, and then gauge the $\mathbb{Z}_2^{\tilde{s}}$-symmetry. For instance, one can consider the diagonal $\Spin(8)_1$ WZW model which contains an anomaly free $SD_{16}$ symmetry embedded in the diagonal $\Spin(8)$ global symmetry. Gauging the corresponding $\mathbb{Z}_2^{\tilde{s}}$ symmetry naturally gives the $\Rep SD_{16}$ symmetry from the group-theoretical construction. 

\section*{Acknowledgements}
We are grateful to Ken Intriligator, Theo Jacobson, Xingyang Yu, Conghuan Luo, Yichul Choi, John McGreevy, Deepak Aryal, Aiden Sheckler, Yi-Zhuang You, Seolhwa Kim for interesting discussions. Research of D.C.L. after September 2024 is supported by the Simons Collaboration on Ultra-Quantum Matter, which is a grant from the Simons Foundation 651440. Z.S. is supported by the Simons Collaboration on Global Categorical Symmetries. Z.Z. is supported in part by the Simons Collaboration on Global Categorical Symmetries, and by Simons Foundation award 568420.

\newpage

\appendix
\section{Extending $\VEC_{\doubleZ_2}$ with a trivially acting $\doubleZ_2$}\label{app:Z2_example}
In this appendix, we consider a simple example where $\VEC_{\doubleZ_2}$ is extended by a trivially acting $\doubleZ_2$ in the bulk. Since $\doubleZ_2$ acts trivially, this is just a central extension of the group with the trivial action. Despite the example being very simple, we want to demonstrate two interesting phenomena which happen in the case of non-invertible symmetries as well. We also want to point it out there are many literature discuss this, for instance, see \cite{Barkeshli:2014cna,Wang:2017loc}.

First, when the $\doubleZ_2^{[0]}$ symmetry in the bulk fractionalizes on some Abelian anyon that is not condensed on the boundary, it will lead to a non-trivial central extension on the boundary extended fusion category symmetries. Second, for certain choice of the fractionalization class, stacking a $\mathbb{Z}_2$ SPT does not change the SET phase, therefore leads to identical fusion category symmetry on the boundary. In this simple example, we can see this very transparently from the bulk, that is the SPT phase is trivialized to a boundary counter term due to the certain choice of the fractionalization class.

Take $\CD = \VEC_{\doubleZ_2^a}$, and since we are taking the trivially acting $\doubleZ_2$ in the bulk, the $\doubleZ_2$-extended fusion category $\CC = \VEC_{\Gamma}^\omega$ for some finite group $\Gamma$ and $\omega \in H^3(\Gamma,U(1))$. Let's first analyze this from the boundary point of view without using fusion category language. The finite group $\Gamma$ fits into the following short exact sequence
\begin{equation}
    0 \rightarrow \doubleZ_2^a \rightarrow \Gamma \rightarrow \doubleZ_2^b \rightarrow 0 ~,
\end{equation}
which is classified by $H^2(\doubleZ_2^b, \doubleZ_2^a) = \doubleZ_2$. A choice of representatives of $\mathfrak{w} \in H^2(\doubleZ_2^b, \doubleZ_2^a)$ is given by $w_0(b,b) = \dsi$ for the trivial class and $w_1(b,b) = a$ for the non-trivial class (where the unspecified 2-cocycle takes value $\dsi$). When we choose $w_0$, the resulting $G = \doubleZ_2^a \times \doubleZ_2^b$. There are four choices of the anomalies parameterized by $\doubleZ_2^{I,b} \times \doubleZ_2^{II,a,b}$, where the first factor denotes the self-anomaly of $\doubleZ_2^b$ and the second factor denotes the mixed anomaly between $\doubleZ_2^a$ and $\doubleZ_2^b$. When we choose $w_1$ instead, we get $G = \doubleZ_4$ as the group multiplication is modified to be $b^2 = a$. There are only two choices of the anomaly taking values in the $\doubleZ_2$ subgroup of $H^3(\doubleZ_4,U(1)) = \doubleZ_4$, as the $\doubleZ_2^a$ subgroup must remain anomaly free. 

The classification of the bulk SET is also known from \cite{Barkeshli:2014cna}. Since the bulk $\doubleZ_2$-symmetry acts trivially, both fractionalization classes and the discrete torsions admit a canonical trivial choice (where on the boundary it leads to $\VEC_{\doubleZ_2 \times \mathbb{Z}_2}$). The bulk $\mathbb{Z}_2$ symmetry admits four different fractionalization classes valued in $H^2(\doubleZ_2, \doubleZ_2^e\times \doubleZ_2^m)$, effectively parameterized by the Abelian anyons in the bulk. When it fractionalizes on $\dsi,e,m$, changing the discrete torsion by stacking a 3d $\doubleZ_2$-SPT will change the SET (therefore this can change the boundary fusion category symmetries); while if it fractionalizes on the fermion $em$, stacking a 3d $\doubleZ_2$-SPT will not change the SET and its effect can be removed by relabeling the bulk twist defects. It is not hard to establish the connection between bulk and boundary summarized in the following Table \ref{tab:Z2_Z2_bulk_bdy}, and we will demonstrate this concretely following our generic approach described in the main text. 
\begin{table}
    \centering
    \begin{tabular}{|c|c|}
        \hline
         boundary symmetry extension & bulk SET data \\
        \hline
        \hline
         $\doubleZ_2^a \times \doubleZ_2^b$ with trivial anomaly & $\nu_{\dsi}$ and $\omega_0$ \\
        \hline
         $\doubleZ_2^a \times \doubleZ_2^b$ with self-anomaly of $\doubleZ_2^b$ & $\nu_{\dsi}$ and $\omega_1$ \\
        \hline
         $\doubleZ_2^a \times \doubleZ_2^b$ with mixed anomaly between $\doubleZ_2^a$ and $\doubleZ_2^b$ & $\nu_{e}$ and $\omega_0$ \\
        \hline
         $\doubleZ_2^a \times \doubleZ_2^b$ with mixed anomaly and self-anomaly of $\doubleZ_2^b$ & $\nu_{e}$ and $\omega_1$ \\
        \hline
         $\doubleZ_4$ with trivial anomaly & $\nu_{m}$ and $\omega_0$ (or $\omega_1$) \\
        \hline
         $\doubleZ_4$ with non-trivial anomaly & $\nu_{em}$ and $\omega_0$ (or $\omega_1$) \\
        \hline
    \end{tabular}
    \caption{We use $\nu_g \in H^2(\doubleZ_2, \doubleZ_2^e \times \doubleZ_2^m)$ where $g = \dsi, e, m, em$ to denote the choice of the fractionalization. And we use $\omega_0, \omega_1$ to denote two choices in $H^3(\doubleZ_2, U(1))$. Here, our convention is that the symmetry boundary is acquired from condensing $e$. It is interesting to notice that while there are two inequivalent choices of discrete torsions when fractionalizing on $m$, they lead to equivalent extension on the boundary as we will demonstrate explicitly.}
    \label{tab:Z2_Z2_bulk_bdy}
\end{table}

Since for all six cases, the bulk $\doubleZ_2$ symmetry remains the same. Then, by our discussion in Section \ref{sec:transformed_F_symbols}, we can start with $\VEC_{\doubleZ_2 \times \doubleZ_2}$ to derive the rest five fusion categories. 

Let's start with the case where the bulk $\doubleZ_2$ fractionalizes on $\dsi$ or $e$. In this case, since both Abelian anyons correspond to the identity line, the fusion rules will not be modified and only anomaly would be activated. It is obvious that if we activate a non-trivial SPT for the bulk $\doubleZ_2$-symmetry, we will turn on the self-anomaly of $\doubleZ_2^b$. It is also straightforward to see that if we turn on the fractionalization class $\nu_e$ on the electric line $e$, we activate a mixed anomaly directly from the $F$-symbol. Parameterizing the $\doubleZ_2^a \times \doubleZ_2^b$ as $(i,j)$ where $i,j=0,1$, this fractionalization will modify the boundary fusion junctions as (where by $[x]$ we mean $x \mod 2$)
\begin{equation}
 ~.
\end{equation}

The modified $\widetilde{F}$-symbols are given by 
\begin{equation}
   %
 ~,\label{eq:Z2Z2Ftildedef}
\end{equation}
Since the original $F$-symbols are simply trivial, the only non-trivial contribution to $\widetilde{F}$ comes from the braiding between the bulk anyon $e^{\frac{1}{2}(j_1 + j_1 - [j_1 + j_2])}$ with the boundary line $(i_3,j_3)$, which leads to the $F$-symbol
\begin{equation}
     \widetilde{F}^{(i_1,j_1),(i_2,j_2),(i_3,j_3)} = (-1)^{\frac{1}{2}i_3 (j_1 + j_2 - [j_1 + j_2])} ~.
\end{equation}
This is nothing but the mixed anomaly between $\doubleZ_2^a$ and $\doubleZ_2^b$.

Then we consider the fractionalization on the $m$ anyon. Fractionalizing on it will modify the fusion junction 
\begin{equation}
    \begin{tikzpicture}[baseline={([yshift=-1ex]current bounding box.center)},vertex/.style={anchor=base,
    circle,fill=black!25,minimum size=18pt,inner sep=2pt},scale=0.5]
        \draw[black, thick, ->-=.3, line width = 0.4mm] (2,-2) -- (2,1.333);

        \draw[black, line width = 0.4mm] (1.333,2) arc (180:270:0.667);

        \draw[black, line width  = 0.4mm, ->-=0.5] (2.667,2) -- (2.667,4);
        
        \draw[black, line width = 0.4mm, ->-=0.5] (1.333,2) -- (1.333,4);
        
        \draw[black, line width = 0.4mm] (2.667,2) arc (0:-90:0.667);

        \draw[red, line width = 0.4mm, ->-=0.5] (2,0) arc (270:360:2 and 4);
        
        \filldraw[black] (2,1.333) circle (2pt);

        \node[black, above] at (3,4) {\tiny ${}_{(i_2,j_2)}$};
        \node[black, below] at (2,-2) {\tiny ${}_{([i_1+i_2 + \frac{1}{2}(j_1 + j_2 - [j_1 + j_2])],[j_1+j_2])}$}; 
        \node[black, above] at (1.3,4) {\tiny ${}_{(i_1,j_1)}$};

        \node[red, right] at (5,4) {\tiny $m^{\frac{1}{2}(j_1 + j_2 - [j_1 + j_2])}$};
    \end{tikzpicture} ~.
\end{equation}
Notice that the outcome of the fusion is modified as $m$ does correspond to the $\doubleZ_2^a$-symmetry line on the boundary. The transformed $\widetilde{F}$ can be defined analogously as in \eqref{eq:Z2Z2Ftildedef} by replacing $e$ with $m$, and since $m$ has trivial braiding with the boundary line, we conclude that
\begin{equation}
    \widetilde{F}^{(i_1,j_1),(i_2,j_2),(i_3,j_3)} = 1 ~.
\end{equation}
On the other hand, if we stack a non-trivial $\doubleZ_2$ SPT, on the boundary it will modify the $\widetilde{F}$ to be
\begin{equation}
    \widetilde{F}^{(i_1,j_1),(i_2,j_2),(i_3,j_3)} = (-1)^{j_1 j_2 j_3} ~.
\end{equation}
However, as one can verify explicitly that this $\widetilde{F}$ is cohomologically trivial in $H^3(\doubleZ_4,U(1))$, as
\begin{equation}\label{eq:Z4_trivialization}
    \widetilde{F}^{(i_1,j_1),(i_2,j_2),(i_3,j_3)} = (d\phi)((i_1,j_1),(i_2,j_2),(i_3,j_3)) ~, \quad \text{where 
  } \phi((i_1,j_1),(i_2,j_2)) = (-1)^{i_1 j_2} ~,
\end{equation}
hence we conclude that these two SETs, even though are different in the bulk, lead to equivalent symmetry extension on the boundary. One can also see this directly from the bulk. The action of the bulk SET with the non-trivial SPT stacked on is 
\begin{equation}
    i \pi \int  a_1 \cup \delta \widehat{a}_1 + \widehat{a}_1 \cup B_1 \cup B_1 + B_1 \cup B_1 \cup B_1  ~,
\end{equation}
where $B_1$ is the background field for the trivially acting $\doubleZ_2$ $0$-form symmetry in the bulk. We want to show that with the Dirichlet boundary condition $a_1 = A_1$, the bulk SPT term is trivialized to a counter term on the boundary. To see this, notice that from the e.o.m. of $\widehat{a}_1$, we have
\begin{equation}
    \delta a_1 = B_1 \cup B_1 ~,
\end{equation}
which then implies that 
\begin{equation}
    i\pi \int_{\mathcal{M}_3} B_1 \cup B_1 \cup B_1 = i\pi \int_{\mathcal{M}_3} \delta a_1 \cup B_1 = i\pi \int_{\mathcal{M}_3} \delta (a_1 \cup B_1) = i \pi \int_{\partial\mathcal{M}_3} A_1 \cup B_1 ~,
\end{equation}
which precisely matches \eqref{eq:Z4_trivialization}.

Before moving, this phenomenon also has a simple interpretation from the boundary. For those who familiar with the LHS spectral sequence, it simply follows from the fact that the specific differential $d_2: H^1(G,H^1(A,U(1))) \rightarrow H^3(G,H^0(A,U(1)))$ is given by the cup product with the $\mathfrak{w}\in H^2(G,A)$ describing the extension
\begin{equation}
    0 \rightarrow A \rightarrow \Gamma \rightarrow G \rightarrow 0 ~,
\end{equation}
which means the image $d_2(H^1(G,H^1(A,U(1)))) \subset H^3(G,H^0(A,U(1)))$ is trivialized by the extension. For more details, see Appendix D of \cite{Wang:2017loc}. Similar phenomena also appears in the context of a $2$-group \cite{Benini:2018reh}.

Finally, let's consider the fractionalization on the $em$ anyon, which will modify the fusion junction as
\begin{equation}
    \begin{tikzpicture}[baseline={([yshift=-1ex]current bounding box.center)},vertex/.style={anchor=base,
    circle,fill=black!25,minimum size=18pt,inner sep=2pt},scale=0.5]
        \draw[black, thick, ->-=.3, line width = 0.4mm] (2,-2) -- (2,1.333);

        \draw[black, line width = 0.4mm] (1.333,2) arc (180:270:0.667);

        \draw[black, line width  = 0.4mm, ->-=0.5] (2.667,2) -- (2.667,4);
        
        \draw[black, line width = 0.4mm, ->-=0.5] (1.333,2) -- (1.333,4);
        
        \draw[black, line width = 0.4mm] (2.667,2) arc (0:-90:0.667);

        \draw[red, line width = 0.4mm, ->-=0.5] (2,0) arc (270:360:2 and 4);
        
        \filldraw[black] (2,1.333) circle (2pt);

        \node[black, above] at (3,4) {\tiny ${}_{(i_2,j_2)}$};
        \node[black, below] at (2,-2) {\tiny ${}_{([i_1+i_2 + \frac{1}{2}(j_1 + j_2 - [j_1 + j_2])],[j_1+j_2])}$}; 
        \node[black, above] at (1.3,4) {\tiny ${}_{(i_1,j_1)}$};

        \node[red, right] at (5,4) {\tiny $(em)^{\frac{1}{2}(j_1 + j_2 - [j_1 + j_2])}$};
    \end{tikzpicture} ~.
\end{equation}
Similar to the case of fractionalizing on $e$, the only contribution to $\widetilde{F}$ arising from the braiding, which leads to
\begin{equation}
    \widetilde{F}^{(i_1,j_1),(i_2,j_2),(i_3,j_3)} = (-1)^{\frac{1}{2}i_3 (j_1 + j_2 - [j_1 + j_2])} ~.
\end{equation}
which represents a non-trivial, order-$2$ anomaly in $H^3(\doubleZ_4,U(1))$. By the similar argument, one can show that stacking a non-trivial $\doubleZ_2$-SPT in the bulk does not change the $\widetilde{F}$ in a gauge-inequivalent way, therefore leads to equivalent fusion category symmetry on the boundary. We want to quickly point it out that unlike the case of fractionalizing on $m$, here, stacking a SPT does not even change the SET in the bulk, as pointed it in \cite{Barkeshli:2014cna}.

\section{Details of some Drinfeld centers}
\subsection{Computing $\CZ(\VEC_G^\omega)$}\label{app:DW_spectrum}
Below we list the formula required for computing $\CZ(\VEC_G^\omega)$ for the specific situation encountered in this paper. More generic result can be found in \cite{Coste:2000tq}.

The Drinfeld center of $\CZ(\VEC_G^\omega)$ contains simple anyons which are parameterized by $([a],\pi_a)$, where $[a]$ is a conjugacy class of $G$ and $\pi_a$ is an irrep of $C_G(a)$ twisted by some $2$-cocycle $\beta_a(h,k)$. Here,
\begin{equation}
    C_G(a) := \{g\in G: ag=ga\} ~,
\end{equation}
and 
\begin{equation}
    \beta_a(h,k) = \frac{\omega(a,h,k)\omega(h,k,(hk)^{-1}ahk)}{\omega(h,h^{-1}ah, k)} ~.
\end{equation}
Notice that as a $2$-cocycle in $H^2(C_G(a), U(1))$, sometimes $\beta_a$ can be cohomologically trivial, meaning that one can find an 1-cochain $\epsilon_a: C_G(a)\rightarrow U(1)$ satisfying
\begin{equation}
    \beta_a(h,k)=(\delta \epsilon)(h,k)=\epsilon_a(h) \epsilon_a(k)\epsilon_a(hk)^{-1} ~,\quad \epsilon_{x^{-1} a x}(x^{-1}h x)=\frac{\beta_a(x,x^{-1} h x)}{\beta_a(h,x)}\epsilon_a(h) ~.
\end{equation}
where $x\in G$ and $h,k\in C_G(a)$ In this case, upon choosing an $\epsilon_a$, any irrep $V(h)$ twisted by $\beta_a$ (that is, $V(h) V(k) = \beta_a(h,k) V(hk)$) are related to the ordinary irrep $U(h)$ via 
\begin{equation}
    V(h) = U(h) \epsilon_a(h) ~.
\end{equation}
This means we can use the character $\chi_a$ of $C_G(a)$ together with a fixed explicit choice of $\epsilon_a$ to label the anyons.

Consider the special case where every $\beta_g$ is cohomologically trivial, we can then denote the simple anyon as $(a,\chi_a)$ as discussed above, where $a$ is the representative of the conjugacy class $a$ and $\chi_a$ is a character of the irrep. Then the modular matrices $S,T$ are given by
\begin{align}\label{eq:STmatrices}
    S^\omega_{(a,\chi),(a',\chi')} &= \frac{1}{\abs{C_G(a)}\abs{C_G(a')}}\sum_{g\in G(a,a')} \chi^* (g a' g^{-1}) \chi^*(g^{-1} a g) \sigma^*(a|g a' g^{-1}),\\
    T^\omega_{(a,\chi),(a',\chi')} &= \delta_{a,a'}\delta_{\chi,\chi'}\frac{\chi(a)}{\chi(\dsi)}\epsilon_a(a).
\end{align}
where $G(a,a') = \{g\in G| agbg^{-1} =g b g^{-1}a\}$, and $\sigma(h|g) = \epsilon_h(g)\epsilon_g(h)$. 

\subsection{Matching between $\CZ(\VEC_{D_8})$ and $\CZ(\Rep D_8)$}\label{app:VECD8_match}
Let us start with the simple case where the TY is $\Rep D_8$. One can compute its Drinfeld center either as $\CZ(\VEC_{D_8})$ or $\CZ(\Rep D_8)$. Here, we compute it as $\CZ(\VEC_{D_8})$ and provide a concrete anyon spectrum match with the $\CZ(\Rep D_8)$ calculation. Recall that we parameterize $D_8$ as
\begin{equation}
    D_8 =\langle r,s| r^4 =s^2 = \dsi, s r s^{-1} = r^{-1} \rangle ~,
\end{equation}
and we denote the group element as  $r^i s^j \equiv (i,j)$. The anyons in $\CZ(\VEC_{D_8})$ are labeled by the pair $(a,\chi^a)$, where $a$ is the representative of the conjugacy class of $D_8$ and $\chi$ is the irreducible character of $C_{D_8}(a)$. In this case, since there is no twist, all $\beta_a$'s are explicitly $1$ and $\epsilon_a = 1$ as well. The complete set of simple anyons are as follows,
\begin{equation}
\begin{array}{c|c|c|c|c|c}
         & \dsi & r^2 & r & s & rs \\ \hline
        C_{D_8}(a) &   D_8 & D_8 & \IZ_4 = \langle r \rangle  & \IZ_2\times \IZ_2 = \langle s, r^2 \rangle & \IZ_2 \times \IZ_2 = \langle rs, r^2 \rangle \\\hline
        \text{character} & \chi_{i}^{\dsi}, i=1\sim5 &\chi_{i}^{r^2}, i=1\sim 5&  \chi^{r}_j,j\in \IZ_4 & \chi^{s}_{a,b},a,b\in \IZ_2  & \chi^{rs}_{a,b},a,b\in \IZ_2
    \end{array}
\end{equation}
Because $C_{D_8}(\dsi) = C_{D_8}(r^2) = D_8$, they share the same character labels, which are given by the character table of $D_8$ itself
\begin{equation}\label{eq:D8CTab}
    \begin{array}{|c|c|c|c|c|c|}
    \hline
         & \dsi & r^2 & r & s & rs \\ \hline
        \chi_1^{D_8} & 1& 1& 1& 1& 1  \\ \hline
        \chi_2^{D_8} & 1& 1& 1&-1&-1  \\ \hline
        \chi_3^{D_8} & 1& 1&-1& 1&-1  \\ \hline
        \chi_4^{D_8} & 1& 1&-1&-1& 1  \\ \hline
        \chi_5^{D_8} & 2&-2& 0& 0& 0 \\
    \hline
    \end{array} ~.
\end{equation}
therefore, $\chi_{i}^{\dsi} = \chi_{i}^{r^2} = \chi_i^{D_8}, i=1\sim 5$. And the other characters are given by
\begin{equation}
    \chi_j^{r}: r^k \rightarrow \exp(\frac{j k 2\pi \ii }{4}),\quad \chi^{s}_{a,b}:  (s)^k (r^2)^l\rightarrow (-1)^{a k+b l},\quad \chi^{rs}_{a,b}: (rs)^k (r^2)^l  \rightarrow (-1)^{a k+b l} ~.
\end{equation}
Therefore the $D_8$ quantum double has 22 anyons. The modular matrices can be computed using \eqref{eq:STmatrices}, and by matching them with those for the $\scriptZ(\Rep D_8)$, we find
\begin{equation}
\begin{aligned}
&\begin{array}{|c|c|c|c|c|c|c|c|}
\hline
 (\dsi,\chi_1) & (\dsi,\chi_3) & (\dsi,\chi_4) &  (r^2,\chi_1) &  (r^2,\chi_2) &  (r^2,\chi_4) &  (r^2,\chi_3) & (\dsi,\chi_2) \\
\hline
 X_{(\dsi,1)} & X_{(a,1)} & X_{(b,1)} & X_{(ab,1)} & X_{(\dsi,-1)} & X_{(a,-1)} & X_{(b,-1)} & X_{(ab,-1)} \\\hline
\end{array}\\
&\begin{array}{|c|c|c|c|c|c|}
\hline 
 (s,\chi_{0,1}) & (r s,\chi_{0,1}) & (r,\chi_0) & (r,\chi_2) & (r s,\chi_{1,1}) & (s,\chi_{1,1}) \\
\hline
 Y_{\{\dsi,a\}} & Y_{\{\dsi,b\}} & Y_{\{\dsi,ab\}} & Y_{\{a,b\}} & Y_{\{a,ab\}} & Y_{\{b,ab\}} \\\hline
\end{array}\\
&\begin{array}{|c|c|c|c|c|c|c|c|}
\hline
  (\dsi,\chi_5) & (s,\chi_{0,0}) & (r s,\chi_{0,0}) & (r,\chi_1) & \left(r^2,\chi_5\right) & (s,\chi_{1,0}) & (r s,\chi_{1,0}) & (r,\chi_3)\\
\hline
 Z_{(2,1)} & Z_{(3,1)} & Z_{(1,1)} & Z_{(4,-1)} & Z_{(2,-1)} & Z_{(3,-1)} & Z_{(1,-1)} & Z_{(4,1)} \\\hline
\end{array}
\end{aligned} ~.
\end{equation}
Corresponding to the $\Rep D_8$ symmetry boundary, the electric Lagrangian algebra is (notice that the choice here is not unique),
\begin{equation}
    \CL_e = ([1],\chi^{D_8}_1)+ ([r^2],\chi^{D_8}_1)+([r],\chi^r_{0})+([s],\chi^{s}_{0,0})+([rs],\chi_{0,0}^{rs}) ~.
\end{equation}
And there are three magnetic Lagrangian algebras given by
\begin{equation}
\begin{aligned}
    \CL_m^1 &= ([\dsi],\chi^{D_8}_1)+([1],\chi^{D_8}_4)+([r^2],\chi^{D_8}_2)+([r^2],\chi^{D_8}_3)+2([rs],\chi_{0,1}^{rs}) ~, \\
    \CL_m^2 &= ([\dsi],\chi^{D_8}_1)+([1],\chi^{D_8}_3)+([r^2],\chi^{D_8}_2)+([r^2],\chi^{D_8}_4)+2([s],\chi_{0,1}^{s}) ~, \\
    \CL_m^3 &= ([\dsi],\chi^{D_8}_1)+([\dsi],\chi^{D_8}_2)+([1],\chi^{D_8}_3)++([1],\chi^{D_8}_4)+2([1],\chi^{D_8}_5) ~.
\end{aligned}
\end{equation}

\subsection{Matching between $\CZ(\Rep H_8)$ and $\CZ(\VEC_{D_8}^\gamma)$}\label{app:RepH8_VECD8gamma_match}
Notice that $\CZ(\Rep H_8) = \CZ(\VEC_{D_8}^\gamma)$, where we parameterize $D_8$ as
\begin{equation}
    D_8 =\langle r,s| r^4 =s^2 = \dsi, s r s^{-1} = r^{-1} \rangle ~,
\end{equation}
The $3$-cocycle $\gamma$ is 
\begin{equation}
    \gamma(g_1,g_2,g_3) = \exp(\frac{4\pi \ii}{4^2}(-1)^{j_1}i_1(i_2+(-1)^{j_2}i_3-[i_2+(-1)^{j_2}i_3]_4))
\end{equation}
where $g_a=(i_a,j_a)$ and $[\cdot]_4$ is shorthand for mod by $4$. It is straightforward to check that the $\beta_a$'s are all cohomologically trivial, therefore the simple anyons still share the same label as the case without a twist. Our choice of $\epsilon_a$'s are
\begin{equation}
\begin{aligned}
    \epsilon_{r^2}(r^i s^j) &= e^{\frac{2\pi \ii}{2}(2j + (-1)^j i)} ~, \\
    \epsilon_{r}(r^i) &= e^{\frac{5\pi \ii}{4} i} ~, \\
    \epsilon_{s}(r^{2i} s^j) &= (-1)^i ~, \\
    \epsilon_{sr}(r^{2i} (rs)^j) &= (-1)^j e^{\frac{\pi \ii}{2} [i]_2} ~.
\end{aligned}
\end{equation}
Again, computing the modular matrices using \eqref{eq:STmatrices} and match them with those for the $\scriptZ(\Rep H_8)$ up to anyon permutation symmetry, we have,
\begin{equation}
\begin{aligned}
&\begin{array}{|c|c|c|c|c|c|c|c|}
\hline
 (\dsi,\chi_1) & (r^2,\chi_3) & (r^2,\chi_4) &  (\dsi,\chi_4) &  (\dsi,\chi_3) &  (r^2,\chi_1) &  (r^2,\chi_2) & (\dsi,\chi_2) \\
\hline
 X_{(\dsi,+1)} & X_{(a,i)} & X_{(b,i)} & X_{(ab,1)} & X_{(\dsi,-1)} & X_{(a,-i)} & X_{(b,-i)} & X_{(ab,-1)} \\\hline
\end{array}\\
&\begin{array}{|c|c|c|c|c|c|}
\hline 
 (s,\chi_{0,1}) & (s,\chi_{0,0}) & (\dsi,\chi_5) & (r^2,\chi_5) & (s,\chi_{1,0}) & (s,\chi_{1,1}) \\
\hline
 Y_{\{\dsi,a\}} & Y_{\{\dsi,b\}} & Y_{\{\dsi,ab\}} & Y_{\{a,b\}} & Y_{\{a,ab\}} & Y_{\{b,ab\}} \\\hline
\end{array}\\
&\begin{array}{|c|c|c|c|c|c|c|c|}
\hline
  (r,\chi_1) & (r s,\chi_{1,1}) & (r s,\chi_{1,0}) & (r,\chi_2) & (r,\chi_3) & (r s,\chi_{0,1}) & (r s,\chi_{0,0}) & (r,\chi_0)\\
\hline
 Z_{(3,+1)} & Z_{(1,+1)} & Z_{(2,+1)} & Z_{(4,1)} & Z_{(3,-1)} & Z_{(1,-1)} & Z_{(2,-1)} & Z_{(4,-1)} \\\hline
\end{array} ~.
\end{aligned}
\end{equation}

\subsection{$\CZ(\VEC_{D_{16}})$}\label{app:ZD16}
The $D_{16}$ is the dihedral group of order 16, which is parameterized by,
\begin{equation}
    D_{16} = \langle r,s|r^8 =s^2= \dsi, s r s^{-1}=r^{-1}\rangle ~,
\end{equation}
and we denote the group elements as $r^i s^j\equiv (i,j)$. The anyons in $D_{16}$ quantum double are labeled by the pair $(a,\chi^a)$, where $a$ is the representative of the conjugacy class of $D_{16}$ and $\chi$ is the irreducible character of $C_{D_{16}}(a)$. In this case, since there is no twist, all $\beta_a$'s are explicitly $1$ and $\epsilon_a = 1$ as well. The complete set of simple anyons are as follows,
\begin{equation}
    \begin{array}{c|c|c|c|c|c|c|c}
        [a] & \dsi & r^4 & r & r^3 & r^2 & r s & s \\ \hline
        C_{D_{16}}(a) &   D_{16} & D_{16} & \IZ_8 = \langle r \rangle & \IZ_8 = \langle r \rangle & \IZ_8 = \langle r \rangle  & \IZ_2\times \IZ_2 = \langle r s, r^5 s\rangle & \IZ_2 \times \IZ_2 = \langle s, r^4s \rangle \\\hline
        \text{character} & \chi_{i}^{\dsi} &\chi_{i}^{r^4}&  \chi^{r}_j,j\in \IZ_8 & \chi^{r}_j,j\in \IZ_8 &\chi^{r}_j,j\in \IZ_8 &\chi^{rs}_{a,b},a,b\in \IZ_2  & \chi^{s}_{a,b},a,b\in \IZ_2
    \end{array}
\end{equation}
where $\chi_{i}^{\dsi}=\chi_{i}^{r^4}=\chi_i, i=1\sim 7$ follow from the character table of $D_{16}$ as shown below,
\begin{equation}\label{eq:D16CTab}
\begin{array}{|c|c|c|c|c|c|c|c|}
\hline
       &\dsi & r^4& r^2 & r & r^3 &s & rs \\\hline
 \chi_1&1 & 1 & 1 & 1 & 1 & 1 & 1 \\\hline
 \chi_2&1 & 1 & 1 & 1 & 1 & -1 & -1 \\\hline
 \chi_3&1 & 1 & 1 & -1 & -1 & 1 & -1 \\\hline
 \chi_4&1 & 1 & 1 & -1 & -1 & -1 & 1 \\\hline
 \chi_5&2 & 2 & -2 & 0 & 0 & 0 & 0 \\\hline
 \chi_6&2 & -2 & 0 & \sqrt{2} & -\sqrt{2} & 0 & 0 \\\hline
 \chi_7&2 & -2 & 0 & -\sqrt{2} & \sqrt{2} & 0 & 0 \\\hline
\end{array} ~.
\end{equation}
And the other characters are given by
\begin{equation}
    \chi_j^{r}: r^k \rightarrow \exp(\frac{j k 2\pi \ii }{8}),\quad \chi^{rs}_{a,b}:  (rs)^k (r^5s)^l\rightarrow (-1)^{a k+b l},\quad \chi^{s}_{a,b}: (s)^k (r^4s)^l  \rightarrow (-1)^{a k+b l} ~.
\end{equation}

We fix the electric Lagrangian algebra to be,
\begin{equation}\label{eq:RepD16eL}
    \CL_e =(1,\chi _1)+(1,\chi _3)+(1,\chi _5)+(1,\chi _6)+(1,\chi _7)+(s,\chi^s _{0,0})+(s,\chi^s _{0,1}) ~,
\end{equation}
which follows from $\Rep D_{16}$ is equivalent to gauging $\IZ_2^s$ in $D_{16}$. There are 3 magnetic Lagrangian algebras for $\CZ(\Rep D_{16})$,
\begin{equation}
\begin{aligned}
    &\CL_{m}^1=(1,\chi _1)+(1,\chi _4)+(r^4,\chi _2)+(r^4,\chi _3)+(r^2,\chi_0^{r})+(r^2,\chi_4^{r})+2 (r s,\chi^{rs}_{0,1}) ~, \\
    &\CL_{m}^2=(1,\chi _1)+(1,\chi _4)+(r^4,\chi _1)+(r^4,\chi _4)+(r^2,\chi_0^{r})+(r^2,\chi _4^{r})+2 (r s,\chi _{0,0}^{rs}) ~,\\
    &\CL_{m}^3=(1,\chi _1)+(1,\chi _2)+(r^4,\chi _1)+(r^4,\chi _2)+2(r,\chi_0^{r})+2(r^2,\chi_0^{r})+2(r^3,\chi_0^{r}) ~.
\end{aligned}
\end{equation}

\paragraph{$\CZ(\VEC_{D_{16}}^{\omega_{0,1}})\cong \CZ(\scriptE_{nm,I}^{(1,-)}\Rep D_8)$}
We fix the electric Lagrangian algebra the same as \eqref{eq:RepD16eL}. The only magnetic Lagrangian algebra is given by,
\begin{equation}
    \CL_m =(1,\chi _1)+(1,\chi _2)+(r^4,\chi _1)+(r^4,\chi _2)+2 (r,\chi _2^r)+2 (r^2,\chi _2^r) +2 (r^3,\chi _2^r)~.
\end{equation}

\paragraph{$\CZ(\VEC_{D_{16}}^{\omega_{4,0}})\cong \CZ(\scriptE_{nm,I}^{(2,+)}\Rep D_8)$}
We fix the electric Lagrangian algebra the same as \eqref{eq:RepD16eL}\footnote{Notice that in all the $\CZ(\VEC_{D_{16}}^\omega)$ we considered, $\beta_a$'s are cohomologically trivial for every $a \in G$. Hence, we use the same characters to label the irreps as we did in the $\CZ(\VEC_{D_8}^\gamma)$ v.s. $\CZ(\VEC_{D_8})$ case.}. The only magnetic Lagrangian algebra,
\begin{equation}
    \CL_m = (1,\chi _1)+(1,\chi _4)+(r^4,\chi _5)+(r^2,\chi^{r} _3)+(r^2,\chi^{r} _7)+(rs,\chi^{rs}_{1,0})+(rs,\chi^{rs}_{1,1})~.
\end{equation}

There are no magnetic Lagrangian algebra for $\CZ(\VEC_{D_{16}}^{\omega_{2/6,0/1}}),\CZ(\VEC_{D_{16}}^{\omega_{4,1}})$.

\subsection{$\CZ(\Rep SD_{16})$}\label{app:ZSD16}
The $SD_{16}$ is the semi-dihedral (or quasi-dihedral) group of order 16, which is parameterized by,
\begin{equation}
    SD_{16} = \langle r,s|r^8 =s^2=\dsi, s r s^{-1}=r^3\rangle ~,
\end{equation}
and we denote the group elements as $r^i s^j\equiv (i,j)$. The anyons in $SD_{16}$ quantum double are labeled by the pair $(a,\chi^a)$, where $a$ is the representative of the conjugacy class of $SD_{16}$ and $\chi$ is the irreducible character of $C_{SD_{16}}(a)$. In this case, since there is no twist, all $\beta_a$'s are explicitly $1$ and $\epsilon_a = 1$ as well. The complete set of simple anyons are as follows,
\begin{equation}
    \begin{array}{c|c|c|c|c|c|c|c}
        [a] & \dsi & r^4 & r & r^5 & r^2 & r s & s \\ \hline
        C_{SD_{16}}(a) &   SD_{16} & SD_{16} & \IZ_8 = \langle r \rangle & \IZ_8 = \langle r \rangle & \IZ_8 = \langle r \rangle  & \IZ_4 = \langle r^5 s\rangle & \IZ_2 \times \IZ_2 = \langle s, r^4 \rangle \\\hline
        \text{character} & \chi_{i}^{\dsi} &\chi_{i}^{r^4}&  \chi^{r}_j,j\in \IZ_8 & \chi^{r}_j,j\in \IZ_8 &\chi^{r}_j,j\in \IZ_8 &\chi^{rs}_{j},j\in \IZ_4  & \chi^{s}_{a,b},a,b\in \IZ_2
    \end{array}
\end{equation}
where $\chi_{i}^{\dsi}=\chi_{i}^{r^4}=\chi_i, i=1\sim 7$ follows from the character table of $SD_{16}$ as shown below,
\begin{equation}\label{eq:SD16CTab}
\begin{array}{|c|c|c|c|c|c|c|c|}
\hline
& \dsi& r^4& r& r^5&r^2& rs&s \\\hline
\chi_1&1 & 1 & 1 & 1 & 1 & 1 & 1 \\\hline
\chi_2&1 & 1 & -1 & -1 & 1 & 1 & -1 \\\hline
 \chi_3&1 & 1 & -1 & -1 & 1 & -1 & 1 \\\hline
\chi_4& 1 & 1 & 1 & 1 & 1 & -1 & -1 \\\hline
\chi_5& 2 & 2 & 0 & 0 & -2 & 0 & 0 \\\hline
\chi_6& 2 & -2 & \ii \sqrt{2} & -\ii \sqrt{2} & 0 & 0 & 0 \\\hline
\chi_7& 2 & -2 & -\ii \sqrt{2} & \ii \sqrt{2} & 0 & 0 & 0 \\\hline
\end{array}
\end{equation}
And the other characters are given by
\begin{equation}
    \chi_j^{r}: r^k \rightarrow \exp(\frac{j k 2\pi \ii }{8}),\quad \chi^{rs}_{j}:  (r^5s)^k \rightarrow \exp(\frac{j k 2\pi \ii }{4}),\quad \chi^{s}_{a,b}: (s)^k (r^4)^l  \rightarrow (-1)^{a k+b l} ~.
\end{equation}

We choose the electric Lagrangian algebra to be,
\begin{equation}
    \CL_e =(1,\chi _1)+(1,\chi _3)+(1,\chi _5)+(1,\chi _6)+(1,\chi _7)+(s,\chi^s _{0,0})+(s,\chi^{s}_{0,1})~.
\end{equation}
which follows from $\Rep SD_{16}$ is equivalent to gauging $\IZ_2^s$ in $SD_{16}$. There are two magnetic Lagrangian algebras for $\CZ(\Rep SD_{16})$,
\begin{equation}
    \begin{aligned}
    &\CL_m^1=(1,\chi _1)+(1,\chi _2)+(r^4,\chi _1)+(r^4,\chi _2)+(r^2,\chi _0^{r})+(r^2,\chi _4^{r})+2 (r s,\chi _0^{rs}) ~, \\
    &\CL_m^2=(1,\chi _1)+(1,\chi _4)+(r^4,\chi _1)+(r^4,\chi _4)+2(r,\chi _0^{r})+2(r^2,\chi _0^{r})+2(r^5,\chi _0^{r}) ~.
    \end{aligned}
\end{equation}
Furthermore, one can explicitly check that there is no magnetic Lagrangian algebras for any other fusion categories with type II fusion rules and we will skip the details here.

\section{Some details on the induction homomorphism: $\ind:\Aut(\CC) \rightarrow \EqBr(\CZ(\CC))$}\label{app:induction_map}
Given a boundary fusion category symmetry $\CC$, its symmetry is captured by the autoequivalence of the fusion category, denoted as $\Aut(\CC)$. For a generic element $\alpha \in \Aut(\CC)$, it maps a simple TDL $a$ to another simple TDL $\alpha(a)$ with the same quantum dimension. Furthermore, it preserves the fusion rule and also acts on the local splitting junction as
\begin{equation}
 ~,
\end{equation}
such that the $F$-symbols are preserved
\begin{equation}\begin{aligned}
     \sum_{\rho,\sigma}[F^{a,b,c}_d]_{(e,\mu,\nu),(f,\rho,\sigma)} [\alpha^{b,c}_f]_{\rho\rho'} [\alpha^{a,f}_d]_{\sigma\sigma'}=\sum_{\mu'\nu'} [\alpha^{a,b}_e]_{\mu\mu'} [\alpha^{e,c}_d]_{\nu\nu'}[F^{\alpha(a),\alpha(b),\alpha(c)}_{\alpha(d)}]_{(\alpha(e),\mu',\nu'),(\alpha(f),\rho',\sigma')}
\end{aligned} ~.
\end{equation}
When the fusion multiplicities are at most $1$ this simplifies to
\begin{equation}
    \frac{[F^{\alpha(a),\alpha(b),\alpha(c)}_{\alpha(d)}]_{\alpha(e),\alpha(f)}}{[F^{a,b,c}_d]_{e,f}}=\frac{\alpha^{b,c}_f \alpha^{a,f}_d}{\alpha^{a,b}_e \alpha^{e,c}_d} ~.
\end{equation}

Naturally, any element in $\Aut(\CC)$ can be lifted to be a symmetry of its SymTFT $\CZ(\CC)$. This is characterized by the induction homomorphism \cite{2016arXiv160304318M}
\begin{equation}
    \ind: \Aut(\CC) \rightarrow \EqBr(\CZ(\CC)) ~.
\end{equation}
However, this homomorphism is neither injective nor surjective. The kernel of this map is characterized by the following. First, consider the group $\Inv(\CC)$ formed by the invertible TDLs in $\CC$. Then, given $X \in \Inv(\CC)$, conjugation by $X$ is a tensor autoequivalence of $\CC$. When $\CC = \VEC_G^\omega$, this is nothing but the inner automorphism of $G$. Generically, this gives a homomorphism
\begin{equation}
    \conj: \Inv(\CC) \rightarrow \Aut(\CC) ~,
\end{equation}
and as shown in \cite{2016arXiv160304318M}, the kernel of $\ind$ is simply the image $\conj(\Inv(\CC))$.

Strictly speaking, $\ind(\alpha)$ consists of two piece of data, the permutation action on the simple anyons as well as the linear action on the space of junctions. Here, we will focus on the first piece of data and describe how $\ind(\alpha)$ transforms the anyons. Consider an anyon $(x,[R^{a,x}_b]^{i}_{\mu,\nu})$ in $\CZ(\CC)$, then the image of $(x,[R^{a,x}_b]^{i}_{\mu\nu})$ is another anyon specified by $(\alpha(x), [R^{a,\alpha(x)}_b]^{\alpha(i)}_{\mu\nu})$, where the transformed $R$-symbol can be easily derived by applying $\alpha$ to \eqref{eq:R_symbol_basis}. It is useful to notice that $\alpha$ may have a non-trivial linear action when mapping the junction between $X$ and $c$ to the junction between $\alpha(X)$ and $\alpha(c)$, but this can be thought of as a gauge transformation when determining the anyon type and will not play an important role here. The transformed $R$-symbols are determined from:
\begin{equation}\label{eq:r-symbol}
    \sum_{\rho,\sigma}  [R^{a,x}_b]^{i}_{(c,\mu,\nu),(d,\rho,\sigma)} [S^X_d]_{\rho\rho'}[\alpha^{a,d}_b]_{\sigma\sigma'}=\sum_{\mu'\nu'} [S^X_c]_{\mu\mu'}[\alpha^{c,a}_b]_{\nu\nu'}[R^{\alpha(a),\alpha(x)}_{\alpha(b)}]^{\alpha(i)}_{(\alpha(c),\mu',\nu'),(\alpha(d),\rho'\sigma')} ~,
\end{equation}
where $(S_c^X)_{\mu\mu'}$ encodes the action on the junctions between $\alpha(X)$ and $c$ as mentioned above. As one can check, the transformed $R$-symbols do satisfy hexagon equations and therefore specify an anyon. To match the transformed $R$-symbols with a known $R$-symbols exactly (instead of just up to gauge transformation), one must fix $S^X_c$'s accordingly.

As a concrete example, let's consider $\CZ(\VEC_G^\omega)$, where simple anyons are labeled by the pair $([a], \pi_a)$. This means $X$ is the direct sum of $\dim \pi_a$ copies of every group elements in the conjugacy class $[a]$, and the possible $R$-symbols are given by
\begin{equation}\label{eq:Gomega_Rsym}
 ~, \quad a' \in [a] ~, \quad \mu = 1,\cdots, \dim \pi_a ~.
\end{equation}
The problem of finding inequivalent solutions to the pentagon equations is essentially the problem of finding irreducible representation of a group on a graded vector space. A classic example is to find the classification of particles in $4$-dim spacetime, where one would first fix a simple choice of the momentum $4$-vector, and consider the irreducible representation of the little group associated to this momentum. The same strategy works here, with the analogy where the momentum is the $a$'s in the conjugacy classes and the little group is $C_G(a)\equiv \{g\in G: ga =ag\}$. Indeed, if we consider the subset of the hexagon equations \eqref{eq:bbhex} where the two boundary TDLs are inside $C_G(a)$, then
\begin{equation}\label{eq:littlegp_irrep}
        \sum_{\nu} [R^{h,([a],\pi_a)}_{ha}]_{\mu\nu}[R^{g,([a],\pi_a)}_{ga}]_{\nu\rho} = [R^{{hg},([a],\pi_a)}_{hga}]_{\mu\rho} \beta_a(h,g), \quad \forall h,g\in C_G(a) ~,
\end{equation}
where $\beta_a = \frac{F^{a,h,g}F^{h,g,{(hg)}^{-1}ahg}}{F^{h,h^{-1}ah,g}}$ becomes a normalised 2-cocycle and thus defines a projective representation of $C_G(a)$. We see indeed that $[R^{g,([a],\pi_a)}_{ga}]_{ij}$ specifies a (projective) representation of $C_G(a)$. And any irreducible (projective) representation of $C_G(a)$ satisfies \eqref{eq:littlegp_irrep} would specify a simple anyon and any two irreducible representations which are similar to each other correspond to the same simple anyon. 

It is straightforward to apply $F_{\phi,\mu}$ on both sides of \eqref{eq:Gomega_Rsym}, we will get another anyon where the irrep data $[R^{g,([a],\pi_a)}_{ga}]_{ij}$ is mapped to (where we have taken $S^X_c$ to be trivial) 
\begin{equation}
    [R^{\phi(g), ([\phi(a)],\phi(\pi_a))}_{\phi(ga)}]_{\mu\nu} = \frac{\mu(g,a)}{\mu(a,g)} [R^{g, ([a],\pi_a)}_{ga}]_{\mu\nu} ~, 
\end{equation}
which, as one can check explicitly, satisfies
\begin{equation}
    \sum_{\nu} [R^{\phi(h), ([\phi(a)],\phi(\pi_a))}_{\phi(ha)}]_{\mu\nu} [R^{\phi(g), ([\phi(a)],\phi(\pi_a))}_{\phi(ga)}]_{\nu\rho} = [R^{\phi(hg), ([\phi(a)],\phi(\pi_a))}_{\phi(hga)}]_{\mu\rho} \beta_{\phi(a)}(\phi(h),\phi(g)) ~,
\end{equation}
therefore does specify an anyon. And the anyon type can be conveniently determined by considering the character of the new (projective) irrep.

\section{Cohomology of (semi)dihedral group}\label{app:Deh_cohomology}
The order $2\ell$ Dihedral group $D_{2\ell}$ is 
\begin{equation}
    D_{2\ell} := \langle r,s | r^{\ell} = s^2 = 1 ~, srs = r^{-1} \rangle ~.
\end{equation}
When $\ell$ is even, its group cohomologies are
\begin{equation}
    H^n(D_{2\ell},\mathbb{Z}) = \begin{cases} \doubleZ ~, \quad & n = 0,\\
    (\mathbb{Z}_2)^{(n-1)/2} ~, \quad & n = 4k + 1 ~, \\
    (\mathbb{Z}_2)^{(n+2)/2} ~, \quad & n = 4k + 2 ~, \\
    (\mathbb{Z}_2)^{(n-1)/2} ~, \quad & n = 4k + 3 ~, \\
    (\mathbb{Z}_{\ell}) \times (\mathbb{Z}_2)^{n/2} ~, \quad & n = 4k ~ (k \neq 0) ~. \\
    \end{cases}
\end{equation}
When $\ell$ is odd, its group cohomologies are
\begin{equation}
    H^n(D_{2\ell},\mathbb{Z}) = \begin{cases} \doubleZ ~, \quad & n = 0,\\
    \mathbb{Z}_2 ~, \quad & n = 4k + 2 ~, \\
    \mathbb{Z}_{\ell} \times \mathbb{Z}_2 ~, \quad & n = 4k ~ (k\neq 0) ~, \\
    0 ~, \quad & n \, \text{odd} ~. \\
    \end{cases}
\end{equation}

The order $8\ell$ ($\ell \geq 2$) semidihedral group $SD_{8\ell}$ is
\begin{equation}
    SD_{8\ell} := \langle r,s | r^{4\ell} = s^2 = 1 ~, srs = r^{2\ell - 1} \rangle ~.
\end{equation}
As an example of metacyclic groups, its integral cohomology has been computed in various places \cite{hayami2018hochschild,larson1988integral}. For even $\ell$, 
\begin{equation}
    H^n(SD_{8\ell},\mathbb{Z}) = \begin{cases} \doubleZ ~, \quad & n = 0,\\
    0 ~, \quad & n = 1, ~ 3, \\
    \mathbb{Z}_{4\ell} \times (\mathbb{Z}_2)^k ~, \quad & n = 4k ~ (k\neq 0) ~, \\
    (\mathbb{Z}_2)^k ~, \quad & n = 4k + 1 ~ (k \neq 0) ~, \\
    (\mathbb{Z}_2)^{k+2} ~, \quad & n = 4k + 2 ~, \\
    (\mathbb{Z}_2)^k ~, \quad & n = 4k + 3 ~ (k \neq 0) ~, \\
    \end{cases}
\end{equation}
while for odd $\ell$, 
\begin{equation}
    H^n(SD_{8\ell},\mathbb{Z}) = \begin{cases} \doubleZ ~, \quad & n = 0,\\
    0 ~, \quad & n = 1, ~ 3, \\
    \mathbb{Z}_{4\ell} \times (\mathbb{Z}_2)^{2k} ~, \quad & n = 4k ~ (k\neq 0) ~, \\
    (\mathbb{Z}_2)^{2k} ~, \quad & n = 4k + 1 ~ (k \neq 0) ~, \\
    \mathbb{Z}_4 \times (\mathbb{Z}_2)^{2k+1} ~, \quad & n = 4k + 2 ~, \\
    (\mathbb{Z}_2)^{2k+1} ~, \quad & n = 4k + 3 ~ (k \neq 0) ~. \\
    \end{cases} 
\end{equation}

\subsection{Restriction map and corestriction (transfer) map}
Let $H$ be a subgroup of $G$. Naturally, a cocycle $\omega$ in $H^*(G,U(1))$ is a naturally a cocycle in $H$ via restriction. This defines the so-called restriction map
\begin{equation}
    \res_{G\rightarrow H}: H^*(G,U(1)) \rightarrow H^*(H,U(1)) ~.
\end{equation}
On the other hand, there is also a homomorphism $\Cor_{H \rightarrow G}: H^*(H,U(1)) \rightarrow H^*(G,U(1))$, which is called the corestriction map or the transfer map. This map admits a explicit formula at the cochain level (e.g. see \cite{weiss1969cohomology}), therefore allows one to construct explicit representative for certain cocycles in $H^*(G,U(1))$ from those in $H^*(H,U(1))$.

Given a cochain $\omega \in C^n(H,U(1))$, its image under the transfer map is given by
\begin{align}
    (\Cor \omega)&(g_1,g_2,\cdots,g_n) \\
    &= \prod_{C \in H \backslash G} \omega(\overline{C} g_1 \overline{C g_1}^{-1}, \overline{C g_1}^{-1} g_1^{-1} g_2 \overline{C g_2}, \overline{C g_2}^{-1} g_2^{-1} g_3 \overline{C g_3}, \cdots, \overline{C g_{n-1}}^{-1} g_{n-1}^{-1} g_n \overline{C g_n}) ~, \nonumber
\end{align}
where $C$'s are the right $H$-cosets in $G$ and $\overline{C}$ denote a fixed choice of representatives for each $C$. It is straightforward to check that $\Cor$ commutes with the differential, hence it defines a homomorphism from $H^n(H,U(1))$ to $H^n(G,U(1))$ at the cochain level. 

It is then left to determine which cocycle in $H^n(G,U(1))$ the $\omega$ is lifted to via the transfer map. For this, we will assume that $H$ is a normal subgroup in $G$. Then, the following relation
\begin{equation}
    \res_{G \rightarrow H} \Cor_{H \rightarrow G}(\gamma) = \prod_{g \in G/H} g \gamma 
\end{equation}
can help us determine the order of the $\Cor_{H\rightarrow G}\gamma$ in concrete examples. For instance, In the construction of the explicit $3$-cocycle of $G = SD_{16}$ via applying the transfer map to the $3$-cocycle of $H = \doubleZ_8$, we see that the order-$8$ generator of $H^3(\doubleZ_8,U(1))$ is only lifted to an order-$4$ generator inside the $\mathbb{Z}_8$-component of $H^3(SD_{16},U(1))$ from the above relation. 

\section{Extension of a finite group and its 3-cocycle}\label{app:extension}
Consider a generic extension of the finite groups captured by the following short exact sequence 
\begin{equation}
    0 \rightarrow N \rightarrow \Gamma \rightarrow G \rightarrow 0 ~.
\end{equation}
In this appendix, we describe the general procedure of extending a $3$-cocycle $\omega_0 \in H^3(N,U(1))$ to a $3$-cocycle $\omega \in H^3(\Gamma,U(1))$. This is done following the Appendix given by \cite{Etingof:2009yvg} and shares some similarity with the work in \cite{Wang:2021nrp}.

To do so, let's first describe the extension data. Notice that we do not require $N$ to lie in the center of $\Gamma$, therefore the extension discussed here is not a central extension in general. Similar to the latter case, as first studied in \cite{schreier1927untergruppen}, a generic extension is characterized by a group homomorphism 
\begin{equation}
    \phi: G \rightarrow \Aut(N) ~,
\end{equation}
as well as a 2-cochain $\chi \in C^2(G,N)$ satisfying
\begin{equation}\label{eq:chi_cond}
\begin{aligned}
    \phi_{g_1}\circ \phi_{g_2}(n) &=\chi(g_1,g_2)\phi_{g_1g_2}(n)\chi^{-1}(g_1,g_2) ~, \\
    (d_\phi \chi)(g_1,g_2,g_3) &= \phi_{g_1}(\chi(g_2,g_3))\chi(g_1,g_2g_3)\chi^{-1}(g_1g_2,g_3)\chi^{-1}(g_1,g_2) = \dsi~.
\end{aligned}
\end{equation}
Denoting a group element of $\Gamma$ as $(n,g) \in N \times G$, the group multiplication law is then given by 
\begin{equation}
    (n_1,g_1) \times (n_2,g_2) = (n_1\phi_{g_1}(n_2)\chi(g_1,g_2),g_1g_2) ~,
\end{equation}
where \eqref{eq:chi_cond} ensures the associativity of the group multiplication. 

Let $G_0$ be a subgroup of $\Aut(N)$, it is interesting to notice that the first condition in \eqref{eq:chi_cond} generically allows an extension with $G = G_0/G_{0,Inn}$ where $G_{0,Inn}$ denotes the subgroup of $G_0$ consisting of all inner automorphisms of $N$. This is the case discussed in Section \ref{sec:gpt_RepH8}, where $N = D_8$ and $G_0 = \mathbb{Z}_4$ generated by $\rho$: since $\rho^2$ is an inner automorphism, we are allowed to construct extensions of the form
\begin{equation}
    0 \rightarrow D_8 \rightarrow D_{16} \rightarrow \mathbb{Z}_2 \rightarrow 0 ~.
\end{equation}

We can derive the extension of the $3$-cocycle with the help of triple-line notation which we now introduce. As mentioned earlier, any element of $\Gamma$ can be denoted as $(n,g) \in N\times G$ and the fusion of $(n_1,g_1)$ and $(n_2,g_2)$ can be represented pictorially as
\begin{equation}
 ~,
\end{equation}
where we use black lines to denote the group elements in $G$ and red and blue lines to denote group elements in $N$, and specifically blue lines are addressed to the junction to ensure the structure of the extension.
$F$-moves can be represented then be represented as
\begin{equation}\label{eq:F-move}
\begin{aligned}
&%
 ~.
    \end{aligned} 
\end{equation}
The whole $F$-move can be decomposed into several elementary moves:
\begin{equation}
\begin{aligned}
      %
 ~
\end{aligned} \label{eq:moves}
\end{equation}
where
\begin{equation}
    \begin{aligned}
       &\mu\in C^1(G,C^2(N,U(1))),\quad \beta \in C^2(G,C^1(N,U(1))), \\
       &\rho \in C^2(G,H^1(N,U(1))),\quad \Theta\in C^3(G,U(1))
    \end{aligned}
\end{equation}
are unknown functions that we need to solve by imposing consistency condition(pentagon equations of $d\omega = 1$). Notice that the above relations hold with some of the red lines replaced with blue lines as they are both elements in $N$. Furthermore, we will also use the fact that the red line and the blue line as trivial crossed braiding as discussed in \cite{2019arXiv191102633G}.

By combining elementary moves in $(\ref{eq:moves})$ to construct the $F$-move in  $(\ref{eq:F-move})$ we can build an ansatz for the 3-cocyle $\omega$:
\begin{equation}
\begin{aligned}
       &\omega[(n_1,g_1),(n_2,g_2),(n_3,g_3)]=\frac{\omega_0[n_1\phi_{g_1}(n_2),\chi(g_1,g_2),\phi_{g_1g_2}(n_3)]\omega_0[n_1,\phi_{g_1}(n_2),\phi_{g_1}(\phi_{g_2}(n_3))]}{\omega_0[n_1\phi_{g_1}(n_2),\phi_{g_1}(\phi_{g_2}(n_3)),\chi(g_1,g_2)]}\\
       &\frac{\omega_0[n_1\phi_{g_1}(n_2\phi_{g_2}(n_3)),\chi(g_1,g_2),\chi(g_1g_2,g_3)]\omega_0[n_1,\phi_{g_1}(n_2\phi_{g_2}(n_3)),\phi_{g_1}(\chi(g_2,g_3))}{\omega_0[n_1\phi_{g_1}(n_2\phi_{g_2}(n_3)),\phi_{g_1}(\chi(g_2,g_3)),\chi(g_1,g_2g_3)]}\\
       &\frac{\rho_{g_1,g_2}[\phi_{g_1}(\phi_{g_2}(n_3))]}{\beta_{g_1,g_2}[\phi_{g_1}(\phi_{g_2}(n_3))]}\mu_{g_1}[\phi_{g_1}(n_2),\phi_{g_1}(\phi_{g_2}(n_3))]\mu_{g_1}[\phi_{g_1}(n_2\phi_{g_2}(n_3)),\phi_{g_1}(\chi(g_2,g_3))]\Theta[g_1,g_2,g_3] ~.
\end{aligned} 
\end{equation}
Pentagon equation that involves $(n_1,g_1),(n_2,\dsi),(n_3,\dsi),(n_4,\dsi)$ imposes consistency condition:
\begin{equation}
\begin{aligned}
  \frac{\phi_{g}\cdot \omega_0}{\omega_0}  &=\delta\mu_g ~,
\end{aligned}    
\end{equation}
where in component it is
\begin{equation}
   \frac{\omega_0(\phi^{-1}_{g_1}(n_2),\phi^{-1}_{g_1}(n_3),\phi^{-1}_{g_1}(n_2))}{\omega_0(n_2,n_3,n_4)}=\frac{\mu_{g_1}(n_3,n_4)\mu_{g_1}(n_2,n_3n_4)}{\mu_{g_1}(n_2,n_3)\mu_{g_1}(n_2n_3,n_4)} ~.
\end{equation}
This condition simply means that unless the cohomology class of $\omega_0$ is invariant under the $\phi_g$'s action, $\omega_0$ can not be lifted.

Pentagon equation that involves $(n_1,g_1),(n_2,g_2),(n_3,\dsi),(n_4,\dsi)$ imposes consistency condition:
\begin{equation}
 \mu_{g_1}(\phi_{g_1}\cdot\mu_{g_2})\alpha^{-1}_{\chi(g_1,g_2)} (c_{\chi(g_1,g_2)}\cdot \mu^{-1}_{g_1g_2})=\delta\beta_{g_1,g_2}
\end{equation}
where $c_n$ stands for conjugation by $n$, where in components it reads
\begin{equation}
\begin{aligned}
\frac{\mu_{g_1}(n_3,n_4)\mu_{g_2}(\phi_{g_1}^{-1}(n_3),\phi_{g_1}^{-1}(n_4))}{\alpha_{\chi_{(g_1,g_2)}}(n_3,n_4)\mu_{g_1g_2}(\chi^{-1}(g_1,g_2)n_3\chi(g_1,g_2),\chi^{-1}(g_1,g_2)n_4\chi(g_1,g_2))}=\frac{\beta_{g_1,g_2}(n_3) \beta_{g_1,g_2}(n_4)}{\beta_{g_1,g_2}(n_3n_4)}
\end{aligned}
\end{equation}
where 
\begin{equation}
    \alpha_{n_1}(n_2,n_3)=\frac{\omega_0(n_2,n_3,n_1)\omega_0(n_1,n_1^{-1}n_2n_1,n_1^{-1}n_3n_1)}{\omega_0(n_2,n_1,n_1^{-1}n_3n_1)} ~.
\end{equation}
In the language of LHS spectral sequence, this is equivalent to demanding $d_2(\omega_0)=0$ on the $E_2$ page.

Pentagon equation that involves $(n_1,g_1),(n_2,g_2),(n_3,g_3),(n_4,\dsi)$ imposes consistency condition:
\begin{equation}
    \begin{aligned}
        &\frac{\beta_{g_1,g_2}(n_4)\beta_{g_1g_2,g_3}(c_{\chi^{-1}(g_1,g_2)}(n_4))}{\beta_{g1,g_2g_3}(c_{\phi_{g_1}(\chi^{-1}(g_2,g_3))}(n_4))\beta_{g_2,g_3}(\phi^{-1}_{g_1}(n_4))}\frac{\gamma_{\chi(g_1,g_2),\chi(g_1g_2,g_3)}(n_4)\mu_{g_1}(\phi_{g_1}(\chi(g_2,g_3)),c_{\phi_{g_1}(\chi^{-1}(g_2,g_3))}(n_4))}{\gamma_{\phi_{g_1}(\chi(g_2,g_3)),\chi(g_1,g_2g_3)}(n_4)\mu_{g_1}(n_4,\phi_{g_1}(\chi(g_2,g_3)))}\\
        &\frac{\rho_{g1,g_2g_3}(c_{\phi_{g_1}(\chi^{-1}(g_2,g_3))}(n_4))\rho_{g_2,g_3}(\phi^{-1}_{g_1}(n_4))}{\rho_{g_1,g_2}(n_4)\rho_{g_1g_2,g_3}(c_{\chi^{-1}(g_1,g_2)}(n_4))}=1
    \end{aligned}
\end{equation}
where
\begin{equation}
    \gamma_{n_1,n_2}(n_3)=\frac{\omega_0[n_1,n_1^{-1}n_3n_1,n_2]}{\omega_0[n_3,n_1,n_2]\omega_0[n_1,n_2,n_2^{-1}n_1^{-1}n_3n_1n_2]}
\end{equation}
This is equivalent to demanding that $d_3(\omega_0)=0$ on the $E_3$ page. 

Finally, once $\mu,\beta,\rho$ are determined from previous conditions, the pentagon equation that involves generic group elements $(n_1,g_1),(n_2,g_2),(n_3,g_3),(n_4,g_4)$ then imposes:
\begin{equation}
 \begin{aligned}
     &\frac{\omega_0[\chi(g_1,g_2),\phi_{g_1g_2}(\chi(g_3,g_4)),\chi(g_1g_2,g_3g_4)]\omega_0[\phi_{g_1}(\chi(g_2,g_3)),\chi(g_1,g_2g_3),\chi(g_1g_2g_3,g_4)]}{\omega_0[\chi(g_1,g_2),\chi(g_1g_2,g_3),\chi(g_1g_2g_3,g_4)]\omega_0[\phi_{g_1}(\chi(g_2,g_3)),\phi_{g_1}(\chi(g_2g_3,g_4)),\chi(g_1,g_2g_3g_4)]}\\
     &\frac{\omega_0[\phi_{g_1}(\phi_{g_2}(\chi(g_3,g_4))),\phi_{g_1}(\chi(g_2,g_3g_4)),\chi(g_1,g_2g_3g_4)]}{\omega_0[\phi_{g_1}(\phi_{g_2}(\chi(g_3,g_4))),\chi(g_1,g_2),\chi(g_1g_2,g_3g_4)]}\frac{\mu_{g_1}(\phi_{g_1}(\chi(g_2,g_3)),\phi_{g_1}(\chi(g_2g_3,g_4)))}{\mu_{g_1}(\phi_{g_1}(\phi_{g_2}(\chi(g_3,g_4)),\phi_{g_1}(\chi(g_2,g_3g_4))))}\\
     &\frac{\beta_{g_1,g_2}(\phi_{g_1}(\phi_{g_2}(\chi(g_3,g_4)))}{\rho_{g_1,g_2}(\phi_{g_1}(\phi_{g_2}(\chi(g_3,g_4)))}=\frac{\Theta[g_1,g_2,g_3g_4]\Theta[g_1g_2,g_3,g_4]}{\Theta[g_1,g_2,g_3]\Theta[g_1,g_2g_3,g_4]\Theta[g_2,g_3,g_4]} ~,
 \end{aligned}   
\end{equation}
where the L.H.S. defines a class in $H^4(G,U(1))$ which must be cohomologically trivial in order for the pentagon equations to be satisfied. This is precisely the $H^4$ obstruction class in the general $G$-extension theory of fusion categories. And when this obstruction vanishes, $\Theta$'s is a trivialization of such class. 

\section{Details of the diagram calculation}
\subsection{Derivation of other crossing relations}\label{app:crossings}
The relation between the $G$-symbols and the $F$-symbols can be derived by evaluating the following diagrams in two different ways:
\begin{equation}
 ~.  
\end{equation}
First, we can directly remove two bubbles using \eqref{eq:fus_spl_relation}
\begin{equation}
    %
 ~.
\end{equation}
Alternatively, we can consider first performing crossing moves for the upper part and lower part simultaneously using \eqref{eq:F-symbols} and \eqref{eq:G-symbols} and then remove the bubbles, where we find
\begin{equation}
\begin{aligned}
    %
 ~.
\end{aligned}
\end{equation}
Since the two calculations must agree, we find
\begin{equation}
    \sum_{f,\rho,\sigma} \left[G_{abc}^d\right]_{(e',\mu',\nu'),(f,\rho,\sigma)} \left[F^{abc}_d\right]_{(e,\mu,\nu),(f,\rho,\sigma)} = \delta_{e,e'} \delta_{\mu,\mu'} \delta_{\nu,\nu'} ~.
\end{equation}

Next, we derive two crossing relations \eqref{eq:other_two_crossings}. The first relation is derived as follows
\begin{equation}
\begin{aligned}
    %
 ~.
\end{aligned} 
\end{equation}
And the second relation can be derived similarly
\begin{equation}
\begin{aligned}
    %
 ~.
\end{aligned} 
\end{equation}

\subsection{Derivation of the $S,T$ matrices}
Before we derive $S$ and $T$ matrices, it will be useful to first evaluate the following diagram:
\begin{equation}\label{eq:F7}
    %
\end{equation}
where $[F^{ab}_{cd}]_{(e,\mu,\nu)(f,\beta,\rho)}=\sqrt{\frac{d_e d_f}{d_d d_a}} \left[F^{c,e,b}_f\right]^{-1}_{(d,\nu,\rho),(a,\mu,\beta)}$.
Notice that in the diagram
\begin{equation}
    %
\end{equation}
is only non-zero when $e = \dsi$ (otherwise shrinking $b$ would lead to a non-trivial junction between the identity line $\dsi$ and the non-identity object $e$). Thus the only contribution in \eqref{eq:F7} comes from $[F^{ab}_{ab}]^{-1}_{(c,\gamma,\alpha),\dsi} = \delta_{\gamma,\alpha}\sqrt{\frac{d_c}{d_ad_b}}$ which can be verified by the following diagram
\begin{equation}
\begin{aligned}
\delta_{e,\dsi} &= \frac{1}{d_ad_b}%
\\
    &= \sum_{c,\mu,\nu}[F^{ab}_{ab}]_{(e,\alpha,\beta)(c,\mu,\nu)}\sqrt{\frac{d_c}{d_ad_b}}\delta_{\mu\nu} ~.
    \end{aligned} 
\end{equation}
Combining these together leads to
\begin{equation}\begin{aligned}
    %
\\
    &= \sqrt{d_ad_bd_c}\delta_{\gamma,\alpha} ~.
    \end{aligned}
\end{equation}
We can now proceed to compute T matrix:
\begin{equation}\label{eq:ST_matrices_cal}
\begin{aligned}
 \theta_X &= \frac{1}{d_X} %
\\
    &=\sum_{a,b,\mu,\alpha}\frac{d_b}{d_X}[R^{a,X}_{b}]_{(a,\mu,\alpha),(a,\mu,\alpha)} ~. \\
    \end{aligned} 
\end{equation}
$S_{XY}$ can be computed using similar techniques:
\begin{equation}
    \begin{aligned}
         S_{XY}&=\frac{1}{d_{\mathcal{Z}(\mathcal{C})}}
    %
\\
    &=\sum_{a,b,c,\mu,\nu,\alpha,\beta}\frac{d_c}{d_{\mathcal{Z}(\mathcal{C})}}[R^{b,X}_c]_{(a,\mu,\alpha),(a,\mu,\beta)}[R^{a,Y}_c]_{(b,\nu,\beta),(b,\nu,\alpha)} ~.
    \end{aligned} 
\end{equation}

\subsection{Derivation of the spin selection rules}
\subsubsection{Non-maximal gauging}\label{app:spin_selection}
To relate $e^{-4\pi is}$ of the states in the defect Hilbert space $\CH_{\CD_i}$ to the action of the symmetry operators, we need to simplify the following diagram:
\begin{equation}\label{eq:der_ss_tp1}
\begin{aligned}
    & %
 ~.
\end{aligned}
\end{equation}
Recall that the symmetry operator is defined as 
\begin{equation}
    %
 ~.
\end{equation}
For simplicity, we will drop the label $\scriptD_i$ and all the black lines denote $\scriptD_i$. The algebraic relations are derived as follows
\begin{equation}\label{eq:tube_alge_der}
\begin{aligned}
    %
 ~.
\end{aligned} 
\end{equation}
This leads to the algebras of the symmetry operators acting on the defect Hilbert space $\CH_{\CD_i}$:
\begin{equation}
    \scriptU[a,b] \scriptU[c,d] = \sum_{e,f} \sqrt{\frac{d_f d_{\CD_i}}{d_b d_d}} \sqrt{\frac{d_a d_c}{d_{e}}} \left[F^{c\CD_i a}_f\right]_{d,b} \left[F^{ca\CD_i}_f\right]_{b,e}^{-1} \left[F^{\CD_i c a}_f\right]^{-1}_{e,d} \scriptU[e,f] ~.
\end{equation}

\subsubsection{Maximal gauging revisited}\label{app:max_spin_selection}
Below we list the new formula for computing the spin selection rule of $\underline{\CE}_{\doubleZ_2}\Rep H_8$ following exactly the same procedure as above. Recall our convention of the symmetry operators acting on the defect Hilbert space are
\begin{equation}
    \begin{tikzpicture}[scale=0.8,baseline={(0,0.75)}]
    \node[above] at (0,3) {\scriptsize$\CD$};
    \node[left] at (0,0.5) {\scriptsize$\CD$};
    \draw[line width = 0.4mm, black, ->-=.6] (0,0) -- (0,1);
    \draw[line width = 0.4mm, black, ->-=.6] (0,2) -- (0,3);
    \node[below] at (0,0) {\scriptsize$\mathcal{O}$};
    \draw[red, line width = 0.4mm, domain=-90:90] plot ({1*cos(\x)}, {1*sin(\x)});
    \draw [red, line width = 0.4mm, domain=90:270, ->-=0.3] plot ({1.5*cos(\x)}, {0.5+1.5*sin(\x)});
    \draw[line width = 0.4mm, black, ->-=0.7] (0,1) -- (0,2);
    \filldraw[] (0,0) circle (1.5pt);
    \node[red, right] at (1.,0.) {\scriptsize$g$};
    
    \end{tikzpicture} \equiv \scriptU_g  ~, \quad \begin{tikzpicture}[scale=0.8,baseline={(0,0.75)}]
    \node[above] at (0,3) {\scriptsize$\CD$};
    \node[left] at (0,0.5) {\scriptsize$\CD$};
    \draw[line width = 0.4mm, black, ->-=.6] (0,0) -- (0,1);
    \draw[line width = 0.4mm, black, ->-=.6] (0,2) -- (0,3);
    \node[below] at (0,0) {\scriptsize$\mathcal{O}$};
    \draw[red, line width = 0.4mm, domain=-90:90] plot ({1*cos(\x)}, {1*sin(\x)});
    \draw [red, line width = 0.4mm, domain=90:270, ->-=0.3] plot ({1.5*cos(\x)}, {0.5+1.5*sin(\x)});
    \draw[line width = 0.4mm, black, ->-=0.7] (0,1) -- (0,2);
    \filldraw[] (0,0) circle (1.5pt);
    \node[red, right] at (1.,0.) {\scriptsize$g$};
    \filldraw[] (0,1) circle (1.5pt);
    \filldraw[] (0,2) circle (1.5pt);
    \node[black, right] at (0,2) {\scriptsize $\nu$};
    \node[black, left] at (0,1) {\scriptsize $\mu$};
    
    \end{tikzpicture} \equiv \scriptU_{\scriptN,\mu\nu} ~,
\end{equation}
and they satisfy the following algebraic relations:
\begin{equation}
\begin{aligned}
    \scriptU_{g} \cdot \scriptU_h &= F^{h\CD g}_{\CD} [F^{hg\CD}_{\CD}]^{-1} G^{\CD}_{\CD hg} \scriptU_{gh} ~, \\
    \scriptU_g \cdot \scriptU_{\CN,\rho\sigma} &= \sum_{\beta,\delta,\kappa} [F^{\CN\CD g}_{\CD}]_{\sigma\beta} [F^{\CN g \CD}_{\CD}]_{\beta\delta} [G^{\CD}_{\CD\CN g}]_{\rho\kappa} \scriptU_{\CN,\kappa\delta} ~, \\
    \scriptU_{\CN,\mu\nu} \cdot \scriptU_g &= \sum_{\rho,\kappa,\delta} [F^{g\CD\CN}_{\CD}]_{\rho\mu} [F^{g\CN\CD}_{\CD}]_{\nu\delta}^{-1} [G^{\CD}_{\CD g \CN}]_{\rho\kappa} \scriptU_{\CN,\kappa\delta} ~, \\
    \scriptU_{\CN,\mu\nu} \cdot \scriptU_{\CN,\rho\sigma} &= 2 \sum_{\lambda,\beta,g}[F^{\CN\CD\CN}_{\CD}]_{\sigma\lambda,\mu\beta} [F^{\CN\CN\CD}_\CD]^{-1}_{\nu\beta,g} [G^{\CD}_{\CD\CN\CN}]_{\rho\lambda,g} \scriptU_g ~.
\end{aligned}
\end{equation}
And the spin in the defect Hilbert space is related to the symmetry operators as
\begin{equation}
    e^{-4\pi is} = \sum_{\CL,\mu,\nu,\sigma} \frac{1}{\sqrt{d_\CL}} [F^{\CD\CD\CD}_{\CD}]^{-1}_{(\CL,\mu,\nu),(\CL,\mu,\sigma)} \scriptU_{\CN,\nu\sigma} ~.
\end{equation}

\bibliographystyle{utphys}
\bibliography{main}

\providecommand{\href}[2]{#2}\begingroup\raggedright\begin{thebibliography}{100}

\bibitem{gaiotto:generalizedsym}
D.~{Gaiotto}, A.~{Kapustin}, N.~{Seiberg}, and B.~{Willett}, ``{Generalized global symmetries},'' {\em Journal of High Energy Physics} {\bf 2015} (Feb., 2015) 172, \href{http://www.arXiv.org/abs/1412.5148}{{\tt 1412.5148}}.

\bibitem{McGreevy:2022oyu}
J.~McGreevy, ``{Generalized Symmetries in Condensed Matter},'' \href{http://www.arXiv.org/abs/2204.03045}{{\tt 2204.03045}}.

\bibitem{Cordova:2022ruw}
C.~Cordova, T.~T. Dumitrescu, K.~Intriligator, and S.-H. Shao, ``{Snowmass White Paper: Generalized Symmetries in Quantum Field Theory and Beyond},'' in {\em {2022 Snowmass Summer Study}}.
\newblock 5, 2022.
\newblock \href{http://www.arXiv.org/abs/2205.09545}{{\tt 2205.09545}}.

\bibitem{daniel2023review}
T.~D. {Brennan} and S.~{Hong}, ``{Introduction to Generalized Global Symmetries in QFT and Particle Physics},'' {\em arXiv e-prints} (June, 2023) arXiv:2306.00912, \href{http://www.arXiv.org/abs/2306.00912}{{\tt 2306.00912}}.

\bibitem{sakura2023review}
S.~{Sch{\"a}fer-Nameki}, ``{ICTP lectures on (non-)invertible generalized symmetries},'' {\em \physrep} {\bf 1063} (Apr., 2024) 1--55, \href{http://www.arXiv.org/abs/2305.18296}{{\tt 2305.18296}}.

\bibitem{lakshya2023review}
L.~{Bhardwaj}, L.~E. {Bottini}, L.~{Fraser-Taliente}, L.~{Gladden}, D.~S.~W. {Gould}, A.~{Platschorre}, and H.~{Tillim}, ``{Lectures on generalized symmetries},'' {\em \physrep} {\bf 1051} (Feb., 2024) 1--87, \href{http://www.arXiv.org/abs/2307.07547}{{\tt 2307.07547}}.

\bibitem{shuheng2023review}
S.-H. {Shao}, ``{What's Done Cannot Be Undone: TASI Lectures on Non-Invertible Symmetry},'' {\em arXiv e-prints} (Aug., 2023) arXiv:2308.00747, \href{http://www.arXiv.org/abs/2308.00747}{{\tt 2308.00747}}.

\bibitem{runkel2023review}
N.~{Carqueville}, M.~{Del Zotto}, and I.~{Runkel}, ``{Topological defects},'' {\em arXiv e-prints} (Nov., 2023) arXiv:2311.02449, \href{http://www.arXiv.org/abs/2311.02449}{{\tt 2311.02449}}.

\bibitem{Costa:2024wks}
D.~Costa {\em et al.}, ``{Simons Lectures on Categorical Symmetries},''
\newblock 11, 2024.
\newblock \href{http://www.arXiv.org/abs/2411.09082}{{\tt 2411.09082}}.

\bibitem{seiberg2023maj}
N.~{Seiberg} and S.-H. {Shao}, ``{Majorana chain and Ising model - (non-invertible) translations, anomalies, and emanant symmetries},'' {\em SciPost Physics} {\bf 16} (Mar., 2024) 064, \href{http://www.arXiv.org/abs/2307.02534}{{\tt 2307.02534}}.

\bibitem{moradi2023sym}
H.~{Moradi}, {\"O}.~M. {Aksoy}, J.~H. {Bardarson}, and A.~{Tiwari}, ``{Symmetry fractionalization, mixed-anomalies and dualities in quantum spin models with generalized symmetries},'' {\em arXiv e-prints} (July, 2023) arXiv:2307.01266, \href{http://www.arXiv.org/abs/2307.01266}{{\tt 2307.01266}}.

\bibitem{seifnashri2023lieb}
S.~{Seifnashri}, ``{Lieb-Schultz-Mattis anomalies as obstructions to gauging (non-on-site) symmetries},'' {\em SciPost Physics} {\bf 16} (Apr., 2024) 098, \href{http://www.arXiv.org/abs/2308.05151}{{\tt 2308.05151}}.

\bibitem{seiberg2024non}
N.~{Seiberg}, S.~{Seifnashri}, and S.-H. {Shao}, ``{Non-invertible symmetries and LSM-type constraints on a tensor product Hilbert space},'' {\em SciPost Physics} {\bf 16} (June, 2024) 154, \href{http://www.arXiv.org/abs/2401.12281}{{\tt 2401.12281}}.

\bibitem{mana2024nisym}
A.~{Parayil Mana}, Y.~{Li}, H.~{Sukeno}, and T.-C. {Wei}, ``{Kennedy-Tasaki transformation and noninvertible symmetry in lattice models beyond one dimension},'' {\em \prb} {\bf 109} (June, 2024) 245129, \href{http://www.arXiv.org/abs/2402.09520}{{\tt 2402.09520}}.

\bibitem{Lu:2024ytl}
D.-C. Lu, Z.~Sun, and Y.-Z. You, ``{Realizing triality and $p$-ality by lattice twisted gauging in (1+1)d quantum spin systems},'' {\em SciPost Phys.} {\bf 17} (2024) 136, \href{http://www.arXiv.org/abs/2405.14939}{{\tt 2405.14939}}.

\bibitem{sal2024mod}
S.~D. {Pace}, G.~{Delfino}, H.~T. {Lam}, and {\"O}.~M. {Aksoy}, ``{Gauging modulated symmetries: Kramers-Wannier dualities and non-invertible reflections},'' {\em arXiv e-prints} (June, 2024) arXiv:2406.12962, \href{http://www.arXiv.org/abs/2406.12962}{{\tt 2406.12962}}.

\bibitem{cao2024nisym}
W.~{Cao}, L.~{Li}, and M.~{Yamazaki}, ``{Generating lattice non-invertible symmetries},'' {\em SciPost Physics} {\bf 17} (Oct., 2024) 104, \href{http://www.arXiv.org/abs/2406.05454}{{\tt 2406.05454}}.

\bibitem{sal2024tdual}
S.~D. {Pace}, A.~{Chatterjee}, and S.-H. {Shao}, ``{Lattice T-duality from non-invertible symmetries in quantum spin chains},'' {\em arXiv e-prints} (Dec., 2024) arXiv:2412.18606, \href{http://www.arXiv.org/abs/2412.18606}{{\tt 2412.18606}}.

\bibitem{Kramers1941}
H.~A. {Kramers} and G.~H. {Wannier}, ``{Statistics of the Two-Dimensional Ferromagnet. Part I},'' {\em Physical Review} {\bf 60} (Aug., 1941) 252--262.

\bibitem{Aasen:2016dop}
D.~Aasen, R.~S.~K. Mong, and P.~Fendley, ``{Topological Defects on the Lattice I: The Ising model},'' {\em J. Phys. A} {\bf 49} (2016), no.~35, 354001, \href{http://www.arXiv.org/abs/1601.07185}{{\tt 1601.07185}}.

\bibitem{Aasen:2020jwb}
D.~Aasen, P.~Fendley, and R.~S.~K. Mong, ``{Topological Defects on the Lattice: Dualities and Degeneracies},'' \href{http://www.arXiv.org/abs/2008.08598}{{\tt 2008.08598}}.

\bibitem{Fendley2018Susy}
E.~{O'Brien} and P.~{Fendley}, ``{Lattice Supersymmetry and Order-Disorder Coexistence in the Tricritical Ising Model},'' {\em \prl} {\bf 120} (May, 2018) 206403, \href{http://www.arXiv.org/abs/1712.06662}{{\tt 1712.06662}}.

\bibitem{Lu2021SelfDualDQCP}
D.-C. {Lu}, C.~{Xu}, and Y.-Z. {You}, ``{Self-duality protected multicriticality in deconfined quantum phase transitions},'' {\em \prb} {\bf 104} (Nov., 2021) 205142, \href{http://www.arXiv.org/abs/2104.05147}{{\tt 2104.05147}}.

\bibitem{Nahum2021SelfdualZ2}
A.~M. {Somoza}, P.~{Serna}, and A.~{Nahum}, ``{Self-Dual Criticality in Three-Dimensional Z$_{2}$ Gauge Theory with Matter},'' {\em Physical Review X} {\bf 11} (Oct., 2021) 041008, \href{http://www.arXiv.org/abs/2012.15845}{{\tt 2012.15845}}.

\bibitem{omer2024selfdual}
{\"O}.~M. {Aksoy}, C.~{Mudry}, A.~{Furusaki}, and A.~{Tiwari}, ``{Lieb-Schultz-Mattis anomalies and web of dualities induced by gauging in quantum spin chains},'' {\em SciPost Physics} {\bf 16} (Jan., 2024) 022, \href{http://www.arXiv.org/abs/2308.00743}{{\tt 2308.00743}}.

\bibitem{arkya2024selfdual}
A.~{Chatterjee}, {\"O}.~M. {Aksoy}, and X.-G. {Wen}, ``{Quantum phases and transitions in spin chains with non-invertible symmetries},'' {\em SciPost Physics} {\bf 17} (Oct., 2024) 115, \href{http://www.arXiv.org/abs/2405.05331}{{\tt 2405.05331}}.

\bibitem{sinha2024selfdual}
M.~{Sinha}, F.~{Yan}, L.~{Grans-Samuelsson}, A.~{Roy}, and H.~{Saleur}, ``{Lattice realizations of topological defects in the critical (1+1)-d three-state Potts model},'' {\em Journal of High Energy Physics} {\bf 2024} (July, 2024) 225, \href{http://www.arXiv.org/abs/2310.19703}{{\tt 2310.19703}}.

\bibitem{tambara1998tensor}
D.~Tambara and S.~Yamagami, ``Tensor categories with fusion rules of self-duality for finite abelian groups,'' {\em Journal of Algebra} {\bf 209} (1998), no.~2, 692--707.

\bibitem{tambara2000representations}
D.~Tambara, ``Representations of tensor categories with fusion rules of self-duality for abelian groups,'' {\em Israel Journal of Mathematics} {\bf 118} (2000), no.~1, 29--60.

\bibitem{Choi:2023vgk}
Y.~Choi, D.-C. Lu, and Z.~Sun, ``{Self-duality under gauging a non-invertible symmetry},'' {\em JHEP} {\bf 01} (2024) 142, \href{http://www.arXiv.org/abs/2310.19867}{{\tt 2310.19867}}.

\bibitem{Perez-Lona:2023djo}
A.~Perez-Lona, D.~Robbins, E.~Sharpe, T.~Vandermeulen, and X.~Yu, ``{Notes on gauging noninvertible symmetries. Part I. Multiplicity-free cases},'' {\em JHEP} {\bf 02} (2024) 154, \href{http://www.arXiv.org/abs/2311.16230}{{\tt 2311.16230}}.

\bibitem{Diatlyk:2023fwf}
O.~Diatlyk, C.~Luo, Y.~Wang, and Q.~Weller, ``{Gauging non-invertible symmetries: topological interfaces and generalized orbifold groupoid in 2d QFT},'' {\em JHEP} {\bf 03} (2024) 127, \href{http://www.arXiv.org/abs/2311.17044}{{\tt 2311.17044}}.

\bibitem{Thorngren:2019iar}
R.~Thorngren and Y.~Wang, ``{Fusion Category Symmetry I: Anomaly In-Flow and Gapped Phases},'' \href{http://www.arXiv.org/abs/1912.02817}{{\tt 1912.02817}}.

\bibitem{Thorngren:2021yso}
R.~Thorngren and Y.~Wang, ``{Fusion Category Symmetry II: Categoriosities at $c$ = 1 and Beyond},'' \href{http://www.arXiv.org/abs/2106.12577}{{\tt 2106.12577}}.

\bibitem{Lu:2022ver}
D.-C. Lu and Z.~Sun, ``{On triality defects in 2d CFT},'' {\em JHEP} {\bf 02} (2023) 173, \href{http://www.arXiv.org/abs/2208.06077}{{\tt 2208.06077}}.

\bibitem{Lu:2024lzf}
D.-C. Lu, Z.~Sun, and Z.~Zhang, ``{Exploring $G$-ality defects in 2-dim QFTs},'' \href{http://www.arXiv.org/abs/2406.12151}{{\tt 2406.12151}}.

\bibitem{Ando:2024hun}
T.~Ando, ``{A journey on self-$G$-ality},'' \href{http://www.arXiv.org/abs/2405.15648}{{\tt 2405.15648}}.

\bibitem{Kaidi:2021xfk}
J.~Kaidi, K.~Ohmori, and Y.~Zheng, ``{Kramers-Wannier-like Duality Defects in (3+1)D Gauge Theories},'' {\em Phys. Rev. Lett.} {\bf 128} (2022), no.~11, 111601, \href{http://www.arXiv.org/abs/2111.01141}{{\tt 2111.01141}}.

\bibitem{Choi:2021kmx}
Y.~Choi, C.~Cordova, P.-S. Hsin, H.~T. Lam, and S.-H. Shao, ``{Non-Invertible Duality Defects in 3+1 Dimensions},'' \href{http://www.arXiv.org/abs/2111.01139}{{\tt 2111.01139}}.

\bibitem{Choi:2022zal}
Y.~Choi, C.~Cordova, P.-S. Hsin, H.~T. Lam, and S.-H. Shao, ``{Non-invertible Condensation, Duality, and Triality Defects in 3+1 Dimensions},'' {\em Commun. Math. Phys.} {\bf 402} (2023), no.~1, 489--542, \href{http://www.arXiv.org/abs/2204.09025}{{\tt 2204.09025}}.

\bibitem{Bhardwaj:2024xcx}
L.~Bhardwaj, T.~D\'ecoppet, S.~Schafer-Nameki, and M.~Yu, ``{Fusion 3-Categories for Duality Defects},'' \href{http://www.arXiv.org/abs/2408.13302}{{\tt 2408.13302}}.

\bibitem{Decoppet:2024moc}
T.~D. D\'ecoppet, ``{Extension Theory and Fermionic Strongly Fusion 2-Categories (with an Appendix by Thibault Didier D\'ecoppet and Theo Johnson-Freyd)},'' {\em SIGMA} {\bf 20} (2024) 092, \href{http://www.arXiv.org/abs/2403.03211}{{\tt 2403.03211}}.

\bibitem{Decoppet:2023bay}
T.~D. D\'ecoppet and M.~Yu, ``{Fiber 2-Functors and Tambara-Yamagami Fusion 2-Categories},'' \href{http://www.arXiv.org/abs/2306.08117}{{\tt 2306.08117}}.

\bibitem{Choi:2024rjm}
Y.~Choi, Y.~Sanghavi, S.-H. Shao, and Y.~Zheng, ``{Non-invertible and higher-form symmetries in 2+1d lattice gauge theories},'' \href{http://www.arXiv.org/abs/2405.13105}{{\tt 2405.13105}}.

\bibitem{Cui:2024cav}
W.~Cui, B.~Haghighat, and L.~Ruggeri, ``{Non-invertible surface defects in 2+1d QFTs from half spacetime gauging},'' {\em JHEP} {\bf 11} (2024) 159, \href{http://www.arXiv.org/abs/2406.09261}{{\tt 2406.09261}}.

\bibitem{Etingof:2009yvg}
P.~Etingof, D.~Nikshych, V.~Ostrik, and w.~a. a. b.~E. Meir, ``{Fusion categories and homotopy theory},'' \href{http://www.arXiv.org/abs/0909.3140}{{\tt 0909.3140}}.

\bibitem{Kitaev:2011dxc}
A.~Kitaev and L.~Kong, ``{Models for Gapped Boundaries and Domain Walls},'' {\em Commun. Math. Phys.} {\bf 313} (2012), no.~2, 351--373, \href{http://www.arXiv.org/abs/1104.5047}{{\tt 1104.5047}}.

\bibitem{symtft2019XGW}
W.~{Ji} and X.-G. {Wen}, ``{Categorical symmetry and non-invertible anomaly in symmetry-breaking and topological phase transitions},'' {\em arXiv e-prints} (Dec., 2019) arXiv:1912.13492, \href{http://www.arXiv.org/abs/1912.13492}{{\tt 1912.13492}}.

\bibitem{symtft2020XGW2}
L.~{Kong}, T.~{Lan}, X.-G. {Wen}, Z.-H. {Zhang}, and H.~{Zheng}, ``{Algebraic higher symmetry and categorical symmetry: A holographic and entanglement view of symmetry},'' {\em Physical Review Research} {\bf 2} (Oct., 2020) 043086, \href{http://www.arXiv.org/abs/2005.14178}{{\tt 2005.14178}}.

\bibitem{symtft2021XGW3}
W.~{Ji} and X.-G. {Wen}, ``{A unified view on symmetry, anomalous symmetry and non-invertible gravitational anomaly},'' {\em arXiv e-prints} (June, 2021) arXiv:2106.02069, \href{http://www.arXiv.org/abs/2106.02069}{{\tt 2106.02069}}.

\bibitem{symtft2023XGW4}
A.~{Chatterjee} and X.-G. {Wen}, ``{Symmetry as a shadow of topological order and a derivation of topological holographic principle},'' {\em \prb} {\bf 107} (Apr., 2023) 155136, \href{http://www.arXiv.org/abs/2203.03596}{{\tt 2203.03596}}.

\bibitem{symtft2022XGW5}
A.~{Chatterjee} and X.-G. {Wen}, ``{Holographic theory for continuous phase transitions -- the emergence and symmetry protection of gaplessness},'' {\em arXiv e-prints} (May, 2022) arXiv:2205.06244, \href{http://www.arXiv.org/abs/2205.06244}{{\tt 2205.06244}}.

\bibitem{symtft2021Gaiotto}
D.~{Gaiotto} and J.~{Kulp}, ``{Orbifold groupoids},'' {\em Journal of High Energy Physics} {\bf 2021} (Feb., 2021) 132, \href{http://www.arXiv.org/abs/2008.05960}{{\tt 2008.05960}}.

\bibitem{symtft2021Sakura}
F.~{Apruzzi}, F.~{Bonetti}, I.~{Garc{\'\i}a Etxebarria}, S.~S. {Hosseini}, and S.~{Schafer-Nameki}, ``{Symmetry TFTs from String Theory},'' {\em arXiv e-prints} (Dec., 2021) arXiv:2112.02092, \href{http://www.arXiv.org/abs/2112.02092}{{\tt 2112.02092}}.

\bibitem{moradi2023topoholo}
H.~{Moradi}, S.~{Faroogh Moosavian}, and A.~{Tiwari}, ``{Topological Holography: Towards a Unification of Landau and Beyond-Landau Physics},'' {\em SciPost Physics Core} {\bf 6} (Oct., 2023) 066, \href{http://www.arXiv.org/abs/2207.10712}{{\tt 2207.10712}}.

\bibitem{symtft2022Apruzzi}
F.~{Apruzzi}, ``{Higher form symmetries TFT in 6d},'' {\em Journal of High Energy Physics} {\bf 2022} (Nov., 2022) 50, \href{http://www.arXiv.org/abs/2203.10063}{{\tt 2203.10063}}.

\bibitem{symtft2022freed2}
D.~S. {Freed}, G.~W. {Moore}, and C.~{Teleman}, ``{Topological symmetry in quantum field theory},'' {\em arXiv e-prints} (Sept., 2022) arXiv:2209.07471, \href{http://www.arXiv.org/abs/2209.07471}{{\tt 2209.07471}}.

\bibitem{Kaidi:2022cpf}
J.~Kaidi, K.~Ohmori, and Y.~Zheng, ``{Symmetry TFTs for Non-Invertible Defects},'' \href{http://www.arXiv.org/abs/2209.11062}{{\tt 2209.11062}}.

\bibitem{symtft2022Kulp}
I.~M. {Burbano}, J.~{Kulp}, and J.~{Neuser}, ``{Duality defects in E$_{8}$},'' {\em Journal of High Energy Physics} {\bf 2022} (Oct., 2022) 186, \href{http://www.arXiv.org/abs/2112.14323}{{\tt 2112.14323}}.

\bibitem{symtft2023kaidi2}
J.~{Kaidi}, E.~{Nardoni}, G.~{Zafrir}, and Y.~{Zheng}, ``{Symmetry TFTs and Anomalies of Non-Invertible Symmetries},'' {\em arXiv e-prints} (Jan., 2023) arXiv:2301.07112, \href{http://www.arXiv.org/abs/2301.07112}{{\tt 2301.07112}}.

\bibitem{Brennan:2024fgj}
T.~D. Brennan and Z.~Sun, ``{A SymTFT for continuous symmetries},'' {\em JHEP} {\bf 12} (2024) 100, \href{http://www.arXiv.org/abs/2401.06128}{{\tt 2401.06128}}.

\bibitem{Bonetti:2024cjk}
F.~Bonetti, M.~Del~Zotto, and R.~Minasian, ``{SymTFTs for Continuous non-Abelian Symmetries},'' \href{http://www.arXiv.org/abs/2402.12347}{{\tt 2402.12347}}.

\bibitem{DelZotto:2024tae}
M.~Del~Zotto, S.~N. Meynet, and R.~Moscrop, ``{Remarks on Geometric Engineering, Symmetry TFTs and Anomalies},'' \href{http://www.arXiv.org/abs/2402.18646}{{\tt 2402.18646}}.

\bibitem{Franco:2024mxa}
S.~Franco and X.~Yu, ``{Generalized Symmetries in 2D from String Theory: SymTFTs, Intrinsic Relativeness, and Anomalies of Non-invertible Symmetries},'' \href{http://www.arXiv.org/abs/2404.19761}{{\tt 2404.19761}}.

\bibitem{Putrov:2024uor}
P.~Putrov and R.~Radhakrishnan, ``{Non-anomalous non-invertible symmetries in 1+1D from gapped boundaries of SymTFTs},'' \href{http://www.arXiv.org/abs/2405.04619}{{\tt 2405.04619}}.

\bibitem{Huang:2024ror}
S.-J. Huang, ``{Fermionic quantum criticality through the lens of topological holography},'' \href{http://www.arXiv.org/abs/2405.09611}{{\tt 2405.09611}}.

\bibitem{Freed:2022qnc}
D.~S. Freed, G.~W. Moore, and C.~Teleman, ``{Topological symmetry in quantum field theory},'' \href{http://www.arXiv.org/abs/2209.07471}{{\tt 2209.07471}}.

\bibitem{Lin:2022dhv}
Y.-H. Lin, M.~Okada, S.~Seifnashri, and Y.~Tachikawa, ``{Asymptotic density of states in 2d CFTs with non-invertible symmetries},'' \href{http://www.arXiv.org/abs/2208.05495}{{\tt 2208.05495}}.

\bibitem{Bhardwaj:2024igy}
L.~Bhardwaj, C.~Copetti, D.~Pajer, and S.~Schafer-Nameki, ``{Boundary SymTFT},'' \href{http://www.arXiv.org/abs/2409.02166}{{\tt 2409.02166}}.

\bibitem{Antinucci:2024ltv}
A.~Antinucci, C.~Copetti, and S.~Schafer-Nameki, ``{SymTFT for (3+1)d Gapless SPTs and Obstructions to Confinement},'' \href{http://www.arXiv.org/abs/2408.05585}{{\tt 2408.05585}}.

\bibitem{Bhardwaj:2024qiv}
L.~Bhardwaj, D.~Pajer, S.~Schafer-Nameki, A.~Tiwari, A.~Warman, and J.~Wu, ``{Gapped Phases in (2+1)d with Non-Invertible Symmetries: Part I},'' \href{http://www.arXiv.org/abs/2408.05266}{{\tt 2408.05266}}.

\bibitem{Bhardwaj:2024qrf}
L.~Bhardwaj, D.~Pajer, S.~Schafer-Nameki, and A.~Warman, ``{Hasse Diagrams for Gapless SPT and SSB Phases with Non-Invertible Symmetries},'' \href{http://www.arXiv.org/abs/2403.00905}{{\tt 2403.00905}}.

\bibitem{Bhardwaj:2023bbf}
L.~Bhardwaj, L.~E. Bottini, D.~Pajer, and S.~Schafer-Nameki, ``{The Club Sandwich: Gapless Phases and Phase Transitions with Non-Invertible Symmetries},'' \href{http://www.arXiv.org/abs/2312.17322}{{\tt 2312.17322}}.

\bibitem{Bhardwaj:2023ayw}
L.~Bhardwaj and S.~Schafer-Nameki, ``{Generalized Charges, Part II: Non-Invertible Symmetries and the Symmetry TFT},'' \href{http://www.arXiv.org/abs/2305.17159}{{\tt 2305.17159}}.

\bibitem{Bhardwaj:2023wzd}
L.~Bhardwaj and S.~Schafer-Nameki, ``{Generalized charges, part I: Invertible symmetries and higher representations},'' {\em SciPost Phys.} {\bf 16} (2024), no.~4, 093, \href{http://www.arXiv.org/abs/2304.02660}{{\tt 2304.02660}}.

\bibitem{Sun:2023xxv}
Z.~Sun and Y.~Zheng, ``{When are Duality Defects Group-Theoretical?},'' \href{http://www.arXiv.org/abs/2307.14428}{{\tt 2307.14428}}.

\bibitem{Zhang:2023wlu}
C.~Zhang and C.~C\'ordova, ``{Anomalies of $(1+1)D$ categorical symmetries},'' \href{http://www.arXiv.org/abs/2304.01262}{{\tt 2304.01262}}.

\bibitem{Copetti:2024onh}
C.~Copetti, ``{Defect Charges, Gapped Boundary Conditions, and the Symmetry TFT},'' \href{http://www.arXiv.org/abs/2408.01490}{{\tt 2408.01490}}.

\bibitem{Argurio:2024oym}
R.~Argurio, F.~Benini, M.~Bertolini, G.~Galati, and P.~Niro, ``{On the symmetry TFT of Yang-Mills-Chern-Simons theory},'' {\em JHEP} {\bf 07} (2024) 130, \href{http://www.arXiv.org/abs/2404.06601}{{\tt 2404.06601}}.

\bibitem{Antinucci:2023ezl}
A.~Antinucci, F.~Benini, C.~Copetti, G.~Galati, and G.~Rizi, ``{Anomalies of non-invertible self-duality symmetries: fractionalization and gauging},'' \href{http://www.arXiv.org/abs/2308.11707}{{\tt 2308.11707}}.

\bibitem{Choi:2024wfm}
Y.~Choi, B.~C. Rayhaun, and Y.~Zheng, ``{Noninvertible Symmetry-Resolved Affleck-Ludwig-Cardy Formula and Entanglement Entropy from the Boundary Tube Algebra},'' {\em Phys. Rev. Lett.} {\bf 133} (2024), no.~25, 251602, \href{http://www.arXiv.org/abs/2409.02806}{{\tt 2409.02806}}.

\bibitem{Choi:2024tri}
Y.~Choi, B.~C. Rayhaun, and Y.~Zheng, ``{Generalized Tube Algebras, Symmetry-Resolved Partition Functions, and Twisted Boundary States},'' \href{http://www.arXiv.org/abs/2409.02159}{{\tt 2409.02159}}.

\bibitem{Cordova:2023bja}
C.~Cordova, P.-S. Hsin, and C.~Zhang, ``{Anomalies of Non-Invertible Symmetries in (3+1)d},'' \href{http://www.arXiv.org/abs/2308.11706}{{\tt 2308.11706}}.

\bibitem{Antinucci:2024zjp}
A.~Antinucci and F.~Benini, ``{Anomalies and gauging of U(1) symmetries},'' \href{http://www.arXiv.org/abs/2401.10165}{{\tt 2401.10165}}.

\bibitem{Bhardwaj:2024ydc}
L.~Bhardwaj, K.~Inamura, and A.~Tiwari, ``{Fermionic Non-Invertible Symmetries in (1+1)d: Gapped and Gapless Phases, Transitions, and Symmetry TFTs},'' \href{http://www.arXiv.org/abs/2405.09754}{{\tt 2405.09754}}.

\bibitem{Chen:2023qnv}
J.~Chen, W.~Cui, B.~Haghighat, and Y.-N. Wang, ``{SymTFTs and duality defects from 6d SCFTs on 4-manifolds},'' {\em JHEP} {\bf 11} (2023) 208, \href{http://www.arXiv.org/abs/2305.09734}{{\tt 2305.09734}}.

\bibitem{Chang:2018iay}
C.-M. Chang, Y.-H. Lin, S.-H. Shao, Y.~Wang, and X.~Yin, ``{Topological Defect Lines and Renormalization Group Flows in Two Dimensions},'' {\em JHEP} {\bf 01} (2019) 026, \href{http://www.arXiv.org/abs/1802.04445}{{\tt 1802.04445}}.

\bibitem{Barkeshli:2014cna}
M.~Barkeshli, P.~Bonderson, M.~Cheng, and Z.~Wang, ``{Symmetry Fractionalization, Defects, and Gauging of Topological Phases},'' {\em Phys. Rev. B} {\bf 100} (2019), no.~11, 115147, \href{http://www.arXiv.org/abs/1410.4540}{{\tt 1410.4540}}.

\bibitem{Gelaki:2009blp}
S.~Gelaki, D.~Naidu, and D.~Nikshych, ``{Centers of graded fusion categories},'' \href{http://www.arXiv.org/abs/0905.3117}{{\tt 0905.3117}}.

\bibitem{Teo:2015xla}
J.~C.~Y. Teo, T.~L. Hughes, and E.~Fradkin, ``{Theory of Twist Liquids: Gauging an Anyonic Symmetry},'' {\em Annals Phys.} {\bf 360} (2015) 349--445, \href{http://www.arXiv.org/abs/1503.06812}{{\tt 1503.06812}}.

\bibitem{vercleyen2024low}
G.~Vercleyen, ``On Low-Rank Multiplicity-Free Fusion Categories,'' {\em arXiv preprint arXiv:2405.20075} (2024).

\bibitem{Ruben2021igSPT}
R.~{Thorngren}, A.~{Vishwanath}, and R.~{Verresen}, ``{Intrinsically gapless topological phases},'' {\em \prb} {\bf 104} (Aug., 2021) 075132, \href{http://www.arXiv.org/abs/2008.06638}{{\tt 2008.06638}}.

\bibitem{huang2023igspt}
S.-J. {Huang} and M.~{Cheng}, ``{Topological holography, quantum criticality, and boundary states},'' {\em arXiv e-prints} (Oct., 2023) arXiv:2310.16878, \href{http://www.arXiv.org/abs/2310.16878}{{\tt 2310.16878}}.

\bibitem{wen2023igSPT}
R.~{Wen} and A.~C. {Potter}, ``{Classification of 1+1D gapless symmetry protected phases via topological holography},'' {\em arXiv e-prints} (Oct., 2023) arXiv:2311.00050, \href{http://www.arXiv.org/abs/2311.00050}{{\tt 2311.00050}}.

\bibitem{sakura2024igspt}
L.~{Bhardwaj}, D.~{Pajer}, S.~{Schafer-Nameki}, and A.~{Warman}, ``{Hasse Diagrams for Gapless SPT and SSB Phases with Non-Invertible Symmetries},'' {\em arXiv e-prints} (Mar., 2024) arXiv:2403.00905, \href{http://www.arXiv.org/abs/2403.00905}{{\tt 2403.00905}}.

\bibitem{Bhardwaj:2017xup}
L.~Bhardwaj and Y.~Tachikawa, ``{On finite symmetries and their gauging in two dimensions},'' {\em JHEP} {\bf 03} (2018) 189, \href{http://www.arXiv.org/abs/1704.02330}{{\tt 1704.02330}}.

\bibitem{Fuchs:2002cm}
J.~Fuchs, I.~Runkel, and C.~Schweigert, ``{TFT construction of RCFT correlators 1. Partition functions},'' {\em Nucl. Phys. B} {\bf 646} (2002) 353--497, \href{http://www.arXiv.org/abs/hep-th/0204148}{{\tt hep-th/0204148}}.

\bibitem{shuheng2024cluster}
S.~{Seifnashri} and S.-H. {Shao}, ``{Cluster State as a Noninvertible Symmetry-Protected Topological Phase},'' {\em \prl} {\bf 133} (Sept., 2024) 116601, \href{http://www.arXiv.org/abs/2404.01369}{{\tt 2404.01369}}.

\bibitem{Inamura:2024jke}
K.~Inamura and S.~Ohyama, ``{1+1d SPT phases with fusion category symmetry: interface modes and non-abelian Thouless pump},'' \href{http://www.arXiv.org/abs/2408.15960}{{\tt 2408.15960}}.

\bibitem{gu2024nispt}
C.~{Meng}, X.~{Yang}, T.~{Lan}, and Z.~{Gu}, ``{Non-invertible SPTs: an on-site realization of (1+1)d anomaly-free fusion category symmetry},'' {\em arXiv e-prints} (Dec., 2024) arXiv:2412.20546, \href{http://www.arXiv.org/abs/2412.20546}{{\tt 2412.20546}}.

\bibitem{Choi:2023xjw}
Y.~Choi, B.~C. Rayhaun, Y.~Sanghavi, and S.-H. Shao, ``{Comments on Boundaries, Anomalies, and Non-Invertible Symmetries},'' \href{http://www.arXiv.org/abs/2305.09713}{{\tt 2305.09713}}.

\bibitem{ostrik2002module}
V.~Ostrik, ``{Module categories over the Drinfeld double of a finite group},'' \href{http://www.arXiv.org/abs/math/0202130}{{\tt math/0202130}}.

\bibitem{Lou:2020gfq}
J.~Lou, C.~Shen, C.~Chen, and L.-Y. Hung, ``{A (dummy\textquoteright{}s) guide to working with gapped boundaries via (fermion) condensation},'' {\em JHEP} {\bf 02} (2021) 171, \href{http://www.arXiv.org/abs/2007.10562}{{\tt 2007.10562}}.

\bibitem{teo2015theory}
J.~C.~Y. {Teo}, T.~L. {Hughes}, and E.~{Fradkin}, ``{Theory of Twist Liquids: Gauging an Anyonic Symmetry},'' {\em arXiv e-prints} (Mar., 2015) arXiv:1503.06812, \href{http://www.arXiv.org/abs/1503.06812}{{\tt 1503.06812}}.

\bibitem{Davydov:2013xov}
A.~Davydov, ``{Bogomolov multiplier, double class-preserving automorphisms, and modular invariants for orbifolds},'' {\em J. Math. Phys.} {\bf 55} (2014) 092305, \href{http://www.arXiv.org/abs/1312.7466}{{\tt 1312.7466}}.

\bibitem{Kobayashi:2025ykb}
R.~Kobayashi and M.~Barkeshli, ``{Soft symmetries of topological orders},'' \href{http://www.arXiv.org/abs/2501.03314}{{\tt 2501.03314}}.

\bibitem{Benini:2018reh}
F.~Benini, C.~C\'ordova, and P.-S. Hsin, ``{On 2-Group Global Symmetries and their Anomalies},'' {\em JHEP} {\bf 03} (2019) 118, \href{http://www.arXiv.org/abs/1803.09336}{{\tt 1803.09336}}.

\bibitem{Brennan:2022tyl}
T.~D. Brennan, C.~Cordova, and T.~T. Dumitrescu, ``{Line Defect Quantum Numbers \& Anomalies},'' \href{http://www.arXiv.org/abs/2206.15401}{{\tt 2206.15401}}.

\bibitem{Delmastro:2022pfo}
D.~G. Delmastro, J.~Gomis, P.-S. Hsin, and Z.~Komargodski, ``{Anomalies and symmetry fractionalization},'' {\em SciPost Phys.} {\bf 15} (2023), no.~3, 079, \href{http://www.arXiv.org/abs/2206.15118}{{\tt 2206.15118}}.

\bibitem{Barkeshli:2019vtb}
M.~Barkeshli and M.~Cheng, ``{Relative Anomalies in (2+1)D Symmetry Enriched Topological States},'' {\em SciPost Phys.} {\bf 8} (2020) 028, \href{http://www.arXiv.org/abs/1906.10691}{{\tt 1906.10691}}.

\bibitem{brown2012cohomology}
K.~S. Brown, {\em Cohomology of groups}, vol.~87.
\newblock Springer Science \& Business Media, 2012.

\bibitem{Cordova:2018cvg}
C.~C\'ordova, T.~T. Dumitrescu, and K.~Intriligator, ``{Exploring 2-Group Global Symmetries},'' {\em JHEP} {\bf 02} (2019) 184, \href{http://www.arXiv.org/abs/1802.04790}{{\tt 1802.04790}}.

\bibitem{2016arXiv160304318M}
I.~{Marshall} and D.~{Nikshych}, ``{On the Brauer-Picard groups of fusion categories},'' {\em arXiv e-prints} (Mar., 2016) arXiv:1603.04318, \href{http://www.arXiv.org/abs/1603.04318}{{\tt 1603.04318}}.

\bibitem{schreier1927untergruppen}
O.~Schreier, ``Die untergruppen der freien gruppen,'' in {\em Abhandlungen aus dem Mathematischen Seminar der universit{\"a}t Hamburg}, vol.~5, pp.~161--183, Springer.
\newblock 1927.

\bibitem{Chang:2020imq}
C.-M. Chang and Y.-H. Lin, ``{Lorentzian dynamics and factorization beyond rationality},'' {\em JHEP} {\bf 10} (2021) 125, \href{http://www.arXiv.org/abs/2012.01429}{{\tt 2012.01429}}.

\bibitem{MMAfile}
Mathematica files of the solutions for the fiber functors are uploaded as Ancillary files on the arXiv.

\bibitem{Wang:2017loc}
J.~Wang, X.-G. Wen, and E.~Witten, ``{Symmetric Gapped Interfaces of SPT and SET States: Systematic Constructions},'' {\em Phys. Rev. X} {\bf 8} (2018), no.~3, 031048, \href{http://www.arXiv.org/abs/1705.06728}{{\tt 1705.06728}}.

\bibitem{Coste:2000tq}
A.~Coste, T.~Gannon, and P.~Ruelle, ``{Finite group modular data},'' {\em Nucl. Phys. B} {\bf 581} (2000) 679--717, \href{http://www.arXiv.org/abs/hep-th/0001158}{{\tt hep-th/0001158}}.

\bibitem{hayami2018hochschild}
T.~Hayami, ``On Hochschild cohomology ring and integral cohomology ring for the semidihedral group,'' {\em International Journal of Algebra and Computation} {\bf 28} (2018), no.~02, 257--290.

\bibitem{larson1988integral}
D.~S. Larson, {\em The integral cohomology rings of split metacyclic groups}.
\newblock University of Minnesota, 1988.

\bibitem{weiss1969cohomology}
E.~Weiss, {\em Cohomology of Groups: Cohomology of Groups}.
\newblock Academic Press, 1969.

\bibitem{Wang:2021nrp}
Q.-R. Wang, S.-Q. Ning, and M.~Cheng, ``{Domain Wall Decorations, Anomalies and Spectral Sequences in Bosonic Topological Phases},'' \href{http://www.arXiv.org/abs/2104.13233}{{\tt 2104.13233}}.

\bibitem{2019arXiv191102633G}
J.~{Green} and D.~{Nikshych}, ``{On the braid group representations coming from weakly group-theoretical fusion categories},'' {\em arXiv e-prints} (Nov., 2019) arXiv:1911.02633, \href{http://www.arXiv.org/abs/1911.02633}{{\tt 1911.02633}}.

\end{thebibliography}\endgroup

\end{document}